# AI4X Roadmap: Artificial Intelligence for the advancement of scientific pursuit and its future directions

Stephen G. Dale[1,2,*], Nikita Kazeev[1], Alastair J. A. Price[3,5], Victor Posligua[1], Stephan Roche[6,7], O. Anatole von Lilienfeld[3,4,5], Konstantin S. Novoselov[1], Xavier Bresson[8], Gianmarco Mengaldo[9, 10, 11], Xudong Chen[12], Terence J. O'Kane[13], Emily R. Lines[14], Matthew J. Allen[14], Amandine E. Debus[14], Clayton Miller[15,16], Jiayu Zhou[17], Hiroko H. Dodge[18], David Rousseau[19], Andrey Ustyuzhanin[1,20,21], Ziyun Yan[2], Mario Lanza[1,2], Fabio Sciarrino[22], Ryo Yoshida[23,24], Zhidong Leong[25], Teck Leong Tan[25], Qianxiao Li[1,10], Adil Kabylda[26], Igor Poltavsky[26], Alexandre Tkatchenko[26], Sherif Abdulkader Tawfik[27], Prathami Divakar Kamath[28,29], Théo Jaffrelot Inizan[29,30,34,35], Kristin A. Persson[28,29,34,35], Bryant Y. Li[28,29], Vir Karan[28,29], Chenru Duan[31], Haojun Jia[31], Qiyuan Zhao[31], Hiroyuki Hayashi[32], Atsuto Seko[32], Isao Tanaka[32,33], Omar M. Yaghi[29,34,35,36], Tim Gould[37], Bun Chan[38], Stefan Vuckovic[39], Tianbo Li[40], Min Lin[40], Zehcen Tang[41], Yang Li[41], Yong Xu[41,42], Amrita Joshi[43], Xiaonan Wang[44], Leonard W.T. Ng[43], Sergei V. Kalinin[45], Mahshid Ahmadi[45], Jiyizhe Zhang[46,47,48], Shuyuan Zhang[46,47], Alexei Lapkin[46,47], Ming Xiao[49], Zhe Wu[49], Kedar Hippalgaonkar[1,43,50], Limsoon Wong[8], Lorenzo Bastonero[51], Nicola Marzari[51,52,53,54,55], Dorye Luis Esteras Córdoba[6], Andrei Tomut[6], Alba Quiñones Andrade[6] and José-Hugo Garcia[6]

[1] Institute of Functional Intelligent Materials, National University of Singapore, 4 Science Drive 2, 117544, Singapore
[2] Department of Material Science and Engineering, National University of Singapore, 9 Engineering Drive 1, 117575, Singapore
[3] Department of Chemistry, University of Toronto, St. George campus, Toronto, ON, Canada
[4] Vector Institute, Toronto, Canada
[5] Acceleration Consortium, University of Toronto, 80 St George St, Toronto, ON M5S 3H6, Canada
[6] Catalan Institute of Nanoscience and Nanotechnology (ICN2), CSIC and BIST, Campus UAB, Bellaterra, 08193 Barcelona, Spain
[7] ICREA, Institució Catalana de Recerca i Estudis Avançats, 08070 Barcelona, Spain
[8] Department of Computer Science, National University of Singapore, Singapore
[9] Department of Mechanical Engineering, National University of Singapore, Singapore
[10] Department of Mathematics, National University of Singapore, Singapore
[11] Honorary Research Fellow, Department of Aeronautics, Imperial College London, London, United Kingdom
[12] Department of Electrical and Computer Engineering, National University of Singapore, Singapore
[13] Environment, CSIRO, Battery Point, Australia
[14] Department of Geography, University of Cambridge, Downing Site, Cambridge CB23EN, United Kingdom
[15] College of Integrative Studies, Singapore Management University, Singapore
[16] Urban Institute, Singapore Management University, Singapore
[17] School of Information, University of Michigan, Ann Arbor, U.S.A
[18] Harvard Medical School, Charlestown, U.S.A
[19] Université Paris-Saclay, CNRS/IN2P3, IJCLab, 91405 Orsay, France
[20] Constructor Knowledge Labs, Bremen, Campus Ring 1, 28759, Germany
[21] Constructor University, Bremen, Campus Ring 1, 28759, Germany
[22] Dipartimento di Fisica, Sapienza Università di Roma, Piazzale Aldo Moro 5, I-00185 Roma, Italy
[23] The Institute of Statistical Mathematics, Research Organization of Information and Systems, Tachikawa, Tokyo 190-8562, Japan
[24] Advanced General Intelligence for Science Program (AGIS), TRIP Headquarters, RIKEN, Wako, Saitama 351-0198, Japan
[25] Institute of High Performance Computing (IHPC), Agency for Science, Technology and Research (A*STAR), 1 Fusionopolis Way, #16-16 Connexis, Singapore 138632, Republic of Singapore
[26] Department of Physics and Materials Science, University of Luxembourg, L-1511 Luxembourg City, Luxembourg
[27] Applied Artificial Intelligence Institute, Deakin University, Geelong, Victoria 3216, Australia
[28] Department of Materials Science and Engineering, University of California, Berkeley, CA, USA
[29] Materials Sciences Division, Lawrence Berkeley National Laboratory, Berkeley, CA, USA
[30] AIMATX Inc., Berkeley, CA, USA
[31] Deep Principle, Inc., Cambridge, MA, USA
[32] Department of Materials Science and Engineering, Kyoto University, Kyoto, Japan
[33] Nano Research Laboratory, Japan Fine Ceramics Center (JFCC), Nagoya, Japan
[34] Department of Chemistry, University of California, Berkeley, CA, USA
[35] Bakar Institute of Digital Materials for the Planet, University of California, Berkeley, CA, USA
[36] Kavli Energy NanoScience Institute, Berkeley, CA, USA
[37] Queensland Micro- and Nanotechnology Centre, Griffith University, Nathan, Qld 4111, Australia

[38] Graduate School of Engineering, Nagasaki University, Bunkyo 1-14, Nagasaki 852-8521, Japan
[39] Department of Chemistry, University of Fribourg, Fribourg, Switzerland
[40] SEA AI Lab, 1 Fusionopolis Place, 138522, Singapore
[41] State Key Laboratory of Low Dimensional Quantum Physics and Department of Physics, Tsinghua University, Beijing, 100084, China
[42] Frontier Science Center for Quantum Information, Beijing, China
[43] School of Materials Science and Engineering, Nanyang Technological University, Singapore, 639798, Singapore
[44] Department of Chemical Engineering, Tsinghua University, Beijing, China, 100084
[45] Dept. of Materials Science and Engineering, University of Tennessee, Knoxville, TN 37923 USA
[46] Department of Chemical Engineering and Biotechnology, University of Cambridge, Cambridge, United Kingdom
[47] Innovation Centre in Digital Molecular Technologies, Yusuf Hamied Department of Chemistry, University of Cambridge, Cambridge, United Kingdom
[48] Department of Chemical Engineering, University of Manchester, Oxford Road, Manchester, United Kingdom
[49] Department of Chemical and Biomolecular Engineering, National University of Singapore, 117585, Singapore
[50] Institute of Materials Research and Engineering, A*STAR (Agency for Science), Singapore, Singapore
[51] U Bremen Excellence Chair, Bremen Centre for Computational Materials Science, and MAPEX Center for Materials and Processes, University of Bremen, 28359 Bremen, Germany
[52] Theory and Simulation of Materials (THEOS), École Polytechnique Fédérale de Lausanne, 1015 Lausanne, Switzerland
[53] National Centre for Computational Design and Discovery of Novel Materials (MARVEL), École Polytechnique Fédérale de Lausanne, 1015 Lausanne, Switzerland
[54] PSI Center for Scientific Computing, Theory and Data (CSD), Paul Scherrer Institut, 5232 Villigen PSI, Switzerland
[55] Cavendish Laboratory (TCM), University of Cambridge, Cambridge CB3 0US, United Kingdom

∗Author to whom any correspondence should be addressed.

E-mail: sdale@nus.edu.sg

**Abstract:**

Artificial intelligence and machine learning are reshaping how we approach scientific discovery, not by replacing established methods but by extending what researchers can probe, predict, and design. In this roadmap we provide a forward-looking view of AI-enabled science across biology, chemistry, climate science, mathematics, materials science, physics, self-driving laboratories and unconventional computing. Several shared themes emerge: the need for diverse and trustworthy data, transferable electronic-structure and interatomic models, AI systems integrated into end-to-end scientific workflows that connect simulations to experiments and generative systems grounded in synthesisability rather than purely idealised phases. Across domains, we highlight how large foundation models, active learning and self-driving laboratories can close loops between prediction and validation while maintaining reproducibility and physical interpretability. Taken together, these perspectives outline where AI-enabled science stands today, identify bottlenecks in data, methods and infrastructure, and chart concrete directions for building AI systems that are not only more powerful but also more transparent and capable of accelerating discovery in complex real-world environments.

Contents

# AI4X: Artificial Intelligence for the advancement of scientific pursuit and its future directions

Stephen G. Dale[1−2] , Nikita Kazeev[1], Alastair J. A. Price[3,5], Victor Posligua[1], Stephan Roche[6−7], O. Anatole von Lilienfeld[3−6], Konstantin S. Novoselov[1]

[1]Institute of Functional Intelligent Materials, National University of Singapore, 4 Science Drive 2, Singapore 117544
[2]Department of Material Science and Engineering, National University of Singapore, 9 Engineering Drive 1, Singapore 117575
[3]Department of Chemistry, University of Toronto, St. George campus, Toronto, ON, Canada
[4]Vector Institute, Toronto, Canada
[5]Acceleration Consortium, University of Toronto, 80 St George St, Toronto, ON M5S 3H6, Canada
[6]Catalan Institute of Nanoscience and Nanotechnology (ICN2), CSIC and BIST, Campus UAB, Bellaterra, 08193 Barcelona, Spain
[7]ICREA, Institució Catalana de Recerca i Estudis Avançats, 08070 Barcelona, Spain
*Author to whom any correspondence should be addressed.

**E-mail:** sdale@nus.edu.sg

**Introduction**

Artificial Intelligence (AI) techniques, particularly machine learning (ML) and related methods, have become increasingly prominent across a wide range of scientific disciplines. Since the release of ChatGPT in 2022, interest in AI has surged dramatically, expanding well beyond its roots in computer science and natural language processing.

The extraordinary capacity of AI to interpret, predict, and model phenomena across fields naturally raises a pivotal question: *Can AI techniques that have succeeded in domains outside my own be leveraged to accelerate progress in my research?*

In response to this interdisciplinary momentum, the AI4X conference (https://ai4x.cc/, short for Artificial Intelligence for Biology, Chemistry, Climate, Finance, Mathematics, Materials Science, Physics, Self-driving Labs, and Unconventional Computing) was convened in Singapore. The event gathered AI researchers from around the world and across disciplines to share developments, exchange insights, and foster new collaborations. This convergence of perspectives reflects a broader global trend: many governments have designated AI research as a national strategic priority, as evidenced by policy frameworks and funding initiatives worldwide.[1, 2, 3, 4, 5] This has been accompanied by the creation of dedicated AI research and regulatory institutions, many of which are represented in the affiliations of this article's authors.

Industry engagement has also been substantial by including companies such as NVIDIA, that has put in perspective the issue of accessing computational resources to support AI research,[6, 7] while major technology leaders including Google,[8, 9] Microsoft,[10, 11] Meta,[12, 13] and OpenAI[14, 15] are actively shaping the development and deployment of AI systems. This ecosystem is further enriched by specialized companies advancing targeted AI applications across scientific domains including Sea AI Labs(SAIL),[16, 17, 18, 19] MaxNet,[20] Insilico Medicine AI Ltd,[21] Constructor,[22] Qognative,[23] among many others.

Given the scale and diversity of AI's adoption, it is crucial for researchers to adopt a collaborative and interdisciplinary approach, not only to remain at the forefront of innovation but to accelerate the impact of their own work. The future of AI in science is promising, but it is also evolving rapidly, demanding ongoing engagement and adaptability from the scientific community.

This perspective highlights key contributions from AI4X and reflects on emerging themes that are likely to define the future of AI-driven research and innovation. With AI4X set to return to Singapore in 2026, in collaboration with the Acceleration Consortium (Canada), we hope this document will serve as both a reference point and a catalyst for continued interdisciplinary exploration and scientific advancement.

**References**
[1] National Research Foundation, Singapore. Research, innovation and enterprise 2025 plan, 2020.
[2] National Science Foundation. Request for information on the development of a 2025 national artificial intelligence (ai) research and development (rd) strategic plan, 2025.

[3] Government of Canada. Pan-canadian artificial intelligence strategy, 2024.

[4] European Commission. European approach to artificial intelligence, 2025.

[5] Ministry of Foreign Affairs People's Republic of China.

[6] Wen Jie Ong, Piero Altoè, Dallas Foster, Melisa Alkan, and Harry Petty. Revolutionizing ai-driven material discovery using nvidia alchemi.

[7] Piero Altoe. Processing units in the age of ai. In *Proceedings of the 21st ACM International Conference on Computing Frontiers*, pages 2–2, 2024.

[8] James Kirkpatrick, Brendan McMorrow, David HP Turban, Alexander L Gaunt, James S Spencer, Alexander GDG Matthews, Annette Obika, Louis Thiry, Meire Fortunato, David Pfau, et al. Pushing the frontiers of density functionals by solving the fractional electron problem. *Science*, 374(6573):1385–1389, 2021.

[9] Amil Merchant, Simon Batzner, Samuel S Schoenholz, Muratahan Aykol, Gowoon Cheon, and Ekin Dogus Cubuk. Scaling deep learning for materials discovery. *Nature*, 624(7990):80–85, 2023.

[10] Claudio Zeni, Robert Pinsler, Daniel Zügner, Andrew Fowler, Matthew Horton, Xiang Fu, Zilong Wang, Aliaksandra Shysheya, Jonathan Crabbé, Shoko Ueda, et al. A generative model for inorganic materials design. *Nature*, 639(8055):624–632, 2025.

[11] Giulia Luise, Chin-Wei Huang, Thijs Vogels, Derk P Kooi, Sebastian Ehlert, Stephanie Lanius, Klaas JH Giesbertz, Amir Karton, Deniz Gunceler, Megan Stanley, et al. Accurate and scalable exchange-correlation with deep learning. *arXiv preprint arXiv:2506.14665*, 2025.

[12] Brandon M Wood, Misko Dzamba, Xiang Fu, Meng Gao, Muhammed Shuaibi, Luis Barroso-Luque, Kareem Abdelmaqsoud, Vahe Gharakhanyan, John R Kitchin, Daniel S Levine, et al. Uma: A family of universal models for atoms. *arXiv preprint arXiv:2506.23971*, 2025.

[13] Daniel S Levine, Muhammed Shuaibi, Evan Walter Clark Spotte-Smith, Michael G Taylor, Muhammad R Hasyim, Kyle Michel, Ilyes Batatia, Gábor Csányi, Misko Dzamba, Peter Eastman, et al. The open molecules 2025 (omol25) dataset, evaluations, and models. *arXiv preprint arXiv:2505.08762*, 2025.

[14] Hunter Lightman, Vineet Kosaraju, Yuri Burda, Harrison Edwards, Bowen Baker, Teddy Lee, Jan Leike, John Schulman, Ilya Sutskever, and Karl Cobbe. Let's verify step by step. In *The Twelfth International Conference on Learning Representations*, 2023.

[15] Zehao Dou and Yang Song. Diffusion posterior sampling for linear inverse problem solving: A filtering perspective. In *The Twelfth International Conference on Learning Representations*, 2024.

[16] Kunhao Zheng and Min Lin. Jax-xc: Exchange correlation functionals library in jax. In *Workshop on" Machine Learning for Materials" ICLR*, volume 2023, 2023.

[17] Min Lin. Automatic functional differentiation in jax. *arXiv preprint arXiv:2311.18727*, 2023.

[18] Tianbo Li, Min Lin, Zheyuan Hu, Kunhao Zheng, Giovanni Vignale, Kenji Kawaguchi, AH Neto, Kostya S Novoselov, and Shuicheng Yan. D4ft: A deep learning approach to kohn-sham density functional theory. *arXiv preprint arXiv:2303.00399*, 2023.

[19] Tianbo Li, Min Lin, Stephen G Dale, Zekun Shi, AH Castro Neto, Kostya S Novoselov, and Giovanni Vignale. Diagonalization without diagonalization: A direct optimization approach for solid-state density functional theory. *Journal of Chemical Theory and Computation*, 21(9):4730–4741, 2025.

[20] Ye Wei, Bo Peng, Ruiwen Xie, Yangtao Chen, Yu Qin, Peng Wen, Stefan Bauer, Po-Yen Tung, and Dierk Raabe. Deep active optimization for complex systems. *Nature Computational Science*, 5(9):801–812, Sep 2025.

[21] Hongfu Lu, Deheng Sun, Haoyu Zhang, Sujing Shi, Huaxing Yu, Hui Cui, Xin Cai, Xiao Ding, Shan Chen, Man Zhang, et al. Discovery of novel, potent, selective, and orally available dgk$\alpha$ inhibitors for the treatment of tumors. *Journal of Medicinal Chemistry*, 2025.

[22] Nikita Kazeev, Wei Nong, Ignat Romanov, Ruiming Zhu, Andrey Ustyuzhanin, Shuya Yamazaki, and Kedar Hippalgaonkar. Wyckoff transformer: Generation of symmetric crystals. *arXiv preprint arXiv:2503.02407*, 2025.

[23] Joshua Rosaler, Luca Candelori, Vahagn Kirakosyan, Kharen Musaelian, Ryan Samson, Martin T Wells, Dhagash Mehta, and Stefano Pasquali. Supervised similarity for high-yield corporate bonds with quantum cognition machine learning. *arXiv preprint arXiv:2502.01495*, 2025.

# Graph Transformers: Overcoming the Barriers of Graph Representation Learning

Xavier Bresson

Department of Computer Science, National University of Singapore

**E-mail:** xaviercs@nus.edu.sg

## 1  Status

Graph-structured data is ubiquitous, representing networks from molecules and materials to social connections and knowledge graphs. Graph Neural Networks (GNNs) have emerged as the standard toolkit for representation learning on such data, leveraging a message-passing paradigm where node representations are iteratively updated by aggregating information from their local neighborhoods [6]. This approach inherently respects the graph topology and is equivariant to node permutations.

The development of GNNs has progressed through several key stages. Early GNNs established the core principles of recurrence and weight sharing but were hampered by issues like the vanishing gradient problem [1]. The subsequent development of Spectral GNNs [2], including the highly influential Graph Convolutional Network (GCN) [3], provided a more stable foundation by defining graph convolutions via spectral graph theory. However, these models were largely isotropic, treating all neighbors equally, which is a significant limitation as most real-world graph data is anisotropic. This led to the creation of Anisotropic GNNs, such as the Graph Attention Network (GAT) [4], which introduced mechanisms to weigh the importance of different neighbors, drawing inspiration from the attention mechanism that revolutionized natural language processing [5]. Despite these advances, this entire class of message-passing GNNs (MP-GNNs) shares fundamental limitations in their expressive power and their ability to capture long-range dependencies, which has motivated the development of a new generation of architectures.

## 2  Current and Future Challenges

Two primary challenges have defined the research frontier for GNNs: limited expressive power and the information bottleneck known as over-squashing.

The expressivity of a GNN refers to its ability to distinguish between non-isomorphic graphs. The Weisfeiler-Lehman (WL) test for graph isomorphism provides a formal hierarchy to measure this capability [7]. It has been proven that the majority of MP-GNNs are, at most, as powerful as the 1-WL test [8, 9]. This theoretical ceiling means that they are incapable of distinguishing between certain simple, non-isomorphic graph structures, and they struggle to perform tasks that require identifying elementary sub-structures like cycles or cliques [10]. Overcoming this limitation is crucial for applications in chemistry and materials science, where molecular function is intimately tied to such structural motifs.

The second major challenge is over-squashing. Because MP-GNNs propagate information locally, capturing dependencies between distant nodes requires stacking many layers. As the number of layers $L$ increases, the receptive field of a node can grow exponentially, $O(2^{L+1})$ for tree-like structures. This exponentially large amount of information must be compressed into a fixed-size node embedding vector at each step. This process creates an information bottleneck, effectively "squashing" long-range signals and preventing the model from learning dependencies beyond a small local neighborhood [11]. This severely limits the effective depth of GNNs and their applicability to large graphs where global context is important.

## 3  Advances in Science and Technology

Recent breakthroughs have directly targeted these fundamental challenges, leading to more powerful and scalable graph learning architectures.

### 3.1  Positional and Structural Encodings

A key strategy to enhance GNN expressivity is to break the symmetry of nodes by providing them with unique identifiers, or Positional Encodings (PE). By augmenting node features with PEs, MP-GNNs can be made provably more expressive than the 1-WL test [12]. A powerful and widely adopted approach is the Laplacian Positional Encoding (LapPE), which uses the low-frequency eigenvectors of the graph Laplacian matrix [13]. These eigenvectors provide a smooth, permutation-equivariant coordinate system for the nodes that captures both local and global structural information, serving as a direct generalization of the sinusoidal encodings used in the original Transformer [5].

*3.2　Graph Transformers*

The immense success of the Transformer architecture prompted its generalization to graphs. A Graph Transformer (GT) applies the self-attention mechanism, but sparsifies it according to the graph's topology; attention is computed only over a node's local neighborhood [13]. This formulation elegantly incorporates the graph structure as an inductive bias. When combined with LapPEs and edge features, GTs have demonstrated superior performance, leading to a new class of expressive models including GraphiT, SAN, and GPS [14, 15, 16]. However, while expressive, these models still rely on a local message-passing scheme and thus remain susceptible to over-squashing on large graphs. Furthermore, models that attempt to use full-graph attention are constrained by a prohibitive $O(N^2)$ complexity.

*3.3　Graph ViT: Scalable Long-Range Learning*

To address both over-squashing and scalability, the ideas from the Vision Transformer (ViT) [17] and MLP-Mixer [18] were generalized to graphs [19]. The Graph ViT/MLP-Mixer architecture introduces a novel hierarchical approach:

1. **Graph Partitioning:** The input graph is first partitioned into a set of smaller subgraphs, or "patches," using a fast graph clustering algorithm like Metis. This process is made stochastic via edge dropout to prevent overfitting.

2. **Local Patch Encoding:** A standard MP-GNN is applied independently to each patch. Since patches are small, this GNN can effectively learn local structural representations without suffering from over-squashing. The output is a single embedding vector for each patch.

3. **Global Mixing:** The sequence of patch embeddings is then processed by a global mixer layer (e.g. a Transformer or MLP-Mixer). This allows for all-to-all communication between patches, enabling the model to learn long-range dependencies across the entire graph with a complexity that is linear in the number of nodes.

This hybrid architecture combines the efficiency of local MP-GNNs with the global reach of Transformers. Empirically, it has been shown to mitigate over-squashing on synthetic benchmarks [11] and achieve state-of-the-art results on large molecular datasets from benchmarks like LRGB, while maintaining linear scalability. It successfully distinguishes graphs that fail 1-WL, 2-WL, and even 3-WL tests, demonstrating a high degree of practical expressivity [19].

## 4　Concluding Remarks

The roadmap for graph representation learning has seen a clear and rapid progression: from simple local aggregators to expressive, attention-based architectures, and now to scalable, hierarchical models that can capture long-range dependencies. The evolution from Vanilla GNNs to Graph Transformers and now Graph ViT reflects a systematic effort to overcome the core limitations of expressivity and scalability. The current state-of-the-art provides a robust and powerful framework for a wide range of graph learning tasks.

　　Looking forward, the next frontier appears to be the incorporation of geometric inductive biases. For graphs that represent objects in 3D space, such as molecules or point clouds, equivariance to permutations is not enough. The new generation of models is equivariant to the continuous symmetries of 3D space—translation, rotation, and reflection, i.e., the SE(3) group. Architectures like Tensor Field Networks [20] and other SE(3)-equivariant models are paving the way for GNNs to become foundational models for the physical sciences.

**Acknowledgments**

XB is supported by NUS Grant ID R-252-000-B97-133 and MOE AcRF T1 Grant ID 251RES2423.

**References**

[1] Scarselli, F., Gori, M., Tsoi, A. C., Hagenbuchner, M., & Monfardini, G. (2009). The graph neural network model. *IEEE Transactions on Neural Networks.*

[2] Defferrard, M., Bresson, X., & Vandergheynst, P. (2016). Convolutional neural networks on graphs with fast localized spectral filtering. *Advances in Neural Information Processing Systems.*

[3] Kipf, T. N., & Welling, M. (2017). Semi-supervised classification with graph convolutional networks. *International Conference on Learning Representations.*

[4] Veličković, P., Cucurull, G., Casanova, A., Romero, A., Lio, P., & Bengio, Y. (2018). Graph attention networks. *International Conference on Learning Representations.*

[5] Vaswani, A., Shazeer, N., Parmar, N., Uszkoreit, J., Jones, L., Gomez, A. N., ... & Polosukhin, I. (2017). Attention is all you need. *Advances in Neural Information Processing Systems.*

[6] Gilmer, J., Schoenholz, S. S., Riley, P. F., Vinyals, O., & Dahl, G. E. (2017). Neural message passing for quantum chemistry. *International Conference on Machine Learning.*

[7] Weisfeiler, B., & Leman, A. (1968). A reduction of a graph to a canonical form and an algebra arising during this reduction. *Nauchno-Tekhnicheskaya Informatsiya.*

[8] Xu, K., Hu, W., Leskovec, J., & Jegelka, S. (2019). How powerful are graph neural networks?. *International Conference on Learning Representations.*

[9] Morris, C., Ritzert, M., Fey, M., Hamilton, W. L., Lenssen, J. E., Rattan, G., & Grohe, M. (2019). Weisfeiler and Leman go neural: Higher-order graph neural networks. *AAAI Conference on Artificial Intelligence.*

[10] Chen, Z., Chen, L., Villar, S., & Bruna, J. (2020). Can graph neural networks count substructures?. *Advances in Neural Information Processing Systems.*

[11] Alon, U., & Yahav, E. (2021). On the bottleneck of graph neural networks and its practical implications. *International Conference on Learning Representations.*

[12] Dwivedi, V. P., Joshi, C. K., Laurent, T., Bengio, Y., & Bresson, X. (2020). Benchmarking graph neural networks. *arXiv preprint arXiv:2003.00982.*

[13] Dwivedi, V. P., & Bresson, X. (2021). A generalization of transformer networks to graphs. *AAAI Workshop on Deep Learning on Graphs.*

[14] Mialon, G., Chen, D., Selosse, J., & Mairal, J. (2021). GraphiT: Encoding graph structure in transformers. *arXiv preprint arXiv:2106.05234.*

[15] Ying, C., Cai, T., Luo, S., Zheng, S., Ke, G., He, D., ... & Liu, T. Y. (2021). Do transformers really perform badly for graph representation?. *Advances in Neural Information Processing Systems.*

[16] Kreuzer, D., Beaini, D., Hamilton, W., Létourneau, V., & Tossou, P. (2021). Rethinking graph transformers with spectral attention. *Advances in Neural Information Processing Systems.*

[17] Dosovitskiy, A., Beyer, L., Kolesnikov, A., Weissenborn, D., Zhai, X., Unterthiner, T., ... & Houlsby, N. (2020). An image is worth 16x16 words: Transformers for image recognition at scale. *International Conference on Learning Representations.*

[18] Tolstikhin, I. O., Houlsby, N., Kolesnikov, A., Beyer, L., Zhai, X., Unterthiner, T., ... & Dosovitskiy, A. (2021). Mlp-mixer: An all-mlp architecture for vision. *Advances in Neural Information Processing Systems.*

[19] He, X., Hooi, B., Laurent, T., Perold, J., LeCun, Y., & Bresson, X. (2022). A generalization of ViT/MLP-Mixer to graphs. *International Conference on Machine Learning.*

[20] Thomas, N., Smidt, T., Kearnes, S., Yang, L., Li, L., Kohlhoff, K., & Riley, P. (2018). Tensor field networks: Rotation- and translation-equivariant neural networks for 3D point clouds. *arXiv preprint arXiv:1802.08219.*

# Explainable AI (XAI) for Science

Gianmarco Mengaldo[1−3,∗]

[1]Department of Mechanical Engineering, National University of Singapore, Singapore
[2]Department of Mathematics, National University of Singapore, Singapore
[3]Honorary Research Fellow, Department of Aeronautics, Imperial College London, London, United Kingdom
∗Author to whom any correspondence should be addressed.

**E-mail:** mpegim@nus.edu.sg

**Status**
Explainable artificial intelligence (XAI) has developed over the past decade in response to the widespread adoption of black-box AI models in scientific and engineering workflows. XAI emphasizes the generation of human-understandable explanations of AI model behavior and decision-making process [1–3]. It uses interpretability methods to clarify *what* data the model deemed important and *how* the data was used, and addresses the question *why* the model reached a certain result, thereby linking AI results to domain knowledge.

Initially, most progress in XAI came through post-hoc interpretability methods, such as saliency maps, and feature attribution tools like LIME [4] and SHAP [5], that provided heuristics for linking model inputs to outputs after training. These approaches were widely adopted due to their model-agnostic nature and ease of application across domains, as they primarily respond to the *what* question of which inputs drive model predictions. However, their limitations quickly became apparent: interpretability results often varied with small perturbations, lacked reproducibility across methods, and risked misrepresenting what data models had actually used [6–9]. To address these concerns, the field has increasingly emphasized intrinsically interpretable models. These include generalized additive models, prototype-based networks, and symbolic regression approaches, all designed to provide transparency by construction – see e.g., [10]. While some of these models trade predictive power for interpretability, they offer stronger guarantees that interpretability results correspond to the actual model decision-making process. Parallel to this trend, hybrid models that integrate physical constraints and other known priors with neural networks have gained momentum, aiming to reconcile predictive power with interpretability – see e.g., [11, 12].

In this context, *causality* emerges as a cornerstone of XAI, as it directly addresses the *why* dimension of explanations by moving beyond correlation toward mechanisms and testable drivers of system behavior. Indeed, *causality* has been deemed a central pillar that should always be taken into account when trying to explain black-box model's behavior and outcomes [13]. This, along with accuracy, reproducibility, and human-centric understandability of interpretability results, constitute the necessary ingredients for the successful adoption of XAI in Science, or what we call "XAI for Science". Accuracy is a necessary (yet not sufficient) requirement for predictions and interpretability results to reflect genuine data–phenomenon relationships; reproducibility requires that interpretability results remain stable across methods and experiments; and understandability emphasizes the alignment of interpretability results with domain-relevant concepts, so that human experts can interpret and explain them. This framing reflects a growing recognition that XAI is not limited to enhancing trust in AI models, but that interpretability should be embedded within the broader epistemic cycle of hypothesis generation, validation, and inductive reasoning [14]. Building on this perspective, XAI for Science can be understood as serving three complementary purposes:

(i) confirming existing knowledge (trustworthy AI),

(ii) discovering new knowledge (AI for scientific discovery),

(iii) shaping model behavior.

In turn, discovering new knowledge can lead to actionable insights that can be applied to e.g., better design of engineering systems [15].

However, it can also lead to the generation of (iv) *false or misleading knowledge* if the fundamental pillars of XAI for Science are not upheld, a risk that could undermine scientific progress and trust in the field. Figure 1 shows a simple schematics of XAI for Science, wit the minimum desiderata, namely causality, accuracy, reproducibility, and understandability, as well as the potential outcomes, and the main areas of research that needs to be further developed. The latter are discussed next.

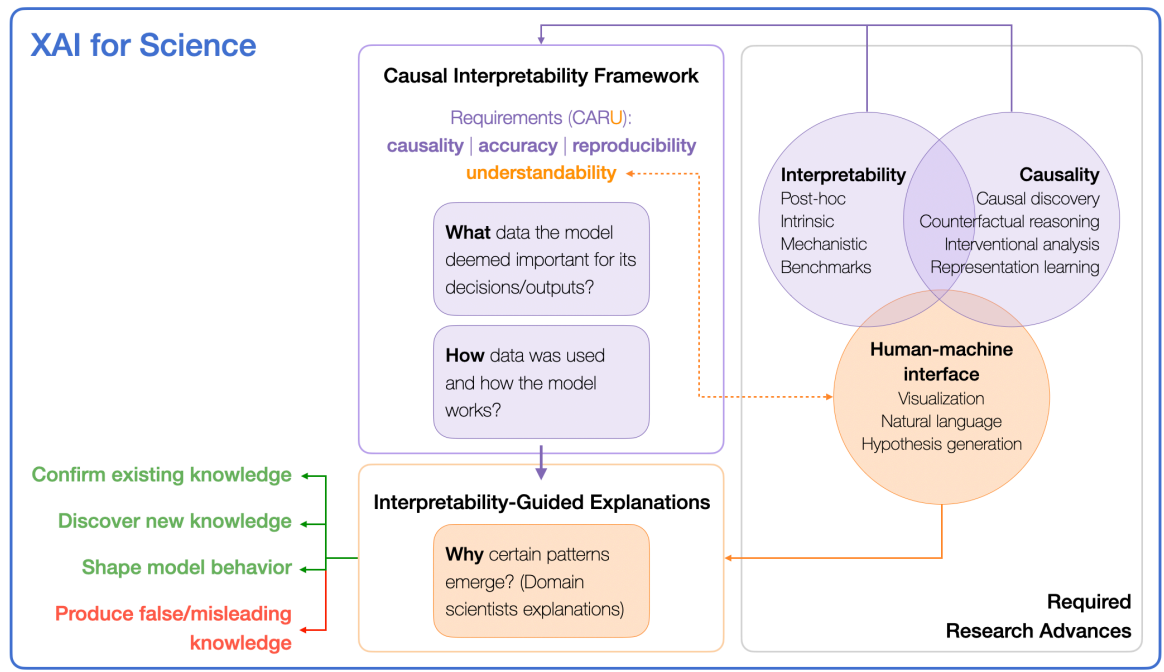

**Figure 1.** XAI for Science provides a unifying framework that integrates interpretability, causality, and human–machine interfaces to advance scientific discovery. The causal interpretability framework (CARU: causality, accuracy, reproducibility, understandability) structures the 'what,' and 'how,' of model interpretability, that are used to provide interpretability-guided answers by human domain experts to 'why' the model reached certain results. This workflow can guide both knowledge confirmation (trustworthy AI) and knowledge discovery, while highlighting the risks of creating misleading or false knowledge. It can also shape model behavior and e.g., correct model biases. Required research advances link methodological innovation in interpretability, causality and human-machine interface.

**Challenges and required advances**
Today, XAI is applied across a broad spectrum of scientific domains. Yet, the criteria required to verify that interpretability outcomes capture genuine systematic relationships, both between input variables and model outputs, and within the model's inner behavior, rather than spurious correlations, are frequently not met. Causality is often not discussed altogether, while accuracy and reproducibility of interpretability results are commonly taken at face value – i.e., the results of interpretability methods are assumed correct, and reproducible – when this is hardly the case, especially in the context of post-hoc interpretability methods. Therefore, human-centric understandability, the basis for XAI, becomes a rather fragile endeavor built on shaky foundations, and potentially leading to *illusionary explanations*.

Given this premise, progress in explainable artificial intelligence (XAI) for scientific applications requires advances at the interface of at least three areas, namely: 1) *interpretability methodology*, 2) *causality research*, and 3) *human-machine knowledge integration*.

First, advances in *interpretability methodology* are needed. Post-hoc approaches remain widely used, but their lack of robustness and faithfulness highlights the need for systematic evaluation frameworks that can assess when and how such explanations should be trusted [8]. In parallel, intrinsically interpretable models offer an alternative path by embedding transparency directly into the architecture, thereby avoiding the pitfalls of post-hoc explanations and providing more faithful insights into the decision-making process [16]. A complementary direction is the development of hybrid frameworks that integrate physical laws, or other domain priors into machine learning models. Such approaches can reduce the risk of spurious correlations and better align explanations with mechanistic understanding and causal reasoning [11, 12]. In addition to model design, there is also a pressing need for new benchmarks and metrics that extend beyond predictive performance to quantify the robustness and reproducibility of interpretability results [3]. Synthetic datasets with ground-truth explanations, complemented by domain-specific testbeds, will be critical to evaluate whether methods truly capture systematic relationships learned by the model.

Second, *causality research* must be more explicitly integrated into XAI. Causal discovery, counterfactual reasoning, interventional analysis, and causal representation learning provide the means to move beyond correlation and to identify drivers of system behavior that can be tested through experiments or simulations [17]. Recent advances in causal machine learning highlight how combining observational data with causal structures can uncover mechanisms that are not accessible through purely predictive models [18]. Embedding such approaches within interpretability pipelines will not only improve the epistemic validity of explanations but also clarify the conditions under which these explanations hold. Without a causal layer, explanations risk being superficially plausible but scientifically misleading, thereby reinforcing the danger of generating *illusionary* knowledge.

Third, advances are required in human–machine knowledge integration. Even when interpretability methods are technically sound, their outputs may not be meaningful to domain experts, either because they lack alignment with domain concepts or because they are presented in inaccessible forms. Developing human-centered interfaces that allow interactive exploration of explanations through visualization, natural-language narration, or hypothesis-oriented workflows will be crucial to make XAI practically useful in scientific settings [19]. Such tools should facilitate iterative dialogue between human reasoning and machine outputs through explanations, enabling scientists to probe assumptions, test hypotheses, and refine models. In this way, XAI can be embedded into the broader epistemic cycle of hypothesis generation, validation, and inductive reasoning, transforming explanations from static outputs into active instruments of discovery.

Finally, interdisciplinary collaboration will remain essential. These advances in XAI cannot be achieved solely within the computer science domain but must be co-designed with domain experts, including physicists, biologists, climate scientists, and engineers, to ensure that explanatory tools reflect the epistemic needs of each discipline.

**Concluding remarks**

Explainable artificial intelligence (XAI) is rapidly evolving, having moved beyond its early emphasis on post-hoc methods that provide heuristic insights into input–output relationships towards a more comprehensive mechanistic understanding of AI model behavior and decision-making process.

This evolution is providing two opportunities. On the one hand, explanations can be used for building trust on the AI model behavior and decision-making process: when explanations are aligned with domain-expert understanding, they can reassure human users and can potentially be used to attribute responsibility [20]. On the other hand, explanations can be used as potential instruments for advancing scientific knowledge and shaping model development and behavior; for instance, explanations can complement domain expertise by uncovering previously hidden mechanisms, generating new hypotheses, and facilitating knowledge discovery. Explanations can also serve as diagnostic tools for identifying spurious correlations, removing unwanted biases, and steering models toward more faithful, physically or biologically consistent behaviors. In this sense, XAI can be seen as an epistemic safeguard that ensures AI models do not mislead, and as a catalyst for discovery and innovation.

These two opportunities of XAI are particularly attractive, as they highlight its dual role as a trust-building mechanism and a driver of discovery.

Yet, we should not be fooled: alongside these opportunities lie two critical risks. At best, explanations may be deemed uninformative, hence discarded, thereby adding little value beyond what predictive performance already provides. At worst, explanations can actively mislead, creating the illusion of understanding and fostering the generation of false knowledge. This danger is especially high when interpretability results are inaccurate, unstable across methods, or not reproducible—conditions under which explanations should be deemed unusable. Unfortunately, such safeguards are not always applied in practice, as interpretability outputs are often assumed to be both correct and reproducible by default. Equally problematic, interpretability results that are accurate but misaligned with domain concepts may fail to resonate with experts, rendering them ineffective in real scientific workflows. In both cases, the epistemic value of XAI breaks down, undermining its promise to contribute to trustworthy and discovery-driven science.

In all the above, causality plays a decisive role. Without causal grounding, even stable and reproducible explanations risk being superficially plausible while scientifically misleading. Embedding causal reasoning into interpretability pipelines through counterfactuals, or mechanistic priors provides the necessary structure to distinguish genuine drivers from spurious correlations. In this sense, causality constitutes a central safeguard that preserves the epistemic value of XAI, ensuring that its outputs serve as reliable instruments for both trust and discovery rather than conduits for illusionary knowledge.

**Acknowledgments**
The author acknowledges support from Ministry of Education, Singapore (MOE) Tier 1 grant number 22-4900-A0001-0: 'Discipline-Informed Neural Networks for Interpretable Time-Series Discovery'.

**References**

[1] Doshi-Velez F and Kim B 2017 *arXiv preprint arXiv:1702.08608*

[2] Gilpin L H, Bau D, Yuan B Z, Bajwa A, Specter M and Kagal L 2018 Explaining explanations: An overview of interpretability of machine learning *2018 IEEE 5th International Conference on Data Science and Advanced Analytics (DSAA)* (IEEE) pp 80–89

[3] Samek W, Montavon G, Lapuschkin S, Anders C J and Müller K R 2021 *Proceedings of the IEEE* **109** 247–278

[4] Ribeiro M T, Singh S and Guestrin C 2016 "why should i trust you?": Explaining the predictions of any classifier *Proceedings of the 22nd ACM SIGKDD International Conference on Knowledge Discovery and Data Mining* (ACM) pp 1135–1144

[5] Lundberg S M and Lee S I 2017 A unified approach to interpreting model predictions *Advances in Neural Information Processing Systems (NeurIPS)* vol 30 pp 4765–4774

[6] Lipton Z C 2018 *Queue* **16** 31–57

[7] Rudin C, Chen C, Chen Z, Huang H, Semenova L and Zhong C 2022 *Statistic Surveys* **16** 1–85

[8] Turbé H, Bjelogrlic M, Lovis C and Mengaldo G 2023 *Nature Machine Intelligence* **5** 250–260

[9] Wei J, Turbé H and Mengaldo G 2024 *arXiv preprint arXiv:2407.19683*

[10] Turbé H, Bjelogrlic M, Mengaldo G and Lovis C 2025 *arXiv preprint arXiv:2502.19577*

[11] Karniadakis G E, Kevrekidis I G, Lu L, Perdikaris P, Wang S and Yang L 2021 *Nature Reviews Physics* **3** 422–440

[12] Wang X, Chen J, Yang J, Adie J, See S, Furtado K, Chen C, Arcomano T, Maulik R, Xue W and Mengaldo G 2025 *arXiv preprint arXiv:2502.13185*

[13] Mengaldo G 2024 *arXiv preprint arXiv:2406.10557*

[14] Popper K 2005 *The logic of scientific discovery* (Routledge)

[15] Cremades A, Hoyas S, Deshpande R, Quintero P, Lellep M, Lee W J, Monty J P, Hutchins N, Linkmann M, Marusic I and Vinuesa R 2024 *Nature Communications* **15** 3864

[16] Rudin C 2019 *Nature Machine Intelligence* **1** 206–215

[17] Pearl J and Mackenzie D 2019 *The Book of Why: The New Science of Cause and Effect* (New York: Basic Books) ISBN 9780465097609

[18] Schölkopf B, Locatello F, Bauer S, Ke N R, Kalchbrenner N, Goyal A and Bengio Y 2021 *Proceedings of the IEEE* **109** 612–634

[19] Hong S R, Hullman J and Bertini E 2020 *Proceedings of the ACM on Human-Computer Interaction* **4** 1–26

[20] Wei J, Verona E, Bertolini A and Mengaldo G 2025 *arXiv preprint arXiv:2509.17334*

# Physics-Assisted Machine Learning for Wave Sensing and Imaging

**Xudong Chen[1]**

[1] Department of Electrical and Computer Engineering, National University of Singapore, Singapore

E-mail: elechenx@nus.edu.sg

**Status**

In recent years, machine learning (ML) has garnered significant attention for addressing challenges in wave imaging and sensing, which involve determining the properties of an unknown target (e.g., shape, position, and material characteristics) from measurements of the scattered fields. By employing electromagnetic or acoustic waves to probe obscured or remote regions, computational imaging techniques enable a wide range of important applications, including nondestructive evaluation, road-target classification, geophysics, security screening, biomedical imaging, remote sensing, and the integration of sensing and communication [1].

**Current and future challenges**

Many approaches treat ML as a black box, overlooking decades of insight grounded in wave physics and rigorous mathematical analysis. In black-box ML, the inputs and outputs of the neural network (NN) are directly defined as the measured data and the parameters to be reconstructed, with the main effort placed on selecting and tuning the NN. However, in wave sensing and imaging, certain physical laws yield well-established mathematical properties (and in some cases even analytical formulas) that do not need to be relearned through data-driven training. Moreover, decades of research have produced valuable domain-specific knowledge, including effective rules of thumb, approximate direct inversion models, and physically meaningful interpretations. To disregard these physical and mathematical insights by relying solely on black-box ML would be a significant waste of established understanding and tools.

From a mathematical perspective, the input–output relationship in black-box ML is fixed, since the inputs and outputs of the NN are directly defined as the measured data and the parameters to be reconstructed. If this input-output relationship is inherently complex, the NN is burdened with a difficult task. Simply designing or fine-tuning a more powerful NN does not change the intrinsic difficulty of the input-output mapping. A more effective approach is to redefine the inputs and/or outputs of the NN using domain-specific knowledge, thereby assigning the network a simpler and more tractable task.

Researchers often ask: Why does the black-box use of ML, which has achieved great success in computer vision, often fall short in wave sensing and imaging? In computer vision, the governing physical law is essentially Fourier optics, which is relatively simple because the wavelength is much smaller than the size of the targets. By contrast, in wave sensing and imaging, particularly in the microwave and millimeter-wave ranges, where the target size is comparable to the wavelength, the governing physical laws are far more complex. In essence, compared with computer vision, wave sensing and imaging are mathematically more challenging problems, regardless of whether ML is applied.

**Advances in science and technology to meet challenges**

Training a NN can, in fact, be viewed as solving an inverse problem, where the unknown filters of the NN are reconstructed from training data. It is therefore crucial to thoroughly understand the corresponding forward problem that underlies these inherently inverse tasks. To this end, many studies have incorporated domain-specific knowledge, i.e., mathematical, physical, and engineering intuition, into NNs, leading to more efficient and elegant solutions.

Extensive efforts have been devoted to leveraging domain-specific knowledge to generate approximate inverses, i.e., reasonably good estimates of the contrast, which are then paired with the ground-truth contrast as input–output data for training NNs. For example, the dominant current scheme in [2] produces more accurate contrast estimates than the back-propagation (BP) method by exploiting the concept of dominant contrast current. In [3], the connection between conventional iterative inverse-scattering algorithms and NNs was explored to develop DeepNIS, an architecture based on three cascaded convolutional neural networks (CNNs) that refine a BP-generated image into a higher-resolution estimate of the dielectric distribution. In [4], an induced current learning method was introduced, incorporating multiple strategies inspired by traditional iterative algorithms into the NN architecture, reformulating the task so that the NN estimates the contrast current rather than the contrast itself. In [5], a deep learning approach for millimeter-wave short-range imaging was proposed, where the NN input is derived from far-field approximation results, leveraging the computational efficiency of the fast Fourier transform (FFT).

There are many approaches to synergizing domain-specific knowledge with ML. In [6], a CNN is trained to learn the complex mapping from magnetic resonance (MR) T1 images to dielectric images, with the predicted dielectric images then serving as the starting point for microwave inverse scattering imaging as a physics-based refinement step. In [7], the supervised descent method is applied to microwave imaging, where inversion models are updated using descent directions obtained during training. For integrated sensing and communication (ISAC), [8] employs beam scanning to perform local imaging at each scanning angle using compressive sensing algorithms, followed by deep learning to fuse the local images into a complete reconstruction. In automotive radar, [9] first processes measured range-Doppler data using wave propagation theory and extracts four physical features that capture key kinematic and geometric characteristics of targets. For classifying road targets into five categories, the proposed NN performs a simple 4-to-5 mapping, significantly outperforming a black-box CNN that requires a 2500-to-5 mapping. For computational imaging problems more broadly, numerous strategies for combining inverse-scattering-problem (ISP) knowledge with ML are reviewed in [10].

**Concluding remarks**

From a mathematical perspective, where a NN maps inputs to outputs, the key to an effective learning approach is to design the NN's inputs and outputs based on domain-specific knowledge, so that the network is assigned a simpler task. Attempting to solve an intrinsically hard problem by merely increasing the NN's complexity is not recommended. From a physical perspective, we should ask: what rule is the NN learning? It is not merely fitting input-output pairs, but capturing the underlying physical laws governing the training data. In sensing and imaging problems, training an NN is, in effect, learning the laws of scattering. The network's generalization can be understood in this context: once these scattering laws are sufficiently learned from the training data, the NN can naturally handle unseen scenarios, such as scatterers with different shapes. Regarding the balance between human and NN effort, it is advisable to extract as much as possible using low-cost computations, leaving the remaining task to the NN. Assigning everything to the NN results in a black-box approach, while doing everything by ourselves corresponds to a traditional non-learning method. To achieve this balance, a thorough understanding of the original problem, including the physical and mathematical insights underlying the training data, is essential.

**References**

[1] Chen X 2018 *Computational Methods for Electromagnetic Inverse Scattering* (Wiley).
[2] Wei, Z. and Chen, X., 2019. Deep-learning schemes for full-wave nonlinear inverse scattering problems. *IEEE Transactions on Geoscience and Remote Sensing*, **57(4)**, pp.1849-1860.
[3] Li, L., Wang, L.G., Teixeira, F.L., Liu, C., Nehorai, A. and Cui, T.J., 2019. DeepNIS: Deep neural network for nonlinear electromagnetic inverse scattering. *IEEE Transactions on Antennas and Propagation*, **67(3)**, pp.1819-1825.
[4] Wei, Z. and Chen, X., 2019. Physics-inspired convolutional neural network for solving full-wave inverse scattering problems. *IEEE Transactions on Antennas and Propagation*, **67(9)**, pp.6138-6148.
[5] Yin, T., Tan, K. and Chen, X., 2023. Far-field approximation learning method for millimeter-wave short-range imaging. *IEEE Transactions on Antennas and Propagation*, **71(4)**, pp.3441-3449.
[6] Chen, G., Shah, P., Stang, J. and Moghaddam, M., 2019. Learning-assisted multimodality dielectric imaging. *IEEE Transactions on Antennas and Propagation*, **68(3)**, pp.2356-2369.
[7] Guo, R., Song, X., Li, M., Yang, F., Xu, S. and Abubakar, A., 2019. Supervised descent learning technique for 2-D microwave imaging. *IEEE Transactions on Antennas and Propagation*, **67(5)**, pp.3550-3554.
[8] Wen, L., Zhao, S., Yin, T., Guo, Y. and Chen, X., 2025. Inverse scattering approach to integration of communication and sensing. *IEEE Journal of Selected Topics in Electromagnetics, Antennas and Propagation*, Accepted.
[9] Tan, K., Yin, T., Ruan, H., Balon, S. and Chen, X., 2022. Learning approach to FMCW radar target classification with feature extraction from wave physics. *IEEE Transactions on Antennas and Propagation*, **70(8)**, pp.6287-6299.
[10] Chen, X., Wei, Z., Maokun, L. and Rocca, P., 2020. A review of deep learning approaches for inverse scattering problems (invited review). *Progress In Electromagnetics Research*, **167**, pp.67-81.

# Dynamics-Informed AI for Complex Systems

Gianmarco Mengaldo[1,2,3,*]

[1]Department of Mechanical Engineering, National University of Singapore, Singapore
[2]Department of Mathematics, National University of Singapore, Singapore
[3]Honorary Research Fellow, Department of Aeronautics, Imperial College London, London, United Kingdom
[*]Author to whom any correspondence should be addressed.

**E-mail:** mpegim@nus.edu.sg

**Status**

Artificial intelligence (AI) has rapidly advanced the forecasting – i.e., the prediction of the future evolution – of complex systems across various domains, from fluid dynamics to weather and climate, among others. Complex systems are characterized by high dimensionality, nonlinear interactions, and multiscale dynamics that often give rise to emergent phenomena such as chaos, intermittency, abrupt regime transitions, and self-organization [1–4]. These properties make them intrinsically difficult to predict.

In this context, AI has shown promising results in forecasting the evolution of such systems over time, and is being adopted to bypass the solution of the expensive deterministic or stochastic equation-based models that typically describe complex systems – e.g., [5, 6]. An important requirement when forecasting complex systems is the accuracy of the forecasts. Yet, accuracy can come in different flavors, with standard error metrics such as mean squared error (MSE) and mean absolute error (MAE) only capturing pointwise discrepancies but failing to quantify whether forecasts remain dynamically (and physically) faithful. Indeed, this aspect is crucial to understand whether AI forecasts respect fundamental physical principles, or they lead to unphysical solutions that do not respect the underlying dynamics of the system being forecasted.

Why the dynamical consistency of AI-based forecasts is important? The answer lies in the fact that AI forecasts which are statistically accurate but dynamically inconsistent may quickly diverge from the true system trajectory, providing misleading information and unreliable results. In chaotic, multiscale systems, pointwise accuracy over short horizons does not guarantee that the model preserves essential dynamical properties. Without such guarantees, AI forecasts risk violating conservation laws, missing regime transitions, or misrepresenting the likelihood of rare and extreme events, especially in long-term forecasts.

The evaluation of dynamical properties in complex systems has been historically conducted via global metrics, such as Lyapunov exponents [7] dimension, and information-theoretic metrics such as entropy — which capture average features of the system's predictability and complexity. While foundational, these global approaches are limited in their ability to capture local, state-dependent dynamical properties and predictability, which are critical in many applications such as weather and climate forecasting. Some extensions were proposed to overcome these limitations. Local or finite-time Lyapunov exponents [8], and later finite-size [9] and nonlinear local Lyapunov exponents [10], were introduced to assess predictability over finite horizons and finite perturbations, but they remain constrained by their reliance on linearized dynamics or the assumption of exponential error growth.

These shortcomings have motivated the development of alternative, data-driven approaches that can more flexibly capture state-dependent predictability, leading to several promising recent advances in dynamical systems theory. Local dynamical indices – such as instantaneous dimension and extremal index [11–13] – offer quantitative, data-driven state-dependent measures of dynamical properties, namely complexity and persistence, and they can provide some glimpses into the predictability of dynamical systems. More recently, the introduction of a new metric, namely time-lagged recurrence (TLR) [14], has enabled the characterization of state-dependent predictability in large, real-world datasets, underlying high-dimensional complex systems. These advances, among many others, could transform the way we understand and build AI forecasting tools. In particular, AI systems can be evaluated, constrained, and informed by data-driven dynamical indices, paving the way for what we term "*dynamics-informed AI*" (*DIAI*) or, in analogy with physics-informed neural networks (PINNs) [15], "dynamics-informed neural networks" (DINNs) (Figure 1). Indeed, DINNs can be a complementary tool to PINNs [15], as DINNs constrain learning through empirical dynamical indices derived solely from data, rather than embedding the system's governing equations into the loss function. Indeed, it could be seen as a data-driven approach to PINNs. Yet, some challenges remain to deliver effective DINNs, and critical advances are required. We briefly discuss this next.

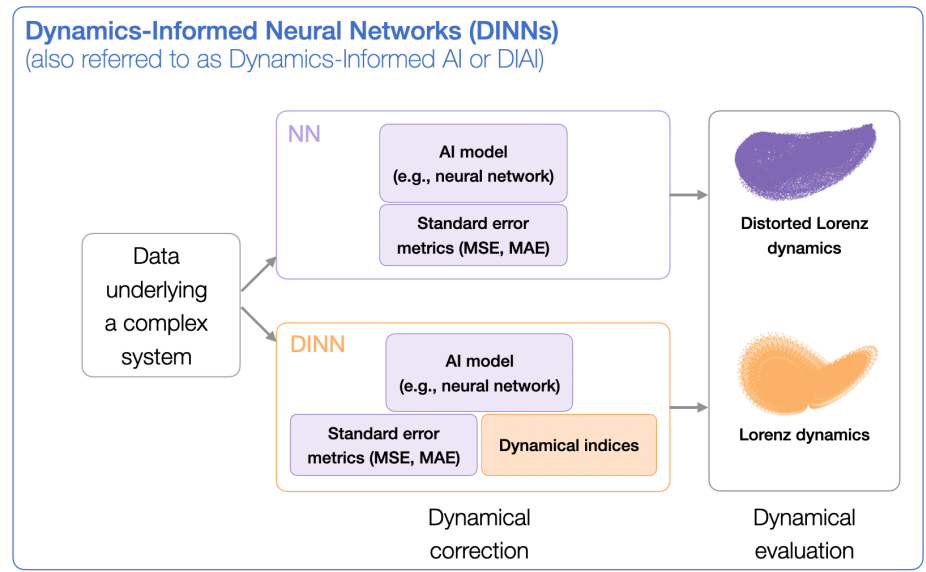

**Figure 1.** Comparison between dynamically consistent and inconsistent forecasts. Top panels: standard workflow, without embedding dynamical constraints in the AI (i.e., neural network) architectures, leading to e.g., distortion of the dynamics for the Lorenz system. Bottom panels: enforcing dynamical constraints in the neural networks architectures, yielding dynamics-informed neural networks (DINNs), that may produce a more consistent dynamical behavior for e.g., the Lorenz system. The results can be further evaluated with dynamical metrics to asses whether they reproduce the true dynamics of the underlying complex system.

**Challenges and required advances**
AI forecasts of complex systems often degrade when extended over long time horizons, as small errors accumulate and eventually cause trajectories to diverge from the true system. This challenge is particularly accentuated in chaotic and multiscale settings, where sensitivity to initial conditions amplifies inaccuracies and may lead to dynamically inconsistent outcomes. Classical error metrics such as MSE and MAE provide little insight into the origin of these failures. By embedding dynamical indices into evaluation frameworks, one can begin to identify where and potentially why a forecast fails. Indeed, recent work shows that AI forecasts tend to perform poorly in regions of high instantaneous dimension, which reflect locally complex dynamics, and low persistence, which correspond to states that quickly lose memory of their neighborhood [16].

This dynamical perspective reveals three characteristic failure modes of AI forecasts:

- *incomplete dynamical representation*, where forecast trajectories do not fully cover the system's variability;

- *systematic shifts in dynamical indices*, indicating biased representation of complexity or persistence; and

- *breakdown of phase-space structure*, where the geometry of the dynamics itself becomes distorted.

These failure modes highlight that even forecasts with low statistical error can be dynamically unfaithful, underscoring the importance of metrics that assess physical and dynamical fidelity.

Beyond using them as evaluation tools, dynamical indices provide opportunities for guiding model development. Forecast errors defined on indices (for example, MSE on instantaneous dimension or persistence) can complement traditional loss functions, encouraging AI models to reproduce state trajectories and the underlying dynamical behavior. This approach shifts the training objective from pointwise accuracy toward dynamical consistency, potentially improving long-horizon accuracy. Similarly, incorporating time-lagged recurrence (TLR) into training pipelines can highlight states where forecasts are intrinsically less predictable. Such state-dependent diagnostics enable models to quantify uncertainty in a principled manner, analogous to ensemble spread in numerical weather prediction, and could guide adaptive allocation of computational resources in regions of low predictability. Several key challenges must be addressed to make these advances feasible.

A first challenge is *scalability*: computing dynamical indices reliably in high-dimensional phase spaces, such as those arising from atmospheric reanalysis or coupled Earth system models, is computationally demanding. Efficient algorithms, approximate surrogates, or GPU-accelerated workflows will be necessary to make such diagnostics practical at scale.

A second challenge is *integration*: while indices can be computed a posteriori to evaluate forecasts, incorporating them directly into training requires differentiable formulations or robust approximations, so that gradient-based optimization remains feasible. Developing loss functions that balance statistical error and dynamical fidelity without destabilizing training is an open research frontier.

A third challenge is *interpretability*. While indices such as instantaneous dimension or persistence provide intuitive measures of complexity, their values may depend sensitively on methodological choices such as thresholds, neighborhood sizes, or distance metrics. Establishing standardized protocols for computing and interpreting dynamical indices will be crucial if they are to become reliable components of AI forecasting workflows. Furthermore, linking index-based diagnostics to physical mechanisms — for example, associating high dimension with multiscale interactions or low persistence with imminent regime transitions — will help ensure that AI models are more physically interpretable.

Finally, there is the broader challenge of *hybridization*: blending dynamical diagnostics with physically constrained AI architectures. Purely data-driven models, even when evaluated with dynamical indices, may still generate unphysical outputs. Embedding conservation laws [15], balance relations, or other domain-specific constraints [17] into network architectures, while simultaneously using indices to monitor and enforce dynamical consistency, represents a promising pathway. This hybrid perspective aligns with the emerging paradigm of hybrid physics-AI modeling (e.g., [17, 18]), where models are trained and evaluated for statistical accuracy and for their adherence to the dynamical principles of the systems they aim to predict.

**Concluding remarks**

AI has shown strong promise in forecasting complex dynamical systems, its long-term reliability is limited by the accumulation of errors and the risk of dynamical inconsistency. Embedding dynamical systems theory — through indices such as instantaneous dimension, persistence, and time-lagged recurrence — offers both diagnostic power and avenues for model improvement. Meeting the challenges of scalability, integration, interpretability, and hybridization will be essential to realize the full potential of dynamics-informed AI, and to deliver forecasts that are accurate and dynamically faithful.

**Acknowledgments**
The author acknowledges support from Ministry of Education, Singapore (MOE) Tier 2 grant number T2EP50221-0017: 'Prediction-to-Mitigation with Digital Twins of the Earth's Weather'. We also thank Prof Adriano Gualandi for fruitful discussions.

**References**

[1] Poincaré H 1890 *Acta mathematica* **13** A3–A270

[2] Lorenz E N 1963 *Journal of Atmospheric Sciences* **20** 130–141

[3] Mandelbrot B 1967 *science* **156** 636–638

[4] Scheffer M, Carpenter S, Foley J A, Folke C and Walker B 2001 *Nature* **413** 591–596

[5] Pathak J, Subramanian S, Harrington P, Raja S, Chattopadhyay A, Mardani M, Kurth T, Hall D, Li Z, Azizzadenesheli K, Hassanzadeh P, Kashinath K and Anandkumar A 2022 *arXiv preprint arXiv:2202.11214*

[6] Lam R, Sanchez-Gonzalez A, Willson M, Wirnsberger P, Fortunato M, Alet F, Ravuri S, Ewalds T, Eaton-Rosen Z, Hu W, Merose A, Hoyer S, Holland G, Vinyals O, Stott J, Pritzel A, Mohamed S and Battaglia P 2023 *Science* **382** 1416–1421

[7] Oseledec V I 1968 *Transactions of the Moscow Mathematical Society* **19** 197–231

[8] Nese J M 1989 *Physica D: Nonlinear Phenomena* **35** 237–250

[9] Aurell E, Boffetta G, Crisanti A, Paladin G and Vulpiani A 1997 *Journal of physics A: Mathematical and general* **30** 1

[10] Ding R and Li J 2007 *Physics Letters A* **364** 396–400

[11] Lucarini V, Faranda D, de Freitas J M M, Holland M, Kuna T, Nicol M, Todd M, Vaienti S *et al.* 2016 *Extremes and recurrence in dynamical systems* (John Wiley & Sons)

[12] Faranda D, Messori G and Yiou P 2017 *Scientific reports* **7** 41278

[13] Dong C, Messori G, Faranda D, Gualandi A, Lucarini V and Mengaldo G 2024 *arXiv preprint arXiv:2412.10069*

[14] Dong C, Faranda D, Gualandi A, Lucarini V and Mengaldo G 2025 *Proceedings of the National Academy of Sciences* **122** e2420252122

[15] Karniadakis G E, Kevrekidis I G, Lu L, Perdikaris P, Wang S and Yang L 2021 *Nature Reviews Physics* **3** 422–440

[16] Fang Z and Mengaldo G 2025 *arXiv preprint arXiv:2504.11074*

[17] Wang X, Chen J, Yang J, Adie J, See S, Furtado K, Chen C, Arcomano T, Maulik R, Xue W and Mengaldo G 2025 *arXiv preprint arXiv:2502.13185*

[18] Rasp S, Pritchard M S and Gentine P 2018 *Proceedings of the national academy of sciences* **115** 9684–9689

# On neural methods, normalizing flows and Bayesian inference in application to climate teleconnections

Terence J. O'Kane[1]*

[1]Environment, CSIRO, Battery Point, Australia

*Author to whom any correspondence should be addressed.

**E-mail:** terence.okane@csiro.aug

**Status**

Chen et al. (2018) [1] demonstrated the utility of posing deep neural networks within the framework of dynamical systems theory - specifically bijective mappings between tangent spaces. They showed that, given a sufficient number of layers, one could parameterize the continuous dynamics of hidden states using an ODE specified by a neural network with few restrictions on the model $f$ (i.e., residual or recurrent networks, normalizing flows, etc ) used to build transformations to a given hidden state $\boldsymbol{h}$

$$\frac{d\boldsymbol{h}(t)}{dt} = f(\boldsymbol{h}(t), t, \theta) \qquad (1)$$

where

$$\boldsymbol{h}_{t+1} = \boldsymbol{h}_t + f(\boldsymbol{h}_t, \theta_t), \ \ t \in \{0, \ldots, T\}, \ \ \boldsymbol{h}_t \in \Re^D. \qquad (2)$$

Training continuous-depth networks relies on reverse-mode differentiation (i.e., back propagation) through the ordinary differential equation (ODE) solver to compute the gradients. Optimizing the loss requires computation of the gradients w.r.t. the model parameters $\theta$ and is reliant on evaluation of vector-Jacobian products via auto-differentiation. In this approach time series are represented as trajectories between latent states $\boldsymbol{z}_{t_0}$ i.e., $\boldsymbol{z}_{t_1}, \boldsymbol{z}_{t_2}, \ldots, \boldsymbol{z}_{t_N} = ODESolve(\mathbf{z}_{t_0}, f, \theta_f, t_0, \ldots, t_N)$ given observations at times $t_0, t_1, \ldots, t_N$. Moving from a discrete set of layers to a continuous transformation of the hidden state i.e., discretizing Eq. 2, can be simplified using methods from normalizing flows. Specifically, normalizing flows are applied to transform an observed data distribution into a Gaussian distribution via a sequence of invertible transformations with a computationally tractable Jacobian.

Complimentary dynamical systems approaches have enabled understand coherent atmospheric weather patterns directly from observational data in combination with machine learning algorithms by 1: reduce the dimensionality of observational data using nonstationary clustering [2] or alternate approaches to normalizing flows [3] thereby allowing 2: direct calculation of transversality and hyperbolicity characteristics of local in space and time chaotic attractors associated with the emergence and lifecycle of metastable states of the flow [4, 5].

Subsequent efforts, in particular at the leading operational centers for weather prediction, have largely focused on learning systems of partial differential equations (PDEs). Noting stability issues in global weather simulations arising at the poles when Fourier Neural Operators [6] - defined on the doubly periodic Euclidean domain - Bonev et al. (2023)[7] introduced Spherical Fourier Neural Operators. This approach enables mappings between function spaces on the sphere where using the appropriate spherical harmonic basis i.e., the spectral representation is learned using a global convolution kernel.

**Current and future challenges**

At the present time, spherical Fourier neural operators (SFNOs) form the basic architecture of many operational artificial intelligence-machine learning (AI-ML) forecast systems with temporal forecast horizons spanning days to decades. Many of these systems are directly descended from the original SFNO architecture [7], for example Ai2 Climate Emulator version 2 (ACE2) [8] from the joint collaboration between the Allen Institute for Artificial Intelligence (AI2) and NVIDIA. SFNOs have also been applied to long-term climate emulation [9] and ensemble prediction [10]. A notable additional example of FNO based weather and climate systems is FourCastNet3 [11]. Some other approaches to atmospheric prediction include Graphcast [12] based on graph neural networks (GNNs), and Aardvark Weather [13], both utilizing an "encoder-processor-decoder" configuration.

However, key challenges remain including, but not limited to,

- the development of fully coupled atmosphere-ocean-sea ice models potentially employing the application of efficient graph neural operator methods to learning the solution operator of large-scale partial differential equations with varying geometries i.e. transforming irregular grids to regular latent grids on which Fourier neural operators can be efficiently applied [14].

- emulating future climates for emission pathways not yet observed or simulated beyond those few available forcing scenarios (Representative Concentration Pathways (RCPs) & Shared Socioeconomic Pathways (SSPs)) prescribed in climate projection studies carried out by the international community as part of the Coupled Model Inter-comparison Project (CMIP). There is an active effort to develop methods that, once trained on projection data from dynamical models forced by available emissions trajectories, can interpolate and extraoplate climate variables given arbitrary emissions pathways [15, 16].

- understanding and addressing the tendency for neural methods (transformers, SFNOs, etc) to act as low pass filters producing overly dispersive energy spectra has been a recognised challenge for some time [17]. Recently, some noteable advances in rectifying the inferred spectra of autoregressive atmospheric predictions has been reported [11].

- interpretability - can attention in model weights for given architectures be translated to physical understanding? A related question is - do AI model outputs reliably reproduce the observed dynamics and the causal relationships between the major weather and climate modes of variability in inference? A recent study suggests AI models do produce reliable dynamical properties of the atmospheric flow [18] while Bayesian inference has been shown to be a powerful tool to infer uncertainty estimates of Granger causal rlationships of climate teleconnections [19].

- due to the spatio-temporal scales involved and the relatively sparse observational record, many of the problems in climate science manifest as "small-data" where there is insufficient data available for training. This is particularly evident when considering the ocean. To address this problem novel prediction methods are being developed specifically to handle the dual challenges of nonstationarity and sparsity of data [20]

**Advances in science and technology to meet challenges**
A diverse array of methods and models are being developed at a fast pace to meet the aforementioned challenges. In operational weather prediction AI models are already outperforming dynamical models with sophisticated data assimilation systems for a number of key metrics. In application to longer term climate predictions AI models are able to emulate a limited number of variables but require vast training data sets from dynamical (CMIP) model projections. For climate model emulation, how to train a model without inheriting biases from the model training data is always an issue. Coupling the atmosphere to the ocean is particularly challenging for AI systems given the ocean has complex boundaries, small spatial scales of the mesoscale eddies together with long e-folding times, which is in stark contrast to the dynamics of the atmosphere.

However, it is exciting to see advances are being made in AI weather and climate prediction almost daily with ever increasing resourcing and scientific excellence. This will only continue as mathematicians, data scientists and experts from weather and climate combine forces to tackle the particular challenges posed by anthropogenic warming and climate change.

**Acknowledgments**
TJO acknowledges support from the Commonwealth Scientific and Industrial Organization (CSIRO) and funding support from the National University of Singapore (NUS)

**References**

[1] Chen R T Q, Rubanova Y, Bettencourt J and Duvenaud D 2018 Neural ordinary differential equations *Proceedings of the 32nd International Conference on Neural Information Processing Systems* NIPS'18 (Red Hook, NY, USA: Curran Associates Inc.) p 6572–6583

[2] Horenko I 2010 *Journal of the Atmospheric Sciences* **67** 1559–1574

[3] Guan S, He Z, Ma S and Gao M 2023 *ISA transactions* **143** 231–243

[4] Axelsen A R, O'Kane T J, Quinn C R and Bassom A P 2025 *Journal of Advances in Modeling Earth Systems* **17** e2024MS004834 URL https://agupubs.onlinelibrary.wiley.com/doi/abs/10.1029/2024MS004834

[5] Badza A and Froyland G 2024 *Chaos: An Interdisciplinary Journal of Nonlinear Science* **34** 123153 ISSN 1054-1500 (*Preprint* https://pubs.aip.org/aip/cha/article-pdf/doi/10.1063/5.0225848/20305595/123153_1_5.0225848.pdf) URL https://doi.org/10.1063/5.0225848

[6] Li Z, Kovachki N, Azizzadenesheli K, Liu B, Bhattacharya K, Stuart A and Anandkumar A 2020 *arXiv:2010.08895v3*

[7] Bonev B, Kurth T, Hundt C, Pathak J, Baust M, Kashinath K and Anandkumar A 2023 Spherical fourier neural operators: learning stable dynamics on the sphere *Proceedings of the 40th International Conference on Machine Learning* ICML'23 (JMLR.org)

[8] Watt-Meyer O, Henn B, McGibbon J, Clark S K, Kwa A, Perkins W A, Wu E, Harris L and Bretherton C S 2025 *npj Climate and Atmospheric Science* **8** 205

[9] Guan H, Arcomano T, Chattopadhyay A and Maulik R 2025 *arXiv:2509.02061*

[10] Mahesh A, Collins W D, Bonev B, Brenowitz N, Cohen Y, Elms J, Harrington P, Kashinath K, Kurth T, North J, O'Brien T, Pritchard M, Pruitt D, Risser M, Subramanian S and Willard J 2025 *Geoscientific Model Development* **18** 5575–5603 URL https://gmd.copernicus.org/articles/18/5575/2025/

[11] Bonev B, Kurth T, Mahesh A, Bisson M, Kossaifi J, Kashinath K, Anandkumar A, Collins W D, Pritchard M S and Keller A 2025 *arXiv:2507.12144*

[12] Lam R, Sanchez-Gonzalez A, Willson M, Wirnsberger P, Fortunato M, Alet F, Ravuri S, Ewalds T, Eaton-Rosen Z, Hu W, Merose A, Hoyer S, Holland G, Vinyals O, Stott J, Pritzel A, Mohamed S and Battaglia P 2023 *Science* **382** 1416–1421 URL https://www.science.org/doi/abs/10.1126/science.adi2336

[13] Allen A, Markou S, Tebbutt W, Requeima J, Bruinsma W P, Andersson T R, Herzog M, Lane N D, Chantry M, Hosking J S and Turner R E 2025 *Nature* **641** 1172–1179

[14] Li Z, Kovachki N B, Choy C, Li B, Kossaifi J, Otta S P, Nabian M A, Stadler M, Hundt C, Azizzadenesheli K and Anandkumar A 2023 Geometry-informed neural operator for large-scale 3d pdes *Proceedings of the 37th International Conference on Neural Information Processing Systems* NIPS '23 (Red Hook, NY, USA: Curran Associates Inc.)

[15] Kitsios V, O'Kane T and Newth D 2023 *Commun. Earth Environ.* **4** 355

[16] Womack C B, Giani P, Eastham S D and Selin N E 2025 *J. Adv. Mod. Earth Systems* **17** e2024MS004523

[17] Wang P, Zheng W and Chen T Wang Z 2022 *International Conference on Learning Representations (ICLR)* **arXiv:2203.05962**

[18] Hakim G J and Masanam S 2024 *Artificial intelligence for earth systems* **3** 1–11

[19] O'Kane T J, Harries D and Collier M A 2024 *J. Adv. Mod. Earth Systems* **16** e2023MS004034

[20] Groom M, Bassetti D, Horenko I and O'Kane T J 2024 *Artificial intelligence for earth systems* **3** 1–20

# AI for Weather and Climate Modeling

Gianmarco Mengaldo[1,2,3,*]

[1]Department of Mechanical Engineering, National University of Singapore, Singapore
[2]Department of Mathematics, National University of Singapore, Singapore
[3]Honorary Research Fellow, Department of Aeronautics, Imperial College London, London, United Kingdom
[*]Author to whom any correspondence should be addressed.

**E-mail:** mpegim@nus.edu.sg

**Status**
Artificial intelligence (AI) is reshaping the landscape of weather and climate simulations. Broadly speaking, two modeling paradigms have emerged: (i) *AI-only* modeling, and (ii) *hybrid physics-AI* modeling.

The first modeling paradigm uses AI, i.e., neural networks, trained on large-scale reanalysis datasets, such as ERA5 [1], and equation-based model outputs, such as SHiELD [2]. The training strategy adopted commonly focuses on supervised learning, that is: learning direct input–output mappings (e.g., state at time $t \to$ forecast at time $t + \Delta t$) of the overall weather/climate "system", thereby bypassing the explicit solution of partial differential equations used in traditional weather and climate models. Prominent examples of such models include *FourCastNet* [3], *GraphCast* [4], *Pangu-Weather* [5], and AIFS [6], which have demonstrated remarkable skill in short- to medium-range weather forecasts. For certain benchmarks, these models match or exceed the accuracy of operational numerical weather prediction systems while running orders of magnitude faster, highlighting the computational efficiency of AI methods on graphical processing units (GPUs) and many-core architectures. More recently, self-supervised training strategies were adopted. These are closer to the training strategies used for large language models (LLMs), where models are pretrained to reconstruct masked or withheld spatiotemporal fields before being adapted to downstream weather and/or climate prediction tasks. Examples include *ClimaX* [7], *AURORA* [8], and *ORBIT* [9], which aim to provide more general-purpose *foundation models* for the Earth system. A list of some supervised and self-supervised (i.e., foundation) models is summarized in Table 1, noting that many promising recent developments are not included given the short scope of this work.

The second modeling paradigm uses a combination of physics and AI, whereby AI models, i.e., neural networks, are used within general circulation models (GCMs) or Earth system models (ESMs) to emulate unresolved processes [?], such as convection, cloud microphysics, or turbulence, while the large-scale dynamics continue to be represented by equation-based physical models. To this end, recent works include NeuralGCM [10], a differentiable framework that enables end-to-end training of subgrid emulators alongside large-scale dynamics, and CondensNet [11], a residual neural network with adaptive physical constraints designed to emulate cloud-resolving models. In Table 1, we report some recent hybrid physics-AI models.

These two paradigms, namely *AI-only* and *hybrid physics-AI*, represent complementary directions in the use of AI for weather and climate prediction. AI-only models offer unprecedented speed and scalability, while hybrid physics-AI approaches emphasize physical consistency and may generalize better to unseen scenarios. In practice, computational cost, time-to-solution, and physical fidelity are the key criteria used to evaluate the respective advantages and limitations of each modeling approach, with energy-to-solution also becoming a key consideration. We discuss these aspects next.

**Table 1.** Overview of major AI models for weather and climate, grouped by paradigm: AI-only versus hybrid physics–AI. For AI-only models, we also distinguish between supervised and self-supervised (foundation) training strategies. The list presented is not meant to be representative of all recent promising works in this space.

| AI-only Models | Year | Architecture; Data | Target application |
|---|---|---|---|
| **Supervised models** | | | |
| FourCastNet | 2022 | Spectral neural operator; ERA5 | Medium-range forecasting [3]. |
| GraphCast | 2023 | Transformer + graph neural network; ERA5 | Medium-range forecasting [4]. |
| Pangu-Weather | 2023 | 3D convolutional neural network; ERA5 | Medium-range forecasting [5]. |
| AIFS | 2024 | Transformer; ERA5 | ECMWF prototype for medium-range forecasting [6]. |
| FuXi-S2S | 2024 | Transformer + VAE style structure; ERA5 | Subseasonal-to-seasonal forecasting [12]. |
| GenCast | 2024 | Conditional diffusion with graph transformer; ERA5 | Probabilistic medium-range ensemble forecasting [13]. |
| ACE2 | 2025 | Residual convolutional neural network; ERA5 + SHiELD | Subseasonal to decadal atmospheric variability [14]. |
| Aardvark Weather | 2025 | Neural process model; multi-source observations (L1B/L1C, in situ) + reanalysis (training) | End-to-end medium-range global + local forecasts [15]. |
| **Self-supervised (foundation models)** | | | |
| ClimaX | 2023 | Masked autoencoder; ERA5 + CMIP6 | Foundation model; multi-task learning [7]. |
| ORBIT | 2024 | Transformer; diverse multi-source datasets | General-purpose Earth system foundation model [9]. |
| AURORA | 2025 | Transformer; diverse multi-source datasets | Foundation model for atmosphere; forecasting, downscaling, emulation [8]. |
| **Hybrid Physics–AI Models** | **Year** | **Architecture / Data** | **Target application** |
| NNCAM | 2022 | ResNet embedded in CAM; SPCAM/CRM | Climate simulation [16]. |
| NeuralGCM | 2023 | Fully connected feed-forward networks embedded in simple differentiable GCM; reanalysis and high-resolution simulations | Climate simulation [10]. |
| CondensNet | 2025 | ResNet with physical constraints embedded in CAM; SPCAM/CRM | Climate simulation [11]. |

**Challenges and required advances**

Both AI-only and hybrid physics-AI approaches present important scientific, technical, and methodological challenges whose resolution will broaden their role and reliability, particularly for long-term climate applications. Addressing these challenges defines key research directions for the coming years. In Figure 1, we summarize the key concepts discussed in this section, providing a workflow comparison between AI-only and hybrid physics-AI modeling.

For AI-only models, the most immediate challenges concern their ability to generalize to unseen scenarios, particularly under climate change forcing, and to ensure physical fidelity in the absence of explicit governing equations. While models such as FourCastNet [3], GraphCast [4], and Pangu-Weather [5] achieve excellent skill at short to medium forecast horizons, their robustness under unseen regimes (e.g., unprecedented weather extremes) is less obvious. An alternative avenue to potentially improve generalizability has emerged with the development of foundation models such as ClimaX [7], AURORA [8], and ORBIT [9]. By adopting self-supervised training strategies inspired by LLMs, these approaches aim to learn flexible, multi-scale representations of the Earth system. In principle, such representations could enable models to compose future unseen scenarios from smaller building blocks learned during training, thereby extending their applicability beyond the regimes directly represented in the data used for training them. Yet, robust extrapolation under strong climate forcing remains an open debate, particularly in capturing rare extremes and faithfully reproducing physical processes. In addition to generalizability, interpretability remains an important challenge for AI-only models. Although these models can achieve competitive skill in certain benchmarks compared to their equation-based counterpart, they typically function as black boxes, making it difficult to trace why specific results are obtained. Indeed, these models may produce unphysical results – to make an analogy with LLMs, they may yield *physical hallucinations*. Such unphysical results often arise because the models are optimized to reproduce statistical relationships present in historical data, without explicit grounding in conservation laws and physics (albeit some models are introducing constraints to respect some physical principles).

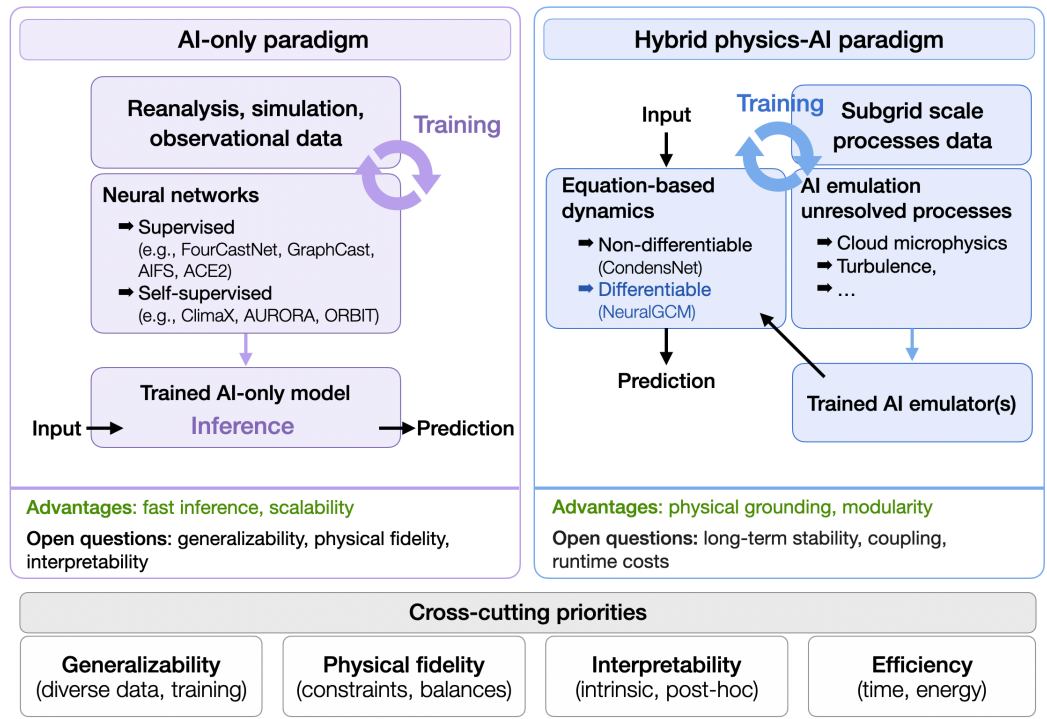

**Figure 1.** Comparison of AI-only and hybrid physics-AI approaches for weather and climate prediction. AI-only models learn direct input–output mappings from data, offering fast inference but facing challenges in generalizability, physical fidelity, and interpretability. Hybrid approaches embed AI emulators of unresolved processes within equation-based dynamics, providing physical grounding and modularity, while presenting challenges in stability, coupling, and runtime costs. Cross-cutting priorities include generalizability, physical fidelity, interpretability, and efficiency.

As a result, AI-only models may rely on correlations that hold in the training regime but do not necessarily correspond to causal physical mechanisms, which can reduce reliability when these correlations no longer apply. This limitation is particularly relevant for high-impact applications, such as forecasting weather extremes or assessing climate change scenarios. In these cases, understanding possible failure modes and/or biases in the models is crucial. Advances in interpretability, encompassing both intrinsically interpretable models [17] and post-hoc interpretability methods [18], may prove essential to better understand and correct failure modes and physical hallucinations. Hybrid physics-AI models address some of the limitations of AI-only approaches, but they introduce a distinct set of challenges of their own. For instance, hybrid physics-AI models offer advantages in terms of generalizability, by grounding simulations in well-established physics-based equations (within GCMs or ESMs) for the dynamics and confining the AI emulators to specific unresolved physical processes, such as cloud microphysics and turbulence [19]. However, they introduce non-trivial coupling challenges: interactions between the AI emulators and the equation-based dynamics can lead to mismatches in certain key variables, which may accumulate over time and manifest as drifts, biases, or instabilities in long-term simulations; the latter leading to simulation failure. Indeed, numerical stability and physical consistency at the interface between AI emulators and the equation-based dynamics remains a central research focus in the context of hybrid approaches. To this end, recent developments, such as CondensNet [11], have shown that embedding adaptive physical constraints can improve stability, while differentiable hybrid frameworks, such as NeuralGCM [10], make the entire model – dynamical core and AI-learned parametrizations – trainable end to end. More specifically, differentiability allows gradients of long-horizon objective functions to be backpropagated through time, so that AI emulators of unresolved processes and their coupling to resolved equation-based dynamics are optimized jointly rather than in isolation. This enables a principled route to align the AI-learned unresolved physics with the needs of the equation-based dynamics across the full integration window, improving physical fidelity, such as the conservations of key variables, and potentially leading to better compatibility of AI emulation–dynamics coupling. Indeed, the two together, namely adaptive physical constraints plus differentiable equation-based dynamics provide a promising way forward. Yet, hybrid physics-AI models remain computationally slower at inference than AI-only approaches, since they must still advance the equation-based dynamical core, which can involve potentially small integration time steps and computationally costly operations, such as the inversion of large sparse matrices. While these solver costs dominate the runtime of every simulation, it is worth noting that the overall expense of training large AI foundation models can be significantly higher; the advantage of AI-only approaches lies in their ability to offset this cost at inference. AI-only models, once trained, bypass the timestepping required by equation-based models, and deliver forecasts through a single forward pass (e.g., 6 hours forecast in one forward pass). To put things into perspective, the Integrated Forecasting System of ECMWF [20] at 9km operational resolution has a timestep of 450 seconds; this means that to provide a 6 hours forecast it needs to perform 48 time steps versus the single forward pass of an AI-only model.

The trade-offs just highlighted show that the two paradigms are likely complementary, especially in terms of target application, with AI-only models currently providing competitive result for short- to medium-range weather forecasts, and hybrid physics-AI models potentially providing a pathway for longer-term climate simulations. Indeed, in the future, the two paradigms may ultimately merge. We briefly discuss these aspects next.

**Concluding remarks**

AI-only and hybrid physics-AI approaches have advanced rapidly from proof-of-concept to promising components of the weather–climate toolbox. Rather than competing, they can and should be seen as complementary tools, potentially converging to a common path: AI-only models deliver remarkable speed and scalability for short- to medium-range prediction, while hybrid methods leverage equation-based dynamics to enhance physical fidelity, especially for long-term simulations.

A key question for both paradigms is whether they can generalize well to unseen scenarios, such as unprecedented extremes or climate change forcing.

AI-only models are statistical tools that try to identify patterns in high-dimensional data. If we use Lorenz's paradigm [21], who in the 1950s distinguished between statistical and dynamical forecasting, AI-only models may be effective "when the future resembles the past". "Future resembling the past", however, could have different flavors – if we use supervised learning strategies, it applies directly; if we use self-supervised training strategies, the composition of previous states may lead to an unseen future and partly extrapolate to future (unseen) scenarios, albeit with the risk of physical hallucinations.

Hybrid physics-AI models could be seen as a form of dynamical forecasting tools within Lorenz's framework. They keep a set of governing equations for the large-scale dynamics, while confining AI emulation to unresolved processes. The emulation of unresolved processes via AI encounters the same issues as AI-only models. Yet, the hope is that by learning specific unresolved physical processes (instead of the whole climate system as done in AI-only models), AI emulators of unresolved processes are able to capture the underlying physics rather than just correlations that may break down under e.g., climate forcing. This could help generalize the description of unresolved processes to climate-system scenarios that were unseen. The addition of differentiability of the equation-based dynamics being put forward in hybrid physics-AI modeling pushes these methods towards a full end-to-end learning strategy. This, in turn, makes hybrid methods closer to AI-only approaches, whereby the overall learning pipeline is learned via AI. However, hybrid approaches remain modularly anchored in physics by construction – the dynamics is still represented via a set of equation eventually corrected via end-to-end learning, and each unresolved process learned separately, thereby keeping the modularity of traditional GCMs and ESMs.

Ultimately, the two paradigms may converge into a single route, where marrying physics and AI will be the primary endeavor. Indeed, AI-only methods will likely require physical constraints to increase physical fidelity and generalizability, while hybrid approaches would need to improve their computational performance, limited by the timestepping of the equation-based dynamics. Progress in both directions will depend on the availability of diverse training data and on advances in interpretability, whether through intrinsically transparent models or post-hoc diagnostics that expose and correct failure modes. The field will also benefit from open benchmarks across weather and climate timescales, ensuring reproducibility through shared data, code, and trained weights.

Sustaining such progress, however, requires equal attention to computing efficiency. In this context, hardware–algorithm co-design, mixed-precision training, and model and data compression can help reduce the carbon footprint of large-scale training and achieve green(er) computing.

**Acknowledgments**
The author acknowledges support from Ministry of Education, Singapore (MOE) Tier 2 grant number T2EP50221-0017: 'Prediction-to-Mitigation with Digital Twins of the Earth's Weather'.

**References**

[1] Hersbach H, Bell B, Berrisford P, Hirahara S, Horányi A, Muñoz-Sabater J, Nicolas J, Peubey C, Radu R, Schepers D, Simmons A, Soci C, Abdalla S, Abellan X, Balsamo G, Bechtold P, Biavati G, Bidlot J R, Bonavita M, Chiara G, Dahlgren P, Dee D, Diamantakis M, Dragani R, Flemming J, Forbes R, Fuentes M, Geer A, Haimberger L, Healy S, Hogan R J, Hólm E, Janisková M, Keeley S, Laloyaux P, Lopez P, Lupu C, Radnoti G, de Rosnay P, Rozum I, Vamborg F, Villaume S and Thépaut J N 2020 *Quarterly Journal of the Royal Meteorological Society* **146** 1999–2049

[2] Harris L, Zhou L, Lin S J, Chen J H, Chen X, Gao K, Morin M, Rees S, Sun Y, Tong M, Xiang B, Bender M, Benson R, Cheng K Y, Clark S, Elbert O D, Hazelton A, Huff J J, Kaltenbaugh A, Liang Z, Marchok T, Shin H H and Stern W 2020 *Journal of Advances in Modeling Earth Systems* **12** e2020MS002223

[3] Pathak J, Subramanian S, Harrington P, Raja S, Chattopadhyay A, Mardani M, Kurth T, Hall D, Li Z, Azizzadenesheli K, Hassanzadeh P, Kashinath K and Anandkumar A 2022 *arXiv preprint arXiv:2202.11214*

[4] Lam R, Sanchez-Gonzalez A, Willson M, Wirnsberger P, Fortunato M, Alet F, Ravuri S, Ewalds T, Eaton-Rosen Z, Hu W, Merose A, Hoyer S, Holland G, Vinyals O, Stott J, Pritzel A, Mohamed S and Battaglia P 2023 *Science* **382** 1416–1421

[5] Bi K, Xie L, Zhang H, Chen X, Gu X and Tian Q 2023 *Nature* **619** 533–538

[6] Lang S, Alexe M, Chantry M, Dramsch J, Pinault F, Raoult B, Clare M C A, Lessig C, Maier-Gerber M, Magnusson L, Ben Bouallègue Z, Prieto Nemesio A, Dueben P D, Brown A, Pappenberger F and Rabier F 2024 *arXiv preprint arXiv:2406.01465*

[7] Nguyen T, Brandstetter J, Kapoor A, Gupta J K and Grover A 2023 *arXiv preprint arXiv:2301.10343*

[8] Bodnar C, Bruinsma W P, Lucic A, Stanley M, Allen A, Brandstetter J, Garvan P, Riechert M, Weyn J A, Dong H, Gupta J K, Thambiratnam K, Archibald A T, Wu C C, Heider E, Welling M, Turner R E and Perdikaris P 2025 *Nature* **641** 1180–1187

[9] Wang X, Liu S, Tsaris A, Choi J Y, Aji A M, Fan M, Zhang W, Yin J, Ashfaq M, Lu D and Balaprakash P 2024 Orbit: Oak ridge base foundation model for earth system predictability *SC24: International Conference for High Performance Computing, Networking, Storage and Analysis* (IEEE) pp 1–11

[10] Kochkov D, Yuval J, Langmore I, Norgaard P, Smith J, Mooers G, Klöwer M, Lottes J, Rasp S, Düben P, Hatfield S, Battaglia P, Sanchez-Gonzalez A, Willson M, Brenner M P and Hoyer S 2024 *Nature* **632** 1060–1066

[11] Wang X, Chen J, Yang J, Adie J, See S, Furtado K, Chen C, Arcomano T, Maulik R, Xue W and Mengaldo G 2025 *arXiv preprint arXiv:2502.13185*

[12] Chen L, Zhong X, Li H, Wu J, Lu B, Chen D, Xie S P, Wu L, Chao Q, Lin C, Hu Z and Qi Y 2024 *Nature Communications* **15** 6425

[13] Price I, Sanchez-Gonzalez A, Alet F, Andersson T R, El-Kadi A, Masters D, Ewalds T, Stott J, Mohamed S, Battaglia P, Lam R and Willson M 2025 *Nature* **637** 84–90

[14] Watt-Meyer O, Henn B, McGibbon J, Clark S K, Kwa A, Perkins W A, Wu E, Harris L and Bretherton C S 2025 *npj Climate and Atmospheric Science* **8** 205

[15] Allen A, Markou S, Tebbutt W, Requeima J, Bruinsma W P, Andersson T R, Herzog M, Lane N D, Chantry M, Hosking J S and Turner R E 2025 *Nature* **641** 1172–1179

[16] Wang X, Han Y, Xue W, Yang G and Zhang G J 2022 *Geoscientific Model Development* **15** 3923–3940

[17] Turbé H, Bjelogrlic M, Mengaldo G and Lovis C 2025 *arXiv preprint arXiv:2502.19577*

[18] Turbé H, Bjelogrlic M, Lovis C and Mengaldo G 2023 *Nature Machine Intelligence* **5** 250–260

[19] Bonan G B and Doney S C 2018 *Science* **359** eaam8328

[20] Kühnlein C, Deconinck W, Klein R, Malardel S, Piotrowski Z P, Smolarkiewicz P K, Szmelter J and Wedi N P 2019 *Geoscientific Model Development* **12** 651–676

[21] Lorenz E N 1956 Empirical orthogonal functions and statistical weather prediction Tech. Rep. Scientific Report No. 1 Massachusetts Institute of Technology, Department of Meteorology, Statistical Forecasting Project Cambridge, MA

# AI for useful forest monitoring with remote sensing

**Authors**:
Emily R. Lines[1*], Matthew J. Allen[1] and Amandine E. Debus[1]
**Affiliation**:
[1]Department of Geography, University of Cambridge, Downing Site, Cambridge CB23EN, United Kingdom.
**Communication**:
*corresponding author, erl27@cam.ac.uk

*Status*

Forests are critical ecosystems in the mitigation of, and adaptation to, the twin threats of climate change and biodiversity loss, but traditional forest monitoring data has restricted spatiotemporal coverage and often comprises only simple measurements (e.g. tree height, trunk diameter, species, and forest health). Such data are expensive to collect and frequently rare in the most carbon and biodiversity rich ecosystems. Data are also not always openly available, and not always accompanied with geospatial location information. High resolution remote sensing data, including ground, airborne and spaceborne 2D imagery and 3D LiDAR have revolutionised our ability to capture and analyse forest structure and change. Such technologies have been shown to be capable of generating improved estimates of key tree structural properties, including biomass, leaf area, height, crown shape and canopy cover (Lines et al., 2022). High resolution remote sensing data also offer direct and proxy measurements for a wide variety of aspects of forest ecosystem structure, function and diversity, habitat provision, and tree health (Lines et al., 2022). Despite their demonstrated value, the use of high resolution remote sensing data to monitor forests presents significant challenges. Data are complex, very large, and require complex processing before informative environmental information can be extracted. Further, 3D datasets such as are generated by LiDAR sensors are frequently noisy, have highly spatially variable point densities and suffer vertical occlusion determined by sensor platform location. Tasks such as individual tree and material segmentation, habitat mapping and land use classification, are extremely time intensive when performed manually, and processing tools developed to automate tasks are frequently highly context and sensor specific. Artificial intelligence (AI) provides an attractive alternative.

From satellite aerial crown segmentation and species monitoring to 3D virtual forests, environmentally-minded computer scientists and computationally-minded environmental scientists are embracing AI to tackle data volume and create new monitoring methods in forests. Successes have been found in individual tree (instance) and material (semantic) segmentation, species identification, and land use and deforestation monitoring (e.g. Brandt et al., 2020; Debus et al., 2025; Krisanski et al., 2021; Puliti et al., 2024). But the field is held back by a scarcity of independently ground-verified datasets and subsequent poor quality training and validation datasets, as well as a lack of universal diverse benchmarking datasets. These challenges are compounded by a current focus of chasing performance metrics, at the expense of rigorous tests of *whether* and *where* the task at hand can be achieved.

*Current and future challenges*

The performance of AI tools to extract information about forests must be verified by measurements taken within forests. However, many AI tools have been built using labels generated by manual annotation labelling 2D and 3D imagery on computers, and so are tested against what a human *thinks* the information is, rather than any grounded truth.

A problematic example of this arises from tree crown identification and delineation in 2D and 3D aerial data. The ability to count and measure trees from the air is a desirable one, allowing the calculation of carbon storage and sequestration potential, and the assessment of forest age, health and habitat value. High profile initiatives have claimed success for this task using AI, with precision and F1 scores ranging from 0.6-0.9 (Hao et al., 2021; Weinstein et al., 2020). But tools are often trained and validated on data labels manually generated by visual inspection of imagery, with minimal or no ground verification. Ours and others' work using ground verified crown segmentations, including using high fidelity terrestrial laser scanning data, has demonstrated that such performance claims are substantially inflated (Allen et al., 2025; Cao et al., 2023). Subcanopy trees are impossible to see in aerial data, and manual labels are likely biased by merging smaller crowns to create incorrect larger trees, with, at best, between 10-36% of canopy trees are well identified in

2D aerial imagery. Performance drops further if accurate crown delineation is required, as delineation success is often judged using intersection over union values as low as 0.5.

Reliance on visually assessed labels is also widespread in the assessment of task success by geospatial foundation models, where human-annotated evaluation datasets are often used to identify land cover or ecosystem classes in satellite imagery (e.g. Kerner et al., 2024). Annotators covering large, diverse regions are unlikely to have sufficient context specific knowledge to accurately interpret the imagery. And there is good reason to believe that this approach will generate poor datasets, since the same land cover class can produce widely different spatio-temporal-spectral patterns depending on local socio-economic and ecological contexts (Dalimier et al., 2022). This approach further reinforces existing inequalities, with training data often taken from well-studied regions in the Global North or is generated by annotators situated within the Global North, while expertise and data from the Global South, where the monitoring need is most urgent, remain scarce or ignored (Nakalembe et al., 2024).

### *Advances in science and technology to meet challenges*

The discipline must tackle reliability and usefulness to generate meaningful information and impact. However, current directions of travel may reproduce blind spots and geographical biases, offering outputs that are too generic to be useful for policy and which underperform compared to task specific models (Rolf et al., 2024). Scalability and transferability to diverse forest types, and usability and trustworthiness for end users, remain underemphasised in both model building and results reporting. Such tasks are challenging and require re-evaluation of our frameworks and approaches.

For many AI tasks in forest monitoring, sample size for model training is a major bottleneck, and human-in-the-loop approaches have promise to leverage expertise without overwhelming manual requirements. Active learning presents as a promising solution; here, the model itself identifies the most informative or uncertain samples for annotation, directing human labelling to where it has the greatest impact, and has been demonstrated to improve satellite deforestation monitoring in data poor regions (Masolele et al., 2024). Iterative labelling, where model prediction labels are accepted or rejected by humans during training, and the model refined accordingly, shows promise in applied computer vision tasks (e.g. Zhang et al., 2023). For 3D forest monitoring there are significant practical barriers to labelling datasets and training models in parallel due to different software norms. New efforts such as TreeAIBox (Xi and Degenhardt, 2025) demonstrate that unified platforms for data labelling and machine learning model training and inference are possible. But it is not always true that suitable data do not exist; too often AI modellers are disconnected from their applied contexts. Even in so-called data poor regions, collaboration with local data collectors and owners can generate comprehensive, ground-truthed data to train high performing models (e.g. Debus et al., 2025).

Despite the clear value of remote sensing data, there is a long-standing, persistent gap between technological advances and actual impact (Fritz et al., 2024), and the complexity and opacity of AI methods and foundation models risk widening this. Model evaluation on standard accuracy-based metrics, and lack of uncertainty and confidence estimation, remain prevalent (Singh et al., 2024), worsening both understanding and trust of outputs. Key to any model's value is its demonstrated *usefulness*, and evaluation based on discipline specific metrics and downstream task performance (e.g. Owen et al., 2025) can offer interpretable performance guidelines to help the user understand where, and where not, to apply it. Communication of model performance via confidence scoring can further improve interpretability, for example by calibrating model performance to reflect true likelihood of correctness (e.g. Guo et al., 2017).

### *Concluding remarks*

The value of AI to unlock the power of remote sensing data for forest monitoring is clear. As data volumes from sensors and sensor networks from the ground to satellites explodes, AI methods are becoming invaluable for extracting information. High quality remote sensing data with which to train, test and validate AI models remains a bottleneck (Schmitt et al., 2023). Poor quality 'benchmark' datasets and reliance on visually interpreted labels result in overpromising for downstream tasks, with real risks of providing incorrect or even harmful information. Yet, better ways forward are available and are effective, including efficient model tuning through human-in-the-loop approaches, and collating labels for historically data poor regions. These depend on expert knowledge and so must be approached with a spirit of equal and open collaboration, as well as understanding and support for the substantial effort and funds often required to generate ground data. For all their promise, remote sensing data in forests have well known information limitations that no data processing method cannot overcome, and – having demonstrated the potential of AI methods – the community must pivot to demonstrating their usefulness through rigorous evaluation and transparent communication.

**Funding acknowledgements**
E.R.L. was supported by the UKRI Future Leaders Fellowship programme (MR/Y033981/1). M.J.A. was supported by the UKRI Centre for Doctoral Training in Application of Artificial Intelligence to the study of Environmental Risks (EP/S022961/1). A.E.D. was supported by the NERC C-CLEAR doctoral training programme (NE/S007164/1).

Allen, M.J., Owen, H.J.F., Grieve, S.W.D., Lines, E.R., 2025. Manual Labelling Artificially Inflates Deep Learning-Based Segmentation Performance on RGB Images of Closed Canopy: Validation Using TLS. https://doi.org/10.48550/arXiv.2503.14273

Brandt, M., Tucker, C.J., Kariryaa, A., Rasmussen, K., Abel, C., Small, J., Chave, J., Rasmussen, L.V., Hiernaux, P., Diouf, A.A., Kergoat, L., Mertz, O., Igel, C., Gieseke, F., Schöning, J., Li, S., Melocik, K., Meyer, J., Sinno, S., Romero, E., Glennie, E., Montagu, A., Dendoncker, M., Fensholt, R., 2020. An unexpectedly large count of trees in the West African Sahara and Sahel. Nature 587, 78–82. https://doi.org/10.1038/s41586-020-2824-5

Cao, Y., Ball, J.G.C., Coomes, D.A., Steinmeier, L., Knapp, N., Wilkes, P., Disney, M., Calders, K., Burt, A., Lin, Y., Jackson, T.D., 2023. Benchmarking airborne laser scanning tree segmentation algorithms in broadleaf forests shows high accuracy only for canopy trees. Int. J. Appl. Earth Obs. Geoinformation 123, 103490. https://doi.org/10.1016/j.jag.2023.103490

Dalimier, J., Achard, F., Delhez, B., Desclée, B., Bourgoin, C., Eva, H., Gourlet-Fleury, S., Hansen, M., Kibambe, J.-P., Mortier, F., Ploton, P., Réjou-Méchain, M., Vancutsem, C., Jungers, Q., Defourny, P., 2022. Distribution of forest types and changes in their classification. CIFOR-ICRAF. URL https://www.cifor-icraf.org/knowledge/publication/8701/ (accessed 9.29.25).

Debus, A., Beauchamp, E., Kamga, J., Verhegghen, A., Zébazé, C., Lines, E.R., 2025. Evaluating satellite data and deep learning for identifying direct deforestation drivers in Cameroon. Remote Sens. Appl. Soc. Environ. 39, 101653. https://doi.org/10.1016/j.rsase.2025.101653

Fritz, S., Milenkovic, M., Georgieva, I., Guzman, K.P., 2024. From Earth Observation data to policy impact. IIASA - Int. Inst. Appl. Syst. Anal. URL http://iiasa.ac.at/blog/mar-2024/from-earth-observation-data-to-policy-impact (accessed 10.6.25).

Guo, C., Pleiss, G., Sun, Y., Weinberger, K.Q., 2017. On Calibration of Modern Neural Networks. https://doi.org/10.48550/arXiv.1706.04599

Hao, Z., Lin, L., Post, C.J., Mikhailova, E.A., Li, M., Chen, Y., Yu, K., Liu, J., 2021. Automated tree-crown and height detection in a young forest plantation using mask region-based convolutional neural network (Mask R-CNN). ISPRS J. Photogramm. Remote Sens. 178, 112–123. https://doi.org/10.1016/j.isprsjprs.2021.06.003

Kerner, H., Nakalembe, C., Yang, A., Zvonkov, I., McWeeny, R., Tseng, G., Becker-Reshef, I., 2024. How accurate are existing land cover maps for agriculture in Sub-Saharan Africa? Sci. Data 11, 486. https://doi.org/10.1038/s41597-024-03306-z

Krisanski, S., Taskhiri, M.S., Gonzalez Aracil, S., Herries, D., Turner, P., 2021. Sensor Agnostic Semantic Segmentation of Structurally Diverse and Complex Forest Point Clouds Using Deep Learning. Remote Sens. 13, 1413. https://doi.org/10.3390/rs13081413

Lines, E.R., Fischer, F.J., Owen, H.J.F., Jucker, T., 2022. The shape of trees: Reimagining forest ecology in three dimensions with remote sensing. J. Ecol. 110, 1730–1745. https://doi.org/10.1111/1365-2745.13944

Masolele, R.N., Marcos, D., De Sy, V., Abu, I.-O., Verbesselt, J., Reiche, J., Herold, M., 2024. Mapping the diversity of land uses following deforestation across Africa. Sci. Rep. 14, 1681. https://doi.org/10.1038/s41598-024-52138-9

Nakalembe, C., Devereux, T., Ginsburg, A., 2024. Whose Priorities? Examining Inequities in Earth Observation Advancements Across Africa. Perspect. Earth Space Sci. 5, e2023CN000220. https://doi.org/10.1029/2023CN000220

Owen, H.J.F., Allen, M.J.A., Grieve, S.W.D., Wilkes, P., Lines, E.R., 2025. PointsToWood: A deep learning framework for complete canopy leaf-wood segmentation of TLS data across diverse European forests. https://doi.org/10.48550/arXiv.2503.04420

Puliti, S., Lines, E.R., Müllerová, J., Frey, J., Schindler, Z., Straker, A., Allen, M.J., Winiwarter, L., Rehush, N., Hristova, H., Murray, B., Calders, K., Terryn, L., Coops, N., Höfle, B., Junttila, S., Krůček, M., Krok, G., Král, K., Levick, S.R., Luck, L., Missarov, A., Mokroš, M., Owen, H.J.F., Stereńczak, K., Pitkänen, T.P., Puletti, N., Saarinen, N., Hopkinson, C., Torresan, C., Tomelleri, E., Weiser, H., Astrup, R., 2024. Benchmarking tree species classification from proximally-sensed laser scanning data: introducing the FOR-species20K dataset. https://doi.org/10.48550/arXiv.2408.06507

Rolf, E., Gordon, L., Tambe, M., Davies, A., 2024. Contrasting local and global modeling with machine learning and satellite data: A case study estimating tree canopy height in African savannas. https://doi.org/10.48550/arXiv.2411.14354

Schmitt, M., Ahmadi, S.A., Xu, Y., Taşkin, G., Verma, U., Sica, F., Hänsch, R., 2023. There Are No Data Like More Data: Datasets for deep learning in Earth observation. IEEE Geosci. Remote Sens. Mag. 11, 63–97. https://doi.org/10.1109/MGRS.2023.3293459

Singh, G., Moncrieff, G., Venter, Z., Cawse-Nicholson, K., Slingsby, J., Robinson, T.B., 2024. Uncertainty quantification for probabilistic machine learning in earth observation using conformal prediction. Sci. Rep. 14, 16166. https://doi.org/10.1038/s41598-024-65954-w

Weinstein, B.G., Marconi, S., Aubry-Kientz, M., Vincent, G., Senyondo, H., White, E.P., 2020. DeepForest: A Python package for RGB deep learning tree crown delineation. Methods Ecol. Evol. 11, 1743–1751. https://doi.org/10.1111/2041-210X.13472

Xi, Z., Degenhardt, D., 2025. A new unified framework for supervised 3D crown segmentation (TreeisoNet) using deep neural networks across airborne, UAV-borne, and terrestrial laser scans. ISPRS Open J. Photogramm. Remote Sens. 15, 100083. https://doi.org/10.1016/j.ophoto.2025.100083

Zhang, Z., Kaveti, P., Singh, H., Powell, A., Fruh, E., Clarke, M.E., 2023. An iterative labeling method for annotating marine life imagery. Front. Mar. Sci. 10. https://doi.org/10.3389/fmars.2023.1094190

# Cities of silos: Urban AI barriers and opportunities

Clayton Miller[1,2]

[1]College of Integrative Studies, Singapore Management University, Singapore
[2]Urban Institute, Singapore Management University, Singapore
*Author to whom any correspondence should be addressed.

E-mail: cmiller@smu.edu.sg

**Status**
The vision of a *smart city* has been evolving for decades, aiming to strike a balance between conserving resources, such as energy and materials, and satisfying human needs. The explosion of digitization and artificial intelligence (AI) capabilities, accompanied by the continuous reduction in the cost of sensors, imaging equipment, data storage, computing capabilities, and human-generated information from smart devices, is fueling the potential for this field (1). Geospatial data sharing and development by researchers, technology companies, and even average citizens have created a renaissance of potential in the urban-scale mapping community, with large open data sets becoming mainstream and crowdsourcing becoming ubiquitous (2; 3). The concept of a *digital twins* of cities has exploded, especially due to the proliferation of geospatial data. The definition of digital twins has been solidifying around an interactive model that has a real-time link to reality towards continuous optimization (4).

However, despite the rapid development of urban digital twins, there is a lack of progress towards practical utilization of these data beyond traditional mapping, wayfinding, and urban context characterization. The fields of district and building design, construction optimization and automation, and real-time support for human decision-making for both citizens and public services providers have been underdeveloped due to impending challenges. One poignant example is building system optimization where market penetration is low despite decades of innovation (5). Several categories of such challenges (and potential solutions) exist that relate to the fragmentation, inaccessibility, and privacy concerns of potential model training data. Many of these concepts are summarized by the descriptor *data silos*.

**Current and future challenges**
Data silos in the urban context impede the ability of advanced AI to capture context-specific and actionable insights. *Silos* exist due to the spatial, technological, and organizational barriers in which most human-generated data are created. Several challenges for Urban AI related to these silos include:

- *Disparate data-generating activities mostly within proprietary systems* – Beyond geospatial data, which is often shared openly in the public domain, a majority of data generated in cities are created and contained within proprietary systems (6). Examples of these systems are sensor networks related to transportation volume, energy consumption, image and video data collection, and air and water quality. The urban context differs in this way from the virtual, internet-based realm, where most AI-driven innovations have occurred.

- *Collection of tangible prediction objective labels* – For prediction-driven AI to be useful, models need to be developed using training information that includes objective labels. An example of such labels could be data related to the tendency of urban occupants to feel uncomfortable with the levels of heat in urban spaces (7). Simply measuring temperature and predicting thermal perception based on thresholds is not AI.

- *Privacy and security of sensitive and personal data* – Protection of individuals' data is essential to reduce the risk of specific harm to a person due to malicious actors using those data to target them for scams, attacks, or other types of crimes (8). Therefore, the use of spatial or temporal data from humans must be handled carefully to minimize these risks. This scenario creates several barriers to accessing and using personal data due to policy and permission-based restrictions.

- *Tactful capture of the time and attention of stakeholders* – AI methods and tools can be developed, but without the dedicated time and attention of decision makers to use them, there will be no value added. Understanding how predictive or generative tools can be utilized in workflows and processes in various industries, including design, construction, and city and building management, is a significant unsolved challenge (9).

**Advances in science and technology to meet challenges**

Potential solutions to the challenges outlined must differ from the conventional AI-driven innovations that have been pioneered in online, internet-based text and image environments. Several key innovations in the technology pipelines could improve these challenges, including:

- *Advancements in data fusion modeling, storage, and retrieval* – The urban context contains numerous disparate silos, different formats of data contained in various systems. Innovations focused on the low-touch fusion of these data sources with minimal expert interaction would be transformational (10). An example of this type of technology is the use of contextual mapping between urban data and user perspectives to create enriched geospatial data that combine information from both sources (11). Another critical innovation towards solving this challenge is the development of data exchange schemas that facilitate the scalability of mapping raw data sources to analytics and modeling platforms (12).

- *Data liberation, sharing incentivization, crowdsourcing, and open data management* – Innovations focused on the generation of scalable crowd-sourced data from humans in the urban context can drive the development of tagged data with prediction objectives (13). Incentives for individuals and companies to provide these types of data for the purpose of objectives that support the general public's well-being would be a game-changer. Innovation in policy, business model development, and behavioral science would push technological innovation even further (14; 15).

- *Addressing privacy with synthetic data and agentic modeling techniques* – When the detailed personal information of an individual is not possible to share, several emerging model-driven techniques preserve privacy through surrogate modeling or synthetic population analysis. An example of this kind of analysis is the use of district-scale Wi-Fi data to train agent-based modeling approaches that emulate the use of a university campus in different ways according to flexible utilization strategies (16; 17).

- *Human-in-the-Loop generative AI for design and planning* – Urban AI models will not penetrate practical implementation and use without innovation related to interaction with human stakeholders, such as architects and urban planners. The significant capital and cultural impact of the design of cities can probably never be fully automated and thus must work to interact and support the design process (18).

- *Social, cultural, and trust-building with citizens and decision-makers* – Beyond the design phase of cities and buildings, there are significant opportunities for Urban AI to influence the decisions of everyday individuals. The field of human-computer interaction is beginning to be expanded through the interface with the *building* and *city* (19).

**Concluding remarks**

Cities and buildings are not the only field that suffers from the data silo effect. Medicine, public health, finance, and education are examples of fields that have struggled with similar issues and have successfully utilized technology, policy, and incentives to drive progress (20). The urban context can learn from these experiences and technological methods. Addressing these silo-related issues can drive the adoption and practical utilization of Urban AI techniques and tools.

**References**

[1] Lai Y and Zhao H 2025 *Nat. Cities* 1–9

[2] Hou Y, Quintana M, Khomiakov M, Yap W, Ouyang J, Ito K, Wang Z, Zhao T and Biljecki F 2024 *ISPRS J. Photogramm. Remote Sens.* **215** 216–238

[3] Milojevic-Dupont N, Wagner F, Nachtigall F, Hu J, Brüser G B, Zumwald M, Biljecki F, Heeren N, Kaack L H, Pichler P P and Creutzig F 2023 *Sci. Data* **10** 147

[4] Abdelrahman M, Macatulad E, Lei B, Quintana M, Miller C and Biljecki F 2025 *Build. Environ.* **274** 112748

[5] Heimar Andersen K, Pommerencke Melgaard S, Johra H, Marszal-Pomianowska A, Lund Jensen R and Kvols Heiselberg P 2024 *Energy Build.* **303** 113801

[6] Gil J, Petrova-Antonova D and Kemp G J L 2024 *Environ. Plan. B Urban Anal. City Sci.*

[7] Quintana M, Abdelrahman M, Frei M, Tartarini F and Miller C 2021 Longitudinal personal thermal comfort preference data in the wild *Proceedings of the 19th ACM Conference on Embedded Networked Sensor Systems* SenSys '21 (New York, NY, USA: Association for Computing Machinery) pp 556–559

[8] Sanchez T W, Brenman M and Ye X 2024 *J. Am. Plann. Assoc.* 1–14

[9] Zhang W, Quintana M and Miller C 2025 *Energy Build.* **329** 115247

[10] Miller C, Abdelrahman M, Chong A, Biljecki F, Quintana M, Frei M, Chew M and Wong D 2021 *J. Phys. Conf. Ser.* **2042** 012041

[11] Zhu Y, Zhang Y and Biljecki F 2025 *Cities* **156** 105535

[12] Balaji B, Bhattacharya A, Fierro G, Gao J, Gluck J, Hong D, Johansen A, Koh J, Ploennigs J, Agarwal Y, Bergés M, Culler D, Gupta R K, Kjærgaard M B, Srivastava M and Whitehouse K 2018 *Appl. Energy* **226** 1273–1292

[13] Miller C, Quintana M, Frei M, Chua Y X, Fu C, Picchetti B, Yap W, Chong A and Biljecki F 2023 Introducing the cool, quiet city competition: Predicting smartwatch-reported heat and noise with digital twin metrics *Proceedings of the 10th ACM International Conference on Systems for Energy-Efficient Buildings, Cities, and Transportation* BuildSys '23 (New York, NY, USA: Association for Computing Machinery) pp 298–299

[14] Mora L, Gerli P, Batty M, Binet Royall E, Carfi N, Coenegrachts K F, de Jong M, Facchina M, Janssen M, Meijer A, Pasi G, Perrino M, Raven R, Sagar A, Sancino A, Santi P, Sharp D, Trencher G, van Zoonen L, Westerberg P, Woods O, Zhang X and Ziemer G 2025 *Nat. Cities* **2** 110–113

[15] Bunnell T, Spicer Z, Miller B, Abbruzzese T, Cardullo P, Chae S, Chang I C C, Charnock G, Chung M K, Heo K, Jou S C, Karvonen A, Kordas O, Kong L, Ribera-Fumaz R, Shin H and Woods O 2025 *Urban Geogr.* 1–26

[16] Mosteiro-Romero M, Miller C, Quintana M, Chong A and Stouffs R 2023 *J. Phys. Conf. Ser.* **2600** 132008

[17] Mosteiro-Romero M, Quintana M, Stouffs R and Miller C 2024 *Build. Environ.* **257** 111479

[18] Ang Y Q, Berzolla Z M, Letellier-Duchesne S, Jusiega V and Reinhart C 2022 *Sustain. Cities Soc.* **77** 103534

[19] Miller C, Chua Y X, Quintana M, Lei B, Biljecki F and Frei M 2025 *Build. Environ.* **284** 113388

[20] Gregorio G D and Ranchordás S 2020 *Legal Challenges of Big Data*

# Language Biomarkers for Dementia: Current Advances and Future Directions

Jiayu Zhou[1,*] and Hiroko H. Dodge[2]

[1]School of Information, University of Michigan, Ann Arbor, U.S.A.
[2]Harvard Medical School, Charlestown, U.S.A.
[*]Author to whom any correspondence should be addressed.

**E-mail:** jiayuz@umich.edu

**Status**

Language abilities are closely linked to cognitive health, and changes in language can serve as early indicators of neurodegeneration [2]. Longitudinal findings show that low linguistic complexity in early adulthood is associated with poorer cognitive function and higher risk of Alzheimer's disease (AD) decades later [15]. Clinically, patients with AD often exhibit language deficits early in the disease, for example, lexical-semantic impairments, simpler grammar, and reduced comprehension of nuanced language [16]. Such communication difficulties are evident even at the mild cognitive impairment (MCI) stage, and performance on verbal tasks is an important diagnostic indicator for both MCI and early AD [16]. Thus, speech and language changes may emerge alongside (or even before) overt memory symptoms, underscoring the potential of language as a window into incipient cognitive decline.

Building on these insights, researchers have made significant advances in automated language analysis for dementia detection. By applying natural language processing (NLP) and machine learning, subtle linguistic and acoustic markers of cognitive decline can be quantified objectively. An early computational study showed that features extracted from short narrative speech samples could differentiate AD patients from healthy older adults with over 80% accuracy [5, 10]. Characteristic speech differences in AD span multiple domains: affected individuals tend to use simpler vocabulary and syntax, produce more pauses or filler words, and convey less information (lower idea density) [3], even in the pre-symptomatic phase. In recent years, shared tasks like the ADReSS challenge have provided benchmark data for comparing algorithms, accelerating progress toward robust speech-based classifiers [12]. Many studies now report high accuracy in detecting mild cognitive impairment from speech [1]. For example, Asgari *et al.* (2017) and Chen *et al.* (2021) achieved high MCI and AD classification accuracy from speech input [1, 4], while Tang et al. (2020) developed an AI dialogue agent that adaptively screens for MCI [18]. These language-based digital biomarkers are non-invasive and low-cost, offering an attractive alternative to expensive neuroimaging or biofluid tests for early detection [7]. Moreover, linguistic markers from short interviews correlate well with standard cognitive assessments and can distinguish MCI/AD from healthy aging [16, 19]. Augmenting traditional neuropsychological evaluation with automated speech analysis could thus enable more scalable screening of populations for subtle signs of dementia, allowing earlier interventions when they are most beneficial.

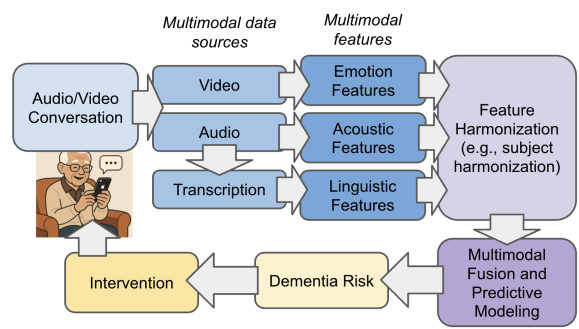

**Figure 1.** Language markers integrated within a multimodal pipeline for dementia risk prediction and monitoring. Conversations are collected in everyday settings and processed to derive multimodal features (acoustic, linguistic, and auxiliary signals). These features are harmonized to reduce subject-specific biases, then combined through multimodal fusion and predictive modeling to generate accessible risk estimates, supporting affordable in-home monitoring and early intervention.

**Current and future challenges**
*Data diversity and generalization:* Many speech-based dementia models are trained on relatively small, homogeneous cohorts, which limits their generalizability. High variability in how individuals speak, due to personal, cultural, or educational differences, makes it hard for language features to generalize to new speakers [7]. A classifier might perform well on its training dataset but fail when applied to a broader patient population that differs in age, background, or dialect.
*Cross-linguistic generalizability:* Because most language-biomarker research has been conducted in English [14], it is uncertain how well these methods extend to other languages. Cognitive-speech patterns of impairment may differ in different languages, so models developed on English speech could lose accuracy in other linguistic contexts. This presents a challenge for making screening tools that are globally applicable to diverse populations.
*Model interpretability and privacy:* Many state-of-the-art AI models output a dementia risk score with no explanation. Clinicians may be reluctant to trust an AI system without understanding its reasoning. There is a need for more transparent models that highlight salient features of a patient's speech (e.g., slow speech rate or frequent pauses) as evidence for their predictions [14]. Additionally, recording patients' speech raises ethical concerns around privacy and consent. Ensuring secure data handling and maintaining patient confidentiality is an ongoing challenge for real-world deployment.
*Multimodal integration and biological validation:* Another challenge is linking speech-derived markers with established biological indicators of dementia. Language changes are useful on their own, but they should ideally be validated against neuroimaging or fluid biomarkers to confirm that they reflect underlying neurodegeneration. Moreover, cognitive assessment is multifaceted, and combining speech analysis with other data (such as imaging) could improve diagnostic accuracy. Determining how to effectively integrate multiple modalities and proving that speech markers correspond to specific neuropathological changes remains difficult, partly due to the lack of data which have both speech-derived markers and established biological markers.
*Clinical validation and deployment:* Translating AI speech tools into clinical practice will require extensive validation and user-friendly implementation. To date, few models have been evaluated on independent cohorts or in prospective trials. For example, a recent review of 86 AI models for AD found that only 2 had undergone external validation [13]. This lack of rigorous validation raises concerns about overfitting and reproducibility. In addition, practical deployment issues must be addressed: memory clinics need tools that are fast, reliable, and easy to use. Initial pilot studies (e.g., telephone-based automated assessments) are promising, but larger multi-site trials are needed to demonstrate real-world utility.

**Advances in science and technology to meet challenges**
*Enhancing data diversity and reach:* Subject harmonization networks, for example, adjust speech feature distributions across individuals and have boosted detection accuracy on new patients [7]. Additionally, data augmentation techniques (for example, generating synthetic training samples) are being used to boost model performance on limited data [11]. Furthermore, multilingual pre-trained models are a promising direction: using speech representations trained on many languages can help a classifier handle non-English input [14, 6]. Early results are encouraging: a Mandarin Chinese speech classifier achieved promising accuracy in detecting cognitive impairment, comparable to results in English [9].
*Improving interpretability and privacy:* The field is also prioritizing trustworthy AI. Some diagnostic systems now provide explanations alongside their predictions. For example, indicating that a patient's slow speech rate and long pauses were key factors in an "at-risk" result, instead of outputting a cryptic score. Even complex neural models can be probed with tools that highlight which features of speech most influenced the decision. Equally important is safeguarding privacy: newer platforms employ encryption and on-device processing so that sensitive audio never leaves the user's device, and patients give explicit consent for any data use. These steps help build confidence that language-based AI tools are both transparent and secure.
*Multimodal integration and validation:* To lend biological credibility to speech biomarkers, efforts are underway to connect them with traditional dementia indicators. Combining different speech features has already proven useful. For example, integrating acoustic and linguistic markers yielded higher MCI detection accuracy than using either alone [17]. Meanwhile, researchers are correlating linguistic metrics with neuroimaging and EEG results to verify that changes in communication mirror changes in the brain. Such multimodal studies are mapping how specific neurodegenerative patterns (e.g., temporal lobe atrophy) correspond to speech and language deficits, providing validation that speech-based markers reflect underlying disease processes.
*Clinical translation and deployment:* Pilot projects have brought automated speech analysis into clinical practice. For example, an automated telephone cognitive test successfully distinguished MCI patients from healthy controls in a trial [19]. Likewise, an "AI virtual interviewer" has been tested for remote dementia screening, demonstrating that speech assessments can be scaled up to reach many people [18]. To encourage adoption, developers are streamlining these tools to be user-friendly. Ongoing validation in diverse patient groups – and training clinicians to interpret AI outputs – will be crucial for making language biomarkers a routine part of dementia care.

**Concluding remarks**
Language provides a unique window into the mind, and leveraging it as a biomarker can transform how we detect dementia. In recent years, research has moved from observational studies to sophisticated AI systems that can flag cognitive impairment from a simple conversation. These advances have demonstrated that subtle features of language and voice often reveal the early footprints of neurodegenerative disease, potentially even before traditional clinical symptoms are obvious. While challenges remain in ensuring these tools are accurate, fair, and trusted, the trajectory of progress is promising. Ongoing efforts to diversify data, expand to many languages, explain AI decisions, and validate speech markers against established biomarkers will be critical for translating prototypes into practical diagnostics. If successful, automated language analysis could become a routine part of check-ups: a quick, non-invasive screen to catch cognitive decline in the earliest stages. Moreover, the same technology may support patients beyond diagnosis: AI chatbots could engage older adults in cognitively stimulating dialogues, providing companionship and cognitive support [8, 20]. By continuing to advance language-focused research, we move closer to a future where a simple spoken exchange can lead to earlier detection, more personalized care, and improved quality of life.

**Acknowledgments**
This material is based in part upon work supported by the National Science Foundation under Grant IIS-2212174, National Institute of Aging (NIA) 1RF1AG072449, R01AG056102, R01AG051628, National Institute of General Medical Sciences (NIGMS) 1R01GM145700.

**References**

[1] M. Asgari, J. Kaye, and H. Dodge. Predicting mild cognitive impairment from spontaneous spoken utterances. *Alzheimer's & Dementia: Translational Research & Clinical Interventions*, 3(2):219–228, 2017.

[2] V. Boschi, E. Catricala, M. Consonni, C. Chesi, A. Moro, and S. F. Cappa. Connected speech in neurodegenerative language disorders: a review. *Frontiers in psychology*, 8:269, 2017.

[3] A. Bose, S. Ahmed, Y. Cheng, and A. Suárez-Gonzalez. Connected speech features in non-english speakers with alzheimer's disease: protocol for scoping review. *Systematic reviews*, 13(1):40, 2024.

[4] J. Chen, J. Ye, F. Tang, and J. Zhou. Automatic detection of alzheimer's disease using spontaneous speech only. In *Interspeech*, volume 2021, page 3830, 2021.

[5] K. C. Fraser, J. A. Meltzer, and F. Rudzicz. Linguistic features identify alzheimer's disease in narrative speech. *Journal of Alzheimer's disease*, 49(2):407–422, 2015.

[6] B. Hoang, Y. Pang, H. Dodge, and J. Zhou. Translingual language markers for cognitive assessment from spontaneous speech. *Interspeech 2024*, pages 977–981, 2024.

[7] B. Hoang, Y. Pang, H. H. Dodge, and J. Zhou. Subject harmonization of digital biomarkers: Improved detection of mild cognitive impairment from language markers. In *Pacific Symposium on Biocomputing. Pacific Symposium on Biocomputing*, volume 29, page 187, 2024.

[8] J. Hong, W. Zheng, H. Meng, S. Liang, A. Chen, H. H. Dodge, J. Zhou, and Z. Wang. A-conect: Designing ai-based conversational chatbot for early dementia intervention. In *ICLR 2024 Workshop on Large Language Model (LLM) Agents*, 2024.

[9] L. Huang, H. Yang, Y. Che, and J. Yang. Automatic speech analysis for detecting cognitive decline of older adults. *Frontiers in Public Health*, 12:1417966, 2024.

[10] A. König, A. Satt, A. Sorin, R. Hoory, O. Toledo-Ronen, A. Derreumaux, V. Manera, F. Verhey, P. Aalten, P. H. Robert, et al. Automatic speech analysis for the assessment of patients with predementia and alzheimer's disease. *Alzheimer's & Dementia: Diagnosis, Assessment & Disease Monitoring*, 1(1):112–124, 2015.

[11] G. Liu, Z. Xue, L. Zhan, H. H. Dodge, and J. Zhou. Detection of mild cognitive impairment from language markers with crossmodal augmentation. In *PACIFIC symposium on biocomputing 2023: Kohala Coast, Hawaii, USA, 3–7 January 2023*, pages 7–18. World Scientific, 2022.

[12] S. Luz, F. Haider, S. de la Fuente Garcia, D. Fromm, and B. MacWhinney. Alzheimer's dementia recognition through spontaneous speech, 2021.

[13] W. Qi, X. Zhu, B. Wang, Y. Shi, C. Dong, S. Shen, J. Li, K. Zhang, Y. He, M. Zhao, et al. Alzheimer's disease digital biomarkers multidimensional landscape and ai model scoping review. *npj Digital Medicine*, 8(1):366, 2025.

[14] A. Shakeri and M. Farmanbar. Natural language processing in alzheimer's disease research: Systematic review of methods, data, and efficacy. *Alzheimer's & Dementia: Diagnosis, Assessment & Disease Monitoring*, 17(1):e70082, 2025.

[15] D. A. Snowdon, S. J. Kemper, J. A. Mortimer, L. H. Greiner, D. R. Wekstein, and W. R. Markesbery. Linguistic ability in early life and cognitive function and alzheimer's disease in late life: Findings from the nun study. *Jama*, 275(7):528–532, 1996.

[16] V. Taler and N. A. Phillips. Language performance in alzheimer's disease and mild cognitive impairment: a comparative review. *Journal of clinical and experimental neuropsychology*, 30(5):501–556, 2008.

[17] F. Tang, J. Chen, H. H. Dodge, and J. Zhou. The joint effects of acoustic and linguistic markers for early identification of mild cognitive impairment. *Frontiers in digital health*, 3:702772, 2022.

[18] F. Tang, I. Uchendu, F. Wang, H. H. Dodge, and J. Zhou. Scalable diagnostic screening of mild cognitive impairment using ai dialogue agent. *Scientific reports*, 10(1):5732, 2020.

[19] D. Ter Huurne, N. Possemis, L. Banning, A. Gruters, A. König, N. Linz, J. Tröger, K. Langel, F. Verhey, M. De Vugt, et al. Validation of an automated speech analysis of cognitive tasks within a semiautomated phone assessment. *Digital biomarkers*, 7(1):115–123, 2023.

[20] Z. Yang, J. Hong, Y. Pang, J. Zhou, and Z. Zhu. Chatwise: A strategy-guided chatbot for enhancing cognitive support in older adults. *arXiv preprint arXiv:2503.05740*, 2025.

# AI for Particle Physics

David Rousseau[1]

[1]Université Paris-Saclay, CNRS/IN2P3, IJCLab, 91405 Orsay, France

**E-mail:** rousseau@ijclab.in2p3.fr

**Status**

Particle physics is at a crossroads concerning AI. The Higgs boson, first postulated in 1964, is central to the Standard Model's explanation of elementary particle mass. Decades later, its existence was experimentally confirmed in 2012 by the ATLAS and CMS collaborations at CERN's Large Hadron Collider (LHC), validating a long-standing theoretical prediction. After the discovery was established, the focus shifted from discovery mode to precision physics mode. Measuring the Higgs boson's properties isn't just about studying an elusive particle; it's about probing the Higgs field itself, a fundamental component of the vacuum that has existed everywhere since the beginning of time (the Big Bang).

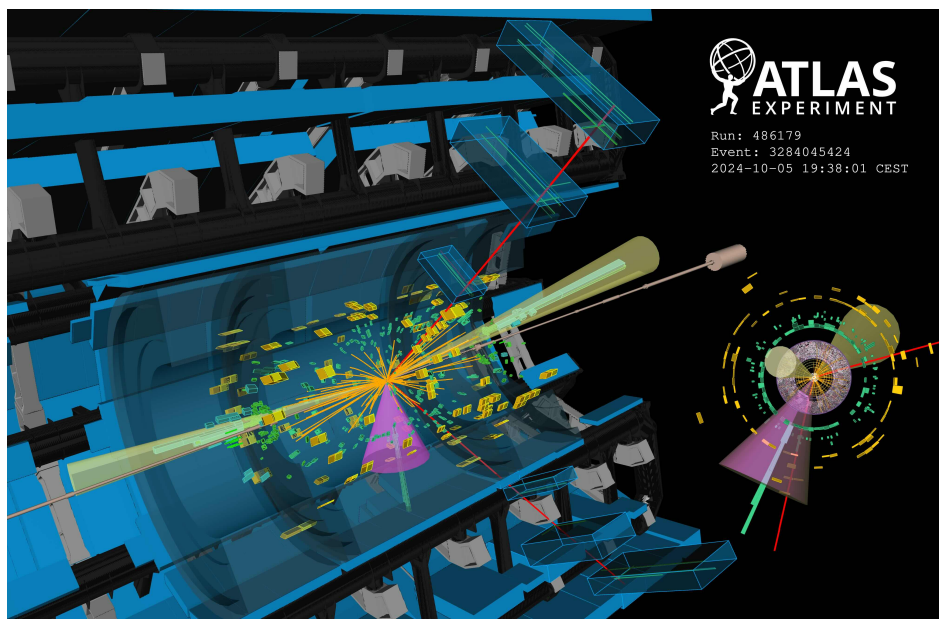

**Figure 1.** Event display of an event recorded by the ATLAS experiment on the 5$^{th}$ October 2024, featuring a decay to a photon (purple cone) and a Z boson, with the Z subsequently decaying to two muons (red lines), accompanied by two jets (yellow cones). Parts of the detector and raw measurements from which the final state is reconstructed are represented. From [1].

The primary subject of investigation in high-energy particle physics is proton-proton collisions, which generate fireworks of particles whose properties are recorded by advanced detector systems. The fundamental unit of analysis is an "event", defined as the complete set of measurements associated with a single collision (see Figure 1). Given the LHC's collision rate of approximately 1 billion events per second, only about 100 events per second are retained by high-performance trigger and data acquisition systems after real-time selection. Over a year, this results in roughly 1 billion recorded events, with each event comprising 1 MB of semi-structured data. The dataset reaches the petabyte scale, presenting substantial storage, processing, and analysis challenges.

AI has already had some role in the 2012 discovery of the Higgs boson [2], with classifiers based on Boosted Decision Tree and simple dense Neural Networks. The field has then grown exponentially: the curated HEPML living review [3] started with a dozen papers, and has now more than 1600 papers (see also [4] for a 2021 status report).

A primary challenge arises from the inherent complexity of event data, which is neither tabular nor image-like but instead semi-structured. Recent advances have been achieved through the adoption of progressively sophisticated machine learning architectures, each offering incremental improvements over prior approaches: Recurrent Neural Networks (RNNs), Graph Neural Networks (GNNs) [5], and recently transformer-based models [6].

**Current and future challenges**
A key field asset is the tradition of developing digital twins combining first-principles calculations with extensive calibration data. They provide billions of labelled simulation events to be used for model training. Although these simulators are relatively accurate, they are computationally expensive. The community has initiated a broad effort to offload resource-intensive components to generative surrogates [7]. Initially based on GANs, VAEs, then Normalising Flows and diffusion models, proofs of concept achieve orders-of-magnitude speedups, but their accuracy has not yet met the requirements for final analyses.

Uncertainty is a central challenge. Experimental measurements are always reported with both statistical (aleatoric) and systematic (epistemic) uncertainties. Statistical uncertainty can be reduced using recent advances in Neural Simulation-Based Inference (NSBI) [8, 9], in which the parameter of interest is estimated via an unbinned maximum-likelihood fit that combines event-level likelihood surrogates derived from neural density–ratio estimators. Systematic (epistemic) uncertainty arises from diverse auxiliary inputs to the experiment, including complex theoretical calculations, detector effects (e.g., drifts in electronics calibration), and assumptions encoded in the digital twin. Key questions for an AI-driven analysis pipeline include (see [10]): Can we continue to report epistemic uncertainties transparently as AI methods proliferate? Can we incorporate these uncertainties via "uncertainty-aware" training during model development rather than only a posteriori? And can we convincingly demonstrate to the community that these uncertainties are being handled rigorously? Addressing these questions links directly to the AI fields of Uncertainty Quantification and explainability [11].

Separately, online filtering and data reduction are natural applications for AI. The objective is to select interesting events at high throughput and to infer particle properties directly from raw detector readouts under tight latency constraints. Relatively simple, reduced-precision architectures deployable on field-programmable gate arrays (FPGAs) have been shown to outperform traditional, non-AI baselines [12]. This progress has prompted a complete re-thinking of the online processing pipeline [13].

The current LHC experiments were conceived in the 1990s, built in the 2000s, and are planned to operate for roughly three decades, through about 2040. AI was not part of the original design and entered the reconstruction and analysis pipelines only later. A new generation of experiments, envisioned for the Future Circular Collider and other frontiers, is being designed this decade. Two central questions now guide community efforts (see, e.g. [14] and subsequent MODE collaboration work): (i) How does treating AI as a first-class design element reshape the architecture and capabilities of future detectors? (ii) Can reinforcement learning or generative AI assist in optimising detector layouts or even proposing novel experimental concepts?

**Advances in science and technology to meet challenges**
The development of AI models for particle physics has so far not been limited by compute, as the required resources are only a small fraction of those routinely available for the field and for other AI applications; scaling is therefore not a primary concern.

A more important key driver of progress in AI for particle physics is sustained collaboration between physicists and computer scientists, experts in two intrinsically complex domains. Open data and standardised benchmarks have been crucial enablers, even if they challenge longstanding traditions around data access in high-energy physics. The 2014 HiggsML competition [15] alerted the community to the transformative potential of machine learning. The 2018–2019 TrackML competitions [16, 17] established a widely used benchmark for final-state particle reconstruction. The 2020 LHC Olympics [18] spurred the adoption of anomaly detection algorithms for data-driven, model-agnostic new physics searches at colliders. The 2021 CaloChallenge [7] catalysed the development of generative surrogates for detector simulation. Looking ahead, the 2024 Fair Universe HiggsML Uncertainties competition [10] promises to become a key benchmark for "uncertainty-aware" modelling.

**Concluding remarks**
While this discussion has centred on Higgs-boson physics at colliders, where the challenges are particularly acute, the underlying methodology is broadly applicable. Even if many models are task-specific, the same principles translate to other scientific domains: rapidly processing massive datasets, rigorously handling diverse sources of uncertainty, and anchoring inference with first-principles calculations.

**References**

 [1] The ATLAS collaboration, *Search for the Higgs boson decay to a Z boson and a photon in pp collisions at $\sqrt{s} = 13$ TeV and 13.6 TeV with the ATLAS detector*, arXiv:2507.12598 [hep-ex].

 [2] A. Radovic, M. Williams, D. Rousseau, M. Kagan, D. Bonacorsi, A. Himmel, A. Aurisano, K. Terao, and T. Wongjirad, *Machine learning at the energy and intensity frontiers of particle physics*, Nature **560** (2018) 41.

 [3] HEP ML Community, "A Living Review of Machine Learning for Particle Physics." https://iml-wg.github.io/HEPML-LivingReview/.

 [4] P. Calafiura, D. Rousseau, and K. Terao, eds., *Artificial Intelligence for High Energy Physics*. World Scientific, 2020. http://dx.doi.org/10.1142/12200.

 [5] J. Duarte and J.-R. Vlimant, *Graph Neural Networks for Particle Tracking and Reconstruction*. In: *Artificial Intelligence for High Energy Physics*, p. 387–436. World Scientific, Feb., 2022. arXiv:2012.01249 [hep-ph]. http://dx.doi.org/10.1142/9789811234033_0012.

 [6] H. Qu, C. Li, and S. Qian, *Particle transformer for jet tagging*, in *Proceedings of the 39th International Conference on Machine Learning*, K. Chaudhuri, S. Jegelka, L. Song, C. Szepesvari, G. Niu, and S. Sabato, eds. PMLR, 17–23 Jul, 2022. https://proceedings.mlr.press/v162/qu22b.html.

 [7] O. Amram *et al.*, *CaloChallenge 2022: A Community Challenge for Fast Calorimeter Simulation*, arXiv:2410.21611 [physics.ins-det].

 [8] J. Brehmer and K. Cranmer, *Simulation-Based Inference Methods for Particle Physics*. In: *Artificial Intelligence for High Energy Physics*, p. 579–611. World Scientific, Feb., 2022. arXiv:2010.06439 [hep-ph]. http://dx.doi.org/10.1142/9789811234033_0016.

 [9] The ATLAS collaboration, *An implementation of neural simulation-based inference for parameter estimation in ATLAS*, Rept. Prog. Phys. **88** (2025) 067801, arXiv:2412.01600 [physics.data-an].

[10] L. Benato, W. Bhimji, P. Calafiura, R. Chakkappai, P.-W. Chang, Y.-T. Chou, S. Diefenbacher, J. Dudley, I. Elsharkawy, S. Farrell, A. Ghosh, C. Giordano, I. Guyon, C. Harris, Y. Hashizume, S.-C. Hsu, E. E. Khoda, C. Krause, A. Li, B. Nachman, P. Nugent, D. Rousseau, R. Schoefbeck, M. Shooshtari, D. Schwarz, B. Thorne, I. Ullah, D. Wang, and Y. Zhang, *FAIR Universe HiggsML Uncertainty Challenge Competition*, Advances in Neural Information Processing Systems (NeurIPS) 2025, arXiv:2410.02867 [hep-ph]. To appear.

[11] T. Y. Chen, B. Dey, A. Ghosh, M. Kagan, B. Nord, and N. Ramachandra, *Interpretable Uncertainty Quantification in AI for HEP*, in *Snowmass 2021*. 8, 2022. arXiv:2208.03284 [hep-ex].

[12] P. Odagiu, Z. Que, J. Duarte, J. Haller, G. Kasieczka, A. Lobanov, V. Loncar, W. Luk, J. Ngadiuba, M. Pierini, P. Rincke, A. Seksaria, S. Summers, A. Sznajder, A. Tapper, and T. K. Årrestad, *Ultrafast jet classification at the hl-lhc*, Machine Learning: Science and Technology **5** (jul, 2024) 035017. https://doi.org/10.1088/2632-2153/ad5f10.

[13] L. Boggia *et al.*, *Review of Machine Learning for Real-Time Analysis at the Large Hadron Collider experiments ALICE, ATLAS, CMS and LHCb*, arXiv:2506.14578 [hep-ex].

[14] MODE, T. Dorigo, A. Giammanco, P. Vischia, M. Aehle, M. Bawaj, A. Boldyrev, P. de Castro Manzano, D. Derkach, J. Donini, A. Edelen, F. Fanzago, N. R. Gauger, C. Glaser, A. G. Baydin, L. Heinrich, R. Keidel, J. Kieseler, C. Krause, M. Lagrange, M. Lamparth, L. Layer, G. Maier, F. Nardi, H. E. S. Pettersen, A. Ramos, F. Ratnikov, D. Röhrich, R. R. de Austri, P. M. R. del Árbol, O. Savchenko, N. Simpson, G. C. Strong, A. Taliercio, M. Tosi, A. Ustyuzhanin, and H. Zaraket, *Toward the end-to-end optimization of particle physics instruments with differentiable programming*, Rev. Phys. **10** (2023) 100085, arXiv:2203.13818 [physics.ins-det].

[15] C. Adam-Bourdarios, G. Cowan, C. Germain, I. Guyon, B. Kégl, and D. Rousseau, *The Higgs boson machine learning challenge*, in *Proceedings of the NIPS 2014 Workshop on High-energy Physics and Machine Learning*, G. Cowan, C. Germain, I. Guyon, B. Kégl, and D. Rousseau, eds. PMLR, Montreal, Canada, 13 Dec, 2015. http://proceedings.mlr.press/v42/cowa14.html.

[16] S. Amrouche, L. Basara, P. Calafiura, V. Estrade, S. Farrell, D. R. Ferreira, L. Finnie, N. Finnie, C. Germain, V. V. Gligorov, T. Golling, S. Gorbunov, H. Gray, I. Guyon, M. Hushchyn, V. Innocente, M. Kiehn, E. Moyse, J.-F. Puget, Y. Reina, D. Rousseau, A. Salzburger, A. Ustyuzhanin, J.-R. Vlimant, J. S. Wind, T. Xylouris, and Y. Yilmaz, *The Tracking Machine Learning Challenge: Accuracy Phase*, in *The NeurIPS 2018 Competition*, pp. 231–264. Springer International Publishing, Nov., 2019. arXiv:1904.06778 [hep-ex].

[17] S. Amrouche, L. Basara, P. Calafiura, D. Emeliyanov, V. Estrade, S. Farrell, C. Germain, V. V. Gligorov, T. Golling, S. Gorbunov, H. Gray, I. Guyon, M. Hushchyn, V. Innocente, M. Kiehn, M. Kunze, E. Moyse, D. Rousseau, A. Salzburger, A. Ustyuzhanin, and J.-R. Vlimant, *The Tracking Machine Learning Challenge: Throughput Phase*, Comput. Softw. Big Sci. **7** (2023) 1, arXiv:2105.01160 [cs.LG].

[18] G. Kasieczka, B. Nachman, D. Shih, O. Amram, A. Andreassen, K. Benkendorfer, B. Bortolato, G. Brooijmans, F. Canelli, J. H. Collins, B. Dai, F. F. De Freitas, B. M. Dillon, I.-M. Dinu, Z. Dong, J. Donini, J. Duarte, D. A. Faroughy, J. Gonski, P. Harris, A. Kahn, J. F. Kamenik, C. K. Khosa, P. Komiske, L. Le Pottier, P. Martín-Ramiro, A. Matevc, E. Metodiev, V. Mikuni, C. W. Murphy, I. Ochoa, S. E. Park, M. Pierini, D. Rankin, V. Sanz, N. Sarda, U. Seljak, A. Smolkovic, G. Stein, C. M. Suarez, M. Szewc, J. Thaler, S. Tsan, S.-M. Udrescu, L. Vaslin, J.-R. Vlimant, D. Williams, and M. Yunus, *The lhc olympics 2020 a community challenge for anomaly detection in high energy physics*, Reports on Progress in Physics **84** (Dec., 2021) 124201. `http://dx.doi.org/10.1088/1361-6633/ac36b9`.

# From Pattern Matching to Emergence Detection: A New Benchmark for Artificial Intelligence

Andrey Ustyuzhanin[1,2,3]

1 Constructor Knowledge Labs, Bremen, Campus Ring 1, 28759, Germany.
2 Constructor University, Bremen, Campus Ring 1, 28759, Germany.
3 Institute for Functional Intelligent Materials, National University of Singapore, 4 Science Drive 2, Singapore 117544, Singapore

**E-mail**: andrey.ustyuzhanin@constructor.org

## Status

The current AI landscape, characterized by rapid yet incremental advancements, has become a hyper-competitive "AI bloodbath." While modern AI excels at pattern matching within existing datasets, it fundamentally lacks the ability to detect and reason about emergent, "life-like" phase transitions [12, 17]—the spontaneous creation of new structures and behaviors in complex systems. This creates an "Intelligence Gap," [1] where machines fail to grasp the "sparks of life" that drive open-ended evolution [4].

To address this, the concept of a "mini-bang" has been introduced. A mini-bang is defined as a spontaneous, localized event of emergence that initiates a multi-scale causal cascade [14], leading to irreversible qualitative shifts in a system's dynamics. A well-understood theoretical model for this phenomenon comes from prebiotic chemistry with the formation of Reflexively Autocatalytic and Food-generated (RAF) networks [7]. These networks demonstrate how a system can undergo a sharp phase transition from a simple collection of molecules to a complex, self-sustaining web, providing a concrete example of a mini-bang. This established model serves as a foundational case study for developing a broader science of emergence detection.

## Current and Future Challenges

The primary challenge is to equip machines with the ability to sense, interpret, and leverage mini-bangs. This overarching goal presents several specific and formidable research hurdles:

- **Mathematical Unification:** A need for mathematical frameworks that can rigorously describe open-ended, distribution-independent growth, moving beyond traditional models that assume convergence.
- **Cross-Domain Benchmarking:** Creating standardized, computationally tractable benchmarks to assess emergence comprehension is difficult. These benchmarks must be able to generate comparable emergent scenarios across diverse domains like chemistry, social dynamics, and technology.
- **Measuring True Comprehension:** Developing quantitative metrics to adequately assess an agent's multi-level understanding is a major challenge. We need to measure not just what an agent observes, but its understanding of *what could be*—its grasp of the new possibilities opened by an emergent event [1].
- **Architectural Innovation:** Current AI architectures, including neural and neurosymbolic models,

are not inherently designed to handle the self-organizing feedback loops and bidirectional causality [10] that define mini-bangs. New architectures are required [6, 11].
- **Robust Detection:** Identifying the subtle onset of mini-bangs amidst the high noise levels of real-world data streams requires new pattern recognition approaches sensitive to emergent causal structures [18, 19].
- **Defining Objective Functions:** It remains an open question how to define a loss function for an agent whose goal is to detect and understand such open-ended, qualitative transitions [4].

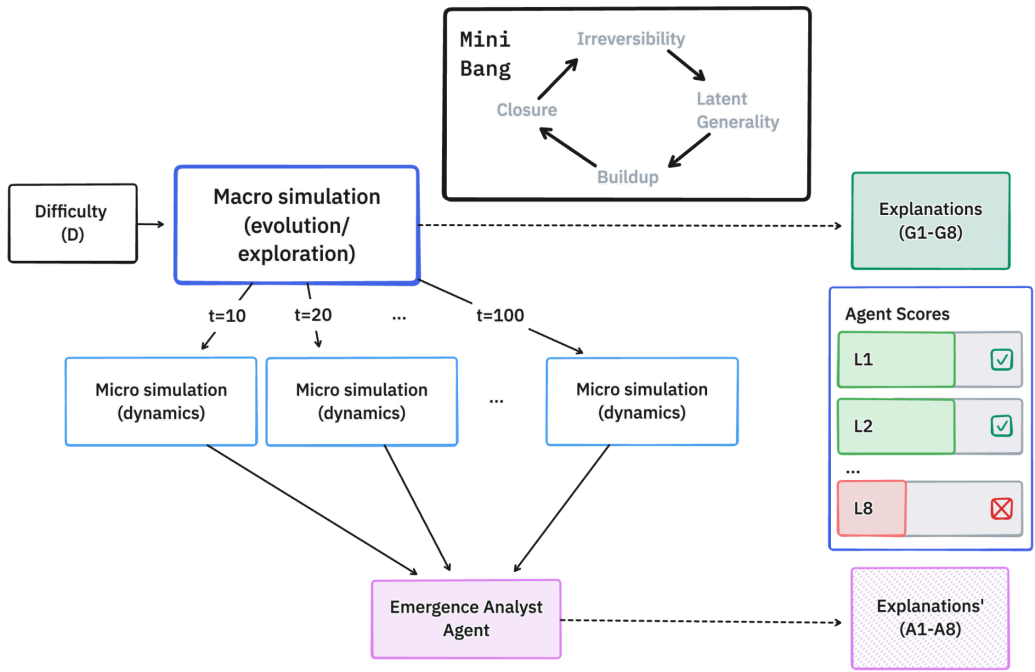

**Figure 1.** Emergence detection evaluation framework.

## Advances in Science and Technology to Meet Challenges

To address these challenges, a novel benchmark-first approach is proposed, centered on a comprehensive framework for generating and evaluating emergence detection. This framework represents a significant technological and scientific advance, built on three key components:

1. **A Formal Model of Emergence:** The "mini-bang" is formalized as a process with four distinct, sequential stages, providing a clear structure for analysis:
    - **Buildup:** A preparatory phase of rising informational or structural complexity [3].
    - **Closure:** The critical moment where a self-sustaining, bidirectional feedback loop emerges [5, 8, 10].
    - **Irreversibility:** The system locks into a new dynamic regime, leading to a divergence of possible future paths.
    - **Latent Generality:** The new structure becomes a platform for future innovations, opening up new "adjacent possibles." [2]
2. **A Hierarchy of Comprehension:** An multilevel hierarchy is introduced to systematically measure

and score an AI's understanding, moving from basic signal detection to deep generative insight [1]. This allows for granular assessment, from identifying an anomaly (Level 1) to reconstructing its structure (Level 3), explaining its mechanism (Level 5), and using that knowledge to induce new mini-bangs (Level 8).
3. **Simulation and Scoring Environment:** The framework utilizes a two-tiered simulation system. A **Macro simulation** evolves a system to the brink of a mini-bang, while sampled **Micro simulations** provide detailed dynamic trajectories for analysis. An "Emergence Analyst Agent" then examines obfuscated code from these simulations and produces explanations, which are scored against ground-truth answers.

**Potential Practical Applications and Novelty:** Beyond a theoretical construct, the ability to detect and understand mini-bangs unlocks vast novelty potential across numerous domains. This framework can be applied to:

- **Novel Knowledge Discovery:** As exemplified by **HypoFinder**, a system designed to induce "scientific mini-bangs." By structuring fragmented knowledge, it aims to transform scattered literature into coherent, actionable research hypotheses and guide research funding toward breakthrough areas.
- **Efficient AI Architectures:** Designing new AI models that can self-organize or undergo "grokking"-like phase transitions [9] intentionally, leading to more efficient and capable systems.
- **Novel Drug Design:** Moving beyond brute-force screening to identify the emergent properties of molecular interactions that lead to therapeutic effects, enabling the design of entirely new classes of drugs.
- **New Computational Paradigms:** Developing novel computational architectures based on physical principles, where computation is an emergent property of the system's dynamics rather than a programmed instruction set.
- **Socio-Economic Policy:** Creating more balanced and resilient policy in politics and economics by identifying the critical thresholds for cascades [14] or market shifts [16] *before* they become irreversible, catastrophic events.

## Concluding Remarks

The study of mini-bangs offers a compelling path forward for AI, aiming to transform models from being simple "pattern matchers to emergence detectives." This research agenda is a direct response to the current state of incrementalism in large language model development, offering a path beyond the "LLM research hell" [4] by focusing on the fundamental principles of creation and open-ended dynamics.

By developing AI that can understand and anticipate these critical transitions [13, 15, 16], we can unlock interpretable and proactive insights into a wide array of complex systems, from chemical evolution to social and technological tipping points [12, 14]. The proposed benchmarking framework provides a concrete roadmap for community progress. Ultimately, this work is a call to multidisciplinary action, urging researchers to collaborate on developing the theory, architectures, and experiments needed to build the next generation of truly intelligent systems—those that can not only analyze the world as it is, but comprehend how it comes to be [5, 8, 10].

## References

1. Chollet, F. (2019). On the measure of intelligence.

2. Kauffman, S. (2000). Investigations.
3. Marshall, S. M., et al. (2021). Assembly Theory.
4. Lehman, J., et al. (2024). Beyond objectives: reconsidering evaluation in open-endedness.
5. Maturana, H., & Varela, F. (1980). Autopoiesis and cognition.
6. Bacon, P. L., Harb, J., & Precup, D. (2017). The Option-Critic Architecture. *Proceedings of the AAAI Conference on Artificial Intelligence*, 31(1).
7. Kauffman, S. A. (1993). *The Origins of Order: Self-Organization and Selection in Evolution*. Oxford University Press.
8. Pattee, H. H. (1995). Evolving a functional epistemology for artificial intelligence. In T. W. Ryan & L. D. Prior (Eds.), *Artificial Intelligence, Expert Systems, and Scientific Computing* (pp. 265-276). Chapman & Hall.
9. Power, A., et al. (2022). Grokking: Generalization Beyond Overfitting on Small Datasets. *arXiv preprint arXiv:2201.02177*.
10. Rosen, R. (1991). *Life Itself: A Comprehensive Inquiry Into the Nature, Origin, and Fabrication of Life*. Columbia University Press.
11. Vezhnevets, A. S., et al. (2017). FeUdal Networks for Hierarchical Reinforcement Learning. *Proceedings of the 34th International Conference on Machine Learning*, PMLR 70, 3591-3600.
12. D'Souza R.M., Arenas A., et al. "Explosive phenomena in complex networks." Advances in Physics (2019).
13. Hagstrom, G. I., & Levin, S. A. (2023). 18 Phase Transitions and the Theory of Early Warning Indicators for Critical Transitions. *How Worlds Collapse: What History, Systems, and Complexity Can Teach Us About Our Modern World and Fragile Future*, 358.
14. Gross B., Havlin S., et al. "Dynamics of cascades in spatial interdependent networks." Chaos (2023).
15. Peters D., Havstad K., et al. "Cross-scale interactions, nonlinearities, and forecasting catastrophic events." PNAS (2004).
16. Sornette D. "Predictability of catastrophic events: Material rupture, earthquakes, turbulence, financial crashes, and human birth." PNAS (2001).
17. Boccaletti S., Zou Y., et al. "Explosive transitions in complex networks' structure and dynamics: Percolation and synchronization." Physics Reports (2016).
18. Lehnertz K. "Time-series-analysis-based detection of critical transitions in real-world non-autonomous systems." Chaos (2024).
19. Liu Z., Yan G., et al. "Early Predictor for the Onset of Critical Transitions in Networked Dynamical Systems." Physical Review X (2024)

# Unconventional Computing

**Ziyun Yan[1], Mario Lanza[1-2]***

[1] Department of Materials Science and Engineering, National University of Singapore, 9 Engineering Drive 1, Building EA, Singapore 117575, Singapore
[2] Institute for Functional Intelligent Materials, National University of Singapore, 4 Science Drive 2, Building S9, Singapore 117544, Singapore

*E-mail: mlanza@nus.edu.sg

### Status

The energy consumed by artificial intelligence (AI) applications is very large because these systems require performing vast amounts of mathematical operations, and for each of them the data has to be transferred between the computational unit and the memory (that is the so-called von Neumann bottleneck) [1]. That is slow and consumes up to 60% of the energy of the entire operation. Currently, AI is implemented in graphics processing units (GPUs) in the cloud, meaning that additional energy is wasted on data transfer to/from the cloud. Extrapolating current trends, by 2030 AI will consume around 20% of the world energy demand [2]. Moreover, such high energy consumption impedes integrating AI directly on edge devices that cannot risk having internet connection interruptions, such as autonomous driving. Under these circumstances, multiple researchers are trying to find alternative ways of computing information, tackling the issue from different angles. Each of these methodologies propose different approaches at the materials, device and system level, and foresee different paths and periods for their implementation in commercial products (if that ends up happening). Among them, the most relevant are:

i) In-memory computing [3]. This is a type of computation that realizes mainly vector-matrix multiplications, a kind of mathematical operation massively used by AI systems, within memory nodes by Kirchhoff's law and Ohm's law, without the need of transferring data.
ii) Neuromorphic computing [4]. This is an asynchronous event-based way of computing information that resembles the functioning of spiking neural networks, which are energetically more efficient.
iii) Quantum computing [5]. This is a radically new computing paradigm that uses the principles of quantum mechanics, such as superposition, entanglement, and interference, to process information in ways that classical computers cannot.
iv) Intelligent materials [6]. The scientific goal of this approach is creating matter that can learn, where its behaviour depends on both its present and history. This matter would have long-term memory, enabling autonomous interaction with its environment and self-regulation of actions.

### Current and future challenges

In the field of in-memory computing, the data can be mapped as the conductance values of memristive devices and computation can be directly performed in-memory. Specifically, by converting input activations into voltage pulses, vector-matrix multiplications can be performed in analogue domain, in place and in parallel, thus achieving high energy efficiency during operation. The main challenge in this direction is that these memory nodes need to exhibit multilevel capacity. Implementation of such memory nodes using Flash memory, dynamic random access memory (DRAM), and static random access memory (SRAM), have been proposed; however, these devices have not been designed for such application and they struggle to exhibit multiple states [7-8]. In this domain, memristive devices have shown potential to enable efficient computing architectures, although they suffer from low reliability and high variability. Still, some small prototypes have shown promising energy efficiencies up to 53 trillions of operations per second per Watt (TOPS/W) [7].

In the field of neuromorphic computing, several groups have proposed the implementation of electronic neurons with leaky integrate-and-fire capability using complementary metal oxide semiconductor (CMOS) circuits. However, the number of transistors and the total area consumed for each electronic neuron is still too large and this prevent upscaling to compete with stat-of-the-art GPUs-based system. Intel's Loihi2 [9] and IBM's NorthPole [10] are two examples of industrial developments, with Loihi2 implementing about 1 million neurons, while NorthPole is realized as a 16×16 core array comprising 22 billion transistors. Using this approach, efficiencies up to 19 TOPS/W have been achieved [11].

In the field of quantum computing, a critical issue is related to the implementation and stability of the qubits. Instead of classical bits (0 or 1), it relies on quantum bits, or qubits, which can exist in multiple states simultaneously, enabling powerful parallelism for certain problems [12], which are: i) Error rates and decoherence. Qubits are fragile and lose information quickly, making long computations unreliable. ii) Scalability. Moving from tens or hundreds of qubits to the millions needed for fault-tolerant quantum computing is extremely difficult. iii) Error correction. Quantum error correction requires large overhead, with thousands of physical qubits needed to form a single "logical" qubit. iv) Hardware diversity. Competing platforms (superconducting circuits, trapped ions, photonics, spin qubits) each face unique engineering hurdles. v) Cryogenics and infrastructure. Many systems need ultra-cold environments and complex support hardware. vi) Algorithms and applications. Identifying practical, near-term use cases where quantum systems clearly outperform classical ones remains an open challenge.

In the field of intelligent materials, some fundamental properties such as intrinsic nonlinearity and reconfigurability [13], history-dependent memory [14], and analogue feature extraction [15] have been demonstrated in academic publications. In combination with simulations, they have been able to claim applicability to handwritten number recognition. However, hardware implementation at the large scale is significantly more challenging and envision of a clear path forward needs to be clarified.

**Advances in science and technology to meet challenges**

In the field of in-memory computing, advances in controlling the physical, chemical, thermal and electronic mechanisms that produce the memristive effect are necessary. Phase-change, ion migration, and spin change have demonstrated good performance as electronic memory and commercial products have been developed [16], but achieving multilevel switching is more complex. In this domain, phase-change and ion migration are well suited because the maximum resistance ratio that can be achieved is larger, which gives space for intermediate states that enable vector matrix multiplication directly in the memory. For phase-change memristors, potential alternatives are three-dimensional stacking, doped chalcogenide alloys, and superlattice-confined structures [16]. For metal-ion based memristor, new strategies to control the diffusion of metal ions into the insulator, including filament confinement, are promising, including the use of molecular crystals [17] and vertical two-dimensional materials [18].

In the field of neuromorphic computing, new device and circuit architectures to implement leaky integrate-and-fire neurons are necessary. Some studies have proposed the use of threshold-type memristors, although scalability seems challenging due to yield issues. A more straightforward approach could be the use of silicon transistors operating in punch through impact ionization regime (using an unconventional biasing protocol), which has shown outstanding reliability and stability [19]. This radically new approach only employs materials and devices that are 100% compatible with the CMOS technology and its supply chain, and now it is only a matter of time that circuit designers employ it to build artificial neural networks.

In the field of quantum computing, progress has focused on improving the scalability, coherence, and operational fidelity of diverse quantum hardware platforms. Algorithmic advances demonstrate that quantum machine learning has potential for computational advantage: neural-network states can efficiently encode entanglement [20,21], and quantum generative adversarial learning has been experimentally realized in superconducting circuits [22]. Theoretical breakthroughs such as the classical shadow method [23] and systematic analyses of challenges and opportunities in quantum machine learning [24] further clarify how quantum devices may process information in ways inaccessible to classical systems. These advances outline a pathway toward establishing quantum computing as a practical unconventional computing paradigm.

In the field of intelligent materials, recent studies have demonstrated that reconfigurable nonlinear processing units (RNPUs) and disordered silicon networks can directly exploit intrinsic material dynamics for computation [13]. Demonstrations in handwritten digit classification and real-time speech recognition at room temperature illustrate their potential for analogue feature extraction [15]. Future progress will depend on the scalable integration of these units into arrays and their co-design with in-memory computing architectures, thereby mitigating the limitations of data movement. Equally important is the establishment of formal theoretical models that capture history-dependent and disordered material responses, providing a principled route to transform variability into reproducible computational functions [14].

## Concluding remarks

Unconventional computing is a term that includes multiple approaches aimed to tackle the limitations of the von Neumann computing architecture, and reduce the energy consumption and increase the performance of current GPU-based AI systems. Some approaches like in-memory computing and neuromorphic computing are closer to realization and require important efforts in terms of upscaling to industrial systems. Quantum computing and intelligent materials are still in an embryonic stage and envisage a longer-term implementation. In this regard, strategies for development need to be pragamatic and consider fundamental implementation challenges at the research stage.

## References

[1] Backus J 1978 Can programming be liberated from the von Neumann style? a functional style and its algebra of programs *Commun. ACM* **21** 613–641 (doi: 10.1145/359576.359579)
[2] Jones N 2018 How to stop data centres from gobbling up the world's electricity *Nature* **561** 163–166 (doi: 10.1038/d41586-018-06610-y)
[3] Yang J J, Strukov D B and Stewart D R 2013 Memristive devices for computing *Nat. Nanotechnol.* **8** 13–24 (doi: 10.1038/nnano.2012.240)
[4] Marković D, Mizrahi A, Querlioz D and Grollier J 2020 Physics for neuromorphic computing *Nat. Rev. Phys.* **2** 499–510 (doi: 10.1038/s42254-020-0208-2)
[5] Gyongyosi L and Imre S 2019 A Survey on quantum computing technology *Comput. Sci. Rev.* **33** 100305 (doi: 10.1016/j.cosrev.2018.11.002)
[6] Qian C, Zhu W, Li X, Shi H, Xu W, Sun L, Chen Y, Zhang K, Wang H, Liu Y et al. 2025 A guidance to intelligent metamaterials and metamaterials intelligence *Nat. Commun.* **16** 492 (doi: 10.1038/s41467-025-56122-3)
[7] MAX78002 Datasheet and Product Info | Analog Devices. https://www.analog.com/en/products/max78002.html.
[8] Mythic M1076 Analog Matrix Processor. https://mythic.ai/products/m1076-analog-matrix-processor/.
[9] Intel 2021 Taking neuromorphic computing with Loihi 2 to the next *Intel Technology Brief*. https://download.intel.com/newsroom/2021/new-technologies/neuromorphic-computing-loihi-2-brief.pdf
[10] Modha D S, Arthur J V, Cassidy A S, Debole M V, Esser S K, Imam N, Lien J W, Moyer B, Naumov M, Paulovicks B, Preissl R, Sawada J, Seo J, Taba B, Tse J S, Williams R D and Andreopoulos A 2023 Neural inference at the frontier of energy, space, and time *Science* **382** 795–803 (doi: 10.1126/science.adg6639)
[11] Gwennap L 2021 Expedera redefines AI acceleration for the edge *The Linley Group White Paper*. https://www.linleygroup.com
[12] de Leon N P, Itoh K M, Kim D, Mehta K K, Northup T E, Paik H, Palmer B S, Samarth N, Sangtawesin S and Schuster D I et al. 2021 Materials challenges and opportunities for quantum computing *Science* **372** eabb2823 (doi: 10.1126/science.abb2823)
[13] Chen T, van Gelder J, van de Ven B, Ahmed I, Banerjee S, Wang S, Famili A, van Nieuwenburg E P L, Singh S, Yuan S et al. 2020 Classification with a disordered dopant-atom network in silicon *Nature* **577** 341–345 (doi: 10.1038/s41586-019-1901-0)
[14] van der Wiel W G, Akhriev A, Akhmerov A R, Aono M, Bandyopadhyay S, Benjamin S C, Bertolazzi S, Bronstein M M, Chen Y, Di Ventra M et al. 2023 Toward a formal theory for computing machines made out of whatever physics offers *Nat. Commun.* **14** 3070 (doi: 10.1038/s41467-023-38674-0)
[15] Zhang Y, Qian C, Nugraha A C, Zhang Q, Li H, Xie L, Gao L, Shi H, Ji Y, Zhang K et al. 2025 Analogue speech recognition based on physical computing *Nature* **625** 263–269 (doi: 10.1038/s41586-024-07892-3)
[16] Lanza M, Villena M A, Yin X, Xie L, Zhou Z, Regan B C, Wang H, Chen X, Ielmini D and Wong H S P et al. 2022 Memristive technologies for data storage, computation, encryption, and radio-frequency communication *Science* **376** eabj9979 (doi: 10.1126/science.abj9979)

[17] Qin L, Guan P, Shao J, Xiao Y, Yu Y, Su J, Zhang C, Li Y, Liu S, Li P et al. 2025 Molecular crystal memristors *Nat. Nanotechnol.* (published 17 Sept 2025) (doi: 10.1038/s41565-025-02013-z)

[18] Xie J, Yekta A E, Al Mamun F, Zhu K, Chen M, Pazos S, Zheng W, Zhang X, Tongay S A, Li X et al. 2025 On-chip direct synthesis of boron nitride memristors *Nat. Nanotechnol.* (published 31 Jul 2025) (doi: 10.1038/s41565-025-01912-5)

[19] Lanza M, Pazos S, Aguirre F, Sebastian A, Le Gallo M, Alam S M, Ikegawa S, Yang J J, Vianello E, Chang M-F et al. 2025 The growing memristor industry *Nature* **640** 613–622 (doi: 10.1038/s41586-025-07489-1)

[20] Carleo G and Troyer M 2017 Solving the quantum many-body problem with artificial neural networks *Science* **355** 602–606 (doi: 10.1126/science.aag2302)

[21] Deng D-L, Li X and Das Sarma S 2017 Quantum entanglement in neural network states *Phys. Rev. X* **7** 021021 (doi: 10.1103/PhysRevX.7.021021)

[22] Hu L, Wu S-T, Cai W, Ma Y, Mu X, Xu Y, Wang H, Song Y, Deng D-L, Zhong Y-P, Xu H and Sun L 2019 Quantum generative adversarial learning in a superconducting quantum circuit *Sci. Adv.* **5** eaav2761 (doi: 10.1126/sciadv.aav2761)

[23] Huang H-Y, Kueng R and Preskill J 2020 Predicting many properties of a quantum system from very few measurements *Nat. Phys.* **16** 1050–1057 (doi: 10.1038/s41567-020-0932-7)

[24] Cerezo M, Arrasmith A, Babbush R, Benjamin S C, Endo S, Fujii K, McClean J R, Mitarai K, Yuan X, Cincio L and Coles P J 2021 Variational quantum algorithms *Nat. Rev. Phys.* **3** 625–644 (doi: 10.1038/s42254-021-00348-9)

# Quantum machine learning via integrated photonics

Fabio Sciarrino[1]

Dipartimento di Fisica, Sapienza Università di Roma, Piazzale Aldo Moro 5, I-00185 Roma, Italy

**E-mail:** fabio.sciarrino@uniroma1.it

**Status**
The current state of the art in photonic quantum computing features advanced techniques for manipulation and exploitation of quantum states of light to perform computational tasks and facilitate quantum communication [5, 17, 18]. Information is encoded into the properties of single photons, such as polarization, phase, and temporal modes, enabling the implementation of various quantum algorithms, particularly using Photonic Integrated Circuits (PICs), combining multiple reconfigurable optical components on a single chip to enhance scalability and performance [1, 11]. Recent technological developments include efficient methods for generating entangled photons, such as quantum dot sources and four-wave mixing, alongside reliable superconducting nanowire single-photon detectors, leading to high-fidelity measurement and manipulation of photonic quantum states [9, 12, 4, 18]. Photonic quantum platforms have successfully demonstrated computational tasks, like Boson Sampling [2], providing a pathway toward quantum advantage [22]. Despite these advancements, challenges such as reducing noise, improving scalability, establishing appropriate hardware architectures, and identifying relevant use cases, remain critical for taking the full benefit from large scale optical quantum computing systems.

There are two distinct frameworks of interest for photonics quantum computing. On the one hand, Boson Sampling is a restricted quantum computation model that uses bosons (photons) evolving through a linear optical network to sample from an output probability distribution that is believed to be hard to do using classical means [2]. Its quantum advantage stems from the inherent complexity of calculating the probabilities related to boson interference patterns, with the downside of limited direct applications mostly in certified randomness generation. Universal photonic quantum computing, on the other hand, aims to create a general-purpose quantum computer using photons as qubits, capable of executing any quantum algorithm through a combination of linear optics, single-photon sources/detectors, and nonlinear elements or adaptive measurements to achieve universality [1, 13, 19]. However, the number of photons and components required for a fault-tolerant quantum computer is out of reach in the short-to-medium-term.

**Current and future challenges**
Current and future challenges for quantum machine learning (QML) with integrated photonics lie at the intersection of several innovative yet complex areas, each offering unique opportunities for exploration and development [11]. Presently, photonic technologies have yet to achieve the full functionality of a universal photonic quantum computer. However, significant achievements have been made towards non-universal models of photon-based computation, notably through architectures based on the Boson Sampling (BS) paradigm. BS models exploit indistinguishable bosons in linear optical networks, generating complex probability distributions that are challenging to sample classically, thus demonstrating potential quantum advantages [2]. Variants like Scattershot Boson Sampling and Gaussian Boson Sampling are built on this principle, though they remain limited in application due to the restricted nature of linear-optical interference. To advance toward universality, the integration of optical nonlinearities becomes essential. This can be accomplished either directly through nonlinear media or indirectly via auxiliary resources coupled with adaptive optical evolution, each presenting distinct technological and resource-related challenges.

In bridging the gap between linear-optical BS and universally applicable photonic computation, intermediate approaches incorporating moderate nonlinearities have emerged. Recent focus has shifted towards enhancing the BS paradigm with elements capable of achieving universality, giving rise to innovations such as the Adaptive Boson Sampling (ABS) paradigm. ABS extends BS by incorporating adaptivity, wherein optical evolution is dynamically reconfigured following intermediate measurements, a concept exemplified by early experimental realizations [7]. Thus, ABS represents a promising pathway toward more versatile quantum computations using photonics.

Parallel to these advances, QML seeks to exploit these quantum capabilities to enhance machine learning tasks, reflected in a proliferation of diverse architectural models [7, 14]. Quantum Extreme

Learning Machines and Quantum Reservoir Computing are approaches introduced recently that map input data into higher-dimensional spaces efficiently, facilitating complex data processing [10]. These techniques illustrate the potential of fixed, random quantum systems in handling time-dependent information and performing sophisticated computations [16, 21]. Quantum Kernel Methods offer another avenue, employing quantum circuits for classification by computing effective kernel functions [7, 20]. Beyond these, photonic quantum systems also contribute to memory-related applications, with quantum simulations of Hopfield memory models and photonic quantum generative adversarial networks highlighting potential applications in data synthesis and recall efficiencies. Specialized quantum classifiers and the emergence of photonic quantum memristors, which propose adaptive neuromorphic circuits, further diversify the optical approach to Quantun Machine Learning. Advanced quantum sampling techniques, such as those used in enhanced image recognition via Gaussian Boson Sampling, demonstrate how interference properties inherent in photonic systems can be harnessed for efficient data processing. Collectively, these innovations signify the rapidly evolving landscape of photonic quantum information processing, charting a course for future advancements and applications in quantum machine learning.

**Advances in science and technology to meet challenges**
Significant efforts are required to address the challenges of realizing effective quantum machine learning (QML) with integrated photonics, and recent developments identify a pathway toward that goal by combining hardware innovation, system-level integration, and theoretical frameworks that target adaptive quantum protocols. Central to these advances is the development of adaptive photonic quantum systems that bridge the gap between linear-optics-limited models and universal quantum computation: progress in source engineering—high-brightness, highly indistinguishable quantum-dot single-photon emitters and polarization-entangled photon sources—provides the on-demand inputs necessary for multi-photon experiments at scale, while fast time-to-space demultiplexers and resonant electro-optic modulators convert temporal resources into spatial modes compatible with large interferometric networks. Fabrication methods such as femtosecond-laser-written waveguide geometries enable reconfigurable, low-loss photonic circuits that can be cascaded into multi-layer architectures combining large number of modes, slowly reconfigurable layers with smaller, fast-reconfigurable circuits; this multi-chip approach would enable processing of many photons across many modes and permits realistic testing of Adaptive Boson Sampling variants, variational algorithms [3, 8], and quantum optical spin machines. Implementing adaptivity requires fast, low-noise switching: phase-preserving fiber-loop delay lines, integrated fast switches, and coherent optical delay infrastructures allow selective mode holding and measurement-triggered reconfiguration, enabling measurement-based control flows and feedback loops inspired by neural networks. Detector and readout technology advances, notably superconducting nanowire single-photon detector arrays with high efficiency and timing resolution, close the loop for real-time adaptive protocols and high-fidelity sampling. Large-scale three-dimensional femtosecond-laser-written photonic circuits has the potential for photonic quantum reservoir computing with a large number of spatial optical modes [6], and coupling selected outputs to fast adaptive modules offers an experimentally accessible route to adaptive quantum reservoir computing. Together, these advances—coordinated across source, modulation, switching, delay, detection, fabrication, and theory— would lead to an integrated photonics platform capable of testing, validating, and exploiting adaptive strategies that enhance expressiveness, mitigate linear-optical limits, and expand the practical reach of quantum machine learning on photonic platforms.

**Concluding remarks**
Photonic platforms have advanced toward practical quantum information processing, but several technological and conceptual challenges need to be addresses. Adaptive integrated photonics offers that route by adding on-demand sources, fast routing and buffering, coherent delays, high-performance detectors, and multilayer reconfigurable chip architectures to linear-optical networks. This intermediate paradigm would enable tailored algorithms, enhanced feature encoding, and feedback-driven training loops that increase computational richness while remaining experimentally accessible [7, 15]. Coordinated progress in device engineering, hybrid integration, control electronics, and theoretical characterization will be essential to validate advantages, identify application niches, and guide scalable designs. Focusing on adaptive architectures that leverage current capabilities while enabling scalable improvements positions the field to move from proof-of-concept to deployable photonic quantum learning systems.

**Acknowledgments**
F.S. acknowledges the ERC Advanced Grant QU-BOSS (QUantum advantage via non-linear BOSon Sampling, Grant Agreement No. 884676), the PNRR MUR project PE0000023-NQSTI and the European Union's Horizon Europe research and innovation program under EPIQUE Project (Grant Agreement No. 101135288).

**References**

[1] J. M. Arrazola, et al., "Quantum circuits with many photons on a programmable nanophotonic chip", Nature 591, 54 (2021).

[2] D. G. Brod, E. F. Galvao, A. Crespi, R. Osellame, N. Spagnolo, F. Sciarrino, "Photonic implementation of boson sampling: a review", Advanced Photonics 1, 034011 (2019).

[3] V. Cimini, M. Valeri, S. Piacentini, F. Ceccarelli, G. Corrielli, R. Osellame, N. Spagnolo, F. Sciarrino, "Variational quantum algorithm for experimental photonic multiparameter estimation", npj Quantum Information 10, 26 (2024).

[4] C. Couteau, S. Barz, T. Durt, T. Gerrits, J. Huwer, R. Prevedel, J. Rarity, A. Shields, G. Weihs, "Applications of single photons to quantum communication and computing", Nat. Rev. Phys. 5, 326 (2023).

[5] F. Flamini, N. Spagnolo, F. Sciarrino, "Photonic quantum information processing: a review", Rep. Progr. Phys. 82, 016001 (2019).

[6] F. Hoch, S. Piacentini, T. Giordani, Z. N. Tian, M. Iuliano, C. Esposito, A. Camillini, G. Carvacho, F. Ceccarelli, N. Spagnolo, A. Crespi, F. Sciarrino, R. Osellame, "Reconfigurable continuously-coupled 3D photonic circuit for Boson Sampling experiments", npj Quantum Information 8, 55 (2022).

[7] F. Hoch, ..., U. Chabaud, R. Osellame, M. Dispenza, F. Sciarrino, "Quantum machine learning with Adaptive Boson Sampling via post-selection", Nature Communications 16, 902 (2025).

[8] F. Hoch, G. Rodari, T. Giordani, P. Perret, N. Spagnolo, G. Carvacho, C. Pentangelo, S. Piacentini, A. Crespi, F. Ceccarelli, R. Osellame, F. Sciarrino, "Variational approach to photonic quantum circuits via the parameter shift rule", Physical Review Research 7, 023227 (2025).

[9] N. Maring, et al., "A versatile single-photon-based quantum computing platform", Nat. Photon. 18, 603–609 (2024).

[10] A. Mehonic, et al., "Roadmap to neuromorphic computing with emerging technologies", APL Mater. 12, 109201 (2024).

[11] E. Pelucchi, et al., "The potential and global outlook of integrated photonics for quantum technologies", Nature Reviews Physics 4, 194 (2022).

[12] M. Pont, et al., "High-fidelity four-photon GHZ states on chip", npj Quantum Inf. 10, 50 (2024).

[13] PsiQuantum team, "A manufacturable platform for photonic quantum computing", Nature 641, 876 (2025).

[14] A. Sakurai, A. Hayashi, W. J. Munro, K. Nemoto, "Quantum optical reservoir computing powered by boson sampling", Optica Quantum 3, 238 (2025).

[15] A. Salavrakos, N. Maring, P.-E. Emeriau, S. Mansfield, "Photon-native quantum algorithms", Mater. Quantum Technol. 5, 023001 (2025).

[16] A. Suprano, D. Zia, L. Innocenti, S. Lorenzo, V. Cimini, T. Giordani, I. Palmisano, E. Polino, N. Spagnolo, F. Sciarrino, G. M. Palma, A. Ferraro, M. Paternostro, "Experimental property reconstruction in a photonic quantum extreme learning machine", Phys. Rev. Lett. 132, 160802 (2024).

[17] J. Wang, F. Sciarrino, A. Laing, M. G. Thompson, "Integrated photonic quantum technologies", Nat. Photonics 14, 273 (2020).

[18] H. Wang, T.C. Ralph, J.J. Renema, C.-Y. Lu, J.-W. Pan, "Scalable photonic quantum technologies", Nature Materials (2025), https://www.nature.com/articles/s41563-025-02306-7.

[19] H. Aghaee Rad, et al., "Scaling and networking a modular photonic quantum computer", Nature 638, 912 (2025).

[20] Z. Yin, I. Agresti, G. de Felice, D. Brown, A. Toumi, C. Pentangelo, S. Piacentini, A. Crespi, F. Ceccarelli, R. Osellame, B. Coecke, P. Walther, "Experimental quantum-enhanced kernel-based machine learning on a photonic processor", Nature Photonics (2025), https://doi.org/10.1038/s41566-025-01682-5.

[21] D. Zia, L. Innocenti, G. Minati, S. Lorenzo, A. Suprano, R. Di Bartolo, N. Spagnolo, T. Giordani, V. Cimini, G. M. Palma, A. Ferraro, F. Sciarrino, M. Paternostro, Quantum extreme learning machines for photonic entanglement witnessing", arXiv:2502.18361 (2025).

[22] H. S. Zhong, et al., "Quantum computational advantage using photons", Science 370, 1460 (2020).

# 7 simple rules for machine learning in chemical space with efficiency, accuracy, transferability, and scalability (EAST)

O. Anatole von Lilienfeld[1,2,3]

[1]Departments of Chemistry, Materials Science and Engineering, Physics, University of Toronto, Toronto, Canada
[2]Vector Institute, Toronto, Canada
[3]Acceleration Consortium, University of Toronto, Toronto, Canada

**E-mail:** anatole.vonlilienfeld@utoronto.ca

**Status**
Chemical compound space (CCS) can be defined as the set of all conceivable stable compounds, i.e. compositions and atomic configurations which correspond to free energy minima surrounded by barriers sufficiently large for life times not to vanish [14]. Notwithstanding recent contributions by researchers including employees from IT giants such as Meta [21], Google [8], Microsoft [5], or Nvidia [1], to simulate certain molecular regions and properties, it is so vast that even just simulated data is scarce in general. Its vastness is due to the combinatorial scaling with size for all the different elements one can draw from the periodic table (compositional isomers), for all the different bonding graphs (constitutional isomers), for all the different internal degrees of freedom (conformational isomers), for all electronic states (spectrum). Addition of further variables such as reaction and activation energies involving other reactants, mixtures thereof, or different external influences (fields, radiation, solvents, temperature, etc) obviously only worsens the combinatorial explosion. While machine learning based generalization within certain molecular regions and properties was shown to succeed beyond any experimentally meaningful precision [26, 10], obvious extrapolation tasks such as the prediction of other observables such as reaction maps, upgrading label quality to experimental (or higher level theory, e.g. explicitly correlated electronic structure methods), or generalizing towards new compounds, have remained elusive. Consequently, one can either resort to sample CCS at random [9], or to restrict it such that exhaustive enumeration is possible [7, 17, 12].

Alternatively—and more straightforwardly— training data needs can be reduced through use of first principles physics based thinking which can be embedded in order to increase data efficiency, accuracy, scalability, and transferability (EAST). In particular, and as we have argued before when faced with the daunting, formally combinatorially, scaling of chemical space, a first principles view on CCS is crucial for the successful development of EAST (efficient, accurate, scalable, transferable) models [14, 16].

**Current and future challenges**
Clearly, it would be desirable to know how to generally obtain machine learning models that exhibit superior EAST and without "cheating", i.e. without increasing the training set size. Strategies to best navigate the trade-off between cost and EAST are not without precedent within the astomistic simulation communities. For example, when performing electronic structure calculations, useful strategies emerged thanks to a clear and (usually) reliable trade-off between computational simulation cost and predictive accuracy. More specifically, Pople's model chemistry [24] as well as Jacob's ladder proposed by Perdew [23] have served as useful guidance for theoretically oriented chemists for many years. In complete analogy, a conceptual map would also be desirable for designing and developing machine learning models applicable throughout chemical space in order to identify and distinguish the more useful strategies from less successful attempts. When it comes to the assessment of machine learning models, learning curves (scaling laws) were introduced in the past [4, 22]. Fig. 1 (Left) illustrates graphically how they can be put to good use for generically assessing the relative performance of different models trained and tested on the same data, see e.g. Ref. [6]. In particular, and as discussed below, the generalization error's offset and slope with respect to training set size depend dramatically on the model's architecture and set-up. We note that it has remained unclear in general, however, what factors will determine off-set and slope, i.e. which factors generally govern the overall data needs, as well as training and inference efficiency.

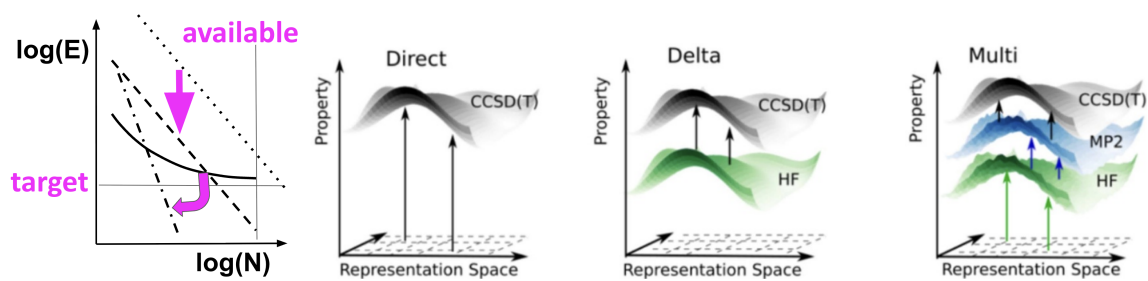

**Figure 1.** Left: Off-set and slope of learning curves (prediction error vs. training set size) can be improved using the 7 simple rules. Right: Illustration how training data needs can be mitigated through rule No. IV by going from direct learning to $\Delta$ learning to multi-level learning (reproduced from Ref. [11]).

**Advances in science and technology to meet challenges**
Following the strategic goal to measure progress in terms of learning curves, seven simple rules (dubbed 'the magnificent seven') have emerged from our work over the course of the last 14 years, typically resulting in improved EAST. More specifically, and in no particular order, we have found significant improvements in learning curves in terms of the following 7 rules:

I input representations, a.k.a. feature engineering or descriptor selection, see e.g. Ref. [19].

II similarity measures, e.g. when selecting distance metrics, or kernel or activation functions, see e.g. Ref. [2]

III loss functions, can be used to incorporate known constraints, such as differential relationships between labels [3]

IV $\Delta$-learning [25], multi-fidelity learning [29], or transfer learning [27] to benefit from heterogeneous data sets with labels of varying fidelity. Direct learning, $\Delta$-learning and multi-fidelity learning are illustrated in Fig. 1 (Right).

V Amon, or fragment based learning, enabling scalable models [13]

VI training set selections geared towards known inference tasks, e.g. Ref. [20]

VII adaptive learning of model parameters, previously and conventionally fixed to globally fit an entire reference data set, e.g. Ref. [18].

Note that changes in learning rate (slope) have been observed for rules V and VI, while all other rules merely result in a reduction of off-set. Also note that some of these rules for sampling chemical compound space with EAST were already known before. In particular, the importance of rule No. I for building predictive machine learning models has already been well known for decades, as summarized in the Handbook of molecular descriptors in 2009 [28].

**Concluding remarks**
We have briefly motivated and described the emergence of seven rules which improve EAST when training machine learning models and using them to predict properties across chemical space. Future work might elucidate to which degree these seven rules can be extended, or simply combined to benefit from synergies. For example, fragment based learning has recently been combined with $\Delta$ ML in order to infer quantum Monte Carlo energies for large molecules using DFT energies of small amons for training [15].

**Acknowledgments**
We acknowledge the support of the Natural Sciences and Engineering Research Council of Canada (NSERC), [funding reference number RGPIN-2023-04853]. Cette recherche a été financée par le Conseil de recherches en sciences naturelles et en génie du Canada (CRSNG), [numéro de référence RGPIN-2023-04853]. This research was undertaken thanks in part to funding provided to the University of Toronto's Acceleration Consortium from the Canada First Research Excellence Fund, grant number: CFREF-2022-00042. O.A.v.L. has received support as the Ed Clark Chair of Advanced Materials, and as a Canada CIFAR AI Chair. O.A.v.L. has received funding from the European Research Council (ERC) under the European Union's Horizon 2020 research and innovation programme (grant agreement No. 772834).

# References

[1] Alice EA Allen, Rui Li, Sakib Matin, Xing Zhang, Benjamin Nebgen, Nicholas Lubbers, Justin S Smith, Richard Messerly, Sergei Tretiak, Garnet Kin Chan, et al. Reactive chemistry at unrestricted coupled cluster level: High-throughput calculations for training machine learning potentials. *arXiv preprint arXiv:2509.10872*, 2025.

[2] Onur Çaylak, O Anatole von Lilienfeld, and Björn Baumeier. Wasserstein metric for improved quantum machine learning with adjacency matrix representations. *Machine Learning: Science and Technology*, 1(3):03LT01, 2020.

[3] Anders S Christensen, Felix A Faber, and O Anatole von Lilienfeld. Operators in quantum machine learning: Response properties in chemical space. *The Journal of Chemical physics*, 150(6):064105, 2019.

[4] Corinna Cortes, Lawrence D Jackel, Sara A Solla, Vladimir Vapnik, and John S Denker. Learning curves: Asymptotic values and rate of convergence. In *Advances in Neural Information Processing Systems*, pages 327–334, 1993.

[5] Sebastian Ehlert, Jan Hermann, Thijs Vogels, Victor Garcia Satorras, Stephanie Lanius, Marwin Segler, Derk P Kooi, Kenji Takeda, Chin-Wei Huang, Giulia Luise, et al. Accurate chemistry collection: Coupled cluster atomization energies for broad chemical space. *arXiv preprint arXiv:2506.14492*, 2025.

[6] Felix A Faber, Luke Hutchison, Bing Huang, Justin Gilmer, Samuel S Schoenholz, George E Dahl, Oriol Vinyals, Steven Kearnes, Patrick F Riley, and O Anatole von Lilienfeld. Prediction errors of molecular machine learning models lower than hybrid DFT error. *J. Chem. Theory Comput.*, 13:5255–5264, 2017.

[7] T. Fink, H. Bruggesser, and J.-L. Reymond. Virtual exploration of the small-molecule chemical universe below 160 Daltons. *Angew. Chem. Int. Ed.*, 44:1504, 2005.

[8] Stefan Ganscha, Oliver T Unke, Daniel Ahlin, Hartmut Maennel, Sergii Kashubin, and Klaus-Robert Müller. The qcml dataset, quantum chemistry reference data from 33.5 m dft and 14.7 b semi-empirical calculations. *Scientific Data*, 12(1):406, 2025.

[9] Thomas Gasevic, Marcel Müller, Jonathan Schöps, Stephanie Lanius, Jan Hermann, Stefan Grimme, and Andreas Hansen. Chemical space exploration with artificial "mindless" molecules. *Journal of Chemical Information and Modeling*, 2025.

[10] Justin Gilmer, Samuel S. Schoenholz, Patrick F. Riley, Oriol Vinyals, and George E. Dahl. Neural message passing for quantum chemistry. In *Proceedings of the 34th International Conference on Machine Learning, ICML 2017*, 2017.

[11] Stefan Heinen, Danish Khan, Guido Falk von Rudorff, Konstantin Karandashev, Daniel Jose Arismendi Arrieta, Alastair JA Price, Surajit Nandi, Arghya Bhowmik, Kersti Hermansson, and O Anatole von Lilienfeld. Reducing training data needs with minimal multilevel machine learning (m3l). *Machine Learning: Science and Technology*, 5(2):025058, 2024.

[12] Bing Huang and O Anatole von Lilienfeld. Dictionary of 140k gdb and zinc derived amons. *arXiv preprint arXiv:2008.05260*, 2020.

[13] Bing Huang and O Anatole von Lilienfeld. Quantum machine learning using atom-in-molecule-based fragments selected on the fly. *Nature chemistry*, 12(10):945–951, 2020.

[14] Bing Huang and O. Anatole von Lilienfeld. Ab initio machine learning in chemical compound space. *Chem. Rev.*, 121:10001, 2021.

[15] Bing Huang, O. Anatole von Lilienfeld, Jaron T. Krogel, and Anouar Benali. Toward DMC Accuracy Across Chemical Space with Scalable $\Delta$-QML. *Journal of Chemical Theory and Computation*, 19(6):1711–1721, March 2023. Publisher: American Chemical Society.

[16] Bing Huang, Guido Falk von Rudorff, and O Anatole von Lilienfeld. The central role of density functional theory in the ai age. *Science*, 381(6654):170–175, 2023.

[17] Danish Khan, Anouar Benali, Scott YH Kim, Guido Falk von Rudorff, and O Anatole von Lilienfeld. Quantum mechanical dataset of 836k neutral closed-shell molecules with up to 5 heavy atoms from c, n, o, f, si, p, s, cl, br. *Scientific Data*, 12(1):1551, 2025.

[18] Danish Khan, Alastair JA Price, Bing Huang, Maximilian L Ach, and O Anatole von Lilienfeld. Adapting hybrid density functionals with machine learning. *Science Advances*, 11(5):eadt7769, 2025.

[19] Danish Khan and O. Anatole von Lilienfeld. Generalized convolutional many-body distribution functional representations. *Proceedings of the National Academy of Sciences*, 122(41), October 2025.

[20] Dominik Lemm, Guido Falk von Rudorff, and O Anatole von Lilienfeld. Improved decision making with similarity based machine learning: applications in chemistry. *Machine Learning: Science and Technology*, 4(4):045043, 2023.

[21] Daniel S Levine, Muhammed Shuaibi, Evan Walter Clark Spotte-Smith, Michael G Taylor, Muhammad R Hasyim, Kyle Michel, Ilyes Batatia, Gábor Csányi, Misko Dzamba, Peter Eastman, et al. The open molecules 2025 (omol25) dataset, evaluations, and models. *arXiv preprint arXiv:2505.08762*, 2025.

[22] K. R. Müller, M. Finke, N. Murata, K. Schulten, and S. Amari. A numerical study on learning curves in stochastic multilayer feedforward networks. *Neural Comp.*, 8:1085, 1996.

[23] John P Perdew and Karla Schmidt. Jacob's ladder of density functional approximations for the exchange-correlation energy. 577(1):1–20, 2001.

[24] John A Pople. Nobel lecture: Quantum chemical models. *Reviews of Modern Physics*, 71(5):1267, 1999.

[25] R. Ramakrishnan, P. Dral, M. Rupp, and O. A. von Lilienfeld. Big Data meets Quantum Chemistry Approximations: The $\Delta$-Machine Learning Approach. *J. Chem. Theory Comput.*, 11:2087, 2015.

[26] M. Rupp, A. Tkatchenko, K.-R. Müller, and O. A. von Lilienfeld. Fast and accurate modeling of molecular atomization energies with machine learning. *Phys. Rev. Lett.*, 108:058301, 2012.

[27] Justin S Smith, Benjamin T Nebgen, Roman Zubatyuk, Nicholas Lubbers, Christian Devereux, Kipton Barros, Sergei Tretiak, Olexandr Isayev, and Adrian E Roitberg. Approaching coupled cluster accuracy with a general-purpose neural network potential through transfer learning. *Nature communications*, 10(1):1–8, 2019.

[28] R. Todeschini and V. Consonni. *Handbook of Molecular Descriptors*. Wiley-VCH, Weinheim, 2009.

[29] Peter Zaspel, Bing Huang, Helmut Harbrecht, and O Anatole von Lilienfeld. Boosting quantum machine learning models with multi-level combination technique: Pople diagrams revisited. *Journal of chemical theory and computation*, 15(3):1546–1559, 2018.

# Strategic Frameworks for Building Data Platforms in AI for Materials Science and Chemistry

**Ryo Yoshida[1,2]**

[1] The Institute of Statistical Mathematics, Research Organization of Information and Systems, Tachikawa, Tokyo 190-8562, Japan  
[2] Advanced General Intelligence for Science Program (AGIS), TRIP Headquarters, RIKEN, Wako, Saitama 351-0198, Japan

E-mail: yoshidar@ism.ac.jp

**Status**

The material space is extraordinarily vast. The central objective of data-driven materials research is to predict and discover novel materials with unprecedented properties and functions within such vast unexplored spaces, with AI serving as a key technical enabler.

The workflow of data-driven materials research is commonly framed in terms of forward and inverse problems [1]. The forward problem aims to predict the output corresponding to a given system input. Typical inputs include compositional and structural features of materials, while the outputs correspond to material properties. By applying machine learning techniques to a dataset of input–output pairs, a forward predictive model is constructed. Then, by exploring its inverse mapping, candidate materials that achieve desired properties are computationally identified. This conceptual framework is broadly applicable and underlies diverse tasks in materials research. This field has been termed materials informatics (MI). Since the mid-to-late 2010s, the paradigm of MI has spread widely across materials science and chemistry. Today, leveraging AI technologies, a wide variety of new materials are being discovered [2-7].

**Current and future challenges**

The amount of data available in materials research remains significantly limited. This data scarcity can be attributed to three primary factors: (a) the high cost of data generation, (b) the diversity of researchers' interests, which hampers the development of a community-wide culture of data sharing, and (c) the strong tendency to protect information from competitors, which reduces incentives for researchers to make their laboratory data publicly available. Since addressing these challenges necessitates a cultural shift, they are unlikely to be resolved in the short to medium term.

To overcome these limitations, computational materials databases, particularly those derived from first-principles calculations, have played a central role in advancing data-driven research. In inorganic materials and organic small molecules, large-scale first-principles property databases have been developed [8-12], comprising on the order of $10^5$–$10^6$ distinct materials. These datasets have served as critical enablers of groundbreaking AI technologies, such as general-purpose machine-learning interatomic potentials [13].

In contrast, the development of databases for polymeric materials has lagged significantly. Polymers, as large-scale and non-periodic systems, are not suited for first-principles calculations. In addition, molecular dynamics (MD) simulations capable of capturing their large structural complexity and long-time behaviour demand prohibitive computational costs. Moreover, automating the modelling of diverse morphologies remains a major challenge. Computational workflows for polymer systems also involve numerous parameters, many of which are determined empirically based on expert knowledge. Such nontrivial parameter tuning constitutes a critical bottleneck for fully automating the entire simulation process.

**Advances in science and technology to meet challenges**

To overcome the challenge of data scarcity in polymer research, we are developing RadonPy, a fully automated pipelining software designed to generate a comprehensive database of polymeric materials through MD-based computational experiments [14]. Given inputs such as polymer repeat units, degrees of polymerization, and temperature conditions, RadonPy executes a fully automated workflow encompassing charge calculations, force-field parameter assignment, initial structure generation, equilibrium and non-equilibrium MD simulations. The latest version implements automated algorithms for 62 distinct properties, across a wide range of polymer systems.

With RadonPy, we are developing PolyOmics, a polymer materials database that spans vast regions of structural and property space for approximately $10^5$ distinct polymers, by operating a supercomputer Fugaku for nearly five years. To accelerate development, we established an academia-industry consortium involving approximately 260 members from 3 national research institutes, 8 universities, and 37 companies, collaborating on the development of automated algorithms for RadonPy and large-scale data production for diverse polymeric systems. This unprecedented, comprehensive database provides crucial scientific insights into previously unexplored regions of polymer materials space.

PolyOmics was developed as a data resource for machine learning. In materials research, simulation-to-real (Sim2Real) transfer learning has played a key role in overcoming the shortage of experimental data [15-18]. In this framework, foundation models trained on the large-scale computational datasets are fine-tuned using limited experimental data to construct predictors for real-world systems, thereby achieving generalization performance unattainable by training from scratch. A crucial aspect of Sim2Real transfer is the presence of scaling laws: as the volume of computational data increases, the generalization performance of fine-tuned models in real-world systems improves monotonically following a power-law relation [19]. Establishing such a scalable Sim2Real transfer workflow provides a clear decision-making strategy—namely, to continue investing computational resources until the target level of generalization is reached. We found that PolyOmics exhibits scalable transfer across a wide range of downstream real-world tasks.

**Concluding remarks**

The most critical barrier in AI for materials science and chemistry lies in the scarcity of data resources. In general, as research approaches the scientific frontier, the cost of data generation increases, thereby exacerbating data scarcity. Even with advances in high-throughput computational experiments or self-driving laboratories, many domains will remain data-poor.

The only viable path to address this challenge is a hierarchical data platform strategy. This approach involves designating foundational domains where high-throughput technologies—such as high-throughput simulations and laboratory automation—can generate abundant data, and then leveraging machine learning to bridge the domain gap with advanced areas where data productivity remains low. As the volume of data in foundational domains increases, the generalization performance of AI across a wide range of downstream tasks improves, following scaling laws. We contend that the construction of such scalable data-generation workflows constitutes the most important challenge for the future of AI in materials science.

**Acknowledgements**

This research was supported by the Ministry of Education, Culture, Sports, Science and Technology (MEXT) through the "Program for Promoting Researches on the Supercomputer Fugaku" (JPMXP1020200314), the Japan Science and Technology Agency (JST) (JPMJCR22O3, JPMJCR2332, JPMJCR2546) and the Japan Society for the Promotion of Science (JSPS) (25H01126).

## References

[1] Ikebata, H., Hongo, K., Isomura, T., Maezono, R., Yoshida, R., Bayesian molecular design with a chemical language model. Journal of Computer-Aided Molecular Design 31, 379–391 (2017).

[2] Wu, S., Kondo, Y., Kakimoto, M., Yang, B., Yamada, H., Kuwajima, I., Lambard, G., Hongo, K., Xu, Y., Shiomi, J., Schick, C., Morikawa, J., Yoshida, R., Machine-learning-assisted discovery of polymers with high thermal conductivity using a molecular design algorithm. npj Computational Materials 5, 66 (2019).

[3] Maeda, H., Wu, S., Marui, R., Yoshida, E., Hatakeyama-Sato, K., Nabae, Y., Nakagawa, S., Ryu, M., Ishige, R., Noguchi, Y., Hayashi, Y., Ishii, M., Kuwajima, I., Jiang, F., Vu, X.T., Ingebrandt, S., Tokita, M., Morikawa, J., Yoshida, R., Hayakawa, T., Discovery of liquid crystalline polymers with high thermal conductivity using machine learning. npj Computational Materials 11, 205 (2025).

[4] Nanjo, S., Arifin, Maeda, H., Hayashi, Y., Hatakeyama-Sato, K., Himeno, R., Hayakawa, T., Yoshida, R., SPACIER: on-demand polymer design with fully automated all-atom classical molecular dynamics integrated into machine learning pipelines. npj Computational Materials 11, 16 (2025).

[5] Liu, C, Fujita. E., Katsura, Y., Inada, Y., Ishikawa, A., Tamura, R., Kimura, K., Yoshida, R., Machine learning to predict quasicrystals from chemical compositions. Advanced Materials 33, 2102507 (2021).

[6] Liu, C, Kitahara, K., Ishikawa, A., Hiroto, T., Singh, A., Fujita, E., Katsura, Y., Inada, Y., Tamura, R., Kimura, K., Yoshida, R., Quasicrystals predicted and discovered by machine learning. Physical Review Materials 7, 093805 (2023).

[7] Uryu, H. et al., Deep learning enables rapid identification of a new quasicrystal from multiphase powder diffraction patterns. Advanced Science 11, 2304546 (2024).

[8] Jain, A. et al., The materials project: a materials genome approach to accelerating materials innovation. APL Materials 1, 011002 (2013).

[9] Curtarolo, S. et al., AFLOW: an automatic framework for high-throughput materials discovery. Computational Materials Science 58, 218–226 (2012).

[10] Merchant, A., Batzner, S., Schoenholz, S.S., Aykol, M., Cheon, G., Cubuk, E.D., Scaling deep learning for materials discovery. Nature 624, 80–85 (2023).

[11] Barroso-Luque, L., Shuaibi, M., Fu, X., Wood, B.M., Dzamba, M., Gao, M., Rizvi, A., Lawrence Zitnick, C., Ulissi, Z.W., Open Materials 2024 (OMat24) inorganic materials dataset and models. arXiv preprint arXiv:2410.12771 (2024).

[12] Ramakrishnan, R., Dral, P. O., Rupp, M., von. Lilienfeld, O. A., Quantum chemistry structures and properties of 134 kilo molecules. Scientific Data 1, 140022 (2014).

[13] Yang, H. et al., MatterSim: A deep learning atomistic model across elements, temperatures and pressures. arXiv preprint arxiv.2405.04967 (2024).

[14] Hayashi, Y., Shiomi, J., Morikawa, J., Yoshida, R., RadonPy: automated physical property calculation using all-atom classical molecular dynamics simulations for polymer informatics. npj Computational Materials 8, 222 (2022).

[15] Yamada, H., Liu, C., Wu, S., Koyama, Y., Ju, S., Shiomi, J., Morikawa, J., Yoshida, R., Predicting materials properties with little data using shotgun transfer learning. ACS Central Science 5, 1717–1730 (2019).

[16] Ju, S., Yoshida, R., Liu, C., Wu, S., Hongo, K., Tadano, T., Shiomi, J., Exploring diamondlike lattice thermal conductivity crystals via feature-based transfer learning. Physical Review Materials 5, 053801 (2021).

[17] Aoki, Y., Wu, S., Tsurimoto, T., Hayashi, Y., Minami, S., Okubo, T., Shiratori, K., Yoshida, R., Multitask machine learning to predict polymer–solvent miscibility using Flory–Huggins interaction parameters. Macromolecules 56, 5446–5456 (2023).

[18] Minami, S., Fukumizu, K., Hayashi, Y., Yoshida, R., Transfer learning with affine model transformation. Advance in Neural Information Processing Systems 36, 17296-17329 (2023).

[19] Minami, S., Hayashi, Y., Wu, S., Fukumizu, K., Sugisawa, H., Ishii, M., Kuwajima, I., Shiratori, K., Yoshida, R., Scaling law of Sim2Real transfer learning in expanding computational materials databases for real-world predictions. npj Computational Materials 11, 146 (2025).

# Generative Models for Solid-State Inorganic Materials: From Perfect Code to Imperfect World

Nikita Kazeev[1]

[1]Institute for Functional Intelligent Materials, University of Singapore, Singapore

**E-mail:** kna@nus.edu.sg

**Status**
Materials science is an applied science – its goal is to create new useful materials. The field as a whole is a generative model, which samples materials conditioned on the target properties. So it's only natural that efforts are underway to make this philosophical preposition a computational reality.

Materials are 3D point clouds of a variable number of atoms. Their properties are invariant towards permutation, rotation and translation. As such, materials present a unique data modality for machine learning models and their representation posed a challenge in the early stage of the field. It has largely been addressed by graph neural networks [XG18] and specialized symmetry-based representations [Goo+22]. They form the core of generative models. Most of them consist of a diffusion or a transformer model trained on a large database produced with Density Functional Theory (DFT) [Hor+25; GVV18; Saa+13]. These models are capable of generating novel thermodynamically stable materials [Zen+25; Oka+25], but not quite "solve" material science. We review the limitations in the next section.

**Current and future challenges**
*Simulated data* Machine learning algorithms have an accuracy ceiling in the form of the quality of the training data – and DFT has systematic errors. They can be classified as functional-driven, due to the approximate nature of the exchange-correlation (XC) functional, and density-driven, when an approximate XC functional produces a qualitatively incorrect electron density [KSB13]. Moreover, most of the important properties, such as mechanical, catalytic, and electron transport, are very expensive to compute even at DFT level of accuracy [Yam+25], resulting in small datasets, and difficulty verifying the results on models' prediction.

Complicating the issue further, real materials are not perfect crystals. Disorder and defects are critically important for conductive, mechanical, optical, magnetic and chemical properties. Modeling them further increases costs by requiring a large simulation cell. Finally, high-throughput modeling of the material synthesis process is totally unfeasible. Processes like crystal growth, sintering, or chemical vapor deposition occur over long timescales and involve complex interactions between multiple phases, which are computationally prohibitive to simulate atomistically. The thermodynamic stability predicted by DFT only indicates that a material will not spontaneously decompose; it says nothing about whether there exists a kinetic pathway to form it from available precursors under achievable laboratory conditions [KSJ25].

*Experimental data* Simulation has a problem in veracity of the causal link between inputs and outputs. Experiment, on the other hand, perfectly links a set of synthesis conditions and a final, tangible material with measurable properties. The problem is, however, that both input and output states are largely not observable. Recorded atomic positions, dynamic reaction pathways, and the local chemical environments are a matter of course for simulation, but are largely not observable in experiments [Vas+19]. The very notable and welcome exception is the determination of atomic structures for pure crystalline materials by X-ray diffraction (XRD), etc. [Zag+19]. Experiments require specialized hardware and are generally expensive, limiting both dataset sizes and the ability to evaluate model predictions.

*Bias towards known material families* Generative models by design learn a probability distribution from the training data; they are very unlikely to generate out-of-distribution novel exotic materials [HM25]. The problem is common for all kinds of heuristic models, but is especially incapacitating for generative models. A mistake by a regression model can be detected by the downstream pipeline, at a cost. But if a new family of materials is never generated, there is no way to learn of its existence.

**Advances in science and technology to meet challenges**
From the applied point of view, generative models currently piggy-back on the existing material design pipeline by supplying candidates to the screening pipelines. Proxy variables, such as chemical system and band gap, are used as property inputs – not because they correspond well to the design targets, but because they are available.

In general, machine learning models are made better by increasing the quality and quantity of the training data. In this respect, there are efforts to combine all the available DFT data into a single dataset [Sir+]; combine simulation data with different fidelity [CRM25].

Active learning, as opposed to training on static datasets, addresses the out of distribution generalization challenge [Ngu+23; Mer+23; Han+25; Sol+24]. The concept is well-known, but the labor and computational costs of a practical implementation mean such works are rare.

The idea of combining all the available data into one multimodal multiscale foundational ML model is fairly obvious and frequently discussed. Doing so involves conceptual algorithmic design challenges, along with a costly and laborious practical implementation. A step into this direction has been taken with LLMs, which are used to combine the structured data in databases with unstructured information from literature [Sri+24], along with heuristic predictions of synthesizability [Son+25]

Conceptually, materials science studies approximate solution of Schrödinger equation for practically useful systems. As such, there is a rough multiscale hierarchy of methods, where more precise low-level computations are used as a basis for faster larger-scale models: quantum chemistry $\rightarrow$ DFT $\rightarrow$ molecular dynamics $\rightarrow$ continuous modeling. AI is being used to improve speed and accuracy of those models, with most prominent successes being electron density prediction [Gri+18] and machine learning interatomic potentials [Jac+25]. These approaches extend the capability of computer simulation to more interesting and useful properties.

Finally, self-driving labs (SDL) are making experiments cheaper and more accessible [Cor+18]. While they don't fundamentally address the core issues forming the gap between experiment and simulations, the orders of magnitude increase in the amount of the data holds a great promise on its own. Crucially, automated laboratories provide a way to quickly evaluate predictions of ML models, enabling active learning. Moreover, an SDL can provide a continuous stream of data, enabling the modeling and control of the synthesis process.

**Concluding remarks**
The development of generative models for solid-state inorganic materials represents a significant leap towards the on-demand creation of materials with desired properties. While challenges persist in addressing the limitations of simulated data, the sparsity of experimental data, and the inherent bias towards known material families, ongoing advancements are promising. The integration of high-quality, diverse training data, the application of active learning strategies, and the emergence of multimodal foundational models are pushing the boundaries of what's possible. Furthermore, the increasing capabilities of AI in accelerating and refining multiscale computational methods, coupled with the proliferation of self-driving labs, are poised to revolutionize material discovery by providing a continuous feedback loop between prediction and experimental verification. These efforts collectively pave the way for a future where novel and exotic materials can be designed and synthesized with unprecedented efficiency.

**Acknowledgments**
This research/project is supported by the Ministry of Education, Singapore, under its Research Centre of Excellence award to the Institute for Functional Intelligent Materials (I-FIM, project No. EDUNC-33-18-279-V12). This research is supported by the National Research Foundation, Singapore under its AI Singapore Programme (AISG Award No: AISG3-RP-2022-028).

We have used LLMs. Namely we have supplied the draft of this chapter with the following prompt: "Critically assess the following roadmap article draft. Fill in in the TODOs. Are the claims there well-substantiated? Are there any works that contradict the presented paradigm? Are there any important works or perspectives which are missing?" to Gemini Deep Research, ChatGPT Deep Research, Qwen Deep Research, Grok, DeepSeek, Elicit, FutureHouse Falcon. We have also used Gemini with the following prompt on the semi-final verison: "Critically assess the following roadmap mini-paper. Cover both the high-level substance and writing details."

**References**

[KSB13]   Min-Cheol Kim, Eunji Sim, and Kieron Burke. "Understanding and Reducing Errors in Density Functional Calculations". In: *Phys. Rev. Lett.* 111 (7 Aug. 2013), p. 073003. DOI: 10.1103/PhysRevLett.111.073003. URL: https://link.aps.org/doi/10.1103/PhysRevLett.111.073003.

[Saa+13]  James E Saal et al. "Materials design and discovery with high-throughput density functional theory: the open quantum materials database (OQMD)". In: *Jom* 65.11 (2013), pp. 1501–1509.

[Cor+18]  Juan-Pablo Correa-Baena et al. "Accelerating materials development via automation, machine learning, and high-performance computing". In: *Joule* 2.8 (2018), pp. 1410–1420.

[GVV18]   Mohammad M Ghahremanpour, Paul J Van Maaren, and David Van Der Spoel. "The Alexandria library, a quantum-chemical database of molecular properties for force field development". In: *Scientific data* 5.1 (2018), pp. 1–10.

[Gri+18]  Andrea Grisafi et al. "Transferable machine-learning model of the electron density". In: *ACS central science* 5.1 (2018), pp. 57–64.

[XG18]    Tian Xie and Jeffrey C. Grossman. "Crystal Graph Convolutional Neural Networks for an Accurate and Interpretable Prediction of Material Properties". In: *Phys. Rev. Lett.* 120 (14 Apr. 2018), p. 145301. DOI: 10.1103/PhysRevLett.120.145301. URL: https://link.aps.org/doi/10.1103/PhysRevLett.120.145301.

[Vas+19]  Rama K Vasudevan et al. "Materials science in the AI age: high-throughput library generation, machine learning and a pathway from correlations to the underpinning physics". In: *MRS communications* 9.3 (2019), pp. 10–1557.

[Zag+19]  Dejan Zagorac et al. "Recent developments in the Inorganic Crystal Structure Database: theoretical crystal structure data and related features". In: *Applied Crystallography* 52.5 (2019), pp. 918–925.

[Goo+22]  Rhys EA Goodall et al. "Rapid discovery of stable materials by coordinate-free coarse graining". In: *Science advances* 8.30 (2022), eabn4117.

[Mer+23]  Amil Merchant et al. "Scaling deep learning for materials discovery". In: *Nature* 624.7990 (2023), pp. 80–85.

[Ngu+23]  Tri Minh Nguyen et al. "Hierarchical gflownet for crystal structure generation". In: *AI for Accelerated Materials Design-NeurIPS 2023 Workshop*. 2023.

[Sol+24]  Mohammad Soleymanibrojeni et al. "An active learning approach to model solid-electrolyte interphase formation in Li-ion batteries". In: *Journal of Materials Chemistry A* 12.4 (2024), pp. 2249–2266.

[Sri+24]  Anuroop Sriram et al. "FlowLLM: Flow matching for material generation with large language models as base distributions". In: *Advances in Neural Information Processing Systems* 37 (2024), pp. 46025–46046.

[CRM25]   Mengnan Cui, Karsten Reuter, and Johannes T Margraf. "Multi-fidelity transfer learning for quantum chemical data using a robust density functional tight binding baseline". In: *Machine Learning: Science and Technology* 6.1 (2025), p. 015071.

[Han+25]  Xiao-Qi Han et al. *InvDesFlow-AL: Active Learning-based Workflow for Inverse Design of Functional Materials*. 2025. arXiv: 2505.09203 [cond-mat.mtrl-sci].

[HM25]    Albertus Denny Handoko and Riko I Made. *Artificial Intelligence and Generative Models for Materials Discovery – A Review*. 2025. arXiv: 2508.03278 [cond-mat.mtrl-sci]. URL: https://arxiv.org/abs/2508.03278.

[Hor+25]  Matthew K Horton et al. "Accelerated data-driven materials science with the Materials Project". In: *Nature Materials* (2025), pp. 1–11.

[Jac+25]  Ryan Jacobs et al. "A practical guide to machine learning interatomic potentials–Status and future". In: *Current Opinion in Solid State and Materials Science* 35 (2025), p. 101214.

[KSJ25]   Seongmin Kim, Joshua Schrier, and Yousung Jung. "Explainable Synthesizability Prediction of Inorganic Crystal Polymorphs Using Large Language Models". In: *Angewandte Chemie International Edition* 64.19 (2025), e202423950.

[Oka+25]   Ryotaro Okabe et al. "Structural constraint integration in a generative model for the discovery of quantum materials". In: *Nature Materials* (2025), pp. 1–8.

[Son+25]   Zhilong Song et al. "Accurate prediction of synthesizability and precursors of 3D crystal structures via large language models". In: *Nature Communications* 16.1 (2025), p. 6530.

[Yam+25]   Shuya Yamazaki et al. *Multi-property directed generative design of inorganic materials through Wyckoff-augmented transfer learning.* 2025. arXiv: `2503.16784 [cond-mat.mtrl-sci]`. URL: `https://arxiv.org/abs/2503.16784`.

[Zen+25]   Claudio Zeni et al. "A generative model for inorganic materials design". In: *Nature* 639.8055 (2025), pp. 624–632.

[Sir+]    Martin Siron et al. "LeMat-Bulk: aggregating, and de-duplicating quantum chemistry materials databases". In: *AI for Accelerated Materials Design-ICLR 2025*.

# Cluster expansion in the AI era: towards scalable inverse alloy design

Zhidong Leong[1] and Teck Leong Tan[1]

[1] Institute of High Performance Computing (IHPC), Agency for Science, Technology and Research (A*STAR), 1 Fusionopolis Way, #16-16 Connexis, Singapore 138632, Republic of Singapore

E-mail: leong_zhidong@a-star.edu.sg; tantl@a-star.edu.sg

**Status**

The rapid advancements of machine learning interatomic potentials (MLIPs)[1] have transformed atomistic modelling, enabling efficient large-scale simulations of complex materials. These MLIPs are trained to be highly versatile, capable of predicting the energies and forces of arbitrary atomic configurations. In contrast, the cluster expansion (CE) method[2][3] tackles a specific subproblem: modelling the energies of atomic configurations largely constrained to a predefined lattice. For over forty years, CE has been a core computational technique in the study of alloyed materials, predicting their ground-state structures[4], phase stability[5], and chemical ordering[6][7]. In the context of today's AI era, it is timely to reexamine CE for possible innovations, by drawing inspirations from related fields.

Based on a generalized Ising model, CE expresses the energy of atomic configurations as a sum of effective interactions among clusters of atoms[3]. Since the set of all possible clusters form a complete basis over the configurational space, CE is sufficiently expressive for modelling the energies (or any properties) within a fixed lattice. The values of these multibody interatomic interactions are determined by fitting to near-first-principles energies. The use of physics-based descriptors in a simple linear model allows CE to be interpretable and accurate even with limited training data.

While originally developed for binary metal alloys[4], CE has in recent years been successfully extended to complex systems such as high-entropy alloys[5][6][7], ionic systems[3][20], adsorbate systems[12], nanomaterials[10], and layered materials[11]. Such extensions are often enabled by using regularization techniques[14][15][20] from machine learning to handle the large number of descriptors in the model.

The continued relevance of CE is also reflected in the growing ecosystem of open-source software packages, from the well-established ATAT[16] to more recent packages like ICET[17] and smol[18]. These tools, together with CE's low computational requirements, make CE accessible for both general users as well as developers interested in integrating modern AI techniques. The next section explores two key challenges CE faces.

**Current and future challenges**

A key challenge of CE is scalability across different lattices and chemical systems. The lattice-based formulation of CE ties each model to a fixed crystallography, requiring separate models for each phase (e.g., face-centred cubic and body-centred cubic) of the same alloy[8]. While CE is highly interpretable, it is difficult to reuse knowledge gained from one system in another[9]. Likewise, even for chemically similar elements on the same lattice, CE must be retrained from scratch for each new material. This is unlike MLIPs, which can generalize across diverse environments[1]. This retraining process becomes increasingly computationally unsustainable as the field shifts towards complex systems for which CE may struggle to model efficiently.

Examples include chemically complex system (e.g., high-entropy alloys), systems with long-range behaviours (e.g., from lattice distortions), and low symmetry systems (e.g., surfaces). To accurately model each of these systems, extensive training data is needed to accommodate a large number of relevant cluster interactions[7] (e.g., which increases combinatorially with the number of elemental species). While various regularization techniques have been developed to improve model sparsity and accuracy, progress remains incremental and tailored to specific systems. A general strategy for scaling CE to chemically diverse or structurally complex systems is still lacking. In particular, the lack of descriptors that describe interatomic distances in CE makes it difficult to model systems with large distortion (e.g., alloys of elements with high atomic size mismatch). The lack of standardized benchmarks also complicates the process of comparing techniques and evaluating their generalizability.

Another major challenge of CE involves inverse design. As a forward predictive model, CE is efficient for evaluating known configurations but less suited for discovering new materials with target properties. For example, searching for ground state structures using CE often relies on brute-force enumeration[4], while discovering compositions with thermodynamically stable phases with a particular chemical order/disorder requires computationally expensive CE-based Monte Carlo simulations across the compositional design space[6][7]. Such approaches often become computationally infeasible for chemically complex systems. An efficient strategy for generating candidate structures or optimizing configurations towards target properties is therefore still lacking.

**Advances in science and technology to meet challenges**

Much of the above challenges can be addressed by adapting state-of-the-art machine learning and AI techniques. Issues such as transferability, scalability, and inverse design have seen significant progress in the development of MLIPs[1], and CE can benefit from these innovations.

For scalability across different chemistries, CE could adapt the rich elemental embeddings and structural representations in MLIPs[1] for use in lattice-based systems. These embeddings and representations could supplement (not replace) the conventional system-specific CE descriptors, enabling knowledge transfer and reducing the need to retrain CE models from scratch for each new material system.

For scalability to complex materials systems, the conventional handcrafted descriptors in CE could be supplemented with machine-learned descriptors that more efficiently capture the many-body and longer ranged physics. Due to the lattice-based formulation of CE, this can be achieved by adapting computer vision techniques, e.g., convolutional neural networks, with on-lattice atomic configurations represented as images.

For inverse design, generative AI offers new possibilities. Image-based generative models such as diffusion models[19], originally developed for conditional image generation, can be repurposed for the inverse design of on-lattice atomic configurations. The generation of atomic configurations can be conditioned on composition, energy, or other target properties. This approach could efficiently produce ground-state structures or equilibrium configurations, bypassing brute-force enumeration and expensive Monte Carlo simulations.

**Concluding remarks**

By adapting recent AI advances, CE can be enhanced to keep up with the demands of modern materials research. Compared to MLIPs, CE has received less attention from the AI community, leaving many opportunities unexplored. These present a fertile ground for innovation, where even modest advances could yield significant impact.

**Acknowledgements**

ZL and TLT acknowledge funding from the MAT-GDT Program at A*STAR via the MTC Programmatic Fund by the Agency for Science, Technology and Research under Grant No. M24N4b0034.

**References**

[1] Zhang, Y.-W.; Sorkin, V.; Aitken, Z. H.; Politano, A.; Behler, J.; P Thompson, A.; Ko, T. W.; Ong, S. P.; Chalykh, O.; Korogod, D.; Podryabinkin, E.; Shapeev, A.; Li, J.; Mishin, Y.; Pei, Z.; Liu, X.; Kim, J.; Park, Y.; Hwang, S.; Han, S.; Sheriff, K.; Cao, Y.; Freitas, R. Roadmap for the Development of Machine Learning-Based Interatomic Potentials. *Modelling Simul. Mater. Sci. Eng.* **2025**, *33* (2), 023301. https://doi.org/10.1088/1361-651X/ad9d63.

[2] Sanchez, J. M.; Ducastelle, F.; Gratias, D. Generalized Cluster Description of Multicomponent Systems. *Physica A: Statistical Mechanics and its Applications* **1984**, *128* (1–2), 334–350. https://doi.org/10.1016/0378-4371(84)90096-7.

[3] Barroso-Luque, L.; Zhong, P.; Yang, J. H.; Xie, F.; Chen, T.; Ouyang, B.; Ceder, G. Cluster Expansions of Multicomponent Ionic Materials: Formalism and Methodology. *Phys. Rev. B* **2022**, *106* (14), 144202. https://doi.org/10.1103/PhysRevB.106.144202.

[4] Blum, V.; Zunger, A. Prediction of Ordered Structures in the Bcc Binary Systems of Mo, Nb, Ta, and W from First-Principles Search of Approximately 3,000,000 Possible Configurations. *Phys. Rev. B* **2005**, *72* (2), 020104. https://doi.org/10.1103/PhysRevB.72.020104.

[5] El Atwani, O.; Vo, H. T.; Tunes, M. A.; Lee, C.; Alvarado, A.; Krienke, N.; Poplawsky, J. D.; Kohnert, A. A.; Gigax, J.; Chen, W.-Y.; Li, M.; Wang, Y. Q.; Wróbel, J. S.; Nguyen-Manh, D.; Baldwin, J. K. S.; Tukac, O. U.; Aydogan, E.; Fensin, S.; Martinez, E. A Quinary WTaCrVHf Nanocrystalline Refractory High-Entropy Alloy Withholding Extreme Irradiation Environments. *Nat Commun* **2023**, *14* (1), 2516. https://doi.org/10.1038/s41467-023-38000-y.

[6] Sobieraj, D.; Wróbel, J. S.; Rygier, T.; Kurzydłowski, K. J.; Atwani, O. E.; Devaraj, A.; Saez, E. M.; Nguyen-Manh, D. Chemical Short-Range Order in Derivative Cr–Ta–Ti–V–W High Entropy Alloys from the First-Principles Thermodynamic Study. *Phys. Chem. Chem. Phys.* **2020**, *22* (41), 23929–23951. https://doi.org/10.1039/D0CP03764H.

[7] Leong, Z.; Ramamurty, U.; Tan, T. L. Microstructural and Compositional Design Principles for Mo-V-Nb-Ti-Zr Multi-Principal Element Alloys: A High-Throughput First-Principles Study. *Acta Materialia* **2021**, *213*, 116958. https://doi.org/10.1016/j.actamat.2021.116958.

[8] Natarajan, A. R.; Dolin, P.; Van der Ven, A. Crystallography, Thermodynamics and Phase Transitions in Refractory Binary Alloys. *Acta Materialia* **2020**, *200*, 171–186. https://doi.org/10.1016/j.actamat.2020.08.034.

[9] Yuge, K. Modeling Configurational Energetics on Multiple Lattices through Extended Cluster Expansion. *Phys. Rev. B* **2012**, *85* (14), 144105. https://doi.org/10.1103/PhysRevB.85.144105.

[10] Cao, L.; Li, C.; Mueller, T. The Use of Cluster Expansions To Predict the Structures and Properties of Surfaces and Nanostructured Materials. *J. Chem. Inf. Model.* **2018**, *58* (12), 2401–2413. https://doi.org/10.1021/acs.jcim.8b00413.

[11] Leong, Z.; Jin, H.; Wong, Z. M.; Nemani, K.; Anasori, B.; Tan, T. L. Elucidating the Chemical Order and Disorder in High-Entropy MXenes: A High-Throughput Survey of the Atomic Configurations in TiVNbMoC3 and TiVCrMoC3. *Chem. Mater.* **2022**, *34* (20), 9062–9071. https://doi.org/10.1021/acs.chemmater.2c01673.

[12] Lerch, D.; Wieckhorst, O.; Hammer, L.; Heinz, K.; Müller, S. Adsorbate Cluster Expansion for an Arbitrary Number of Inequivalent Sites. *Phys. Rev. B* **2008**, *78* (12), 121405. https://doi.org/10.1103/PhysRevB.78.121405.

[13] Yang, J. H.; Chen, T.; Barroso-Luque, L.; Jadidi, Z.; Ceder, G. Approaches for Handling High-Dimensional Cluster Expansions of Ionic Systems. *npj Comput Mater* **2022**, *8* (1), 1–11. https://doi.org/10.1038/s41524-022-00818-3.

[14] Nelson, L. J.; Hart, G. L. W.; Zhou, F.; Ozoliņš, V. Compressive Sensing as a Paradigm for Building Physics Models. *Physical Review B* **2013**, *87* (3). https://doi.org/10.1103/PhysRevB.87.035125.

[15] Leong, Z.; Tan, T. L. Robust Cluster Expansion of Multicomponent Systems Using Structured Sparsity. *Phys. Rev. B* **2019**, *100* (13), 134108. https://doi.org/10.1103/PhysRevB.100.134108.

[16] van de Walle, A. Multicomponent Multisublattice Alloys, Nonconfigurational Entropy and Other Additions to the Alloy Theoretic Automated Toolkit. *Calphad* **2009**, *33* (2), 266–278. https://doi.org/10.1016/j.calphad.2008.12.005.

[17] Ångqvist, M.; Muñoz, W. A.; Rahm, J. M.; Fransson, E.; Durniak, C.; Rozyczko, P.; Rod, T. H.; Erhart, P. ICET – A Python Library for Constructing and Sampling Alloy Cluster Expansions. *Advanced Theory and Simulations* **2019**, *2* (7), 1900015. https://doi.org/10.1002/adts.201900015.

[18] Barroso-Luque, L.; Yang, J. H.; Xie, F.; Chen, T.; Kam, R. L.; Jadidi, Z.; Zhong, P.; Ceder, G. Smol: A Python Package for Cluster Expansions and Beyond. *Journal of Open Source Software* **2022**, *7* (77), 4504. https://doi.org/10.21105/joss.04504.

[19] Ho, J.; Jain, A.; Abbeel, P. Denoising Diffusion Probabilistic Models. In *Advances in neural information processing systems*; Larochelle, H., Ranzato, M., Hadsell, R., Balcan, M. F., Lin, H., Eds.; Curran Associates, Inc., 2020; Vol. 33, pp 6840–6851.

[20] Yang, J. H.; Chen, T.; Barroso-Luque, L.; Jadidi, Z.; Ceder, G. Approaches for Handling High-Dimensional Cluster Expansions of Ionic Systems. *npj Comput Mater* **2022**, *8* (1), 1–11. https://doi.org/10.1038/s41524-022-00818-3.

# AI-driven modelling of dynamical processes

Qianxiao Li[1,2,*]

[1]Department of Mathematics, National University of Singapore, Singapore
[2]Institute for Functional Intelligent Materials, National University of Singapore, Singapore
[*]Author to whom any correspondence should be addressed.

**E-mail:** qianxiao@nus.edu.sg

**Status**
Dynamical processes are ubiquitous in science and engineering, and building models for these processes is of fundamental importance. Although quantum mechanics in principle provides microscopically accurate descriptions for most practical systems, the associated computational cost renders direct simulation infeasible. To enable efficient simulation, modelling, and control, we therefore turn to reduced models that operate on larger spatial and temporal scales.

Two main classical approaches have been developed to achieve this goal. The first is *phenomenological modelling*, where mesoscopic or macroscopic models are formulated by hand to capture the key qualitative properties of the system. A well-known example is the Landau model [1] for continuous phase transitions. While physically well motivated, the main drawback of phenomenological models is their limited accuracy. This is to be expected, as these models are designed to capture essential qualitative features, rather than precise dynamical properties. For instance, the Allen–Cahn equation [2] (which can be viewed as the gradient flow of the Ginzburg–Landau energy functional) fails to capture the true hydrodynamic limit of a kinetic spin system (such as the Ising magnet with Glauber dynamics), where more sophisticated tools are required [3].

The second approach, *multiscale modelling* [4], aims to address these limitations by coupling mesoscopic and macroscopic dynamical models with targeted microscopic computations. In principle, this strategy allows one to construct efficient coarse-grained models that retain the accuracy of microscopic descriptions. Multiscale modelling has yielded many successes, but also faces well-known shortcomings, such as the need for detailed knowledge of physical systems, reliable coarse-graining procedures, and clear scale separation.

With the advent of Artificial Intelligence (AI)-enabled high-throughput computational and experimental workflows, however, a new opportunity has emerged: by fusing machine learning with multiscale modelling, we can begin to rethink how to build accurate yet efficient mesoscopic and macroscopic dynamical models. This in turn enables scalable analysis of complex dynamical processes, and may ultimately transform the way we model and simulate complex systems.

**Current progress and future challenges**
AI-driven modelling and simulation of dynamical processes is an increasingly active area of research. We first give some examples of current progress, and then outline the future challenges that need to be resolved.

The first line of research aims to accelerate classical modelling and simulation workflows for dynamical systems. At the microscopic level, ab-initio molecular dynamics [5] has been a central tool to simulate the dynamics of atomic and molecular systems with quantum accuracy. The most expensive step is the computation of energy gradients (forces), typically using density functional theory (DFT), which hinders simulation of most considerably sized systems required to study macroscopic behaviour. Machine-learning interatomic potentials (MLIPs) [6, 7, 8] have emerged as a promising solution to this problem. MLIPs employ neural networks as function approximators to estimate energies and forces. Unlike DFT-based computations, once trained MLIPs can accelerate force evaluations by orders of magnitude, thereby making molecular dynamics simulations much more tractable. Beyond microscopic simulations, AI is also playing an increasingly important role in augmenting multiscale workflows. An example is the modelling of constitutive relations in continuum fluid models, such as polymeric fluids. Classically, microscopic polymer geometry must be incorporated into Brownian dynamics models to derive constitutive relations, which are valid only for highly idealised fluids. Machine learning offers an alternative, enabling constitutive relations to be learned directly from microscopic computations [9], greatly improving the generality and transferability of multiscale workflows.

The second line of research concerns the construction of mesoscopic and macroscopic models directly from data. Early approaches relied on sparse regression with hand-crafted or learned dictionary sets for system identification [10]. A major issue is that the models identified are highly sensitive to noise and the choice of dictionary. More importantly, for realistic complex dynamical systems, it is difficult to select appropriate model spaces that guarantee both simulation stability and physical interpretability. This motivates combining broad physical insights on structure and symmetry with flexible hypothesis spaces. Examples include Hamiltonian systems [11] and dissipative systems [12], with successful applications to non-equilibrium polymer dynamics [13]. Unlike direct system identification, these models enforce chosen physical principles *a priori*, with stability and interpretability built directly into their architectures. Rather than relying on a sparse regression framework, they aim to learn representations of key components of the dynamics (e.g. free energy functions) without committing to a fixed functional form, such as low-order polynomials.

Looking forward, there are important challenges to be addressed. Below we discuss a subset of them.

*Microscopic data generation.* While MLIPs have significantly advanced the scale at which accurate dynamical simulations can be carried out, they are not without limitations. One issue is that, due to the stiffness of dynamical systems, it remains prohibitive to simulate large systems over long time scales (e.g. typical MLIP models can only output 1–100 ns of simulation time per day of computation [14]) A promising direction is to increase the effective time step in molecular dynamics simulations with MLIPs, either by adapting the learned potentials to solvers capable of taking large time steps, or by employing conditional generative models that replace explicit integration of Newton's equations, where stiffness limits step sizes. Another limitation is that MLIPs are typically trained on DFT-relaxation trajectory data concentrated near local minima of the potential landscape. Using them to explore rare events involving first- or higher-order transition states poses fundamental challenges that remain unresolved. With efficient and reliable microscopic trajectory data from MLIPs, we can better inform a new class of data-driven mesoscopic and macroscopic dynamical models.

*Mesoscopic models.* Although progress has been made in building macroscopic thermodynamic models, learning mesoscopic models (e.g. continuum models, phase-field models), which are central to modelling and simulation, remains under-explored. Mesoscopic models offer the advantage of fast numerical simulation compared to ab-initio methods, while still preserving sufficient spatial and temporal resolution to study detailed dynamical behaviour over a rich set of initial, boundary or ambient conditions. Building accurate yet efficient mesoscopic models is therefore of great importance to both physics and computation. The primary challenge is two-fold. First, one must formulate hypothesis spaces for mesoscopic models that capture essential physics and symmetries, yet allow for flexible learning of complex dynamics. For macroscopic systems there are established methods (e.g. those due to Lars Onsager [15]), but for spatially extended mesoscopic systems there are few general approaches beyond gradient-flow dynamics. Second, one must account for non-smooth behaviour—such as defects—within continuum models. This likely requires rethinking hypothesis spaces and moving beyond the (partial) differential equation framework that dominates classical mesoscale modelling.

*Automated workflows.* Ultimately, we seek unified AI-driven workflows for multiscale modelling of dynamics, coupling machine learning, computational data generation, and experimental validation. Such workflows are far from mature, and their development requires collaboration across multiple disciplines. For a given complex system, an end-to-end workflow would need to first encode broad physical constraints, design model architectures that respect these constraints, generate microscopic training data to inform mesoscopic and macroscopic models, and finally correct for systematic errors through experimental validation. Such an automated orchestrator would fundamentally change the way we model dynamical processes, and enable subsequent analysis and control workflows to achieve both understanding and functionality.

**Concluding remarks**

AI-driven multiscale modelling is poised to deliver efficient, accurate dynamical models that bridge scales from atoms to continuum. Overcoming the remaining challenges in microscopic data generation, mesoscopic hypothesis spaces, and end-to-end automation will require close interplay between theory, computation and experiment. With coordinated effort, these advances can transform our ability to predict, control and design complex dynamical systems.

**Acknowledgments**

This research is supported by the National Research Foundation, Singapore under its AI Singapore Programme (AISG Award No: AISG3-RP-2022-028).

**References**

[1] Lev Davidovich Landau and Evgenii Mikhailovich Lifshitz. *Statistical Physics: Volume 5*. Vol. 5. Elsevier, 2013. ISBN: 0-08-057046-1.

[2] S. M Allen and J. W Cahn. "Ground State Structures in Ordered Binary Alloys with Second Neighbor Interactions". In: *Acta Metallurgica* 20.3 (Mar. 1, 1972), pp. 423–433. ISSN: 0001-6160. DOI: 10.1016/0001-6160(72)90037-5. URL: https://www.sciencedirect.com/science/article/pii/0001616072900375 (visited on 09/26/2025).

[3] A. De Masi, E. Orlandi, E. Presutti, and L. Triolo. "Glauber Evolution with Kac Potentials. I. Mesoscopic and Macroscopic Limits, Interface Dynamics". In: *Nonlinearity* 7.3 (May 1994), p. 633. ISSN: 0951-7715. DOI: 10.1088/0951-7715/7/3/001. URL: https://doi.org/10.1088/0951-7715/7/3/001 (visited on 09/26/2025).

[4] Weinan E. *Principles of Multiscale Modeling*. Cambridge University Press, 2011. URL: https://books.google.com/books?hl=en&lr=&id=9uTjKO2Ix-YC&oi=fnd&pg=PR12&dq=principles+of+multiscale+modeling&ots=vNM8Pp80SH&sig=7JKGjPC0lB3WZPaoYpcKfTQPpbo (visited on 10/02/2024).

[5] R. Car and M. Parrinello. "Unified Approach for Molecular Dynamics and Density-Functional Theory". In: *Phys. Rev. Lett.* 55.22 (Nov. 25, 1985), pp. 2471–2474. DOI: 10.1103/PhysRevLett.55.2471. URL: https://link.aps.org/doi/10.1103/PhysRevLett.55.2471 (visited on 09/26/2025).

[6] Linfeng Zhang, Jiequn Han, Han Wang, Roberto Car, and E. Weinan. "Deep Potential Molecular Dynamics: A Scalable Model with the Accuracy of Quantum Mechanics". In: *Physical review letters* 120.14 (Apr. 2018), p. 143001. DOI: 10.1103/PhysRevLett.120.143001.

[7] Ilyes Batatia, Dávid Péter Kovács, Gregor N C Simm, Christoph Ortner, and Gábor Csányi. "MACE: Higher Order Equivariant Message Passing Neural Networks for Fast and Accurate Force Fields". In: *NeurIPS*. 2022.

[8] Bowen Deng, Peichen Zhong, KyuJung Jun, Janosh Riebesell, Kevin Han, Christopher J. Bartel, and Gerbrand Ceder. "CHGNet as a Pretrained Universal Neural Network Potential for Charge-Informed Atomistic Modelling". In: *Nat Mach Intell* 5.9 (Sept. 2023), pp. 1031–1041. ISSN: 2522-5839. DOI: 10.1038/s42256-023-00716-3. URL: https://www.nature.com/articles/s42256-023-00716-3 (visited on 09/26/2025).

[9] Huan Lei, Lei Wu, and Weinan E. "Machine Learning Based Non-Newtonian Fluid Model with Molecular Fidelity". In: *Phys. Rev. E* 102.4 (Oct. 13, 2020), p. 043309. ISSN: 2470-0045, 2470-0053. DOI: 10.1103/PhysRevE.102.043309. arXiv: 2003.03672 [physics]. URL: http://arxiv.org/abs/2003.03672 (visited on 04/02/2025).

[10] Steven L. Brunton, Joshua L. Proctor, and J. Nathan Kutz. "Discovering Governing Equations from Data by Sparse Identification of Nonlinear Dynamical Systems". In: *Proc. Natl. Acad. Sci. U.S.A.* 113.15 (Apr. 12, 2016), pp. 3932–3937. ISSN: 0027-8424, 1091-6490. DOI: 10.1073/pnas.1517384113. URL: https://pnas.org/doi/full/10.1073/pnas.1517384113 (visited on 11/13/2023).

[11] Tom Bertalan, Felix Dietrich, Igor Mezić, and Ioannis G. Kevrekidis. "On Learning Hamiltonian Systems from Data". In: *Chaos: An Interdisciplinary Journal of Nonlinear Science* 29.12 (Dec. 31, 2019), p. 121107. ISSN: 1054-1500. DOI: 10.1063/1.5128231. URL: https://doi.org/10.1063/1.5128231 (visited on 11/13/2023).

[12] Haijun Yu, Xinyuan Tian, Weinan E, and Qianxiao Li. "OnsagerNet: Learning Stable and Interpretable Dynamics Using a Generalized Onsager Principle". In: *Phys. Rev. Fluids* 6.11 (Nov. 23, 2021), p. 114402. DOI: 10.1103/PhysRevFluids.6.114402. arXiv: 2009.02327. URL: https://link.aps.org/doi/10.1103/PhysRevFluids.6.114402 (visited on 03/27/2022).

[13]  Xiaoli Chen, Beatrice W. Soh, Zi-En Ooi, Eleonore Vissol-Gaudin, Haijun Yu, Kostya S. Novoselov, Kedar Hippalgaonkar, and Qianxiao Li. "Constructing Custom Thermodynamics Using Deep Learning". In: *Nature Computational Science* 4.1 (Jan. 2024), pp. 66–85. URL: https://www.nature.com/articles/s43588-023-00581-5 (visited on 08/04/2024).

[14]  Jianxiong Li, Boyang Li, Zhuoqiang Guo, Mingzhen Li, Enji Li, Lijun Liu, Guojun Yuan, Zhan Wang, Guangming Tan, and Weile Jia. "Scaling Molecular Dynamics with Ab Initio Accuracy to 149 Nanoseconds per Day". In: *SC24: International Conference for High Performance Computing, Networking, Storage and Analysis*. Comment: 11 pages, 11 figures, 3 tables, SC'24. Nov. 17, 2024, pp. 1–15. DOI: 10.1109/SC41406.2024.00036. arXiv: 2410.22867 [cs]. URL: http://arxiv.org/abs/2410.22867 (visited on 09/26/2025).

[15]  Lars Onsager. "Reciprocal Relations in Irreversible Processes. I." In: *Physical review* 37.4 (1931), p. 405.

# Machine learning force fields: Recent advances and remaining challenges II

Adil Kabylda[1], Igor Poltavsky[1] and Alexandre Tkatchenko[1,*]

[1]Department of Physics and Materials Science, University of Luxembourg, L-1511 Luxembourg City, Luxembourg
*Author to whom any correspondence should be addressed.

E-mail: alexandre.tkatchenko@uni.lu

**Status**
Atomistic simulations have become indispensable for exploring chemical and physical processes at the microscopic scale. Advances in graph neural networks (e.g., SchNet, NequIP, MACE, SO3krates) now routinely achieve sub-chemical accuracy across diverse systems [1–4]. A recent major development in machine learning force field (MLFF) research is the emergence of broadly applicable models pre-trained on large, diverse datasets containing millions of structures [5–7]. Often referred to as "general-purpose", "universal", or "foundational" MLFFs, these models represent valuable community achievements, built on collective advances in ML architectures, training strategies, and the aggregation of datasets from many research efforts.

Recent work on general-purpose MLFFs has extended applicability across a broad chemical and structural domain. State-of-the-art models now cover 10–100 elements and apply to solids, liquids, and molecular systems, including organic (bio)molecules, transition-metal oxides, alloys, and electrolytes. They achieve accuracy close to the underlying reference level of theory while scaling to millions of atoms and nanosecond timescales, effectively replacing *ab initio* MD in many regimes. This capability enables predictive studies of conformational exploration, ionic and thermal transport, phase stability, and mechanical response [8].

The SO3LR model is an example of a pre-trained, state-of-the-art MLFF that is broadly applicable to (bio)organic systems. It combines a fast and stable SO(3)-equivariant neural network for semi-local interactions with universal, physically grounded pairwise potentials that describe short-range repulsion, long-range electrostatics, and dispersion [7]. This design reserves network capacity for short- and intermediate-range many-body effects while enforcing correct asymptotic behavior at both extremes, yielding higher efficiency and improved transferability compared with simply increasing the number of message-passing layers. SO3LR scales to 200k atoms at a latency of $\sim 3\,\mu$s/atom/step on a single H100 GPU, approaching sizes and timescales relevant for realistic biomolecules. Pretraining on 4M reference fragments computed at PBE0+MBD level of theory [9-12] confers broad transferability, enabling the simulation of small biomolecular building blocks, polyalanine systems, bulk water, proteins, glycoproteins, and a lipid bilayer with a single MLFF model.

**Current/future challenges and required advances**
Despite the notable progress of modern MLFFs, a number of challenges remain. A reliable model not only requires a robust MLFF architecture but also representative training data at the appropriate level of theory, along with training strategies that capture interactions across different scales, from atoms to extended systems. For practical use, MLFFs must further combine speed, accuracy, and transferability—often conflicting demands that force difficult compromises in model design.

Community efforts are essential to track progress and address these issues. The TEA Challenge 2023 served as such a benchmark, applying diverse MLFF architectures, each trained by their respective developers, to tasks of variable complexity, covering molecules, materials, and interfaces [13, 14]. This extensive initiative revealed both strong agreement in describing molecular dynamics and striking disparities in computational cost, with some comparable architectures exhibiting an order-of-magnitude difference. The study also demonstrated the decisive role of training data quality, both in terms of the theory level and sampling completeness, and highlighted the persistence of systematic errors even in state-of-the-art MLFF models.

For general-purpose MLFFs, similar efforts would be highly beneficial. However, several apparent deficiencies can already be underlined. In particular, the accuracy and robustness of general-purpose MLFFs remain inferior to those of bespoke models tailored to specific systems and conditions. In practice, broadly applicable models typically require fine-tuning on system-specific data, which is nevertheless far more data-efficient than training new models from scratch. Direct applications without fine-tuning are possible, yet their results must be treated with caution and independently verified.

Another limitation of MLFFs is their relatively low robustness compared to traditional computational chemistry methods, which often rely on variational principles and inherently respect physical constraints. While modern MLFF models achieve sub-chemical accuracy across static benchmarks, they can notably violate physical laws and chemical rules [15], leading to unstable MD trajectories or failures in capturing specific bonding motifs, especially when training data are imbalanced [16]. Expanding datasets mitigates but cannot eliminate these issues given the vast conformational and compositional diversity of chemical space.

Enhancing the accuracy and robustness of MLFFs would require not only scaling existing models, but also their careful analysis, ensuring adherence to physically and chemically motivated constraints, and further architectural developments. By balancing the flexibility of deep neural networks with the reliability of physical principles, ML models can achieve stronger generalization, transferability to unseen states and even out-of-domain chemical systems. Examples of such strategies include employing real-space chemical descriptors or predicting physical quantities beyond atomic forces and energies [17].

Key directions for improvement, identified in the TEA Challenge 2023 and other similar works, include: (i) robust treatment of non-local interactions (van der Waals, electrostatics), (ii) mitigating dataset incompleteness via advanced sampling and active learning, (iii) addressing limitations inherited from reference methods, and (iv) incorporating intrinsic quality measures such as uncertainty quantification. Progress in these areas will be critical for advancing MLFFs toward reliable, robust, and efficient atomistic modeling.

**Concluding remarks**

Machine learning force fields have crossed the threshold from "promising" to "practically useful". Reaching truly universal force fields will require hybrid physics-aware architectures, active learning and multi-fidelity data curation, robust uncertainty quantification, and standardized, stress-tested validation. Given the current momentum, general-purpose force fields capable of providing predictive insights in chemistry, biology, and material science seem to be within reach.

**Acknowledgments**

We thank colleagues in the MLFF community for releasing preprints, open-sourcing code, models, and datasets. A.K. acknowledges financial support from the Luxembourg National Research Fund (FNR AFR Ph.D. Grant 15720828). I.P. and A.T. acknowledge the Luxembourg National Research Fund under grant FNR-CORE MBD-in-BMD (18093472) and the European Research Council under ERC-AdG grant FITMOL (101054629).

**References**

[1] Schütt K T, Sauceda H E, Kindermans P J, Tkatchenko A and Müller K. R. 2018 SchNet–a deep learning architecture for molecules and materials J. Chem. Phys. 148(24) (doi: 10.1063/1.5019779)

[2] Batzner S, Musaelian A, Sun L, Geiger M, Mailoa J P, Kornbluth M, Molinari N, Smidt T E and Kozinsky B 2022 E(3)-equivariant graph neural networks for data-efficient and accurate interatomic potentials Nat. Commun. 13 2453 (doi: 10.1038/s41467-022-29939-5)

[3] Batatia I, Kovacs D P, Simm G, Ortner C and Csányi G 2022 MACE: higher order equivariant message passing neural networks for fast and accurate force fields Adv. Neural Inf. Process. Syst. 35 11423

[4] Frank J T, Unke O T, Müller K-R and Chmiela S 2024 A Euclidean transformer for fast and stable machine learned force fields Nat. Commun. 15 6539 (doi: 10.1038/s41467-024-50620-6)

[5] Deng B, Zhong P, Jun K, Riebesell J, Han K, Bartel C J and Ceder G 2023 CHGNet as a pretrained universal neural network potential for charge-informed atomistic modelling Nat. Mach. Intell. 5 1031 (doi: 10.1038/s42256-023-00716-3)

[6] Unke O T, Stöhr M, Ganscha S, Unterthiner T, Maennel H, Kashubin S, Ahlin D, Gastegger M, Medrano Sandonas L, Berryman J T, Tkatchenko A and Müller K-R 2024 Biomolecular dynamics with machine-learned quantum-mechanical force fields trained on diverse chemical fragments Sci. Adv. 10(14) eadn4397 (doi: 10.1126/sciadv.adn4397)

[7] Kabylda A, Frank J T, Dou S S, Khabibrakhmanov A, Sandonas L M, Unke O T, Chmiela S, Müller K-R and Tkatchenko A 2025 Molecular simulations with a pretrained neural network and universal pairwise force fields J. Am. Chem. Soc. ASAP (doi: 10.1021/jacs.5c09558)

[8] Jacobs R, Morgan D, Attarian S, Meng J, Shen C, Wu Z, Xie C Y, Yang J H, Artrith N, Blaiszik B, Ceder G, Choudhary K, Csányi G, Cubuk E D, Deng B, Drautz R, Fu X, Godwin J, Honavar V, Isayev O, Johansson A, Kozinsky B, Martiniani S, Ong S P, Poltavsky I, Schmidt K J, Takamoto S, Thompson A P, Westermayr J, Wood B M 2025 A practical guide to machine learning interatomic potentials–Status and future Curr. Opin. Solid State Mater. Sci. 35 101214 (doi: 10.1016/j.cossms.2025.101214)

[9] Perdew J P, Burke K and Ernzerhof M 1996 Generalized gradient approximation made simple Phys. Rev. Lett. 77 3865 (doi: 10.1103/PhysRevLett.77.3865)

[10] Adamo C and Barone V 1999 Toward reliable density functional methods without adjustable parameters: The PBE0 model J. Chem. Phys. 110 6158 (doi: 10.1063/1.478522)

[11] Tkatchenko A, DiStasio R A Jr, Car R and Scheffler M 2012 Accurate and efficient method for many-body van der Waals interactions Phys. Rev. Lett. 108 236402 (doi: 10.1103/PhysRevLett.108.236402)

[12] Ambrosetti A, Reilly A M, DiStasio R A Jr and Tkatchenko A 2014 Long-range correlation energy calculated from coupled atomic response functions J. Chem. Phys. 140 18A508 (doi: 10.1063/1.4865104)

[13] Poltavsky I, Charkin-Gorbulin A, Puleva M, Fonseca G, Batatia I, Browning N J, Chmiela C, Cui M, Frank J T, Heinen S, Huang B, Käser S, Kabylda A, Khan D, Müller C, Price A, Riedmiller K, Töpfer K, Ko T W, Meuwly M, Rupp M, Csányi G, von Lilienfeld O A, Margraf J T, Müller K-R and Tkatchenko A 2025 Crash testing machine learning force fields for molecules, materials, and interfaces: model analysis in the TEA Challenge 2023 Chem. Sci. 16 3720 (doi: 10.1039/D4SC06529H)

[14] Poltavsky I, Charkin-Gorbulin A, Puleva M, Fonseca G, Batatia I, Browning N J, Chmiela C, Cui M, Frank J T, Heinen S, Huang B, Käser S, Kabylda A, Khan D, Müller C, Price A, Riedmiller K, Töpfer K, Ko T W, Meuwly M, Rupp M, Csányi G, von Lilienfeld O A, Margraf J T, Müller K-R and Tkatchenko A 2025 Crash testing machine learning force fields for molecules, materials, and interfaces: molecular dynamics in the TEA Challenge 2023 Chem. Sci. 16 3738 (doi: 10.1039/D4SC06530A)

[15] Esders M, Schnake T, Lederer J, Kabylda A, Montavon G, Tkatchenko A and Müller K-R 2025 Analyzing atomic interactions in molecules as learned by neural networks J. Chem. Theory Comput. 21 714 (doi: 10.1021/acs.jctc.4c01424)

[16] Charkin-Gorbulin A, Kokorin A, Sauceda H E, Chmiela S, Quarti C, Beljonne D, Tkatchenko A and Poltavsky I 2025 Atomic orbits in molecules and materials for improving machine learning force fields Mach. Learn.: Sci. Technol. 6 035005 (doi: 10.1088/2632-2153/adea0b)

[17] Gallegos M, Vassilev-Galindo V, Poltavsky I, Martín Pendás Á and Tkatchenko A 2024 Explainable chemical artificial intelligence from accurate machine learning of real-space chemical descriptors Nat. Commun. 15(1) 4345 (doi: 10.1038/s41467-024-48567-9)

# Making physics-informed machine learning more physics-y: DIEP

**Sherif Abdulkader Tawfik[1]**

[1]Applied Artificial Intelligence Institute, Deakin University, Geelong, Victoria 3216, Australia

E-mail: s.abbas@deakin.edu.au

**Status**

At the heart of the flourishing field of machine learning potentials (MLPs) are graph neural networks, where deep learning is interwoven with physics-informed machine learning (PIML) architectures. Various PIML models, upon training with density functional theory (DFT) material structure-property datasets, have achieved unprecedented prediction accuracy for a range of molecular and material properties. A critical component in the learned graph representation of crystal structures in PIMLs is how the various fragments of the structure's graph are embedded in a neural network. Several of the state-of-art PIML models apply spherical harmonic functions. Such functions assume that DFT computes the Coulomb potential of atom-atom interactions. However, DFT does not directly compute such potentials, but integrates the electron-atom potentials. We introduce the direct integration of the external potential (DIEP) methods which more faithfully reflects that actual computational workflow in DFT. DIEP integrates the external (electron-atom) potential and uses these quantities to embed the structure graph into a deep learning model. We demonstrate the enhanced accuracy of the DIEP model in predicting the energies of pristine and defective materials. By training DIEP to predict the potential energy surface, we show the ability of the model in predicting the onset of fracture of pristine and defective carbon nanotubes.

**Current and future challenges**

In recent years, the frontiers of materials simulation have been radically redrawn by machine learning potentials (MLPs). These models bridge the precision of *ab initio* methods, particularly density functional theory (DFT) with the computational speed of classical molecular dynamics. Within this new paradigm, graph neural networks (GNNs) have emerged as the backbone for encoding atomic environments, allowing complex interactions to be mapped into neural architectures that can predict forces, energies, and structural evolution with near-quantum-level accuracy.

Yet, despite the proliferation of MLPs, with leading examples such as, but not limited to, SchNet [1], DimeNet [2], M3GNET[3] and MACE [4], a fundamental question has persisted: how faithfully do these neural representations capture the true physics of interatomic interactions, particularly under extreme conditions like fracture?

Fracture, a process that drives material failure through the nucleation and propagation of defects and bond breakage, poses a stringent test for any interatomic potential. Traditional force fields, calibrated for equilibrium conditions, often collapse under the non-linearities of bond rupture. Conversely, DFT provides exquisite accuracy but is computationally prohibitive for large systems and long timescales.

**Advances to meet challenges**

Our work (Tawfik et al. [5,6]) confront this dual challenge head-on, introducing a physics-grounded MLP framework: the Direct Integration of the External Potential (DIEP). We apply it to both defective crystals and fracturing carbon nanotubes (CNTs), and Figure 1 displays a schematic illustration of the DIEP computational workflow. The critical aspect of the workflow is the mathematical embedding function that is used for embedding the graph features (the bonds and the triplets in the figure): the function is inspired by the form of the electron-ion interaction energy, the $E_{ext}$ term in the Kuhn-Sham energy functional. Please refer to Eqs. 5-8 in Ref. [5] for further details.

The DIEP embedding follows the physics that is employed by DFT more faithfully than other MLPs, particularly DimeNet and its successor M3GNET: these MLPs assume the presence of nucleus-nucleus interactions according to the time-independent Schrödinger equation, which is not the case in typical DFT calculation workflows. Typically, the nucleus-nucleus interaction is ignored in the electron density optimisation in DFT, and is only accounted for via the atomic forces. The only term in the Kohn-Sham energy functional that represents the nuclei is the $E_{ext}$ term, which is the focus of the DIEP embedding method.

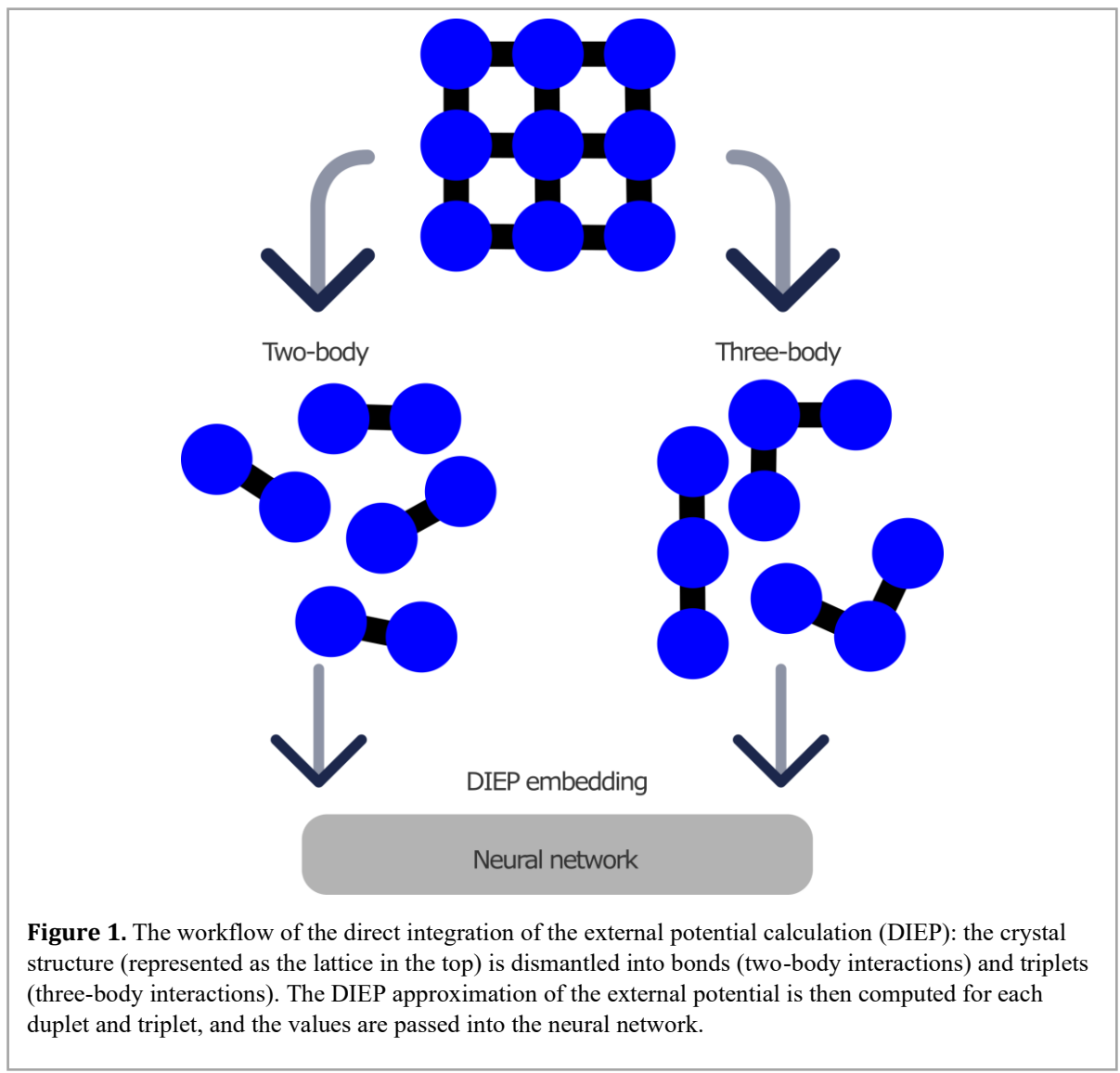

**Figure 1.** The workflow of the direct integration of the external potential calculation (DIEP): the crystal structure (represented as the lattice in the top) is dismantled into bonds (two-body interactions) and triplets (three-body interactions). The DIEP approximation of the external potential is then computed for each duplet and triplet, and the values are passed into the neural network.

Our carbon nanotube (CNT) fracture study in Ref. [6] serves as both a validation and a showcase of DIEP's multiscale reach. Fracture of nanotubes involves interplay between atomic bonding and mesoscopic geometry. The model accurately reproduces chirality-dependent mechanical anisotropy, bridging *ab initio* theory, MD, and experimental data across five orders of magnitude in system size. By predicting trends across 186 chiralities, DIEP transcends the sampling limitations of DFT-based fracture studies (which rarely exceed 10 chiralities). This capacity opens the door to high-throughput mechanical screening of nanostructures, enabling the rational selection of chiral configurations for targeted mechanical performance.

A related matter is: *how do we judge the accuracy of MLPs?* Given that MLPs are trained on DFT-based datasets such as the Materials Project PES datasets, it is tempting to judge the accuracy of an MLP by the mean absolute error (MAE) of the MLP's predictions for a standardised test set. As discussed in our Viewpoint article [7], simply relying on the MAE for a test set might not be a sufficient accuracy measure to judge the generalisability of the MLP to various, and importantly large, simulation structures.

Deciding the accuracy of MLPs should use measures beyond the MAE of a test set, such as the numerous simulations in Ref. [8] from various material science domains and the fracture simulations in our work.

**Concluding remarks**

The Direct Integration of the External Potential (DIEP) concept re-anchors machine learning potentials in first principles, demonstrating that physical correctness and predictive power are not opposing goals. By successfully modelling both the subtle energetics of point defects and the catastrophic non-linearity of fracture, DIEP sets a new benchmark for MLP development. In the domain of material fracture, DIEP's success hints at an exciting prospect: the emergence of machine learning potentials that can truly understand failure as a phenomenon governed by fundamental forces. As materials science advances toward autonomous discovery and design, models like DIEP will play a central role. They will not merely predict when a material breaks; they will help us understand why, uniting the language of deep learning with the logic of quantum mechanics, and transforming fracture from an unpredictable catastrophe into a predictable, designable property of matter.

**References**

[1] Schütt K T, Sauceda H E, Kindermans P-J, Tkatchenko A and Müller K-R 2018 SchNet – A deep learning architecture for molecules and materials *The Journal of Chemical Physics* **148** 241722

[2] Gasteiger J, Groß J and Günnemann S 2022 Directional Message Passing for Molecular Graphs

[3] Chen C and Ong S P 2022 A universal graph deep learning interatomic potential for the periodic table *Nat Comput Sci* **2** 718–28

[4] Batatia I, Kovács D P, Simm G N C, Ortner C and Csányi G 2023 MACE: Higher Order Equivariant Message Passing Neural Networks for Fast and Accurate Force Fields

[5] Tawfik S A, Nguyen T M, Russo S P, Tran T, Gupta S and Venkatesh S 2024 Embedding material graphs using the electron-ion potential: application to material fracture *Digital Discovery* **3** 2618–27

[6] Tawfik S A, Nguyen T M, Tran T, Gupta S and Venkatesh S 2025 Impact of chirality on nanotube fracture strain: comprehensive machine learning potential calculations *Nanoscale* **17** 22802–6

[7] Tawfik S A 2025 Computational Material Science Has a Data Problem *J. Chem. Inf. Model.* **65** 5823–6

[8] Batatia I, Benner P, Chiang Y, Elena A M, Kovács D P, Riebesell J, Advincula X R, Asta M, Avaylon M, Baldwin W J, Berger F, Bernstein N, Bhowmik A, Blau S M, Cărare V, Darby J P, De S, Della Pia F, Deringer V L, Elijošius R, El-Machachi Z, Falcioni F, Fako E, Ferrari A C, Genreith-Schriever A, George J, Goodall R E A, Grey C P, Grigorev P, Han S, Handley W, Heenen H H, Hermansson K, Holm C, Jaafar J, Hofmann S, Jakob K S, Jung H, Kapil V, Kaplan A D, Karimitari N, Kermode J R, Kroupa N, Kullgren J, Kuner M C, Kuryla D, Liepuoniute G, Margraf J T, Magdău I-B, Michaelides A, Moore J H, Naik A A, Niblett S P, Norwood S W, O'Neill N, Ortner C, Persson K A, Reuter K, Rosen A S, Schaaf L L, Schran C, Shi B X, Sivonxay E, Stenczel T K, Svahn V, Sutton C, Swinburne T D, Tilly J, van der Oord C, Varga-Umbrich E, Vegge T, Vondrák M, Wang Y, Witt W C, Zills F and Csányi G 2024 A foundation model for atomistic materials chemistry

# Accelerating the study of Metal-organic Frameworks with Machine Learning Interatomic Potentials

Prathami Divakar Kamath[1,2], Théo Jaffrelot Inizan[2,3] and Kristin A. Persson[1,2,*]

[1] Department of Materials Science and Engineering, University of California, Berkeley, CA, USA
[2] Materials Sciences Division, Lawrence Berkeley National Laboratory, Berkeley, CA, USA
[3] AIMATX Inc., Berkeley, CA, USA
[*] Author to whom any correspondence should be addressed.

**E-mail:** kapersson@lbl.gov

**Status**
Metal-organic Frameworks (MOFs) are widely studied today due to their potential to address important challenges such as Direct Air Capture (DAC) [1]. MOFs are inorganic–organic hybrid materials with high porosity, which gives them excellent gas adsorption abilities. Machine Learning Interatomic Potentials (MLIPs) [2], and especially universal MLIPs (uMLIPs) [3], have emerged as efficient methods to describe complex chemistries such as MOFs. uMLIPs are often large networks with millions of parameters, requiring training on extensive Density Functional Theory (DFT) databases such as inorganic crystals from the Materials Project [4] and Alexandria [5]. This enables them to capture out-of-distribution atomic interactions and chemistries. Further improvements for domain-specific applications can be achieved by fine-tuning these uMLIPs on MOF specific datasets, increasing accuracy in capturing organic interactions and enabling better transferability to diverse chemical environments [6].

However, ready-to-use MOF DFT datasets for training remain very limited. Their vast chemical space and large unit cells make DFT data generation and extensive chemical-space coverage a challenge compared to their inorganic counterparts. As larger MOF DFT datasets become available more MOF-generalist MLIPs are emerging, providing transferable and generalizable potentials. Recent evaluations such as MOFSimBench [7] show that the top-performing uMLIPs surpass classical potentials and fine-tuned ML models in several specific areas for MOFs across diverse chemistries, including tasks such as structure optimization, molecular dynamics stability, and prediction of bulk properties. Their analysis indicates that the superior performance of uMLIPs is mainly correlated with the size, diversity, and quality of the DFT training data rather than the specific model architectures, prompting us to more closely examine the MOF DFT datasets available for training MLIPs.

**Current and future challenges**
Generating high-fidelity DFT datasets is a challenging problem for MOFs, not only because of their large size but also because MOFs are soft and flexible, demonstrating the propensity for structural reorganization and phase changes. Hence, generating a meaningful dataset consisting of strained configurations, equilibrium configurations and transition pathways is essential. This makes choosing the number of snapshots of the potential energy surface (PES) and the chemical diversity of the sampled MOFs an important problem of balancing cost and accuracy in training MLIPs. Although large data sets with millions of DFT data points have emerged they mainly rely on the generalized gradient approximations (GGAs) functionals of PBE [8] or PBE + D3 (BJ) [9, 10] due to their cost effectiveness for large, flexible unit cells typical in MOFs. However, standard PBE + D3 (BJ) can systematically overestimate binding energies and misrepresent electron localization for open-metal sites and correlated centers [11]. MLIPs trained on such data will inherit these approximations, potentially producing erroneous predictions for adsorption properties and magnetic ground states in MOFs with strong correlation. This issue is most problematic for prominent frameworks containing first-transition series elements like Fe, Cr, Cu, etc. which underpin a huge variety of synthetic frameworks and applications. Table 1 summarizes the MOF datasets available today and their key characteristics, highlighting the lack of high-quality DFT. These datasets continue to serve as the foundation for ML training, but expanding coverage with more correlated and dynamic configurations remains an ongoing challenge.

Table 1. Comparison of major MOF computational datasets with their key features and limitations.

| Dataset Name | Size | DFT Functional(s) Used | Key Features | Limitations |
|---|---|---|---|---|
| OpenDAC (ODAC23) [12] | ∼8,412 MOFs, ∼38 million calculations | PBE + D3 (BJ) | Large-scale adsorption and defect data: covers 57 metals and multi-metallic frameworks | Limited spin state sampling |
| QMOF Database [13] | ∼20,000 relaxed MOF crystal structures | PBE + D3 (BJ) | Structural and electronic properties; covers 79 elements with diverse linkers and topologies | Forces and stress tensors not available for training MLIPs; equilibrium structures |
| Phonon-focused MACE-MP-MOF0 dataset [6] | 127 MOFs, ∼5000 calculations | PBE + D3 (BJ) | Efficient PES and diversity coverage via farthest-point sampling; covers 13 metals with diverse linkers and functionalized MOFs | Excludes several spin-polarized MOFs |
| Specialized Active Learning dataset [14] | Small (∼few thousand points) | PBE + D3 (BJ) | High-quality out-of-equilibrium configurations obtained based on uncertainty estimates: covers 3 metals | Smaller scale and only for for specific MOFs |

**Advances in science and technology to meet challenges**
Recent advancements to meet challenges in improving the performance of MLIPs for MOFs have focused on improving phase-space sampling and accuracy of underlying DFT functionals for dataset generation. In order to maximize diversity of PES configurations and sampled phase-space of MOFs, uMLIPs are being used to replace expensive Ab Initio Molecular Dynamics (AIMD) with short MD runs combined with Grand Canonical Monte Carlo simulations to obtain inputs for DFT data generation covering regions of framework flexibility as well gas adsorbate interactions [6, 15]. To further increase the accuracy of MLIPs in areas of poor performance, active learning strategies based on uncertainty estimates are also being used to expand the sampling of training sets [14]. Recent studies on improving DFT accuracy beyond standard PBE functionals for MOFs, have highlighted that other meta-GGA functionals like r2SCAN [16], or including Hubbard U correction [17], offer a better balance in cost and accuracy to capture the lattice dynamics and long-range interactions in spin-polarized MOFs [18]. This motivates the development of MLIPs for MOFs trained on datasets generated at a higher level of theory while also noting that such efforts can be computationally demanding. Hence, cross-functional learning that incorporates cheaper DFT functionals and higher-level r2SCAN calculations offers an alternative to optimize data efficiency and accuracy trade-offs for MLIPs for MOFs [19]. MLIPs trained on these datasets with improved phase-space sampling and meta-GGA functional could provide a better description of key properties from the breathing behavior of MOFs [20], to transition metal-based MOFs and gas adsorption.

**Concluding remarks**
In summary, MLIPs for MOFs have reached a mature and practical level for many key simulation challenges. Overall, DFT datasets of MOFs form a foundational basis for MLIP training but require careful curation and expansion with diverse chemical environments and dynamic states to fully realize their potential for accurate and transferable MLIPs for MOFs.

**Acknowledgments**
P.D.K. acknowledges the National Energy Research Scientific Computing Center (NERSC) financial support from U.S. National Science Foundation's "The Quantum Sensing Challenges for Transformational Advances in Quantum Systems (QuSeC-TAQS)" program. T.J.I. used resources of NERSC award DOE-ERCAP0031751 'GenAI@NERSC'. T.J.I. also thanks the Bakar Institute of Digital Materials for the Planet Fellowship for its support.

**References**

[1] Zhou H.-C., Long J. R., and Yaghi O. M. Introduction to metal–organic frameworks. *Chem. Rev.*, 112:673–674, 2012.

[2] Behler J. and Parrinello M. Generalized neural-network representation of high-dimensional potential-energy surfaces. *Phys. Rev. Lett.*, 98:146401, 2007.

[3] Batatia I., Benner P., Chiang Y., Elena A. M., Kovács D. P., Riebesell J., Advincula X. R., Asta M., Avaylon M., Baldwin W. J., Berger F., Bernstein N., Bhowmik A., Blau S. M., Cărare V., Darby J. P., De S., Della Pia F., Deringer V. L., Elijošius R., El-Machachi Z., Falcioni F., Fako E., Ferrari A. C., Genreith-Schriever A., George J., Goodall R. E. A., Grey C. P., Grigorev P., Han S., Handley W., Heenen H. H., Hermansson K., Holm C., Jaafar J., Hofmann S., Jakob K. S., Jung H., Kapil V., Kaplan A. D., Karimitari N., Kermode J. R., Kroupa N., Kullgren J., Kuner M. C., Kuryla D., Liepuoniute G., Margraf J. T., Magdău I.-B., Michaelides A., Moore J. H., Naik A. A., Niblett S. P., Norwood S. W., O'Neill N., Ortner C., Persson K. A., Reuter K., Rosen A. S., Schaaf L. L., Schran C., Shi B. X., Sivonxay E., Stenczel T. K., Svahn V., Sutton C., Swinburne T. D., Tilly J., van der Oord C., Varga-Umbrich E., Vegge T., Vondrák M., Wang Y., Witt W. C., Zills F., and Csányi G. A foundation model for atomistic materials chemistry. *arXiv preprint arXiv:2401.00096*, 2024.

[4] Jain A., Ong S. P., Hautier G., Chen W., Richards W. D., Dacek S., Cholia S., Gunter D., Skinner D., Ceder G., and Persson K. A. Commentary: The materials project: A materials genome approach to accelerating materials innovation. *APL Mater.*, 1:011002, 2013.

[5] Schmidt J., Wang H.-C., Cerqueira T. F. T., Botti S., and Marques M. A. L. A dataset of 175k stable and metastable materials calculated with the pbesol and scan functionals. *Sci. Data*, 9:64, 2022.

[6] Elena A. M., Kamath P. D., Jaffrelot Inizan T., Rosen A. S., Zanca F., and Persson K. A. Machine learned potential for high-throughput phonon calculations of metal–organic frameworks. *npj Comput. Mater.*, 11:125, 2025.

[7] Kraß H., Huang J., and Moosavi S. M. Mofsimbench: Evaluating universal machine learning interatomic potentials in metal–organic framework molecular modeling. *arXiv preprint*, arXiv:2507.11806, 2025.

[8] Perdew J. P., Burke K., and Ernzerhof M. Generalized gradient approximation made simple. *Phys. Rev. Lett.*, 77:3865–3868, 1996.

[9] Grimme S., Antony J., Ehrlich S., and Krieg H. A consistent and accurate ab initio parametrization of density functional dispersion correction (dft-d) for the 94 elements h–pu. *J. Chem. Phys.*, 132:154104, 2010.

[10] Grimme S., Ehrlich S., and Goerigk L. Effect of the damping function in dispersion corrected density functional theory. *J. Comput. Chem.*, 32:1456–1465, 2011.

[11] Rosen A. S., Notestein J. M., and Snurr R. Q. Comparing gga, gga+u, and meta-gga functionals for redox-dependent binding at open metal sites in metal–organic frameworks. *J. Chem. Phys.*, 152:224101, 2020.

[12] Sriram A., Choi S., Yu X., Brabson L. M., Das A., Ulissi Z., Uyttendaele M., Medford A. J., and Sholl D. S. The open dac 2023 dataset and challenges for sorbent discovery in direct air capture. *ACS Cent. Sci.*, 10:923–941, 2024.

[13] Rosen A. S., Iyer S. M., Ray D., Yao Z., Aspuru-Guzik A., Gagliardi L., Notestein J. M., and Snurr R. Q. Machine learning the quantum-chemical properties of metal–organic frameworks for accelerated materials discovery. *Matter*, 4:1578–1597, 2021.

[14] Wieser S. and Zojer E. Machine learned force-fields for an ab-initio quality description of metal–organic frameworks. *npj Comput. Mater.*, 10:18, 2024.

[15] Tayfuroglu O., Kocak A., and Zorlu Y. Modeling gas adsorption and mechanistic insights into flexibility in isoreticular metal–organic frameworks using high-dimensional neural network potentials. *Langmuir*, 41(11):7323–7335, 2025.

[16] Furness J. W., Kaplan A. D., Ning J., Perdew J. P., and Sun J. Accurate and numerically efficient r2scan meta-generalized gradient approximation. *J. Phys. Chem. Lett.*, 11:8208–8215, 2020.

[17] Anisimov V. I., Zaanen J., and Andersen O. K. Band theory and mott insulators: Hubbard u instead of stoner i. *Phys. Rev. B*, 44:943–954, 1991.

[18] Edzards J., Santana-Andreo J., Saßnick H.-D., and Cocchi C. Benchmarking selected density functionals and dispersion corrections for mof-5 and its derivatives. *J. Chem. Theory Comput.*, 21:7062–7074, 2025.

[19] Huang X., Deng B., Zhong P., Kaplan A. D., Persson K. A., and Ceder G. Cross-functional transferability in universal machine learning interatomic potentials. *arXiv preprint*, arXiv:2504.05565, 2025.

[20] Cockayne E. Thermodynamics of the flexible metal-organic framework material mil-53(cr) from first principles. *J. Phys. Chem. C*, 121:4312–4317, 2017.

# Towards an atomistic understanding of solid-state reaction interfaces using machine-learned interatomic potentials

Bryant Y. Li[1,2,$], Vir Karan[1,2,$] and Kristin A. Persson[1,2*]

[1] Department of Materials Science and Engineering, University of California Berkeley, Berkeley, California 94720, United States
[2] Materials Science Division, Lawrence Berkeley National Laboratory, Berkeley, California 94720, United States
*Author to whom any correspondence should be addressed.  $These authors contributed equally.

**E-mail:** kristinpersson@berkeley.edu

**Status**
Reactions in the solid-state, in particular those driven by bulk ionic motion, affect the synthesis and subsequent performance of functional materials, with applications ranging from traditional ceramic manufacturing and metal processing to modern technologies such as semiconductors and electrochemical interfaces [1]. Figure 1 displays some of the processes observed in the early stages of reactions at the interface between two solids. Unlike the well-defined reaction mechanisms of organic chemistry, our understanding of solid-state reactions has been limited by the sheer complexity of modeling solid-phases undergoing reactions through inter-diffusion, nucleation and growth processes. This has confined our understanding of such reactions primarily to quantities that can be derived from bulk thermodynamics, classical nucleation theory and Fickian diffusion[2]. First-principles methods, particularly density functional theory (DFT), are a key enabler to compute reaction energetics and evaluate thermodynamic feasibility. While such approaches have shown promise in describing energetic pathways for pairwise reactions by Muira et al.[3], they remain incapable of fully capturing the kinetics of the predicted reaction pathways, while being generalizable to both metallic and ceramic solid state systems.

While classical molecular dynamics (MD) is valuable for diffusion and vibrational dynamics, its non-reactive force fields preclude robust treatment of chemistry. Ab initio MD (AIMD) is therefore the standard for reaction studies, but its $O(n^3)$ cost limits it to $10^2 - 10^3$ atoms and sub-nanosecond (ns) trajectories. In this regard, the development of machine-learning interatomic potentials (MLIPs) has transformed the modeling of solid-state reactivity, since they allow for retaining ab-initio accuracy for a fraction of the cost. Modern MLIP–MD can reach timescales of several nanoseconds and $10^2$ nanometer length scales on modest computational resources, providing simulations of early solid–solid interface evolution and the prediction of both mesoscale reactivity morphologies and localized structural transformations. As demonstrated in the recent literature, a plethora of advances in the study of solid-state reactions made possible by the introduction of MLIPs [4, 5, 6].

For example, the evolution of the metal anode/solid state electrolyte battery interface for Li/Li$_7$P$_3$S$_{11}$ was explored with MLIPs. Prior AIMD studies have been restricted to ≈100-atom cells and timescales of up to 20 ps [7, 8], capturing only the earliest stages of interfacial reactivity and the formation of SEI products such as Li$_2$S and Li$_3$P. With MLIPs, Li et al.[9] were able to develop an interatomic potential trained on DFT configurations of dynamic and disordered states, combined with interfacial AIMD data, to represent the Li-P-S reactionary space. This MLIP enabled large-scale MLIP-MD simulations, scaling to 10 ns for ≈ 500, 000 atoms, providing an atomistic resolution of interphase evolution, including the onset of passivation and the emergence of heterogeneous interphase layering. These structural features were consistent with experimental validation from X-ray Photoelectron Spectroscopy (XPS) and Electrochemical Impedance Spectroscopy (EIS) studies [10, 11]. Furthermore, through Onsager analysis, Li et al. [9] was able to capture how interface morphology is impacted by kinetics, particularly correlated ionic transport, thereby revealing solid-state reaction pathways that remain inaccessible to conventional DFT or AIMD approaches.

A related study addressed the broader challenge of modeling the evolution of bulk reaction products in powder synthesis, using the Ba–Ti–O chemical space as a case study [12]. Traditionally, such reactions have been described by advancing along the thermodynamic hull of stability in a pairwise, energetics-only fashion [13]. In contrast, analogous to the Li–P–S system, the full correlated ionic transport tensor was computed via MLIP-MD for several "liquid-like" analogues of thermodynamically feasible intermediates in the $BaO$–$TiO_2$ reaction. These transport coefficients provide estimates of relative product formation rates under diffusion-limited conditions, where transport occurs through a liquid-like interfacial medium. The coefficients were then incorporated into a cellular automaton framework to simulate the evolution of bulk powder phases, thereby bridging atomic-scale transport to macroscopic synthesis outcomes [14]. This work represents the first model for powder synthesis reactions that explicitly accounts for both thermodynamic driving forces and kinetically preferred transport pathways. While the cellular automaton is only one possible route for scale bridging, its flexibility readily accommodates more sophisticated theories of solid–solid interfacial reactivity.

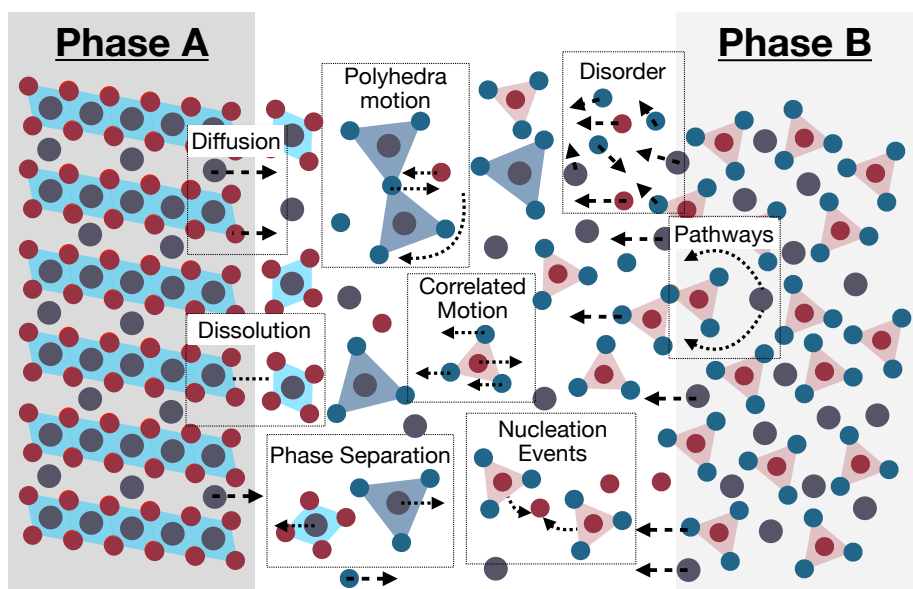

**Figure 1.** Illustration of kinetic phenomena captured at solid-state interfaces, revealed by MLIP-MD simulations, such as phase formation, reaction pathways, and kinetic barriers at solid-state interfaces over extended timescales and length scales that are not accessible with AIMD or DFT alone, providing atomistic insight into interface evolution.

Hence, the introduction of MLIPs into solid-state reactivity research has shown significant promise in bridging the gap between density functional theory (DFT) and molecular dynamics (MD) simulations to learn more about the complex, multiscale phenomena with unprecedented detail and accuracy.

**Current and future challenges**
Despite their successes, most MLIPs currently in use face significant limitations, both architecturally and in terms of training data, that hinder routine and reliable application to complex solid-state interfaces. A key gap, widely recognized in the community, is the explicit incorporation of charge (and charge-related quantities) into model predictions [15]. Existing state-of-the-art models struggle to capture long-range interactions, which are essential for describing charge distributions. Compounding this challenge is the lack of consensus on what "charge" should represent: different charge partitioning schemes yield incompatible partial charges, which often diverge from intuitive notions of formal charges in molecules or oxidation states in solids. These issues become especially consequential in studies such as Li et al. [9], where SEI formation depends critically on redox processes and charge transfer during battery cycling. Without explicitly accounting for charge, modeling solid-state interfaces (and surfaces) with potentially open-shell atoms under applied external potentials is not feasible. Recent work has applied the concept of 'latent' charges in the Cartesian Atomic Cluster Expansion (CACE) framework[16, 17], to capture atomistic charge-like quantities for computing charge-dependent long-range interactions in solids. Large-scale use and validation of this method still awaits.

Another challenge with MLIPs has been designing generally applicable metrics to validate their behavior on tasks outside the distribution of equilibrium and near-equilibrium solids. Most benchmarks focus on properties that are obtained from close-to-equilibrium calculations and associated mean absolute errors in the predictions of energies, forces and stress.[18, 19] However, the suitability of MLIPs, especially for probing interfaces and interphases, cannot be evaluated solely by equilibrium benchmarks, because the prediction task is fundamentally different. As shown by Karan et al. [12], extracting correlated transport from long-timescale MD requires both fidelity to DFT reference data and stable dynamics across equilibrium and non-equilibrium states. Indeed, under such conditions, e.g. elevated temperatures, the simulation can devolve into unphysical behavior such as atoms acquiring ballistic, plasma-like velocities. Such behavior indicates that the learned PES contains discontinuities or non-smooth regions [19]. The most reliable remedy is to smooth the short-range (core-repulsive) portion of the PES and incorporate high-temperature AIMD data into the training set. Secondly, MLIP simulations typically overestimate reaction rates due to the preferential sampling of lower-energy states. Although Li et al.[9] applied active learning over disordered atomistic states to expand out-of-equilibrium coverage, the well-known smoothing of the potential led to the onset of interphase passivation occurring faster than suggested by experimental evidence. These deficiencies point to a clear need for task-specific benchmarks that directly probe a model's ability to predict interfacial energetics and forces, and to maintain MD stability, rather than relying on equilibrium metrics alone.

In summary, as MLIP architectures and data-generation pipelines mature, access to observables long inaccessible to DFT, such as long-time correlated transport and high-temperature interfacial kinetics, can directly serve as 'digital twins' for solid state synthesis and aid in the operation of autonomous laboratories [20]. Self-driving platforms for solid-state synthesis are especially well positioned to exploit these insights: as in the Ba–Ti–O example, one can envision a cross-periodic-table database of high-temperature thermodynamic and transport descriptors to steer automated exploration of under-sampled chemistries. Crucially, such closed-loop experiments will also serve as ground-truth to DFT-trained MLIPs, allowing us to distinguish genuine interfacial reaction mechanisms from artifacts introduced by model bias or PES pathologies.

**Concluding remarks**

MLIPs are redefining how we explore solid-state interfaces and reaction mechanisms, offering first-principles accuracy at scales that reveal phenomena previously beyond reach. By bridging the gap between electronic structure calculations and large-scale dynamical simulations, MLIPs are uncovering pathways, kinetics, and interfacial behaviors inaccessible to DFT-type methods alone. Yet, critical challenges remain, such as integrating robust charge descriptions, ensuring stability under extrapolation, and establishing universal benchmarks for fair and rigorous evaluation. Advances in charge-informed architectures and standardized benchmarking are positioning MLIPs to become foundational tools in the field of solid state chemistry, where MLIPs can operate as digital twins within autonomous laboratories, coupling simulation, synthesis, and characterization in closed-loop discovery pipelines.

**Acknowledgments**

B.Y.L acknowledges the Energy Storage Research Alliance (ESRA), an Energy Innovation Hub funded by the U.S. Department of Energy. V.K acknowledges D2S2, the core program KCD2S2, which is supported by the U.S. Department of Energy, Office of Science, Office of Basic Energy Sciences, Materials Sciences and Engineering Division under contract no. DE-AC02-05-CH11231.

**References**

[1] DiSalvo F J 1990 *Science* **247** 649–655 ISSN 1095-9203 URL
    http://dx.doi.org/10.1126/science.247.4943.649

[2] Schmalzried H 2008 *Chemical kinetics of solids* (John Wiley & Sons)

[3] Miura A, Bartel C J, Goto Y, Mizuguchi Y, Moriyoshi C, Kuroiwa Y, Wang Y, Yaguchi T, Shirai M, Nagao M, Rosero-Navarro N C, Tadanaga K, Ceder G and Sun W 2021 *Advanced Materials* **33** ISSN 1521-4095 URL http://dx.doi.org/10.1002/adma.202100312

[4] Wang C, Aykol M and Mueller T 2023 *Chemistry of Materials* **35** 6346–6356 ISSN 1520-5002 URL http://dx.doi.org/10.1021/acs.chemmater.3c00993

[5] Bienvenu B, Todorova M, Neugebauer J, Raabe D, Mrovec M, Lysogorskiy Y and Drautz R 2025 *npj Computational Materials* **11** ISSN 2057-3960 URL
    http://dx.doi.org/10.1038/s41524-025-01574-w

[6] Holekevi Chandrappa M L, Qi J, Chen C, Banerjee S and Ong S P 2022 *Journal of the American Chemical Society* **144** 18009–18022 ISSN 1520-5126 URL
    http://dx.doi.org/10.1021/jacs.2c07482

[7] Camacho-Forero L E and Balbuena P B 2018 *Journal of Power Sources* **396** 782–790 ISSN 0378-7753 URL http://dx.doi.org/10.1016/j.jpowsour.2018.06.092

[8] Lepley N D and Holzwarth N A W 2015 *Physical Review B* **92** ISSN 1550-235X URL http://dx.doi.org/10.1103/PhysRevB.92.214201

[9] Li B Y, Karan V, Kaplan A D, Wen M and Persson K A 2025 *The Journal of Physical Chemistry C* **129** 16043–16054 ISSN 1932-7455 URL http://dx.doi.org/10.1021/acs.jpcc.5c03589

[10] Wood K N, Steirer K X, Hafner S E, Ban C, Santhanagopalan S, Lee S H and Teeter G 2018 *Nature Communications* **9** ISSN 2041-1723 URL http://dx.doi.org/10.1038/s41467-018-04762-z

[11] Wenzel S, Leichtweiss T, Krüger D, Sann J and Janek J 2015 *Solid State Ionics* **278** 98–105 ISSN 0167-2738 URL http://dx.doi.org/10.1016/j.ssi.2015.06.001

[12] Karan V, Gallant M C, Fei Y, Ceder G and Persson K A 2025 Ion correlations explain kinetic selectivity in diffusion-limited solid state synthesis reactions URL https://arxiv.org/abs/2501.08560

[13] McDermott M J, Dwaraknath S S and Persson K A 2021 *Nature Communications* **12** ISSN 2041-1723 URL http://dx.doi.org/10.1038/s41467-021-23339-x

[14] Gallant M C, McDermott M J, Li B and Persson K A 2024 *Chemistry of Materials* **37** 210–223 ISSN 1520-5002 URL http://dx.doi.org/10.1021/acs.chemmater.4c02301

[15] Deng B, Zhong P, Jun K, Riebesell J, Han K, Bartel C J and Ceder G 2023 *Nature Machine Intelligence* 1–11

[16] Cheng B 2025 *npj Computational Materials* **11** ISSN 2057-3960 URL http://dx.doi.org/10.1038/s41524-025-01577-7

[17] Kim D, King D S, Zhong P and Cheng B 2024 Learning charges and long-range interactions from energies and forces URL https://arxiv.org/abs/2412.15455

[18] Riebesell J, Goodall R E A, Benner P, Chiang Y, Deng B, Ceder G, Asta M, Lee A A, Jain A and Persson K A 2025 *Nature Machine Intelligence* **7** 836–847 ISSN 2522-5839 URL http://dx.doi.org/10.1038/s42256-025-01055-1

[19] Chiang Y, Kreiman T, Weaver E, Kuner M, Zhang C, Kaplan A, Chrzan D, Blau S M, Krishnapriyan A S and Asta M 2025 MLIP arena: Advancing fairness and transparency in machine learning interatomic potentials through an open and accessible benchmark platform *AI for Accelerated Materials Design - ICLR 2025* URL https://openreview.net/forum?id=ysKfIavYQE

[20] Szymanski N J, Rendy B, Fei Y, Kumar R E, He T, Milsted D, McDermott M J, Gallant M, Cubuk E D, Merchant A, Kim H, Jain A, Bartel C J, Persson K, Zeng Y and Ceder G 2023 *Nature* **624** 86–91 ISSN 1476-4687 URL http://dx.doi.org/10.1038/s41586-023-06734-w

# Generative AI towards new chemical reactions

Chenru Duan[1],[*]  Haojun Jia[1] , Qiyuan Zhao[1]

[1] Deep Principle, Inc., Cambridge, MA, USA
[*] Author to whom any correspondence should be addressed.

**E-mail:** duanchenru@gmail.com

**Status**
Transition state (TS) is the key in chemical reaction for elucidating mechanisms and optimizing catalysts and conditions[1]. Locating a TS computationally, however, remains the kinetic bottleneck for building trustworthy reaction networks[2, 3, 4]. Classical methods, such as nudge elastic band (NEB), growing string methods (GSM), and stochastic surface walking, are accurate but can be slow and sensitive to initialization. Despite limited by their scale in chemical reaction exploration, these established approaches were recently used to generate large 3D reactive datasets. For example, Transition1X provides 9.6M DFT single points along approximately 10k NEB pathways at $\omega$B97X/6-31G(d)[5]. Complementary, RGD1 offers approximately 177k validated TSs from graphically defined reaction space, many with multiple conformers, at both GFN2-xTB and DFT levels[6]. These datasets enable chemical reaction learning as a three-dimensional (3D) structure generation task, which can then be addressed by generative AI frameworks[7, 8], most recently denoising diffusion probabilistic models, flow matching, and optimal transport[9, 10].

OA-ReactDiff learns the joint 3D distribution over reactants (R), TS, and products (P) with object-aware SE(3) symmetry that follows all physical symmetries and constraints in chemical reactions[11]. With R and/or P provided, it generates the TS through inpainting, delivering sub-Å TS guesses in seconds and pairing naturally with a confidence model for ranking[12]. Independently, TSDiff generates TS structures from 2D graphs alone, sampling diverse TS conformations efficiently[13]. React-OT reframes double-ended TS generation as optimal transport with flow matching, solving an ordinary differential equation defined by a learned vector field started by a linear interpolation of R and P structures as initial TS guess[14]. Ablation study shows both the optimal transport objective and a good initial TS guess matter. Compared on Transition1X, React-OT reduces both TS RMSD and barrier error versus OA-ReactDiff while being orders of magnitude faster at inference. In addition, pretraining on RGD1-xTB further improves the performance of React-OT and shows transferability to out-of-distribution reactions (e.g., diverse Diels–Alder reaction sets).

**Current and future challenges**
Despite the success in generating accurate elementary reactions with speed, there are still challenges ahead to push generative AI useful for practical chemical reaction network constructions to enable new reaction discovery.

*Limitations of reactive data.-* Widely used datasets remain biased toward small, neutral, organic molecules. Chemistry involving charges, radicals, heteroatoms (S, P, halogens), transition metals, and explicit solvation are underrepresented[15]. Conformational diversity in the vicinity of saddle points is also sparse[16]. Heterogeneous catalysis datasets (e.g., OC22[17]) advanced surface absorption optimizations, but they are task-specific without extensive touch on cost-demanding TS acquisition for surface chemistry.

*Model transferability.-* Performance often degrades under distribution shift across chemistry, for example, unseen reaction families, larger systems, or different electronic-structure levels. Evidence indicates that pretraining on broad, low-level reaction dataset (e.g., RGD1-xTB), followed by light high-level fine-tuning, improves both geometric and barrier accuracy. Committee-based active learning similarly enhances MLIP robustness in reactive regions[18].

*Benchmarks to trust.-* Mean absolute errors (MAE) on energies and forces are inadequate proxies for end-to-end performance for chemical reactions[19]. Yet establishing an end-to-end framework to evaluate the performance of a generative model or MLIP is a non-trial mission where many factors come to play, optimization procedure, definition of success, to name a few.

*Cost of a network.-* Even when one-shot TS inference is sub-second, RP preparation (often via xTB/MLIP optimization) and DFT IRC calculations dominate wall-time[14]. In addition, the combinatorial explosion of the number of potential reactions still poses challenges on constructing complete chemical reaction networks when initial reactants are provided.

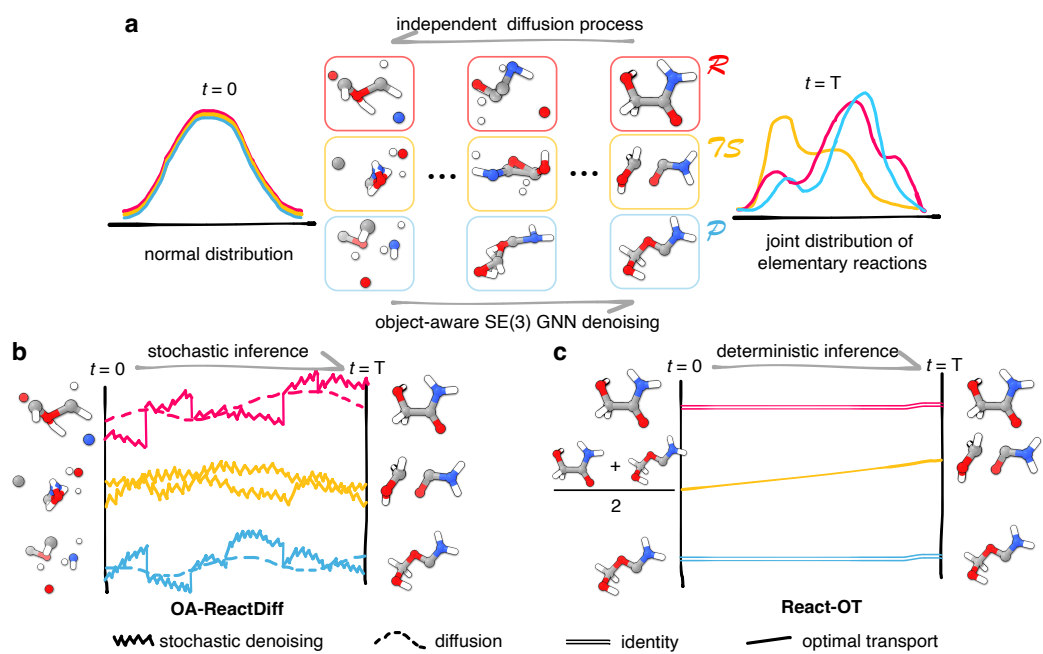

**Figure 1.** Overview of the OA-ReactDiff and React-OT for generating TS. a. Learning the joint distribution of structures in elementary reactions (reactant in red, TS in yellow, and product in blue). b. Stochastic inference with inpainting in OA-ReactDiff. c. Deterministic inference with React-OT. Atoms are colored as follows: C for gray; N for blue, O for red, and H for white. Reproduced with permission from Nature publishing group from React-OT[14].

### Advances in science and technology to meet challenges

To make generative AI-based reaction exploration reliable, transferable, and cost-effective, the field should prioritize the following.

*Broaden chemical coverage.-* Build reactive datasets that systematically include charged and radical species, heavier p-block elements, transition-metal chemistry, and larger systems, with validated TS geometries and energies. Solvent effects, counter-ions, and surface reactions are desired in a controlled manner where they materially influence barriers or TS geometry.

*Evaluate transfer under domain shift.-* Report performance on held-out reaction families, charge states, molecular sizes, and electronic-structure levels, using rigorously designed cross-domain splits.

*Benchmark end-to-end performance.-* Prioritize intended-TS discovery verified by intrinsic reaction coordinate calculations , success of downstream NEB/GSM refinement, TS structure accuracy, and barrier MAE, with confidence intervals and stratification by reaction class and system size. Sustain public, reproducible leaderboards with fixed splits, code, and document evaluation pipelines so progress is attributable to modeling advances rather than evaluation drift[20].

*Engineer for network constructions.-* Modern engineering standard, such as batch inference, cache intermediates, kernel accelerators should be applied for diffusion and flow matching during the growth of chemical reaction networks. For practical applications, the pipeline should be instrumented to attribute time and compute to each stage against a declared accuracy budget.

### Concluding remarks

A pragmatic recipe is emerging for the paradigm shift from intuition-driven to generative AI-powered chemical reaction discovery[8]. Deterministic transport models are well suited for one-shot, accurate TS generation. Stochastic diffusion models remain valuable where diversity is required, enumerating alternative products, conformers, or proposing mechanistic hypotheses. Equivariant MLIPs, ideally trained with second-order Hessian supervision, provide fast and stable refinement and triage before selective DFT/NEB/GSM validation. Stitching these components together, with more diverse reactive data that improves model transferability, proper end-to-end framework that tracks the progress, and necessary engineering effort that accelerates the network growth, generative AI would enable reaction discovery at an unprecedented speed.

### Acknowledgments

We would like to thank our entire team from Deep Principle for helpful discussions and support.

**References**

[1] Jan P. Unsleber and Markus Reiher. The exploration of chemical reaction networks. *Annual Review of Physical Chemistry*, 71(1):121–142, 2020. PMID: 32105566.

[2] Qiyuan Zhao and Brett M Savoie. Simultaneously improving reaction coverage and computational cost in automated reaction prediction tasks. *Nature Computational Science*, 1(7):479–490, 2021.

[3] Qiyuan Zhao and Brett M Savoie. Deep reaction network exploration of glucose pyrolysis. *Proceedings of the National Academy of Sciences*, 120(34):e2305884120, 2023.

[4] Mingjian Wen, Evan Walter Clark Spotte-Smith, Samuel M. Blau, Matthew J. McDermott, Aditi S. Krishnapriyan, and Kristin A. Persson. Chemical reaction networks and opportunities for machine learning. *Nature Computational Science*, 3(1):12–24, Jan 2023.

[5] Mathias Schreiner, Arghya Bhowmik, Tejs Vegge, Jonas Busk, and Ole Winther. Transition1x - a dataset for building generalizable reactive machine learning potentials. *Scientific Data*, 9(1):779, Dec 2022.

[6] Qiyuan Zhao, Sai Mahit Vaddadi, Michael Woulfe, Lawal A. Ogunfowora, Sanjay S. Garimella, Olexandr Isayev, and Brett M. Savoie. Comprehensive exploration of graphically defined reaction spaces. *Scientific Data*, 10(1):145, Mar 2023.

[7] Sunghwan Choi. Prediction of transition state structures of gas-phase chemical reactions via machine learning. *Nature Communications*, 14(1):1168, Mar 2023.

[8] Yuanqi Du, Arian R. Jamasb, Jeff Guo, Tianfan Fu, Charles Harris, Yingheng Wang, Chenru Duan, Pietro Liò, Philippe Schwaller, and Tom L. Blundell. Machine learning-aided generative molecular design. *Nature Machine Intelligence*, 6(6):589–604, Jun 2024.

[9] Yaron Lipman, Ricky T. Q. Chen, Heli Ben-Hamu, Maximilian Nickel, and Matt Le. Flow matching for generative modeling, 2023.

[10] Konstantin Mark, Leonard Galustian, Maximilian P. P. Kovar, and Esther Heid. Feynman-kac-flow: Inference steering of conditional flow matching to an energy-tilted posterior, 2025.

[11] Chenru Duan, Yuanqi Du, Haojun Jia, and Heather J. Kulik. Accurate transition state generation with an object-aware equivariant elementary reaction diffusion model. *Nature Computational Science*, 3(12):1045–1055, 2023.

[12] Gabriele Corso, Hannes Stärk, Bowen Jing, Regina Barzilay, and Tommi Jaakkola. DiffDock: Diffusion steps, twists, and turns for molecular docking. *arXiv:2210.01776*, 2023.

[13] Seonghwan Kim, Jeheon Woo, and Woo Youn Kim. Diffusion-based generative AI for exploring transition states from 2D molecular graphs. *Nature Communications*, 15(1):341, 2024.

[14] Chenru Duan, Guan-Horng Liu, Yuanqi Du, Tianrong Chen, Qiyuan Zhao, Haojun Jia, Carla P Gomes, Evangelos A Theodorou, and Heather J Kulik. Optimal transport for generating transition states in chemical reactions. *Nature Machine Intelligence*, 7(4):615–626, 2025.

[15] Daniel S. Levine, Muhammed Shuaibi, Evan Walter Clark Spotte-Smith, Michael G. Taylor, Muhammad R. Hasyim, Kyle Michel, Ilyes Batatia, Gábor Csányi, Misko Dzamba, Peter Eastman, Nathan C. Frey, Xiang Fu, Vahe Gharakhanyan, Aditi S. Krishnapriyan, Joshua A. Rackers, Sanjeev Raja, Ammar Rizvi, Andrew S. Rosen, Zachary Ulissi, Santiago Vargas, C. Lawrence Zitnick, Samuel M. Blau, and Brandon M. Wood. The open molecules 2025 (omol25) dataset, evaluations, and models, 2025.

[16] Taoyong Cui, Yunhong Han, Haojun Jia, Chenru Duan, and Qiyuan Zhao. Horm: A large scale molecular hessian database for optimizing reactive machine learning interatomic potentials, 2025.

[17] Richard Tran*, Janice Lan*, Muhammed Shuaibi*, Brandon Wood*, Siddharth Goyal*, Abhishek Das, Javier Heras-Domingo, Adeesh Kolluru, Ammar Rizvi, Nima Shoghi, Anuroop Sriram, Zachary Ulissi, and C. Lawrence Zitnick. The open catalyst 2022 (oc22) dataset and challenges for oxide electrocatalysts. *ACS Catalysis*, 2023.

[18] B. Kalita, R. Zubatyuk, D. M. Anstine, M. Bergeler, V. Settels, C. Stork, et al. AIMNet2-NSE: A transferable reactive neural network potential for open-shell chemistry. *ChemRxiv*, 2025. Preprint; not peer-reviewed.

[19] Qiyuan Zhao, Yunhong Han, Duo Zhang, Jiaxu Wang, Peichen Zhong, Taoyong Cui, Bangchen Yin, Yirui Cao, Haojun Jia, and Chenru Duan. Harnessing machine learning to enhance transition state search with interatomic potentials and generative models. *Adv. Sci.*, page e06240, 2025.

[20] Matthew K. Horton, Patrick Huck, Ruo Xi Yang, Jason M. Munro, Shyam Dwaraknath, Alex M. Ganose, Ryan S. Kingsbury, Mingjian Wen, Jimmy X. Shen, Tyler S. Mathis, Aaron D. Kaplan, Karlo Berket, Janosh Riebesell, Janine George, Andrew S. Rosen, Evan W. C. Spotte-Smith, Matthew J. McDermott, Orion A. Cohen, Alex Dunn, Matthew C. Kuner, Gian-Marco Rignanese, Guido Petretto, David Waroquiers, Sinead M. Griffin, Jeffrey B. Neaton, Daryl C. Chrzan, Mark Asta, Geoffroy Hautier, Shreyas Cholia, Gerbrand Ceder, Shyue Ping Ong, Anubhav Jain, and Kristin A. Persson. Accelerated data-driven materials science with the materials project. *Nature Materials*, Jul 2025.

# Recommender system for discovery of inorganic compounds

Hiroyuki Hayashi[1], Atsuto Seko[1] and Isao Tanaka[1,2]

[1] Department of Materials Science and Engineering, Kyoto University, Kyoto, Japan
[2] Nano Research Laboratory, Japan Fine Ceramics Center (JFCC), Nagoya, Japan
E-mail: tanaka@cms.mtl.kyoto-u.ac.jp

**Status**

The discovery of inorganic compounds continues to drive progress in materials science and technology. Two routes are common. One is to reveal new functions in known compounds. The other is to find compounds not yet reported. The second route is difficult because the composition space is very large and cannot be explored well without data-driven support. Large experimental databases such as the ICSD [1] provide a foundation, but growth in new entries has slowed and oxides dominate the records among ionic compounds. This trend suggests that traditional search alone is not sufficient.

The idea of chemically relevant compositions (CRC) has become important. These are compositions that may form stable or metastable compounds under given conditions. First-principles calculations can test stability, but the cost is high when structures are unknown. Machine learning now complements these tools. In particular, recommender systems use existing data to score new compositions and guide synthesis. Descriptor-based [2] and tensor-based [3] models have both shown value. Some predicted compositions have been validated by experiment, and related ideas extend to synthesis conditions as well [4,5]. Notably, experimental groups have already succeeded in discovering new compounds, including lithium-ion conducting oxides and novel nitrides, using recommender systems [6-8].

**Current and future challenges**

Despite the demonstrated potential of recommender systems in materials discovery, several fundamental challenges remain. A primary difficulty lies in the availability and diversity of training data. Existing inorganic compound databases are strongly biased toward stable oxides and ground-state phases, which are easier to synthesize and characterize. As a result, vast regions of the chemical composition space, particularly non-oxides, disordered systems, and metastable compounds, remain underexplored. This imbalance restricts the generalizability of recommender models and limits their ability to propose compositions beyond well-established chemical domains.

Another challenge is the limited success rate of experimental validation. Even highly ranked compositions may fail in synthesis [4], showing that machine learning alone does not provide sufficient guidance for realization. The absence of comprehensive datasets for synthesis conditions further complicates this problem. Text-mined synthesis databases, while valuable, are heavily biased toward successful reports in the literature, and thus lack the negative data that are equally essential for robust machine learning.

There are also computational challenges. Validation of predicted compositions using first-principles calculations remains resource-intensive, especially when candidate structures are unknown. Exhaustive structure searching and convex hull construction are feasible only for a limited number of systems. As recommender systems propose increasingly complex compositions, the associated computational and experimental workload may grow disproportionately.

**Advances in science and technology to meet challenges**

Meeting the challenges outlined above requires advances in both data infrastructure and methodological innovation. The first priority is the systematic expansion of datasets that underpin recommender systems. Current databases are biased toward stable oxides and well-studied systems. To overcome this, new strategies must integrate data from multiple sources: literature mining, high-throughput first-principles calculations, and automated synthesis experiments. Particularly important is the inclusion of unsuccessful synthesis attempts, which can provide critical negative examples for machine learning models. The development of robotic and parallel experimental platforms offers a promising pathway to generate large, unbiased datasets that capture both successes and failures.

On the algorithmic front, hybrid approaches that combine the strengths of descriptor-based and tensor-based methods are needed. Descriptor-based models are effective when physical features correlate with target properties, while tensor-based approaches excel with large amounts of uniformly distributed data. Integrating these strategies can improve predictive performance across diverse domains. Furthermore, the incorporation of uncertainty quantification into recommender scores is a key advance. Confidence estimates, already established in other machine learning applications, can help prioritize candidates for experimental validation and allocate resources more effectively.

Progress will depend on stronger integration between computation and experiment. Recommender systems for synthesis conditions, when combined with automated laboratories, could enable closed-loop discovery in which predictions are continuously refined by feedback from high-throughput experiments. This vision of an autonomous discovery cycle represents a paradigm shift in materials research. By uniting large-scale data generation, machine learning, and physical modeling, the field can move toward a future where the exploration of vast chemical spaces becomes practical and routine.

**Concluding remarks**

Recommender systems have emerged as a powerful and data-efficient strategy for accelerating the discovery of inorganic compounds. By leveraging existing databases, these methods have successfully identified chemically relevant compositions and, in some cases, guided the experimental synthesis of previously unknown materials. Extending the concept to synthesis conditions has further demonstrated the versatility of recommender approaches, showing that machine learning can aid not only in identifying promising compositions but also in suggesting pathways for their realization.

**Acknowledgements**
H.H. was supported by PRESTO, JST, and JSPS KAKENHI (Grant No. JP20H02423). A.S. was supported by PRESTO, JST, and JSPS KAKENHI (Grant Nos. JP18K18942 and JP19H05787). I.T. was supported by JSPS KAKENHI (Grant Nos. JP21H04621 and 23K17835).

**References**

[1] Bergerhoff G and Brown I D 1987 In Crystallographic Databases edited by Allen F H et al International Union of Crystallography Chester

[2] Seko A, Hayashi H and Tanaka I 2018 Compositional descriptor based recommender system for the materials discovery The Journal of Chemical Physics 148 241719

[3] Seko A, Hayashi H, Kashima H and Tanaka I 2018 Matrix and tensor based recommender systems for the discovery of currently unknown inorganic compounds Physical Review Materials 2 013805

[4] Hayashi H, Hayashi K, Kouzai K, Seko A and Tanaka I 2019 Recommender system of successful processing conditions for new compounds based on a parallel experimental data set Chemistry of Materials 31 9984 to 9992

[5] Hayashi H, Seko A and Tanaka I 2022 Recommender system for discovery of inorganic compounds npj Computational Materials 8 217

[6] Suzuki K, Ohura K, Seko A, Iwamizu Y, Zhao G, Hirayama M, Tanaka I and Kanno R 2020 Fast material search of lithium ion conducting oxides using a recommender system Journal of Materials Chemistry A 8 11582 to 11588

[7] Koyama Y, Seko A, Tanaka I, Funahashi S and Hirosaki N 2021 Combination of recommender system and single particle diagnosis for accelerated discovery of novel nitrides The Journal of Chemical Physics 154 224117

[8] Nakayama T, Watanabe K, Iwamizu Y, Suzuki K, Matsui N, Seko A, Tanaka I, Kanno R, Hirayama M 2025 Simultaneous Exploration of Candidates for Electrode and Electrolyte Materials for All-Solid-State Batteries Using Predicted Rating from a Recommender System ACS Applied Energy Materials 8 2260 to 2267

# Autonomous Material Discovery with Generative AI Systems

Theo Jaffrelot Inizan[1,2,3,4], Omar M. Yaghi[1,2,3,5,*] and Kristin A. Persson[1,2,*]

[1] Department of Chemistry, University of California, Berkeley, CA, USA.
[2] Bakar Institute of Digital Materials for the Planet, University of California, Berkeley, CA, USA.
[3] Materials Sciences Division, Lawrence Berkeley National Laboratory, Berkeley, CA, USA.
[4] AIMATX Inc., Berkeley, CA, USA.
[5] Kavli Energy NanoScience Institute, Berkeley, CA, USA.
[*] Author to whom any correspondence should be addressed.

**E-mail:** yaghi@berkeley.edu, kristinpersson@berkeley.edu

**Status**

Traditional material discovery has mainly relied on heuristics, trial-and-error and, more recently, quantum mechanical computational workflows [1]. Generative AI agents, such as large language models (LLMs), have begun to reshape the discovery process by helping experimentalists in designing experiments through data extraction, synthesis condition exploration and molecular design [2, 3]. Although the first use of generative models for crystal structure generation, through Generative Adversarial Networks, dates back to 2018 [4], more recent works have greatly improved robustness employing diffusion-based approaches [5, 6, 7].

Despite these advances, the steps toward synthesizing GenAI-designed materials are still limited, particularly for large systems with thousands of atoms and a large chemical space. Some strategies have been proposed to bypass these difficulties through the use of coarse graining [8] and novel representations [9, 6]. Most recently, such approaches have been applied to generate large crystal structures of metal-organic frameworks (MOFs) [10] with unit cells exceeding 250 atoms [7].

However, scalability remains difficult, as generated crystal structures become less physical when models are scaled to larger systems. This limitation prevents current models from targeting large complex systems such as MOFs or disordered materials, which typically involve larger unit cells and diverse chemistries. In addition, while tens of millions of structures have been generated from diffusion algorithms, only a handful of them have been synthesized so far: an inorganic crystal [11] and five MOFs [7]. Nevertheless, the rapid development of generative methods suggests that further advances in architectures, physical constraints and synthesizability awareness metrics could significantly bridge the gap between GenAI design and experimental feasibility.

**Current and future challenges**

While scalability has been extensively studied in image generation [12], materials generation remains underexplored, particularly how training-set size affects chemical validity. This is directly related to the lack of physical metrics during training, which will be discussed further later. Inorganic crystal databases are relatively large and sufficient for training diffusion models, the database size decreases with system size: from hundred of thousands of inorganic crystals (e.g. Materials Project [13] and AFLOW [14]), to a few hundred thousand MOFs. As an example, MOFGen leverages 300k structures drawn from essentially all major MOF databases [15, 16], yet without augmentation this dataset effectively renders diffusion models data-limited.

Furthermore, a major limitation is the lack of domain-informed metrics to assess the novelty, chemical validity, and synthesizability of a generated compound. Current measures such as uniqueness, validity, and formation energy give some indication of physical meaningfulness, but they lack a mechanistic understanding of synthesizability [17]. While the commonly used 'energy above hull' metric is relevant for assessing an upper bound of synthesizability of inorganic compounds, it is restricted to chemistries with few elemental types and does not apply well to larger and more complex systems with many atom types, such as MOFs. In addition, the generated structures usually require post-processing steps to remove floating atoms, especially light ones such as hydrogen.

Assessing the accuracy and robustness of ML methods remains difficult in the absence of a widely accepted benchmark. While MatBench has advanced standardization [18], its public availability and repeated reuse can introduce model-selection bias and inadvertent overfitting to the benchmark. Thus, when new metrics appear, past best models perform significantly worse on more general metrics. To mitigate bias, test sets should remain blinded and inaccessible during model development. A second gap is the absence of robust generalization benchmarks tied to diverse experimental observables; analogous to the SAMPL challenge.

A current trend is to focus on generating large databases rather than considering experimental constraints. While enumerating new structures, for example through atomic substitution, can be useful for small inorganic compounds, it requires large computational resources and does not scale to large frameworks. In the future, we envision stronger integration of experimentalist-in-the-loop approaches for 3D structure generation, where experimental feasibility guides AI-driven design.

**Advances in science and technology to meet challenges**

Addressing these challenges will require combining advances in modeling, benchmarking, and evaluation. Scalability could be improved by training and testing on benchmark databases of varying size and chemical complexity. Future model improvements in scalability will likely involve physics-inspired architectures or models conditioned under physical constraints, such as pairwise interactions, chemical composition rules, topology-constrained generation, or in-painting strategies. Making models more scalable with training size and computationally less intensive would enable a greater community to tackle problems that currently sometimes need hundreds of GPUs for training, which are not accessible to many labs. This would also enable their use on challenging or novel materials with small databases.

Progress will also require additional evaluation metrics, especially focused on synthesizability. This would enable better out-of-distribution predictions, generalization, and ultimately real discovery. Currently, the tendency of the community is to generate large databases that are often unusable for experimentalists or created with little emphasis on accessibility. A better understanding of what makes a material synthesizable is a key point and would necessitate interdisciplinary discussion. Possible directions include incorporating retrosynthesis rules, quantum physics-based descriptors, or reaction constraints directly into generative workflows, thereby prioritizing experimentally feasible structures from the beginning.

Benchmark databases should be designed not only to test crystal structure position recovery but also to assess out-of-distribution generalization and the ability to generate novel structures. Thus, model predictions should be tested on benchmarks that, if possible, are not publicly accessible to avoid bias. Such benchmarks should consist of sets of crystal structures with chemical diversity, spanning different chemistries and involving experimental observables directly linked to the crystal structure. We think that the community should take inspiration from work in biochemistry. Such studies would also contribute to understanding structure–property relationships and, more generally, help the development of adjacent fields such as machine learning interatomic potentials [19, 20].

Finally, emphasis should not be limited to diffusion models or LLMs, which each face intrinsic issues. We believe that physically aware models, where physical knowledge is embedded directly into the architecture, will define the next generation of 3D structure generators and open new directions for AI in science.

**Concluding remarks**

Generative AI has begun to transform material discovery but still faces challenges of scalability, synthesizability, benchmarking and physical awareness. Most models continue to generate unphysical structures even when trained on very large datasets. Future developments in physically grounded architectures combined with synthesizability metrics and automated experimentation, will be critical.

In the near future, the combination of generative models with autonomous laboratories could dramatically accelerate validation cycles. Their coupling with high-throughput robotics and automated synthesis may reduce discovery timelines from years to days. Generative AI thus holds the potential not only to generate new materials but also to enable their synthesis, reshaping the pace and scope of material discovery.

**Acknowledgments**

T.J.I. acknowledges the National Energy Research Scientific Computing Center (NERSC) financial support from U.S. National Science Foundation's "The Quantum Sensing Challenges for Transformational Advances in Quantum Systems (QuSeC-TAQS)" program. T.J.I. used resources of NERSC award DOE-ERCAP0031751 'GenAI@NERSC'. T.J.I. also thanks the Bakar Institute of Digital Materials for the Planet Fellowship for its support.

T.J.I. and O.M.Y. acknowledges the Bakar Institute of Digital Materials for the Planet (BIDMaP). Additional support on data generation and workflow development was obtained from

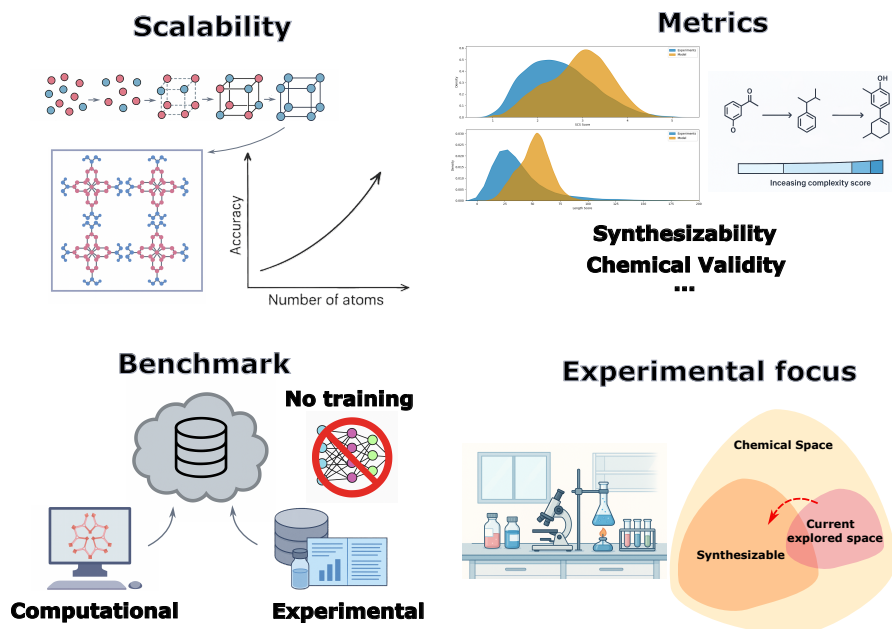

**Figure 1.** Scalability, synthesis-representative metrics, unbias model testing, benchmark diversity and experimental focus are the future challenges in 3D crystal structure generation on the path toward fully autonomous discovery of synthesizable materials. This include (1) improving the scalability of generative models to handle large atomic-level structure, (2) developing metrics that accurately assess chemical validity, diversity, and synthesizability, (3) establishing community-driven test set benchmarks using both computational and experimental data and 4) strengthening the experimental focus through better understanding of experimental constraint and challenges.

**References**

[1] Alex Ganose, Hrushikesh Sahasrabuddhe, Mark Asta, Kevin Beck, Tathagata Biswas, Alexander Bonkowski, Joana Bustamante, Xin Chen, Yuan Chiang, Daryl Chrzan, Jacob Clary, Orion Cohen, Christina Ertural, Max Gallant, Janine George, Sophie Gerits, Rhys Goodall, Rishabh Guha, Geoffroy Hautier, Matthew Horton, Aaron Kaplan, Ryan Kingsbury, Matthew Kuner, Bryant Li, Xavier Linn, Matthew McDermott, Rohith Srinivaas Mohanakrishnan, Aakash Naik, Jeffrey Neaton, Kristin Persson, Guido Petretto, Thomas Purcell, Francesco Ricci, Benjamin Rich, Janosh Riebesell, Gian-Marco Rignanese, Andrew Rosen, Matthias Scheffler, Jonathan Schmidt, Jimmy-Xuan Shen, Andrei Sobolev, Ravishankar Sundararaman, Cooper Tezak, Victor Trinquet, Joel Varley, Derek Vigil-Fowler, Duo Wang, David Waroquiers, Mingjian Wen, Han Yang, Hui Zheng, Jiongzhi Zheng, Zhuoying Zhu, and Anubhav Jain. *Atomate2: Modular workflows for materials science*. Jan. 22, 2025.

[2] Daniil A. Boiko, Robert MacKnight, Ben Kline, and Gabe Gomes. "Autonomous chemical research with large language models". In: *Nature* 624.7992 (Dec. 2023). Publisher: Nature Publishing Group, pp. 570–578.

[3] Zhiling Zheng, Nakul Rampal, Theo Jaffrelot Inizan, Christian Borgs, Jennifer T. Chayes, and Omar M. Yaghi. "Large language models for reticular chemistry". In: *Nature Reviews Materials* (Feb. 1, 2025). Publisher: Nature Publishing Group, pp. 1–13.

[4] Asma Nouira, Nataliya Sokolovska, and Jean-Claude Crivello. *CrystalGAN: Learning to Discover Crystallographic Structures with Generative Adversarial Networks*. May 25, 2019.

[5] Tian Xie, Xiang Fu, Octavian-Eugen Ganea, Regina Barzilay, and Tommi Jaakkola. *Crystal Diffusion Variational Autoencoder for Periodic Material Generation*. Mar. 14, 2022.

[6] Sherry Yang, KwangHwan Cho, Amil Merchant, Pieter Abbeel, Dale Schuurmans, Igor Mordatch, and Ekin Dogus Cubuk. *Scalable Diffusion for Materials Generation*. June 3, 2024.

[7] Theo Jaffrelot Inizan, Sherry Yang, Aaron Kaplan, Yen-hsu Lin, Jian Yin, Saber Mirzaei, Mona Abdelgaid, Ali H. Alawadhi, KwangHwan Cho, Zhiling Zheng, Ekin Dogus Cubuk, Christian Borgs, Jennifer T. Chayes, Kristin A. Persson, and Omar M. Yaghi. *System of Agentic AI for the Discovery of Metal-Organic Frameworks*. Apr. 18, 2025.

[8] Xiang Fu, Tian Xie, Andrew S. Rosen, Tommi Jaakkola, and Jake Smith. *MOFDiff: Coarse-grained Diffusion for Metal-Organic Framework Design*. Oct. 16, 2023.

[9] Olaf Ronneberger, Philipp Fischer, and Thomas Brox. *U-Net: Convolutional Networks for Biomedical Image Segmentation*. May 18, 2015.

[10] Omar M. Yaghi, Michael O'Keeffe, Nathan W. Ockwig, Hee K. Chae, Mohamed Eddaoudi, and Jaheon Kim. "Reticular synthesis and the design of new materials". In: *Nature* 423.6941 (June 2003). Number: 6941 Publisher: Nature Publishing Group, pp. 705–714.

[11] Claudio Zeni, Robert Pinsler, Daniel Zügner, Andrew Fowler, Matthew Horton, Xiang Fu, Zilong Wang, Aliaksandra Shysheya, Jonathan Crabbé, Shoko Ueda, Roberto Sordillo, Lixin Sun, Jake Smith, Bichlien Nguyen, Hannes Schulz, Sarah Lewis, Chin-Wei Huang, Ziheng Lu, Yichi Zhou, Han Yang, Hongxia Hao, Jielan Li, Chunlei Yang, Wenjie Li, Ryota Tomioka, and Tian Xie. "A generative model for inorganic materials design". In: *Nature* (Jan. 16, 2025), pp. 1–3.

[12] Tom Henighan, Jared Kaplan, Mor Katz, Mark Chen, Christopher Hesse, Jacob Jackson, Heewoo Jun, Tom B. Brown, Prafulla Dhariwal, Scott Gray, Chris Hallacy, Benjamin Mann, Alec Radford, Aditya Ramesh, Nick Ryder, Daniel M. Ziegler, John Schulman, Dario Amodei, and Sam McCandlish. *Scaling Laws for Autoregressive Generative Modeling*. Nov. 6, 2020.

[13] Anubhav Jain, Shyue Ping Ong, Geoffroy Hautier, Wei Chen, William Davidson Richards, Stephen Dacek, Shreyas Cholia, Dan Gunter, David Skinner, Gerbrand Ceder, and Kristin A. Persson. "Commentary: The Materials Project: A materials genome approach to accelerating materials innovation". In: *APL Materials* 1.1 (July 18, 2013), p. 011002.

[14] Stefano Curtarolo, Wahyu Setyawan, Gus L. W. Hart, Michal Jahnatek, Roman V. Chepulskii, Richard H. Taylor, Shidong Wang, Junkai Xue, Kesong Yang, Ohad Levy, Michael J. Mehl, Harold T. Stokes, Denis O. Demchenko, and Dane Morgan. "AFLOW: An automatic framework for high-throughput materials discovery". In: *Computational Materials Science* 58 (June 1, 2012), pp. 218–226.

[15] Yongchul G. Chung, Emmanuel Haldoupis, Benjamin J. Bucior, Maciej Haranczyk, Seulchan Lee, Hongda Zhang, Konstantinos D. Vogiatzis, Marija Milisavljevic, Sanliang Ling, Jeffrey S. Camp, Ben Slater, J. Ilja Siepmann, David S. Sholl, and Randall Q. Snurr. "Advances, Updates, and Analytics for the Computation-Ready, Experimental Metal–Organic Framework Database: CoRE MOF 2019". In: *Journal of Chemical & Engineering Data* 64.12 (Dec. 12, 2019), pp. 5985–5998.

[16] Andrew S. Rosen, Shaelyn M. Iyer, Debmalya Ray, Zhenpeng Yao, Alán Aspuru-Guzik, Laura Gagliardi, Justin M. Notestein, and Randall Q. Snurr. "Machine learning the quantum-chemical properties of metal–organic frameworks for accelerated materials discovery". In: *Matter* 4.5 (May 5, 2021), pp. 1578–1597.

[17] Wenhao Gao and Connor W. Coley. "The Synthesizability of Molecules Proposed by Generative Models". In: *Journal of Chemical Information and Modeling* 60.12 (Dec. 28, 2020). Publisher: American Chemical Society, pp. 5714–5723.

[18] Janosh Riebesell, Rhys E. A. Goodall, Philipp Benner, Yuan Chiang, Bowen Deng, Gerbrand Ceder, Mark Asta, Alpha A. Lee, Anubhav Jain, and Kristin A. Persson. "A framework to evaluate machine learning crystal stability predictions". In: *Nature Machine Intelligence* 7.6 (June 2025). Publisher: Nature Publishing Group, pp. 836–847.

[19] Jörg Behler and Michele Parrinello. "Generalized Neural-Network Representation of High-Dimensional Potential-Energy Surfaces". In: *Physical Review Letters* 98.14 (Apr. 2, 2007), p. 146401.

[20] Alin Marin Elena, Prathami Divakar Kamath, Théo Jaffrelot Inizan, Andrew S. Rosen, Federica Zanca, and Kristin A. Persson. *Machine Learned Potential for High-Throughput Phonon Calculations of Metal-Organic Frameworks*. Dec. 18, 2024.

# How do we ensure transferability in artificial intelligence based density functional approximations?

Stephen G. Dale[1−2∗],  Tim Gould[3], Bun Chan[4], Stefan Vuckovic[5]

[1]Institute of Functional Intelligent Materials, National University of Singapore, 4 Science Drive 2, Singapore 117544
[2]Department of Material Science and Engineering, National University of Singapore, 9 Engineering Drive 1, Singapore 117575
[3]Queensland Micro- and Nanotechnology Centre, Griffith University, Nathan, Qld 4111, Australia
[4]Graduate School of Engineering, Nagasaki University, Bunkyo 1-14, Nagasaki 852-8521, Japan
[5]Department of Chemistry, University of Fribourg, Fribourg, Switzerland
∗Author to whom any correspondence should be addressed.

**E-mail:** sdale@nus.edu.sg

**Status**
Density functional theory (DFT) is an exceptional modern tool for modeling the electronic structure of materials, and frequently forms the basis for training data of artificial intelligence (AI) materials discovery methods.[1, 2] Starting from approximation to the Infinite Uniform Electron Gas,[3] which hardly represents realistic electron distributions in real molecules and materials, density functional approximations (DFAs) have shown a remarkable degree of transferability to all chemical problems with minimal changes to modeling methods.[4, 5] This transferability is extremely valuable and permits a high degree of confidence when modeling new and unknown systems, and even where it fails, many DFAs fail predictably.[6]

As AI techniques are introduced to start producing new DFAs which are trained on large data sets it becomes dangerous to assume these AI-DFAs are transferable, with AI methods prone to overfitting, and in some cases an inability to extrapolate beyond the training set, or introducing instabilities to the self-consistent field convergence (SCF) technique.[7, 8, 9] To scrutinize these issues the author and coworkers[10] trained the $XYG_p$ functional on the 55 different subsets of the "General-main group thermochemistry, kinetics and non-covalent interaction" dataset (GMTKN55),[1] and the transition metal dataset TMC151.[11] By training on one subset of data, and then testing on another, we are able to recommend metrics for and then measure the transferability of the generated AI-DFAs. This resulted in both surprising and reassuring results.

1. Unsurprisingly, the chosen training set can have a remarkable impact on transferability.

2. Reassuringly, the training superset is the most transferable set.

3. Surprisingly, the Mindless Benchmark dataset (MB16-43)[12] was shown to be the next most transferable training set.

4. Reassuringly, it was shown that transferability can be optimized for in the T100 dataset, with only 100 datapoints compared to GMTKN55s 1505, also showing data efficiency.

Figure 1 shows the mean absolute deviation for the GMTKN55 superset when trained on all of the different GMTKN55 subsets, and the visual conclusion of points 2 and 3 are very clear. This is particularly alarming as other training sets are specifically curated by seasoned chemists to be representative of a particular aspect of chemistry with which we are familiar. This would seem to suggest that there is significant benefit to machine generated and randomized datasets, with limited curation by a human scientists being ideal.

**Current/future challenges and required advances**
It is clear based on the training of the $XYG_p$ functional on GMTKN55 and its subsets that choice of the training set is critical to maintaining transferability. However, the results discussed above very much simplify the problem in order to make analysis tractable, restricting the functional to 7 parameters or fewer, and the training superset of 1505 datapoints, orders of magnitude smaller that other AI-DFT methods available in the literature. This presents two very clear questions:

1. How do the results discussed above apply to AI-DFAs with orders of magnitude more paramaters and data-points?

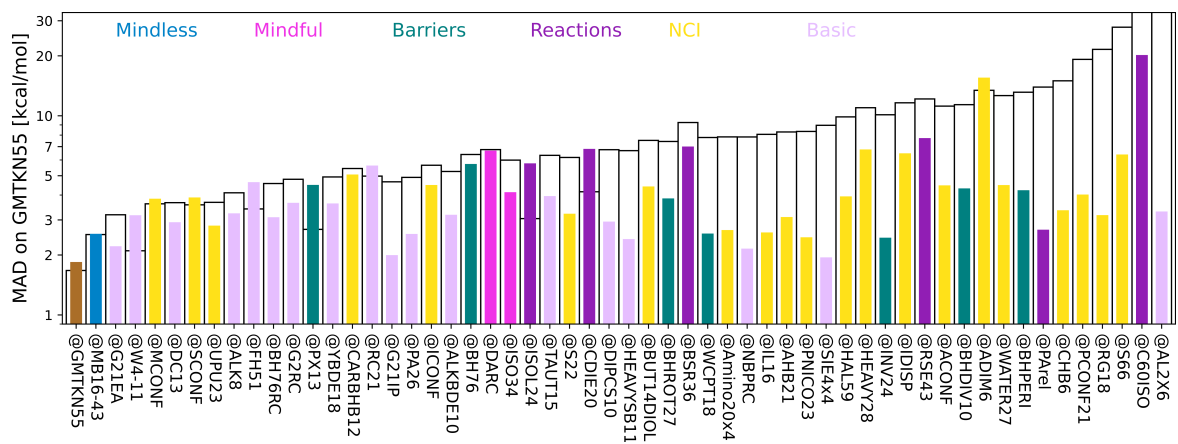

**Figure 1.** Mean absolute deviation (MAD, log scale) for GMTKN55@subset, where the notation [test set]@[training set] is used and subset is a subset of GMTKN55. Hollow bars are the XYG$_7$ functional, and solid bars are the XYG$_3$ functional with the ordering based on the average MAD and difference between MADs of each functional. Colour coding is listed the guide the reader to the type of chemistry represented by each subset of GMTKN55. Reproduced from Ref.[10] with permission from the Royal Society of Chemistry

2. Training a single AI-DFA method is already massively expensive, assessing transferability as discussed above would be similarly more expensive.

The advancement required to answer this question is a method for assessing transferability without retraining of the method. Otherwise, sufficiently complex AI-DFA methods will have to be treated current as gold-standard DFA methods, assumed to be transferable and with extensive testing on a wide range of systems looking for outlier cases. Alternatively, a pragmatic choice could be to limit the complexity of the AI-DFA methods that we generate to enable the type of transferability analysis discussed above. Perhaps it is possible to use the confidence estimations of AI models that have seen success in other fields[13, 14] as a rough measure of where more training is needed in our future AI-DFA models?

It is also clear that datasets with a focus on diversity of chemical representation, that is not necessarily generated by human intuition is required, which certainly goes against my own scientific intuition. It is well understood that the amount of chemical data available is too scarce, evidenced by many materials prediction AI tools generating artificial DFT based training data.[15, 16] There is a growing understanding that the type of chemical data available is also limited, primarily focused on very stable ground state perfectly structured systems. The reality is that chemical systems, particularly solids, exist in disordered,[17] metastable or even transition states[18, 19] which a chemically complete model will require some training in. A significant advance will be the creation of datasets that can reliable represent these disordered and imperfect chemical spaces. Promising advancements are being made with the recent expansion of the mindless dataset.[20]

**Concluding remarks**
Reasonable exasperation was expressed at the conclusion that machine generated and randomized training data gives better results than scientist curated training data. While understandable, I do not think this is a cause for despair, rather evidence that chemical space is VAST, and still not fully understood with potential for new discoveries on the horizon. Similarly, there is still room for the improvement of our existing datasets and AI models used in the DFT and materials modeling fields, and potential solutions beyond those discussed here.

**Acknowledgments**
SGD was supported by the Ministry of Education, Singapore, under its Research Centre of Excellence award to the Institute for Functional Intelligent Materials, with Project No. EDUNC-33-18-279-V12.

**References**
[1] Lars Goerigk, Andreas Hansen, Christoph Bauer, Stephan Ehrlich, Asim Najibi, and Stefan Grimme. A look at the density functional theory zoo with the advanced gmtkn55 database for general main group thermochemistry, kinetics and noncovalent interactions. *Physical Chemistry Chemical Physics*, 19(48):32184–32215, 2017.

[2] Vahe Gharakhanyan, Luis Barroso-Luque, Yi Yang, Muhammed Shuaibi, Kyle Michel, Daniel S Levine, Misko Dzamba, Xiang Fu, Meng Gao, Xingyu Liu, et al. Open molecular crystals 2025 (omc25) dataset and models. *arXiv preprint arXiv:2508.02651*, 2025.

[3] W Kohn and LJ Sham. Self-consistent equations including exchange and correlation effects. *Phys. Rev*, 140(4A):A1133–A1138, 1965.

[4] Axel D Becke. Perspective: Fifty years of density-functional theory in chemical physics. *The Journal of chemical physics*, 140(18), 2014.

[5] Michael G Medvedev, Ivan S Bushmarinov, Jianwei Sun, John P Perdew, and Konstantin A Lyssenko. Density functional theory is straying from the path toward the exact functional. *Science*, 355(6320):49–52, 2017.

[6] Kyle R Bryenton, Adebayo A Adeleke, Stephen G Dale, and Erin R Johnson. Delocalization error: The greatest outstanding challenge in density-functional theory. *Wiley Interdisciplinary Reviews: Computational Molecular Science*, 13(2):e1631, 2023.

[7] James Kirkpatrick, Brendan McMorrow, David HP Turban, Alexander L Gaunt, James S Spencer, Alexander GDG Matthews, Annette Obika, Louis Thiry, Meire Fortunato, David Pfau, et al. Pushing the frontiers of density functionals by solving the fractional electron problem. *Science*, 374(6573):1385–1389, 2021.

[8] Heng Zhao, Tim Gould, and Stefan Vuckovic. Deep mind 21 functional does not extrapolate to transition metal chemistry. *Physical Chemistry Chemical Physics*, 26(16):12289–12298, 2024.

[9] Giulia Luise, Chin-Wei Huang, Thijs Vogels, Derk P Kooi, Sebastian Ehlert, Stephanie Lanius, Klaas JH Giesbertz, Amir Karton, Deniz Gunceler, Megan Stanley, et al. Accurate and scalable exchange-correlation with deep learning. *arXiv preprint arXiv:2506.14665*, 2025.

[10] Tim Gould, Bun Chan, Stephen G Dale, and Stefan Vuckovic. Identifying and embedding transferability in data-driven representations of chemical space. *Chemical Science*, 15(28):11122–11133, 2024.

[11] Bun Chan, Peter MW Gill, and Masanari Kimura. Assessment of dft methods for transition metals with the tmc151 compilation of data sets and comparison with accuracies for main-group chemistry. *Journal of chemical theory and computation*, 15(6):3610–3622, 2019.

[12] Martin Korth and Stefan Grimme. "mindless" dft benchmarking. *Journal of chemical theory and computation*, 5(4):993–1003, 2009.

[13] Yarin Gal and Zoubin Ghahramani. Dropout as a bayesian approximation: Representing model uncertainty in deep learning. In *international conference on machine learning*, pages 1050–1059. PMLR, 2016.

[14] Antonio Loquercio, Mattia Segu, and Davide Scaramuzza. A general framework for uncertainty estimation in deep learning. *IEEE Robotics and Automation Letters*, 5(2):3153–3160, 2020.

[15] Amil Merchant, Simon Batzner, Samuel S Schoenholz, Muratahan Aykol, Gowoon Cheon, and Ekin Dogus Cubuk. Scaling deep learning for materials discovery. *Nature*, 624(7990):80–85, 2023.

[16] Claudio Zeni, Robert Pinsler, Daniel Zügner, Andrew Fowler, Matthew Horton, Xiang Fu, Zilong Wang, Aliaksandra Shysheya, Jonathan Crabbé, Shoko Ueda, et al. A generative model for inorganic materials design. *Nature*, 639(8055):624–632, 2025.

[17] Martin Hoffmann Petersen, Ruiming Zhu, Haiwen Dai, Savyasanchi Aggarwal, Nong Wei, Andy Paul Chen, Arghya Bhowmik, Juan Maria Garcia Lastra, and Kedar Hippalgaonkar. Dis-gen: Disordered crystal structure generation. *arXiv preprint arXiv:2507.18275*, 2025.

[18] Erin R Johnson, Owen J Clarkin, Stephen G Dale, and Gino A DiLabio. Kinetics of the addition of olefins to si-centered radicals: the critical role of dispersion interactions revealed by theory and experiment. *The Journal of Physical Chemistry A*, 119(22):5883–5888, 2015.

[19] Dylan M Anstine, Qiyuan Zhao, Roman Zubatiuk, Shuhao Zhang, Veerupaksh Singla, Filipp Nikitin, Brett M Savoie, and Olexandr Isayev. Aimnet2-rxn: A machine learned potential for generalized reaction modeling on a millions-of-pathways scale. 2025.

[20] Thomas Gasevic, Marcel Müller, Jonathan Schöps, Stephanie Lanius, Jan Hermann, Stefan Grimme, and Andreas Hansen. Chemical space exploration with artificial" mindless" molecules. 2025.

# Adaptive Density Functional Theory in the AI Era

Alastair J. A. Price
*Department of Chemistry, University of Toronto, Toronto, Ontario, Canada*
(Dated: October 27, 2025)

Density Functional Theory (DFT) remains the quantum–mechanical workhorse of chemistry and materials science. Its reproducibility, scalability, and interpretability make it a natural foundation for the emerging fields of machine learning (ML) and artificial intelligence (AI) for science. Yet, when DFT is called upon as the "oracle" in data-driven discovery, its accuracy and therefore the ceiling of predictive fidelity for any model trained upon it, is defined by the exchange–correlation approximations at its core. Here we outline how *adaptive density functional theory*, a general framework that allows functional parameters to respond to the physics and chemistry of a system, provides a principled path forward. Adaptive hybrids, dispersion corrections, double hybrids, and even pseudopotentials exemplify how we can bridge first–principles rigour with trustworthy data-driven modelling. As AI accelerates discovery, DFT must evelove with it; adaptivity provides the mechanism by which physics remains not only relevant, but essential.

## I. THE ROLE OF PHYSICS IN THE AI ERA

The intersection of artificial intelligence and quantum mechanics is reshaping scientific discovery. Machine learning and AI models now appear in every stage of chemical and materials design; from generative molecular assembly and autonomous synthesis to property prediction and inverse optimization. Within this growing ecosystem, DFT remains the de facto foundation. It is the first general quantum–mechanical theory capable of treating systems across the periodic table with a balance of accuracy and cost that no other electronic-structure method has matched.

DFT's reproducibility and grounding in the electron density make it uniquely interpretable among quantum approaches, and thus naturally suited to act as the physics prior for AI workflows in science.[1] However, this same strength exposes a limitation: the physics is only as accurate as the exchange–correlation (XC) functional employed. Decades of developments have produced a hierarchy of approximations, from the local density and generalized gradient approximations (GGAs)[2, 3], to hybrid, range-separated, and double-hybrid approximations.[4–6] Yet, none are universally transferable to all chemical systems. Each approximation embeds fixed assumptions about the balance of exchange and correlation, dispersion and delocalization, or short- and long-range behaviour. When these assumptions fail, as they do for charge transfer, strongly correlated systems, and extended noncovalent networks, DFT's predictive powers lose reliability. In turn, any ML model that treats DFT as its oracle inherits those same blind spots. Like overfitted functionals, purely data-driven models can reproduce the training data with extraordinary precision yet fail catastrophically when asked to extrapolate. As von Neumann warned, with enough parameters one can fit an elephant but not necessarily describe how it walks. Physics provides the priors that allow us to move beyond imitation toward understanding.

## II. THE LIMITATIONS OF FIXED FUNCTIONAL APPROXIMATIONS

The core of DFT's challenge lies in delocalization and static correlation errors, two manifestations of imperfect cancellation between approximate exchange and correlation. In GGAs such as PBE and B86bPBE, delocalization error spuriously stabilizes fractional charges and stretched odd-electron bonds, as seen in systems like $H_2^+$, leading to qualitatively incorrect charge transfer, band gaps, and radical energetics.[7] For regular bonds with paired electrons, GGAs instead predict dissociation limits that are far too high due to static correlation error. Hybrids partially mitigate these effects by mixing a fraction of Hartree–Fock exchange, $a_x$, with GGA exchange:

$$E_X^{\text{hybrid}} = a_x E_X^{\text{HF}} + (1-a_x) E_X^{\text{DFA}}. \tag{1}$$

However, the optimal value of $a_x$ depends strongly on system and property. The canonical PBE0[5] and B3LYP[4, 8] employ a global fixed admixture of 25% and 20%, respectively, which work well for typical covalent molecules but fail for ionic crystals, hydrogen-bonded networks, and low-gap or strongly correlated systems.[9, 10] Increasing HF exchange improves delocalization but over-localizes charge density elsewhere. Similar trade-offs also appear in dispersion corrections, which provide a clear example of how adaptivity improves transferability. In early empirical models such as D2 or D3, the dispersion coefficients are fixed or depend only weakly on the local environment. These schemes perform well within the systems they were parameterized for but often fail when the bonding or charge distribution differs significantly from the training data. By contrast, models such as the Tkatchenko–Scheffler (TS), exchange–hole dipole moment (XDM), and many-body dispersion(MBD) approaches derive their dispersion coefficients directly from the self-consistent electron density, allowing them to respond to changes in molecular geometry, charge state, or chemical composition.[11–15] This density dependence makes them inherently more transferable, since the interaction strength follows the physics of the underlying electronic structure rather than an external fit. In every case, responsiveness to the electronic environment rather than reliance on fixed parameters correlates with broader applicability and reliability. The lesson is universal: fixed global parameters limit transferability. What is required is adaptivity, the ability for the functional to adjust to the physics of the system itself.

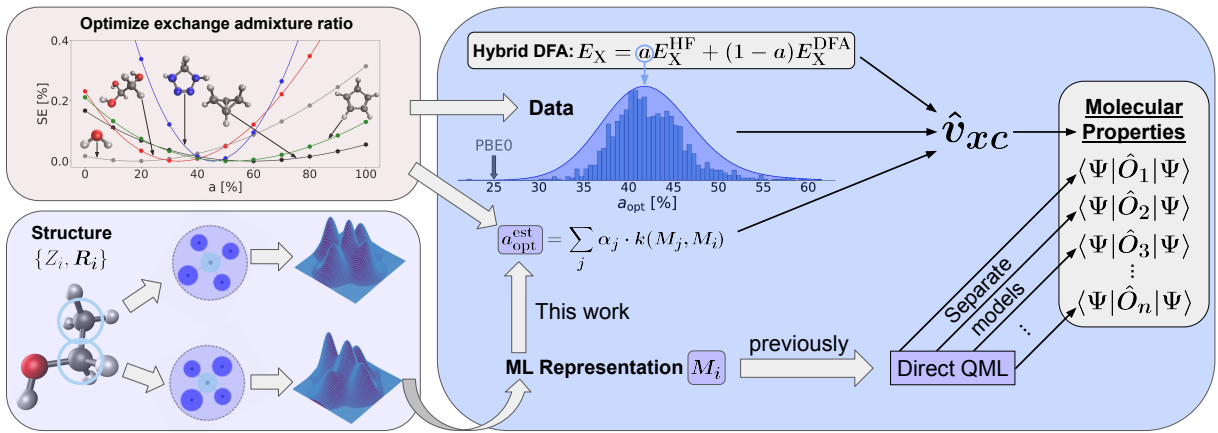

FIG. 1. Schematic of the adaptive hybrid workflow (*aPBE0*). Optimal HF fractions are learned from reference atomization energies and predicted for new systems with negligible cost. Adapted from Khan *et al.*, *Sci. Adv.* **11**, eadt7769 (2025).

## III.  ADAPTIVITY: LEARNING THE RIGHT PHYSICS

By adaptive DFT, we mean an approach in which the functional parameters within the method, whether an exact-exchange fraction, range separation constant, dispersion damping term, or correlation weights, become context sensitive. Adaptivity does not discard Kohn-Sham DFT or replace physics with data; rather, it augments the functional form with learned or inferred dependencies derived from the electronic environment. Adaptive DFT thus preserves variational consistency and interpretability while improving transferability across chemical space.

The adaptive hybrid framework, exemplified by the *aPBE0* functional,[16] illustrates this idea. Here, the optimal exchange fraction $a_\text{opt}$ is learned as a difference from the base 25%, through a smooth, data-efficient function of molecular geometry and composition, predicted once and used self-consistently within the SCF cycle. The cost at runtime is identical to that of a standard hybrid, but the accuracy improves dramatically, the workflow can be seen in Fig. 1. Across benchmarks such as QM5, QM9, and GMTKN55, *aPBE0* reduces atomization-energy errors by nearly a factor of three, improves electron densities by roughly 33%, and reduces HOMO–LUMO gap errors by 75%. Spin-state energetics, long a weak point of DFT, benefit similarly: for 110 carbenes, the mean singlet–triplet gap error drops from 10 to 3 kcal mol$^{-1}$.

These improvements arise not from ad hoc parmeterization but from restoring the missing dependence between the functional form and its local physics. Similar philosophies underlie emerging adaptive enhancement factors from data[?] and adaptive range-separated hybrids that predict system-specific screening parameters. The concept of adaptive double hybrid are also emerging, where perturbative correlation contributions can be modulated dynamically to balance static and dynamical correlation. The same concept can be extended to pseudopotentials, whose transferability suffers from fixed-core definitions. Environmental-aware pseudopotentials that respond to bonding or charge-state context could eliminate many long-standing sources of error between molecular and condensed-phase calculations, as well as offer a route to reducing memory and computational costs for a given calculation.

## IV.  THE PROPOSED PATH FORWARD

We envision adaptive methods as a unifying force that transcends individual functional families. The same logic applies whether we tune exchange functions, damping parameters, or other parameters within quantum mechanical methods. Adaptivity ensures that functionals can evolve with their data and context, something that has long been absent from the fixed landscapes of semiempirical corrections and empirical parameterizations. The next generation of adaptive methods will likely combine physical constraints with lightweight ML models that enforce monotonicity, asymptotic behaviour, and sum rules derived from exact conditions.

For this vision to mature, several practical challenges must be met. Benchmark datasets must capture diverse bonding motifs and include high-level reference energies and densities; ML models must remain compact and uncertainty-aware; and adaptive frameworks must be implemented in mainstream electronic-structure codes to allow self-consistent integration with solid-state and interfacial simulations. But the intellectual direction is clear: we must move beyond fixed forms and let our functionals learn from the systems they describe. Adaptivity is not a replacement for physics, it is how physics adapts to complexity.

As we move deeper into the AI era of scientific discovery, one truth endures: *physics is not optional*. It is the language through which we ensure that the intelligence we build, artificial or otherwise, remains grounded in the laws that govern nature.

### ACKNOWLEDGMENTS

We thank Danish Khan, Maximilian L. Ach, and O. Anatole von Lilienfeld for collaboration and insight and Erin Johnson for her review and comments on this work. Support from the University of Toronto, and the Acceleration Consortium is gratefully acknowledged.

[1] B. Huang, G. F. von Rudorff, and O. A. von Lilienfeld, Science **381**, 170 (2023).
[2] J. P. Perdew, K. Burke, and M. Ernzerhof, Physical review letters **77**, 3865 (1996).
[3] A. D. Becke, Physical review A **38**, 3098 (1988).
[4] A. D. Becke, The Journal of chemical physics **98**, 5648 (1993).
[5] C. Adamo and V. Barone, The Journal of chemical physics **110**, 6158 (1999).
[6] L. Goerigk, A. Hansen, C. Bauer, S. Ehrlich, A. Najibi, and S. Grimme, Physical Chemistry Chemical Physics **19**, 32184 (2017).
[7] K. R. Bryenton, A. A. Adeleke, S. G. Dale, and E. R. Johnson, Wiley Interdisciplinary Reviews: Computational Molecular Science **13**, e1631 (2023).
[8] C. Lee, W. Yang, and R. G. Parr, Physical review B **37**, 785 (1988).
[9] A. J. Price, A. Otero-de-la Roza, and E. R. Johnson, Chemical Science **14**, 1252 (2023).
[10] A. J. Price, R. A. Mayo, A. Otero-de-la Roza, and E. R. Johnson, CrystEngComm **25**, 953 (2023).
[11] A. Tkatchenko and M. Scheffler, Physical review letters **102**, 073005 (2009).
[12] A. D. Becke and E. R. Johnson, The Journal of chemical physics **123** (2005).
[13] E. R. Johnson and A. D. Becke, The Journal of chemical physics **123** (2005).
[14] A. Otero-De-La-Roza and E. R. Johnson, The Journal of chemical physics **138** (2013).
[15] A. Tkatchenko, R. A. DiStasio Jr, R. Car, and M. Scheffler, Physical review letters **108**, 236402 (2012).
[16] D. Khan, A. J. Price, B. Huang, M. L. Ach, and O. A. von Lilienfeld, Science Advances **11**, eadt7769 (2025).

# Differentiable Programming for Density Functional Theory: A Paradigm Shift in Electronic Structure Calculations

Tianbo Li[1] and Min Lin[1]

[1]SEA AI Lab, 1 Fusionopolis Place, Singapore 138522
*Author to whom any correspondence should be addressed.

**Status**

The rapid advancement of artificial intelligence (AI) is quietly transforming quantum chemistry. Increasingly, researchers are leveraging AI to address long-standing challenges in density functional theory (DFT) calculations. A prominent example is the development of sophisticated machine learning-based exchange-correlation (XC) functionals, such as DM21 [3]. These models, often trained on high-fidelity quantum chemical data, demonstrate a remarkable ability to go beyond the approximations of traditional human-designed functionals. By offering a more nuanced and data-driven description of electron correlations, these AI-enhanced functionals significantly improve the accuracy and predictive power of DFT for complex systems.

The development and effective training of these powerful AI-driven models are intrinsically dependent on the modern AI software ecosystem, with differentiable programming emerging as a critical enabler. Frameworks like Jax [1] and PyTorch [7] provide automatic differentiation (AD), which automates the computation of gradients—a fundamental requirement for training neural networks. This capability eliminates the need for researchers to derive and implement complex analytical gradients manually. Consequently, scientists can focus on designing innovative model architectures while relying on AD to handle the intricate optimization process efficiently. The success of complex functionals like DM21 underscores that the continued advancement of AI-in-DFT is inextricably linked to the power and flexibility of these differentiable programming frameworks.

Recognizing this synergy, the field is moving towards fundamentally rebuilding the core infrastructure of DFT around the principle of differentiability. The goal is to move beyond merely embedding AI components within traditional DFT codes and instead create fully differentiable DFT frameworks. This paradigm shift involves re-implementing key computational components, such as the self-consistent field (SCF) procedure, using AD tools [6, 4, 5, 2]. Although these new packages are not yet the most widely used, these early efforts nonetheless signal a future of closer integration between DFT and differentiable programming.

**Current and future challenges**

Unlike traditional numerical approaches, differentiable programming represents computational operations as differentiable graphs, enabling systematic evaluation of gradients through automatic differentiation. This idea has been widely validated in machine learning and deep learning, and its adoption in DFT is now gaining momentum. Following this trend, many researchers are exploring the use of popular AD frameworks from AI research, such as PyTorch and JAX, to accelerate DFT development. These frameworks provide several advantages: (i) automatic computation of gradients without the need for manual derivations; (ii) extensive support for GPUs and other high-performance hardware; and (iii) rich libraries of linear algebra routines and mathematical utilities crucial for electronic structure methods.

Despite the remarkable success of differentiable frameworks in AI research, their application to DFT still faces numerous challenges. We summarize these challenges below.

1. **Numerical Stability and Precision.** DFT calculations require high numerical precision to achieve chemical accuracy. This demands a careful trade-off between precision and efficiency, imposing stringent requirements on the numerical stability of differentiable operations. Critical yet challenging components include non-standard operators like Fourier transforms and matrix diagonalization, which can be sensitive in gradient-based optimization.

2. **High-Dimensional and Higher-Order Derivatives.** Applications such as phonon calculations require higher-order derivatives (e.g., the dynamical matrix, a second energy derivative). Computing these high-dimensional tensors with AD can be prohibitively expensive in terms of memory and computation, and may suffer from numerical instability. While AD is general in principle, practical applications often require specialized algorithms or approximations for efficiency.

3. **Complex Optimization Problems.** Differentiable DFT extends beyond ground-state calculations to optimization-driven tasks, such as training XC functionals against reference data. This leads to bilevel optimization problems: an inner loop for the self-consistent DFT calculation and an outer loop for adjusting functional parameters. Traditional density-functional perturbation theory offers limited flexibility here, and existing AD frameworks are not optimized for such nested tasks. Efficiently solving these optimization problems remains an open challenge.

4. **The Formula-Code Gap.** Most existing DFT software is engineering-oriented, with implementations that diverge from their mathematical formulations. This gap increases development costs and hinders the prototyping of new methods. In differentiable programming, the problem is exacerbated: code must be numerically efficient and seamlessly compatible with AD frameworks. Developing higher-level abstractions that map physical equations directly to differentiable code is essential for improving reproducibility, extensibility, and research agility.

**Advances in science and technology to meet challenges**

To address these challenges, we propose three coordinated directions that fuse advanced methodologies with modern computational practice:

**Gradient-Centric, GPU-Native Algorithm Design.** Traditional SCF loops and eigensolvers, designed for CPU-bound linear algebra, are ill-suited for GPU parallelism. Shifting to gradient-centric optimization algorithms better leverages GPU capabilities through large-batch, fused-kernel execution. A GPU-native design should favor batched, synchronization-light operations and memory-access patterns optimized for throughput. Mixed-precision strategies with rigorous error control can further enhance computational efficiency.

**Tools for High-Dimensional and Higher-Order Differentiation.** Emerging applications, such as synthesis path prediction and inverse materials design, require robust handling of high dimensionality and higher-order derivatives. A unifying computational core should provide memory-efficient automatic differentiation, support operator-level rules for differential operators, and integrate stochastic estimators and dimensionality reduction while maintaining numerical stability.

**Bridging the Path from Theory to Implementation.** A modular software framework is needed, with clear separation between physics models, numerical algorithms, and hardware-specific code. This modularity enables a declarative front end for intuitive problem specification and a plugin system for community contributions. Reproducibility must be a core design goal, achieved through standardized benchmarks, versioned datasets, and provenance tracking. Such a framework would systematically accelerate innovation by closing the gap between theoretical concepts and practical implementation.

**Concluding remarks**

Differentiable programming offers a promising path toward more accurate and efficient DFT calculations. Its success in AI underscores its potential, but realizing this potential in DFT requires addressing unique challenges in numerical stability, high-dimensional differentiation, and complex optimization. By pursuing the outlined directions—GPU-native redesign, unified differentiation frameworks, and modular software design—the community can harness differentiable programming to accelerate discovery in quantum chemistry and materials science. With concerted effort, this paradigm shift will fundamentally advance electronic structure theory.

**Acknowledgments**

This work was supported by SEA AI Lab (SAIL) and Institute of Functional Intelligence Materials (IFIM) of National University of Singapore (NUS). We would like to thank Prof. Giovanni Vignale and Prof. Stephen Gregory Dale for their valuable comments and suggestions.

**References**

[1] James Bradbury, Roy Frostig, Peter Hawkins, Matthew James Johnson, Chris Leary, Dougal Maclaurin, George Necula, Adam Paszke, Jake VanderPlas, Skye Wanderman-Milne, and Qiao Zhang. JAX: composable transformations of Python+NumPy programs, 2018.

[2] Michael F Herbst, Antoine Levitt, and Eric Cancès. Dftk: A julian approach for simulating electrons in solids. In *Proceedings of the JuliaCon conferences*, volume 3, page 69, 2021.

[3] James Kirkpatrick, Brendan McMorrow, David HP Turban, Alexander L Gaunt, James S Spencer, Alexander GDG Matthews, Annette Obika, Louis Thiry, Meire Fortunato, David Pfau,

et al. Pushing the frontiers of density functionals by solving the fractional electron problem. *Science*, 374(6573):1385–1389, 2021.

[4] Tianbo Li, Min Lin, Zheyuan Hu, Kunhao Zheng, Giovanni Vignale, Kenji Kawaguchi, AH Neto, Kostya S Novoselov, and Shuicheng Yan. D4ft: A deep learning approach to kohn-sham density functional theory. *arXiv preprint arXiv:2303.00399*, 2023.

[5] Tianbo Li, Zekun Shi, Stephen Gregory Dale, Giovanni Vignale, and Min Lin. Jrystal: A jax-based differentiable density functional theory framework for materials. In *Machine Learning and the Physical Sciences Workshop at NeurIPS*, 2024.

[6] Pablo A M Casares, Jack S Baker, Matija Medvidović, Roberto dos Reis, and Juan Miguel Arrazola. Graddft. a software library for machine learning enhanced density functional theory. *The Journal of Chemical Physics*, 160(6), 2024.

[7] Adam Paszke, Sam Gross, Francisco Massa, Adam Lerer, James Bradbury, Gregory Chanan, Trevor Killeen, Zeming Lin, Natalia Gimelshein, Luca Antiga, et al. Pytorch: An imperative style, high-performance deep learning library. *Advances in neural information processing systems*, 32, 2019.

# Advancing first-principles electronic structure calculations through deep-learning electronic Hamiltonian methods

Zehcen Tang[1], Yang Li[1] and Yong Xu[1,2,*]

[1]State Key Laboratory of Low Dimensional Quantum Physics and Department of Physics, Tsinghua University, Beijing, 100084, China
[2]Frontier Science Center for Quantum Information, Beijing, China
*Author to whom any correspondence should be addressed.

**E-mail:** yongxu@mail.tsinghua.edu.cn

## Status

First-principles methods based on fundamental principles of quantum mechanics have played a pivotal role in the development of computational physics and materials science. Among them, density functional theory (DFT) is one of the most widely used, offering a favorable balance between computational efficiency and accuracy [1]. Despite its tremendous success, DFT faces great challenges in studying large-size material systems and performing high-throughput simulations for constructing large materials database, as constrained by the expensive computational cost.

The integration of deep learning (DL) offers new opportunities to extend the applicability of DFT. One prominent direction is the development of DL-based force fields, where neural-network methods are applied to model the potential energy surfaces and atomic forces for given material structures, enabling highly efficient atomic-structure calculations, such as geometry optimization and molecular dynamics simulation [2]. These atomic-structure models have already found broad applications, where the electronic-structure information is coarse grained for simplify. In parallel, DL-based electronic-structure calculation methods have emerged to learn and predict electronic structure properties by neural-network models. A representative example is the deep-learning DFT Hamiltonian (DeepH) method, which predicts the DFT Hamiltonian as a function of material structures [3]. Through bypassing the time-consuming self-consistent iterations for solving DFT equations, DeepH enables efficient derivation of electronic properties through post-processing of the Hamiltonian predicted by neural networks, offering substantial speedup compared with conventional DFT.

Two principal features largely define the application prospects of DeepH: First, DeepH leverages Walter Kohn's "quantum nearsightedness principle", suggesting that local electronic properties depends mainly on the neighboring environment [4]. By predicting DFT Hamiltonians under localized atomic bases with graph neural networks, DeepH achieves linear scaling with system size, and it enables models trained on small-sized training structures to reliably generalize to much larger ones, ensuring applicability to large-scale materials [3]. Second, by encoding physical priors and exploiting the abundance of Hamiltonian data, DeepH achieves strong transferability across material classes, opening the door to training high-accurate foundation model of electronic structures, which could be useful for AI-driven material discovery [5].

Currently, mainstream DeepH frameworks predict DFT Hamiltonian matrix elements in localized atomic bases, which allows seamless integration with DFT programs employing such bases. In these networks, a crystal graph neural network is first constructed from the material structure [6]. After several layers of message passing, node and edge features are utilized to construct onsite and hopping Hamiltonian elements, respectively. The model is trained by minimizing a loss function, typically defined as the mean absolute or mean squared error of the Hamiltonian elements. The resulting Hamiltonian can be interpreted as a non-orthogonal tight-binding model with first-principles accuracy, which can then be further post-processed to obtain any electronic properties on the mean-field level.

Although electronic-structure models have received less attention than atomic-structure models, the past five years have witnessed their rapid development, including emergence of several DL Hamiltonian packages [3, 7–9]. Since the first proposal of DeepH in 2021 [10], systematic progress has been made in both methodology and applications. Despite these advances, important challenges remain for DeepH methods to further improve accuracy, efficiency, and transferability, as we outline in the following sections.

**Current and future challenges**
A critical aspect for DeepH is accuracy, which inherently depends on the underlying DFT approximations. For example, (semi-)local DFT underestimates band gaps, and the limitation is partly resolved by hybrid functionals [1]. Thus, coupling DeepH with hybrid functionals is crucial for applications requiring accurate band-gap descriptions, such as optical property predictions [11]. Future research may extend this idea to more advanced frameworks, including post-Hartree–Fock and *GW* theories. Such integration, however, is non-trivial: these methods are not strictly single-electron based, therefore whether and how DL-Hamiltonian models can be incorporated remains uncertain. If these integrations are feasible, the efficiency advantage of DL over conventional computations could be even more apparent, given the much higher computational demands of these methods compared with standard DFT. In addition, embedding DeepH into DFT-based frameworks offers opportunities for computing wider range of properties. For instance, combining DeepH with density functional perturbation theory have enabled predicting electron–phonon couplings [12]. Looking ahead, these efforts could be extended to other couplings, such as electron–magnon interactions or higher-order perturbations.

In DL-accelerated simulations of *large-scale* materials, post-processing often becomes the main bottleneck for DeepH, in some cases exceeding the time required for Hamiltonian prediction itself. Addressing this challenge will require coordinated advances in hardware, software, and algorithmic optimization. As DeepH applications are still in their early stages, specialized post-processing tools for large-scale electronic Hamiltonians remain underdeveloped. We anticipate that the integration of efficient and low-scaling algorithms–such as Wannierization or kernel polynomial methods [1]–could significantly expand the applicability of DeepH.

Another critical application is the integration of DeepH with *high-throughput* materials discovery. Even for small-sized systems, high-throughput DFT simulations are costly due to the vast number of candidate materials in chemical space. To enable DL-DFT research at the database scale for unseen materials, the transferability of models across compositions, structures, and prototypes is of crucial concern. DeepH models with such capability may be regarded as "universal models". Although progress in universal DeepH models has lagged behind universal DL force fields [13], the past two years have already witnessed the emergence of several primary frameworks [5]. In the coming years, it is foreseeable that the development of universal models trained on larger, more diverse datasets, together with explorations of their transferability to increasingly complex materials datasets, such as alloys, defects, and interfaces. Beyond their standalone usage, DeepH also holds promise for accelerating DFT convergence: since predicted electronic Hamiltonians encode underlying electronic structure information, restarting calculations from these predictions could enable faster convergence, offering a pathway to more efficient construction of material databases.

**Advances in science and technology to meet challenges**
Methodological advances of DeepH largely depend on the innovations in neural-network architectures. Many innovations are shared with DL force fields, such as crystal graph convolutional neural networks and equivariant neural networks [6, 14]. However, while DL force fields typically handle only equivariant scalars (energies) and vectors (forces), DeepH must also account for equivariant vectors with higher angular momentum $l$, associated with the angular quantum number of atomic orbitals. For materials with spin–orbit coupling or magnetism, Hamiltonian prediction additionally requires handling half-integer angular momenta. DeepH addresses this challenge by applying tensor decomposition rules that represent half-integer equivariant tensors in terms of integer ones [15, 16]. The intrinsic requirement for high-$l$ features introduces significant computational challenges, particularly in the "tensor product" operations of equivariant features, whose cost grows steeply with $l$. To address this, several efficient tensor-product strategies have recently been proposed, including the eSCN tensor product [17]. Initially developed for DL force fields and later adapted to DeepH, implementation of eSCN tensor product substantially improve computational performance [18]. Such advances illustrate how innovations in neural-network methodology can directly enhance the scalability and applicability of DeepH architectures. Apart from shared architecture development with DL force fields, a distinctive feature of deep-learning electronic structure is its compatibility with the DFT variational principle, which enables physics-informed unsupervised training of DeepH and yields improved descriptions of derived properties from predicted Hamiltonians [19].

Finally, a notable prerequisite for developing DeepH models is the availability of material databases containing Hamiltonian matrix elements in localized atomic-like bases. Unlike plane-wave bases, atomic bases depend strongly on the underlying DFT program and basis-construction parameters, posing major challenges for data interoperability. To address this, future progress will likely depend on several directions, including: 1. Developing high-throughput workflows based on atomic-basis DFT codes to generate and store Hamiltonian data; 2. Establishing data standards for electronic Hamiltonian datasets to ensure better interoperability; and 3. Advancing the "projection and reconstruction" scheme that maps plane-wave Hamiltonians onto localized bases, thereby leveraging the accuracy of plane-wave DFT while retaining the DeepH-compatibility of atomic-basis approaches [20]. These advances collectively expand the prospects for DeepH developments, enabling more consistent framework developments.

**Concluding remarks**

DeepH provides a promising pathway to accelerate DFT simulations by predicting electronic Hamiltonians with high accuracy. Its future development focuses on further improving accuracy through integration with advanced DFT-based methods and enhancing applicability in both large-scale and high-throughput materials studies. Neural network innovations, database development, and low-scaling post-processing will shape the technological progress of DeepH-derived materials research. With continued advances across these fronts, DeepH is poised to evolve from a specialized tool into a broadly applicable framework that complements and extends conventional DFT, enabling computational electronic structure study of materials at scales and complexities previously out of reach. Since DeepH universal models aim to encode most DFT-level electronic structures of vast materials, we may anticipate that, through continued development, DeepH could evolve into a "large materials model", serving as a foundation model in computational materials research [5].

**Acknowledgments**

We acknowledge support from the Basic Science Center Project of NSFC (grant no. 52388201), the National Key Basic Research and Development Program of China (grant nos. 2024YFA1409100 and 2023YFA1406400), the National Natural Science Foundation of China (grants nos. 12334003, 12421004, 12361141826, and 124B2072), and the National Science Fund for Distinguished Young Scholars (grant no. 12025405). Yang Li is funded by the Shuimu Tsinghua Scholar program. We also thank support from Tianhe new generation supercomputer at National Supercomputer Center in Tianjin.

**References**

[1] Martin RM 2020 Electronic structure: basic theory and practical methods (Cambridge university press) (doi:10.1017/CBO9780511805769)

[2] Batzner S, Musaelian A, Sun L, Geiger M, Mailoa JP, Kornbluth M, Molinari N, Smidt T and Kozinsky B 2022 E(3)-equivariant graph neural networks for data-efficient and accurate interatomic potentials *Nat. Commun.* **13** 2453 (doi:10.1038/s41467-022-29939-5)

[3] Li H, Wang Z, Zou N, Ye M, Xu R, Gong X, Duan W and Xu Y 2022 Deep-learning density functional theory Hamiltonian for efficient ab initio electronic-structure calculation *Nat. Comput. Sci.* **2** 36 (doi:10.1038/s43588-024-00723-3)

[4] Kohn W 1996 Density functional and density matrix method scaling linearly with the number of atoms *Phys. Rev. Lett.* **76** 3168 (doi:10.1103/PhysRevLett.76.3168)

[5] Wang Y, Li Y, Tang Z, Li H, Yuan Z, Tao H, Zou N, Bao T, Liang X, Chen Z, Xu S, Bian C, Xu Z, Wang C, Si C, Duan W and Xu Y 2024 Universal materials model of deep-learning density functional theory Hamiltonian *Sci. Bull.* **69** 2514 (doi:10.1016/j.scib.2024.06.011)

[6] Xie T and Grossman JC 2018 Crystal graph convolutional neural networks for an accurate and interpretable prediction of material properties *Phys Rev. Lett.* **120** 145301 (doi:10.1103/PhysRevLett.120.145301)

[7] Unke O, Bogojeski M, Gastegger M, Geiger M, Smidt T and Müller KR 2021 SE(3)-equivariant prediction of molecular wavefunctions and electronic densities *Advances in Neural Information Processing Systems* **34** 14434

[8] Yu H, Xu Z, Qian X, Qian X and Ji S 2023 Efficient and Equivariant Graph Networks for Predicting Quantum Hamiltonian *Proceedings of the 40th International Conference on Machine Learning* 40412

[9] Zhong Y, Yu H, Su M, Gong X and Xiang H 2023 Transferable equivariant graph neural networks for the Hamiltonians of molecules and solids *npj Comput. Mater.* **9** 182 (doi:10.1038/s41524-023-01130-4)

[10] Li H, Wang Z, Zou N, Ye M, Duan W and Xu Y 2021 arXiv:2104.03786

[11] Tang Z, Li H, Lin P, Gong X, Jin G, He L, Jiang H, Ren X, Duan W and Xu Y 2024 A deep equivariant neural network approach for efficient hybrid density functional calculations *Nat. Commun.* **15** 8815 (doi:10.1038/s41467-024-53028-4)

[12] Li H, Tang Z, Fu J, Dong WH, Zou N, Gong X, Duan W and Xu Y 2024 Deep-learning density functional perturbation theory *Phys. Rev. Lett.* **132** 096401 (doi:10.1103/PhysRevLett.132.096401)

[13] Chen C and Ong SP 2022 A universal graph deep learning interatomic potential for the periodic table *Nat. Comput. Sci.* **2** 718 (doi:10.1038/s43588-022-00349-3)

[14] Geiger M and Smidt T 2022 arXiv:2207.09453

[15] Gong X, Li H, Zou N, Xu R, Duan W and Xu Y 2023 General framework for E(3)-equivariant neural network representation of density functional theory Hamiltonian *Nat. Commun.* **14** 2848 (doi:10.1038/s41467-023-38468-8)

[16] Li H, Tang Z, Gong X, Zou N, Duan W and Xu Y 2023 Deep-learning electronic-structure calculation of magnetic superstructures *Nat. Comput. Sci.* **3** 321 (doi:10.1038/s43588-023-00424-3)

[17] Passaro S and Zitnick CL 2023 Reducing SO(3) Convolutions to SO(2) for Efficient Equivariant GNNs *Proceedings of the 40th International Conference on Machine Learning* 27420

[18] Wang Y, Li H, Tang Z, Tao H, Wang Y, Yuan Z, Chen Z, Duan W and Xu Y 2024 arXiv:2401.17015

[19] Li Y, Tang Z, Chen Z, Sun M, Zhao B, Li H, Tao H, Yuan Z, Duan W and Xu Y 2024 Neural-network density functional theory based on variational energy minimization *Phys. Rev. Lett.* **133** 076401 (doi:10.1103/PhysRevLett.133.076401)

[20] Gong X, Louie SG, Duan W and Xu Y 2024 Generalizing deep learning electronic structure calculation to the plane-wave basis *Nat. Comput. Sci.* **4** 752 (doi:10.1038/s43588-024-00701-9)

# Autonomous Materials Discovery: From Automated Screening to Self-Driving Laboratories

Amrita Joshi[1,*] ⦿   Xiaonan Wang[2,*] ⦿   Leonard W.T. Ng[1,*] ⦿

[1] School of Materials Science and Engineering, Nanyang Technological University, Singapore, 639798, Singapore
[2] Department of Chemical Engineering, Tsinghua University, Beijing, China, 100084
[*] Author to whom any correspondence should be addressed.     **E-mail:** leonard.ngwt@ntu.edu.sg

**Status**
Materials discovery has evolved through four paradigms: empirical observation, theoretical understanding, computational prediction, and data-driven approaches [1]. Each advancement accelerated our ability to predict and screen materials, yet the fundamental challenge persists. Translating laboratory discoveries to commercial products typically requires 10-20 years [2]. This timeline has become increasingly untenable as society demands rapid solutions for energy storage, quantum computing, and sustainable manufacturing. The emergence of self-driving laboratories (SDLs) represents a fifth paradigm that promises to collapse this timeline through intelligent automation that integrates synthesis, characterization, and optimization for scientific discovery in closed-loop systems [3].

Current SDL implementations demonstrate significant acceleration factors, reducing experimental cycles from months to days while conducting orders of magnitude more experiments with superior precision [4]. The field has progressed from simple high-throughput screening with predefined protocols to sophisticated platforms employing Bayesian optimization for intelligent experiment planning [5]. Leading examples include autonomous synthesis platforms for inorganic materials achieving 71% success rates [6], roll-to-roll printed photovoltaics[7] with integrated optimization [8], and microfluidic systems for nanoparticle synthesis [9]. These systems combine three critical components: hardware automation for physical experimentation, software algorithms for intelligent decision-making, and domain expertise encoded in constraints and heuristics [10] (Figure 1a).

The transition from automation to true autonomy represents the frontier of materials science (Figure 1b). While high-throughput methods multiply experimental capacity, they remain bound by predetermined experimental designs. Modern SDLs incorporating machine learning achieve parameter optimization within defined spaces [11], but next-generation agentic systems [12] will demonstrate goal-directed reasoning with integrated knowledge representation [13]. This evolution parallels developments in artificial intelligence, where large language models now enable natural language specification of research objectives and automated hypothesis generation [14]. The convergence of these technologies positions materials science at an inflection point where fully autonomous discovery becomes feasible.

**Current and future challenges**
The primary challenge facing SDLs extends beyond automation efficiency and algorithmic sophistication; it lies in bridging the persistent gap between laboratory optimization and industrial translation and scalability. Most current platforms optimize easily measurable laboratory properties while neglecting manufacturing constraints, leading to "champion" materials that prove impossible to scale [2]. This disconnect manifests in multiple dimensions: materials optimized for peak performance often require synthesis conditions incompatible with industrial processes; device-level properties emerge from interfaces and defects not captured in bulk characterization [15]; and long-term stability frequently contradicts initial performance metrics.

Hardware limitations constrain current SDL implementations. Seamless integration between synthesis, processing, and characterization modules remains elusive, with most platforms focusing on single aspects of the materials development pipeline [10]. Experimental viability poses persistent challenges. Even sophisticated platforms like A-Lab report success rates around 71%, with only 37% of the 355 synthesis recipes producing targets, indicating significant room for improvement in handling experimental failures autonomously [6]. The absence of standardized interfaces between equipment from different manufacturers forces custom engineering for each platform, limiting scalability and knowledge transfer between research groups [16].

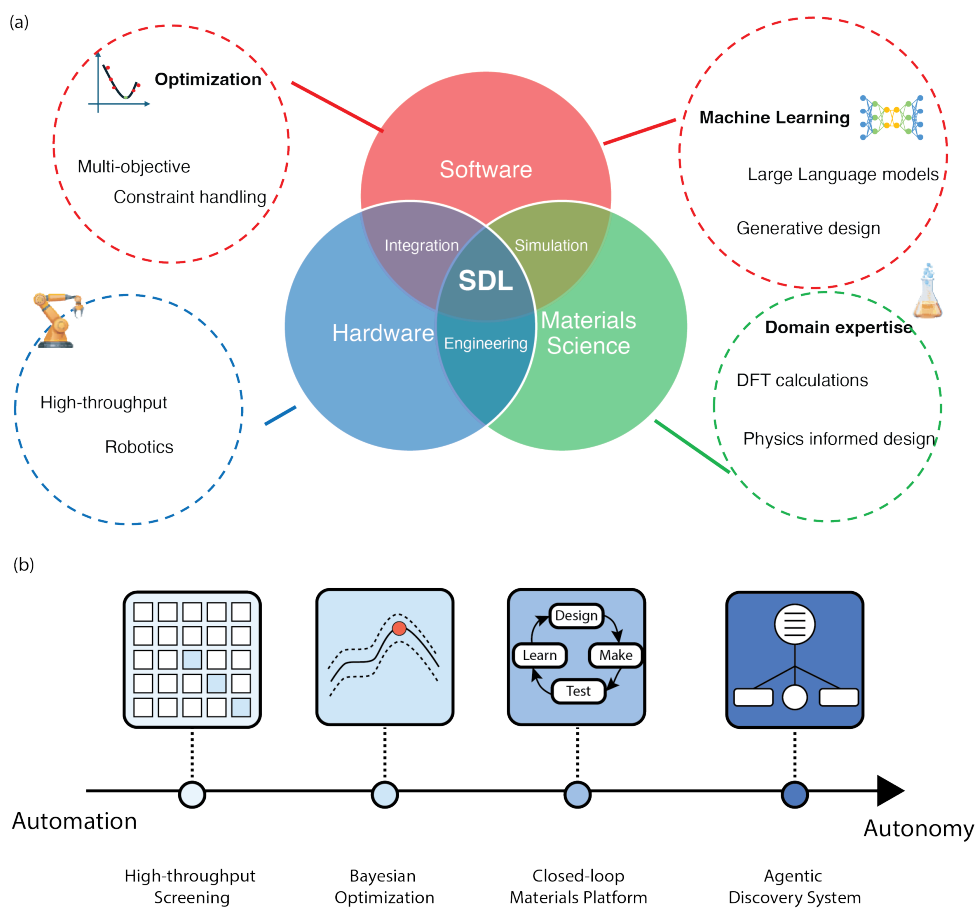

**Figure 1.** (a) The convergence of multiple technologies relating to software, hardware and domain expertise to form SDLs for materials science. Reproduced from ref. [1] with permission from [ACS]. Copyright [2025] [American Chemical Society.] (b)Evolution from automation to autonomy in materials discovery systems, showing the progression from high-throughput screening through Bayesian optimization and closed-loop platforms to agentic discovery systems with corresponding advances in decision-making, knowledge representation, and scientific reasoning capabilities.

The software ecosystem for SDLs faces equally significant challenges. Current optimization algorithms excel in low-data regimes but struggle with multi-objective optimization incorporating manufacturing constraints [17]. The integration of physics-based models with data-driven approaches remains ad hoc, missing opportunities to leverage domain knowledge systematically [18]. Most critically, platforms lack the ability to reason about experimental failures, adjust strategies autonomously, and incorporate lessons from failed experiments into future planning [19]. The emergence of foundation models trained on materials data offers promise but requires careful integration with experimental uncertainty and physical constraints [20].

A persistent gap remains between laboratory optimization and commercial translation. Most SDLs prioritize performance metrics measurable at small scale bandgap, mobility, catalytic rate, while neglecting cross-scale process fidelity. Reproducibility alone is insufficient, manufacturing readiness, cost variance, and yield predictability must become co-optimization objectives [1].

**Advances in science and technology to meet challenges**

Addressing these challenges requires fundamental advances in both technical capabilities and conceptual frameworks. The evolution from automation to autonomy demands systems that integrate decision-making, knowledge representation, and scientific reasoning (Figure 2). Hardware advances must enable modular, interoperable platforms where synthesis, characterization, and testing modules communicate through standardized protocols [16]. Digital twin representations of experimental systems will enable predictive maintenance, uncertainty quantification, and virtual experimentation before committing physical resources [8].

Software innovations center on three areas: optimization algorithms, knowledge representation,

and autonomous reasoning. Multi-fidelity Bayesian optimization incorporating manufacturing constraints enables concurrent optimization across scales, from molecular properties through device performance to production yield [17, 18]. Beyond conventional optimization pipelines, the evolution of SDLs now points toward expert-in-the-loop architectures that couple machine intelligence with human intuition. In this paradigm, large language models (LLMs) and multimodal agents act as cognitive layers that translate experimental intent into executable protocols, interpret unstructured results, and trigger expert intervention when uncertainties exceed acceptable thresholds [14, 21]. Such hybrid frameworks move SDLs beyond blind automation toward context-aware autonomy, where reasoning and decision-making are grounded in both data and domain knowledge. Critical to this evolution is the development of causal models that distinguish correlation from causation in experimental data, enabling genuine scientific discovery rather than mere optimization [13].

The concept of cross-scale optimization represents a paradigm shift from sequential development to concurrent engineering [15]. Rather than optimizing materials properties, then device performance, then manufacturing processes, future SDLs will consider all scales simultaneously. This requires new mathematical frameworks for multi-objective optimization with constraints spanning orders of magnitude in length and time scales [9]. Successful implementation demands close coupling between experimental platforms and simulation tools, with machine learning models serving as surrogate functions to guide experimentation efficiently [5, 22].

Integration of manufacturing concepts into laboratory platforms elevates them from research tools to prototype factories. Implementing asynchronous job scheduling, in-line quality control, and statistical process control in SDLs ensures discoveries translate directly to production [19, 23]. This manufacturing-aware approach influences experimental design from the outset, prioritizing robust processes over peak performance and considering supply chain constraints in materials selection [21].

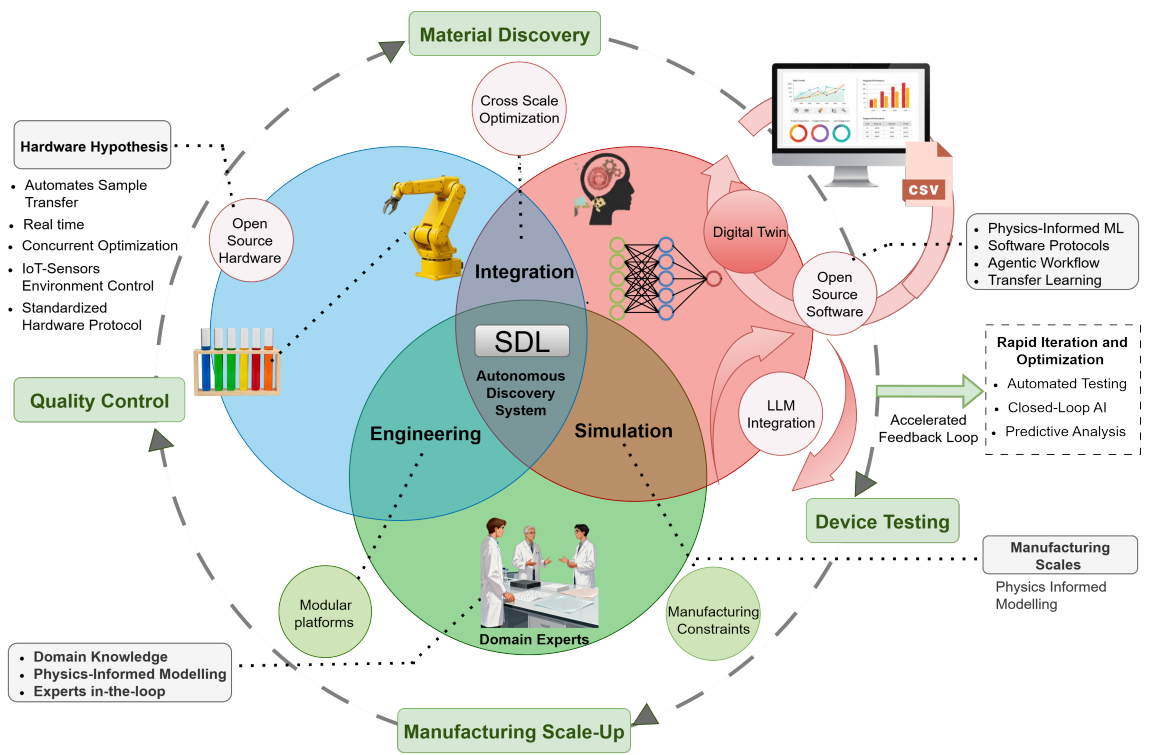

**Figure 2.** Vision for future self-driving laboratories integrating hardware automation, software intelligence, and domain expertise to enable cross-scale optimization from materials discovery through device testing to manufacturing scale-up.

Figure 2 illustrates the conceptual architecture of a fully integrated Autonomous Discovery System-SDLs that unites materials science, engineering, and simulation through digital-twin-enabled intelligence. The three overlapping domains represent how modular hardware platforms, open-source software, and data-driven simulation environments converge to form a closed-loop learning framework. Within this framework, cross-scale optimization bridges laboratory synthesis with factory-scale repeatability, while LLM-driven orchestration and digital twins facilitate the smooth transition between physical experimentation and virtual validation. The involvement of domain experts within this process guarantees that contextual judgment, safety, and manufacturability are prioritized in every iteration. As stated [1], hybrid SDL designs transcend isolated automation by incorporating physics-informed modeling, agentic workflows, and adaptive learning cycles that facilitate rapid iteration, predictive analysis, and scalable process transfer. This integration paradigm underpins the next generation of self-driving autonomous factories, where hypothesis, fabrication, testing, and manufacturing feedback are synchronized within a singular intelligent ecosystem.

**Concluding remarks**
The transformation of materials science from a primarily empirical discipline to an autonomous, AI-driven field is underway. Self-driving laboratories represent not merely automation of existing processes but a fundamental reimagining of how we discover and develop materials [3]. Success requires abandoning the traditional sequential pipeline in favor of integrated platforms that consider manufacturability from day one [15]. The convergence of hardware automation, intelligent algorithms, and encoded domain expertise creates unprecedented opportunities for accelerated materials innovation [10]. As these systems evolve from current closed-loop platforms toward truly agentic discovery systems with advanced scientific reasoning capabilities [13], they promise to collapse the laboratory-to-factory timeline from decades to years. The materials community must embrace this transformation through investment in standardized interfaces [24], open-source tools [25], and educational programs preparing researchers for this new paradigm. The trajectory is clearly defined from automation through optimization to true autonomous discovery and optimisation.

**Acknowledgments**
The authors thank Professor Stephen Dale for the invitation and the AI4X community for valuable discussions. L. N. W. T. Acknowledges funding from the Singapore Ministry of Education Tier 1 (RG86/23) And Tier 2 funding. GenAI (Claude.io) was used to paraphrase and proofread sections (Introduction) of the text.

**References**
[1] A.K.Y. Low, J.J.W. Cheng, K. Hippalgaonkar, L.W.T. Ng, Self-Driving Laboratories: Translating Materials Science from Laboratory to Factory, ACS Omega 10 (2025) 29902–29908.

[2] J.P. Correa-Baena, K. Hippalgaonkar, J. van Duren, S. Jaffer, V.R. Chandrasekhar, V. Stevanovic, C. Wadia, S. Guha, T. Buonassisi, Accelerating Materials Development via Automation, Machine Learning, and High-Performance Computing, Joule 2 (2018) 1410–1420.

[3] J. Wagner, C.G. Berger, X. Du, T. Stubhan, J.A. Hauch, C.J. Brabec, The evolution of Materials Acceleration Platforms: toward the laboratory of the future with AMANDA, J Mater Sci 56 (2021) 16422–16446.

[4] B. Burger, P.M. Maffettone, V.V. Gusev, C.M. Aitchison, Y. Bai, X. Wang, X. Li, B.M. Alston, B. Li, R. Clowes, A mobile robotic chemist, Nature 583 (2020) 237–241.

[5] S. Sun, N.T.P. Hartono, Z.D. Ren, F. Oviedo, A.M. Buscemi, M. Layurova, D.X. Chen, T. Ogunfunmi, J. Thapa, S. Ramasamy, C. Settens, B.L. DeCost, A.G. Kusne, Z. Liu, S.I.P. Tian, I.M. Peters, J.P. Correa-Baena, T. Buonassisi, Accelerated Development of Perovskite-Inspired Materials via High-Throughput Synthesis and Machine-Learning Diagnosis, Joule 3 (2019) 1437–1451.

[6] N.J. Szymanski, B. Rendy, Y. Fei, R.E. Kumar, T. He, D. Milsted, M.J. McDermott, M. Gallant, E.D. Cubuk, A. Merchant, H. Kim, A. Jain, C.J. Bartel, K. Persson, Y. Zeng, G. Ceder, An autonomous laboratory for the accelerated synthesis of novel materials, Nature 624 (2023) 86–91.

[7] X. Xiao, M. Chalh, Z.R. Loh, E. Mbina, T. Xu, R.C. Hiorns, Y. Li, M. Das, K. N'konou, L.W.T. Ng, Strategies to achieve efficiencies of over 19% for organic solar cells, Cell Rep Phys Sci 6 (2025) 102390.

[8] L.W.T. Ng, N.G. An, L. Yang, Y. Zhou, D.W. Chang, J.-E. Kim, L.J. Sutherland, T. Hasan, M. Gao, D. Vak, A printing-inspired digital twin for the self-driving, high-throughput, closed-loop optimization of roll-to-roll printed photovoltaics, Cell Rep Phys Sci (2024).

[9] A.K.Y. Low, F. Mekki-Berrada, A. Gupta, A. Ostudin, J. Xie, E. Vissol-Gaudin, Y.-F. Lim, Q. Li, Y.S. Ong, S.A. Khan, K. Hippalgaonkar, Evolution-guided Bayesian optimization for constrained multi-objective optimization in self-driving labs, NPJ Comput Mater 10 (2024) 104.

[10] R. Vescovi, T. Ginsburg, K. Hippe, D. Ozgulbas, C. Stone, A. Stroka, R. Butler, B. Blaiszik, T. Brettin, K. Chard, M. Hereld, A. Ramanathan, R. Stevens, A. Vriza, J. Xu, Q. Zhang, I. Foster, Towards a modular architecture for science factories, Digital Discovery 2 (2023) 1980–1998.

[11] Y. Jing, A.K.Y. Low, Y. Liu, M. Feng, J.W.M. Lim, S.M. Loh, Q. Rehman, S.A. Blundel, N. Mathews, K. Hippalgaonkar, T.C. Sum, A. Bruno, S.G. Mhaisalkar, Stable and Highly Emissive Infrared Yb-Doped Perovskite Quantum Cutters Engineered by Machine Learning, Advanced Materials 36 (2024) 2405973.

[12] M. Thway, J. Recatala-Gomez, F.S. Lim, K. Hippalgaonkar, L.W.T. Ng, Harnessing GenAI for Higher Education: A Study of a Retrieval Augmented Generation Chatbot's Impact on Learning, J Chem Educ 102 (2025) 3849–3857.

[13] H. Kitano, Nobel Turing Challenge: creating the engine for scientific discovery, NPJ Syst Biol Appl 7 (2021) 29.

[14] M. Thway, A.K.Y. Low, S. Khetan, H. Dai, J. Recatala-Gomez, A.P. Chen, K. Hippalgaonkar, Harnessing GPT-3.5 for text parsing in solid-state synthesis–case study of ternary chalcogenides, Digital Discovery 3 (2024) 328–336.

[15] J. Zhang, J.A. Hauch, C.J. Brabec, Toward Self-Driven Autonomous Material and Device Acceleration Platforms (AMADAP) for Emerging Photovoltaics Technologies, Acc Chem Res 57 (2024) 1434–1445.

[16] C.J. Leong, K.Y.A. Low, J. Recatala-Gomez, P. Quijano Velasco, E. Vissol-Gaudin, J. Da Tan, B. Ramalingam, R. I Made, S.D. Pethe, S. Sebastian, Y.-F. Lim, Z.H.J. Khoo, Y. Bai, J.J.W. Cheng, K. Hippalgaonkar, An object-oriented framework to enable workflow evolution across materials acceleration platforms, Matter 5 (2022) 3124–3134.

[17] A.K.Y. Low, E. Vissol-Gaudin, Y.-F. Lim, K. Hippalgaonkar, Mapping pareto fronts for efficient multi-objective materials discovery, Journal of Materials Informatics 3 (2023) 11.

[18] D. Bash, Y. Cai, V. Chellappan, S.L. Wong, X. Yang, P. Kumar, J. Da Tan, A. Abutaha, J.J.W. Cheng, Y. Lim, Multi-fidelity high-throughput optimization of electrical conductivity in P3HT-CNT composites, Adv Funct Mater 31 (2021) 2102606.

[19] M. Christensen, L.P.E. Yunker, P. Shiri, T. Zepel, P.L. Prieto, S. Grunert, F. Bork, J.E. Hein, Automation isn't automatic, Chem Sci 12 (2021) 15473–15490.

[20] J. Yin, H. Chen, J. Qiu, W. Li, P. He, J. Li, I.A. Karimi, X. Lan, T. Wang, X. Wang, SurFF: a foundation model for surface exposure and morphology across intermetallic crystals, Nature Computational Science (2025).

[21] C. Zeni, R. Pinsler, D. Zügner, A. Fowler, M. Horton, X. Fu, Z. Wang, A. Shysheya, J. Crabbé, S. Ueda, R. Sordillo, L. Sun, J. Smith, B. Nguyen, H. Schulz, S. Lewis, C.-W. Huang, Z. Lu, Y. Zhou, H. Yang, H. Hao, J. Li, C. Yang, W. Li, R. Tomioka, T. Xie, A generative model for inorganic materials design, Nature (2025).

[22] B.A. Koscher, R.B. Canty, M.A. McDonald, K.P. Greenman, C.J. McGill, C.L. Bilodeau, W. Jin, H. Wu, F.H. Vermeire, B. Jin, T. Hart, T. Kulesza, S.C. Li, T.S. Jaakkola, R. Barzilay, R. Gómez-Bombarelli, W.H. Green, K.F. Jensen, Autonomous, multiproperty-driven molecular discovery: From predictions to measurements and back, Science 382 (2023).

[23] A.E. Siemenn, B. Das, E. Aissi, F. Sheng, L. Elliott, B. Hudspeth, M. Meyers, J. Serdy, T. Buonassisi, Archerfish: a retrofitted 3D printer for high-throughput combinatorial experimentation via continuous printing, Digital Discovery (2025).

[24] A. Jain, J. Montoya, S. Dwaraknath, N.E.R. Zimmermann, J. Dagdelen, M. Horton, P. Huck, D. Winston, S. Cholia, S.P. Ong, The materials project: Accelerating materials design through theory-driven data and tools, Handbook of Materials Modeling: Methods: Theory and Modeling (2020) 1751–1784.

[25] F. Häse, M. Aldeghi, R.J. Hickman, L.M. Roch, M. Christensen, E. Liles, J.E. Hein, A. Aspuru-Guzik, Olympus: a benchmarking framework for noisy optimization and experiment planning, Mach Learn Sci Technol 2 (2021) 035021.

# Reward engineering from long-term objectives towards realization of interactive AI workflows

Sergei V. Kalinin[1] and Mahshid Ahmadi[1]

[1] Dept. of Materials Science and Engineering, University of Tennessee, Knoxville, TN 37923 USA
E-mail: sergei2@utk.edu, mahmadi3@utk.edu

**Status**

Machine learning methods are progressively considered as a part of real-world technological solutions, including workflows for materials synthesis and optimization, computation, imaging, and characterization.[1, 2] Building ML/AI-controlled tools, integration of dissimilar tools in geographically-localized self-driving lab (SDL), and operation of multiple geographically-distributed SDL nodes requires algorithms for decision making beyond closed-loop Bayesian Optimization.[1, 3-7] In general, such algorithms can be defined as a search in the space of all executable instrument and SDL operations with known costs and latencies. Methods such as reinforcement learning (RL) that had been shown to be highly effective in environments such as computer games or simulations, are often inadequate for real-world applications both due to large data budget and lack of well-defined reward functions. Similarly, implementation of stochastic optimization methods including A*, Monte Carlo Decision Trees, etc. requires both rewards functions and estimated rewards via roll-out functions. However, for many real-world problems the reward functions available in the end of experimental campaign (or after several steps) are absent; rather the experiments are motivated by the long-term objectives with probabilistic relationship between experiment measure of success (reward) and long-term objectives. Correspondingly, designing reward functions that adequately represents real-world objective and does not lead to reward hacking is a challenge.

**Current and future challenges**

Consider climate change, the problem motivating multi-billion-dollar investments over the globe. Minimizing climate change is a very long-term objective. The lower rank objectives are the development of solar and wind energy and associated grid-level storage and effective energy transport methods. The even lower rank objectives are the development of cheap, environmentally friendly, and stable chemistries for photovoltaic systems. None of these objectives can be directly translated into a reward for a realistic experimental campaign. Rather, these objectives serve as a motivation for experiment planning.

Potential technological solution for solar energy is solution-processable low-cost hybrid perovskites.[8, 9] These materials require joint optimization across a design space involving both the composition space of inorganic framework and chemical space of organic molecular spacers.[10] This exploration is further complicated by the frequent coexistence of multiple phases and complex defect chemistries in synthesized materials, necessitating focus on process optimization. For these, even simple BO materials optimization relies on transformation of measured experimental response (Photoluminescence spectrum, X-Ray Diffraction, Raman) into reward function. For example, X-Ray pattern can be transformed into lattice parameters, used as a measure of crystallinity and disorder, or detect the presence of impurity phase. Optimization and discovery loops based on these reward functions will naturally guide active experiments to different outcomes. This limitations persist to the level of individual measurements of microstructure and local structure-property relationships,[11, 12] where reward functions such as image optimization, minimization of predictive uncertainty of specific hypothesis[13] or structure-property relationships[14-16] can guide the operation of specific instrument. These simple reward functions are often insufficiently selective, and the parts of the search space corresponding to the maximum reward are still very large. Similarly, reward functions can be imperfectly aligned with the target functionalities – for example, while minimum free energy is correlated with thermodynamic phase stability, many metastable phases can be kinetically stable. We refer to such reward functions as weak and mis-aligned, respectively. When multiple characterization and modelling modalities can be realized over the same candidate space (e.g. phase diagram, chemical space, or their direct product), the construction of (joint) reward function becomes the key challenge for automated workflow design.

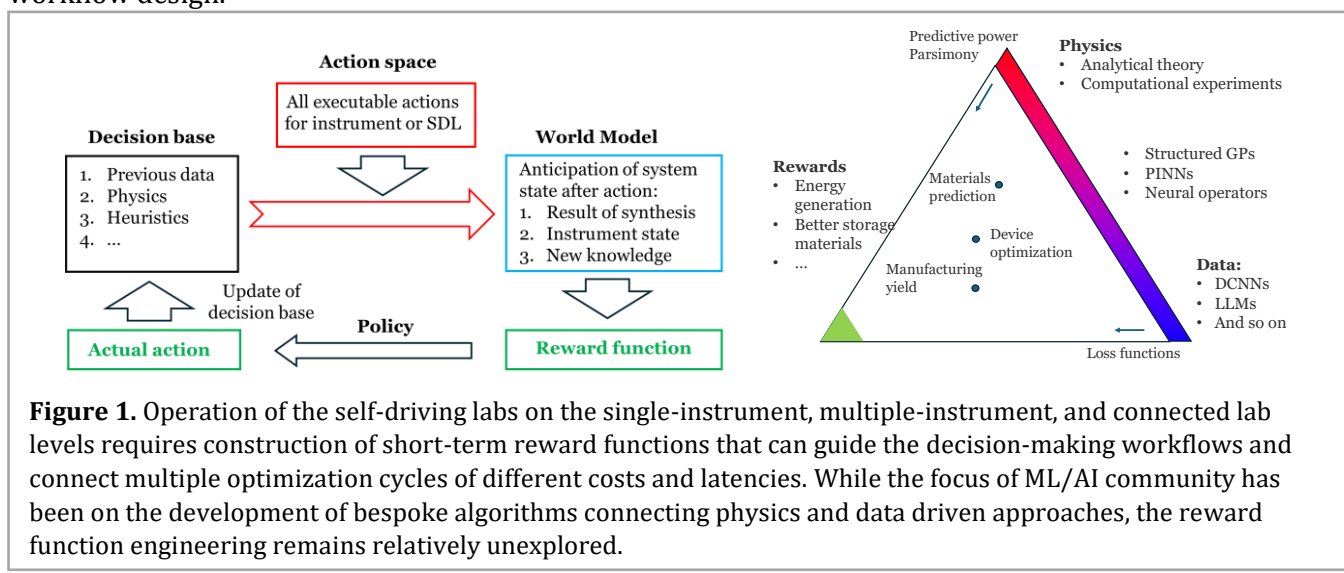

**Figure 1.** Operation of the self-driving labs on the single-instrument, multiple-instrument, and connected lab levels requires construction of short-term reward functions that can guide the decision-making workflows and connect multiple optimization cycles of different costs and latencies. While the focus of ML/AI community has been on the development of bespoke algorithms connecting physics and data driven approaches, the reward function engineering remains relatively unexplored.

**Advances in science and technology to meet challenges**

The discovery of the short-term rewards that can be used for hypothesis making to guide experimental research, and to guide and ascertain the success of experimental campaigns is the missing link required to connect ML to real-world applications. As potential pathways to address this challenge, we consider (a) LLM-based literature mining towards building the DAGs connecting experimental outcomes (rewards) and objectives (motivation), (b) Technoeconomic analysis of past publications outcomes, and (c) Crowdsourcing to the community of experts (aka "what would be the potential of high temperature conductivity to change the world" to "how does the phase separation in cuprates affects peak-effect and losses").

**Concluding remarks**

Reward functions will serve as the primary integrative mechanism for coupling heterogeneous experimental workflows including synthesis, characterization, data analysis, and device integration, into a coherent autonomous framework. By parameterizing reward metrics at multiple temporal scales, the approach will reconcile rapid, short term performance improvements with the pursuit of long-term scientific objectives, thereby facilitating both efficient optimization and strategic discovery. Furthermore, it can be anticipated that a dedicated community of "reward function engineers" will emerge, prioritizing the systematic design of high value reward structures along with the traditional focus on large scale data aggregation The important element of this approach is that humans are the part of the theory-experiment loop – and hence the structure of the rewards can be amended via human feedback.[3, 17, 18]

**Acknowledgements**

This work is supported by NSF DMR proposal Award Number:2523284 and NSF Award Number 2043205.

**References**

(1) MacLeod, B. P.; Parlane, F. G. L.; Morrissey, T. D.; Hase, F.; Roch, L. M.; Dettelbach, K. E.; Moreira, R.; Yunker, L. P. E.; Rooney, M. B.; Deeth, J. R.; Lai, V.; Ng, G. J.; Situ, H.; Zhang, R. H.; Elliott, M. S.; Haley, T. H.; Dvorak, D. J.; Aspuru-Guzik, A.; Hein, J. E.; Berlinguette, C. P. Self-driving laboratory for accelerated discovery of thin-film materials. *Sci Adv* **2020**, *6* (20), eaaz8867. DOI: 10.1126/sciadv.aaz8867 From NLM PubMed-not-MEDLINE.
(2) Flores-Leonar, M. M.; Mejía-Mendoza, L. M.; Aguilar-Granda, A.; Sanchez-Lengeling, B.; Tribukait, H.; Amador-Bedolla, C.; Aspuru-Guzik, A. Materials Acceleration Platforms: On the way to autonomous experimentation. *Current Opinion in Green and Sustainable Chemistry* **2020**, *25*, 100370. DOI: 10.1016/j.cogsc.2020.100370.
(3) Hysmith, H.; Foadian, E.; Padhy, S. P.; Kalinin, S. V.; Moore, R. G.; Ovchinnikova, O.; Ahmadi, M. J. D. D. The future of self-driving laboratories: From Human in the Loop Interactive AI to Gamification. **2024**.
(4) Abolhasani, M.; Kumacheva, E. The rise of self-driving labs in chemical and materials sciences. *Nature Synthesis* **2023**, *2* (6), 483-492.
(5) Kalinin, S. V.; Ziatdinov, M.; Hinkle, J.; Jesse, S.; Ghosh, A.; Kelley, K. P.; Lupini, A. R.; Sumpter, B. G.; Vasudevan, R. K. Automated and Autonomous Experiments in Electron and Scanning Probe Microscopy. *Acs Nano* **2021**, *15* (8), 12604-12627. DOI: 10.1021/acsnano.1c02104.
(6) Soldatov, M. A.; Butova, V. V.; Pashkov, D.; Butakova, M. A.; Medvedev, P. V.; Chernov, A. V.; Soldatov, A. V. Self-driving laboratories for development of new functional materials and optimizing known reactions. *Nanomaterials* **2021**, *11* (3), 619.
(7) Epps, R. W.; Bowen, M. S.; Volk, A. A.; Abdel-Latif, K.; Han, S.; Reyes, K. G.; Amassian, A.; Abolhasani, M. Artificial Chemist: An Autonomous Quantum Dot Synthesis Bot. *Adv Mater* **2020**, *32* (30), e2001626. DOI: 10.1002/adma.202001626 From NLM PubMed-not-MEDLINE.
(8) Grancini, G.; Nazeeruddin, M. K. Dimensional tailoring of hybrid perovskites for photovoltaics. *Nature Reviews Materials* **2019**, *4* (1), 4-22.
(9) Stoumpos, C. C.; Malliakas, C. D.; Kanatzidis, M. G. Semiconducting tin and lead iodide perovskites with organic cations: phase transitions, high mobilities, and near-infrared photoluminescent properties. *Inorg Chem* **2013**, *52* (15), 9019-9038. DOI: 10.1021/ic401215x From NLM PubMed-not-MEDLINE.
(10) Yang, J. H.; Kalinin, S. V.; Cubuk, E. D.; Ziatdinov, M.; Ahmadi, M. Toward self-organizing low-dimensional organic-inorganic hybrid perovskites: Machine learning-driven co-navigation of chemical and compositional spaces. *Mrs Bulletin* **2023**, *48* (2), 164-172. DOI: 10.1557/s43577-023-00490-y.
(11) Kalinin, S. V.; Ziatdinov, M.; Ahmadi, M.; Ghosh, A.; Roccapriore, K.; Liu, Y.; Vasudevan, R. K. J. A. P. R. Designing workflows for materials characterization. **2024**, *11* (1).
(12) Kalinin, S. V.; Liu, Y.; Biswas, A.; Duscher, G.; Pratiush, U.; Roccapriore, K.; Ziatdinov, M.; Vasudevan, R. J. M. T. Human-in-the-Loop: The Future of Machine Learning in Automated Electron Microscopy. **2024**, *32* (1), 35-41.
(13) Liu, Y.; Morozovska, A. N.; Eliseev, E. A.; Kelley, K. P.; Vasudevan, R.; Ziatdinov, M.; Kalinin, S. V. Autonomous scanning probe microscopy with hypothesis learning: Exploring the physics of domain switching in ferroelectric materials. *Patterns (N Y)* **2023**, *4* (3), 100704. DOI: 10.1016/j.patter.2023.100704 From NLM PubMed-not-MEDLINE.
(14) Liu, Y.; Yang, J.; Vasudevan, R. K.; Kelley, K. P.; Ziatdinov, M.; Kalinin, S. V.; Ahmadi, M. Exploring the Relationship of Microstructure and Conductivity in Metal Halide Perovskites via Active Learning-Driven Automated Scanning Probe Microscopy. *J Phys Chem Lett* **2023**, *14* (13), 3352-3359. DOI: 10.1021/acs.jpclett.3c00223 From NLM PubMed-not-MEDLINE.
(15) Liu, Y. T.; Kelley, K. P.; Vasudevan, R. K.; Funakubo, H.; Ziatdinov, M. A.; Kalinin, S. V. Experimental discovery of structure-property relationships in ferroelectric materials via active learning. *Nature Machine Intelligence* **2022**, *4* (4), 341-350. DOI: 10.1038/s42256-022-00460-0.
(16) Roccapriore, K. M.; Kalinin, S. V.; Ziatdinov, M. Physics Discovery in Nanoplasmonic Systems via Autonomous Experiments in Scanning Transmission Electron Microscopy. *Adv Sci (Weinh)* **2022**, *9* (36), e2203422. DOI: 10.1002/advs.202203422 From NLM Medline.
(17) Liu, Y.; Ziatdinov, M. A.; Vasudevan, R. K.; Kalinin, S. V. Explainability and human intervention in autonomous scanning probe microscopy. *Patterns (N Y)* **2023**, *4* (11), 100858. DOI: 10.1016/j.patter.2023.100858 From NLM PubMed-not-MEDLINE.
(18) Biswas, A.; Liu, Y.; Creange, N.; Liu, Y.-C.; Jesse, S.; Yang, J.-C.; Kalinin, S. V.; Ziatdinov, M. A.; Vasudevan, R. K. J. n. C. M. A dynamic Bayesian optimized active recommender system for curiosity-driven partially Human-in-the-loop automated experiments. **2024**, *10* (1), 29.

# Ontological Representation of Chemical Process Knowledge for Agentic AI

Jiyizhe Zhang,[1,2,3] Shuyuan Zhang[1,2] and Alexei Lapkin[1,2]

[1] Department of Chemical Engineering and Biotechnology, University of Cambridge, Cambridge, United Kingdom
[2] Innovation Centre in Digital Molecular Technologies, Yusuf Hamied Department of Chemistry, University of Cambridge, Cambridge, United Kingdom
[3] Department of Chemical Engineering, University of Manchester, Oxford Road, Manchester, United Kingdom

E-mail: aal35@cam.ac.uk

**Status**

Advances in artificial intelligence (AI) are crucial to achieving sustainability goals and net-zero ambitions in the chemical industry. Digitalisation of R&D accelerates discovery and reduces risks/cost of development and scale-up, whereas digital manufacturing is promising unprecedented agility, resilience and efficiency along the complete chemicals value chain [1]. The chemical industry was one of the earliest sectors to adopt digital and AI technologies in the late 1980s [2], rapidly taking advantage of developments in cheminformatics, process simulation software, and expert systems tools. Today we are experiencing fast transformation of chemical R&D and manufacturing, supported by progress in computing and AI.

Unlocking the full potential of AI to attain high predictive accuracy and to achieve higher levels of reasoning in chemistry and chemical engineering remains challenging [3]. A key bottleneck is the current lack of *systematic and extensive encoding of domain knowledge*. Within the domain of chemical engineering this translates into: (*i*) *knowledge representation*, and (*ii*) *knowledge integration and standardisation*.

Chemical processes are inherently complex, involving coupled spatio-temporal phenomena across multiple scales. Representing such knowledge requires capturing phenomena and identifying their relationships. Chemical knowledge often exists in highly heterogeneous formats. Some knowledge is curated in datasets; other types of knowledge are embedded in model equations or further exist in disparate records across literature, laboratory instruments, and proprietary industrial records. Achieving true interoperability demands a shared digital infrastructure that links experimental, modelling, and operational data seamlessly across organisations and along the value chain.

These challenges were recognised early in computer science, where semantic web technologies were proposed with a vision of representing knowledge and their relationship as an ontological graph. Both humans and machines can read, and autonomous agents can act within such graphs for discovery and functional tasks. In chemical engineering, pioneering efforts began in the early 2000s with initiatives such as OntoCAPE [4], a comprehensive ontology of chemical process concepts, and domain-specific frameworks for pharmaceutical manufacturing [5]. More recently, ontologies began to be developed for specific tasks, such as enabling communication between laboratory equipment [6] or supporting digital research workflows [7]. International initiatives such as NFDI4Chem (Germany), PSDI (UK), Allotrope Foundation and Pistoia alliance are actively developing the foundational ontologies and related agents.

**Current and future challenges**

The establishment of a fundamental and standardised knowledge infrastructure would mark a transformative step for chemical research and development. It would not only allow AI agents to automate routine tasks but will also provide deep, cross-domain insights that extend beyond any individual's expertise. Further advances in this area will therefore be pivotal for creating digital workflows that are interoperable, scalable, and supporting the sustainable development of the industry. Already today there are early examples of integrated digital workflows being developed using state-of-the-art language models and agentic AI framework [8].

The most significant challenge in progressing the applications of AI in chemistry and chemical engineering is the availability of good-quality data. Most applications targeting predictions of properties of molecules, materials, formulations, or outcomes of reactions, by-product formation and reaction yield are directly linked to access to data. Historical data is available in proprietary industrial records, in open source and behind paywalls of journals and curated databases. However, quality of historical data is frequently low for ML applications, as datasets are imbalanced, do not contain negative results, or do not systematically sample the potential experimental space [9]. While data foundries would undoubtedly emerge in response to this need, future challenges around new collaborative research models and adjustments of business models towards data and code sharing will be required.

In the challenging area of knowledge representation, the string representations of molecules and reactions, such as the community-adopted canonical identifiers SMILES (Simplified Molecular Input Line Entry System) and InChI (International Chemical Identifier), and similar formats like SMARTS (SMILES Arbitrary Target Specification), became the baseline cheminformatics methods for chemistry-AI applications [10]. These data formats can be natively referenced by chemical process ontologies to enable deterministic and interoperable molecular representations. RXNO (Royal Society of Chemistry name reactions ontology) and CHEMINF (the CHEMical INFormation ontology) are examples of recent ontologies for developing common reaction and property vocabularies, albeit restricted to metadata with limited chemical scope [11]. The remaining significant challenge is standardising reaction ontologies with multi-granular descriptors, from mechanic bond breaking/forming to operational temperature/pressure. This would allow building AI workflows that span not only molecular discovery and property prediction, but also synthesis planning, process design, scale-up and environmental impact prediction.

**Advances in science and technology to meet challenges**

The capabilities of AI models are closely linked to the availability of data. Chemical laboratories are rapidly becoming mixed human-robotic environments geared towards the generation of high-quality AI-ready experimental data [12]. Thus, chemical experimental facilities may take the form of 'data foundries', contributing towards a collaborative effort in data generation for shared high-quality AI models. Certainly, the availability of lower-cost automation and robotic equipment, and access to open-source codes, enables rapid development of automated experimental pipelines that generate data for AI workflows, e.g. [13].

To facilitate data transfer from experimental systems, built-in exporters in ELNs (Electronic Lab Notebooks), LIMSs (Laboratory Information Management Systems), and instrument firmware should be able to provide semantically rich metadata (persistent identifiers, provenance, measured uncertainty) in standardised schema forms. Automated validation pipelines using SHACL (SHApes Constraint Language) checks and unit/quantity verification will reduce manual effort and ensure immediate interoperability of the ingested data.

The increased pace of adoption of FAIR (Findable, Accessible, Interoperable, Reusable) data principles in publishing is contributing to the creation of datasets that are actionable by AI agents [14], and could be mined for the creation of relational data enriched with ontology schemas. It is unclear if initiatives such as the Open Reaction Database [15] and a similar data initiative by NFDI4Chem would significantly alter the quantity/quality of semantically enriched data to threaten the usability of data behind paywalls of commercial publishers and within corporate environments.

We expect that development and maintenance of ontologies will become significantly easier with advancements in symbolic AI and agentic AI systems [16]. NLP (Natural Language Processing) techniques can extract terms, relations, and axioms from unstructured ELNs, SOPs (Standard Operating Procedures), patents, and flow sheets for seeding and populating domain ontologies. Conversely, ontologies provide rich semantic grounding for developing interpretable and trustworthy AI agents, such as graph neural networks, large language models, and symbolic systems, for supporting decision-making in the chemical industry.

Our final comment relates to the environmental cost of computing. The academic community and industrial community focusing on societal challenges (health, climate restoration, mobility, etc.) are highly conscious of the potential environmental impact of the increased use of AI methods [17]. While this is not yet quantified, digitalisation of R&D may lead to the reduction in resource intensity of research through avoidance of unnecessary repeats and better learning on historical data requiring fewer new experiments [18]. At the same time, the rebound effect of the increased ease of performing experiments in automated algorithm-directed facilities may cause an increase in resource use (both chemical inputs and energy). Systems-level analytics and decision support will be required to figure out the most resource-effective deployment of automation and AI, while the foundational layer for multi-domain systems analysis is the ontology.

**Concluding remarks**

The development and adoption of ontologies is supporting the emergent symbolic AI applications and is likely to bring significant future advances over the statistical ML methods that do not have access to semantically enriched data. There is, however, a significant challenge in developing ontologies in chemical sciences and engineering, as these domains have not yet emerged from the chaos of random names for objects and phenomena, a plethora of competing terms, non-standard experimental protocols and a complete lack of benchmarks. A growing scientific community is collaborating on solving the challenges of data and protocol standards, and a growing number of vendors are part of this effort as well. As we emerge from the 'hype' phase of AI in chemistry, we begin to develop semantically rich, contextualised datasets, and take advantage of the reasoning capabilities of advanced AI models. Further connecting contextualised data across chemical sciences and chemical technology domains will enable addressing the challenges of sustainability and climate adaptation.

**Acknowledgements**

We acknowledge public funding from UKRI under grants EP/W031019/1 "Bio-derived and Bio-inspired Advanced Materials for Sustainable Industries (VALUED)" and EP/Z531339/1 "Accelerated Development of Pharmaceutical Processes Through Digitally Coupled Reaction Screening and Process Optimisation (OptiMed)". Innovation Centre in Digital Molecular Technologies is co-funded by its industrial members and the University of Cambridge.

**References**

[1] P. Fantke, C. Cinquemani, P. Yaseneva, J. De Mello, H. Schwabe, B. Ebeling, A.A. Lapkin, Transition to sustainable chemistry through digitalization, Chem 7 (2021) 2866–2882. https://doi.org/10.1016/j.chempr.2021.09.012.

[2] V. Venkatasubramanian, The promise of artificial intelligence in chemical engineering: Is it here, finally?, Aiche J. 65 (2019) 466–478. https://doi.org/10.1002/aic.16489.

[3] V. Venkatasubramanian, Do large language models "understand" their knowledge?, AIChE Journal 71 (2025) e18661. https://doi.org/10.1002/aic.18661.

[4]  J. Morbach, A. Wiesner, W. Marquardt, OntoCAPE-A (re)usable ontology for computer-aided process engineering, Comput. Chem. Eng. 33 (2009) 1546–1556. https://doi.org/10.1016/j.compchemeng.2009.01.019.

[5]  L. Hailemariam, V. Venkatasubramanian, Purdue Ontology for Pharmaceutical Engineering: Part I. Conceptual Framework, Journal of Pharmaceutical Innovation 5 (2010) 88–99. https://doi.org/10.1007/s12247-010-9081-3.

[6]  J. Bai, S. Mosbach, C.J. Taylor, D. Karan, K.F. Lee, S.D. Rihm, J. Akroyd, A.A. Lapkin, M. Kraft, A dynamic knowledge graph approach to distributed self-driving laboratories, Nature Communications 15 (2024) 462. https://doi.org/10.1038/s41467-023-44599-9.

[7]  S.D. Rihm, J. Bai, A. Kondinski, S. Mosbach, J. Akroyd, M. Kraft, Transforming research laboratories with connected digital twins, Nexus 1 (2024) 100004. https://doi.org/10.1016/j.ynexs.2024.100004.

[8]  A.M. Bran, S. Cox, O. Schilter, A.D. White, P. Schwaller, ChemCrow: Augmenting large-language models with chemistry tools, arXiv 2304.05376 (2023). https://doi.org/10.48550/arXiv.2304.05376.

[9]  J. Jiang, C. Zhang, L. Ke, N. Hayes, Y. Zhu, H. Qiu, B. Zhang, T. Zhou, G. Wei, A review of machine learning methods for imbalanced data challenges in chemistry, CHEMICAL SCIENCE 16 (2025) 7637–7658. https://doi.org/10.1039/d5sc00270b.

[10] D.S. Wigh, J.M. Goodman, A.A. Lapkin, A review of molecular representation in the age of machine learning, WIREs Computational Molecular Science n/a (2022) e1603. https://doi.org/10.1002/wcms.1603.

[11] P. Strömert, J. Hunold, A. Castro, S. Neumann, O. Koepler, Ontologies4Chem: the landscape of ontologies in chemistry, 94 (2022) 605–622. https://doi.org/10.1515/pac-2021-2007.

[12] P.Q. Velasco, K. Hippalgaonkar, B. Ramalingam, Emerging trends in the optimization of organic synthesis through high-throughput tools and machine learning, Beilstein J. Org. Chem. 21 (2025) 10–38. https://doi.org/10.3762/bjoc.21.3.

[13] Z.J. Liew, Z. Elkhaiary, A.A. Lapkin, Parameter efficient multi-model vision assistant for polymer solvation behaviour inference, Npj Computational Materials 11 (2025) 161. https://doi.org/10.1038/s41524-025-01658-7.

[14] K.M. Jablonka, L. Patiny, B. Smit, Making the collective knowledge of chemistry open and machine actionable, Nature Chemistry 14 (2022) 365–376. https://doi.org/10.1038/s41557-022-00910-7.

[15] S.M. Kearnes, M.R. Maser, M. Wleklinski, A. Kast, A.G. Doyle, S.D. Dreher, J.M. Hawkins, K.F. Jensen, C.W. Coley, The Open Reaction Database, J. Am. Chem. Soc. 143 (2021) 18820–18826. https://doi.org/10.1021/jacs.1c09820.

[16] A. Maedche, S. Staab, Mining Ontologies from Text, Springer-Verlag, Berlin Heidelberg, 2000. https://doi.org/10.1007/3-540-39967-4_14.

[17] C. Jay, Y. Yu, I. Crawford, S. Archer-Nicholls, P. James, A. Gledson, G. Shaddick, R. Haines, L. Lannelongue, E. Lines, S. Hosking, D. Topping, Prioritize environmental sustainability in use of AI and data science methods, NATURE GEOSCIENCE 17 (2024) 106–108. https://doi.org/10.1038/s41561-023-01369-y.

[18] G. Pesciullesi, P. Schwaller, T. Laino, J.L. Reymond, Transfer learning enables the molecular transformer to predict regio- and stereoselective reactions on carbohydrates, Nature Communications 11 (2020) 8. https://doi.org/10.1038/s41467-020-18671-7.

# Machine Learning in Model Predictive Control of Chemical Processes

Ming Xiao[1], and Zhe Wu[1,*]

[1]Department of Chemical and Biomolecular Engineering, National University of Singapore, 117585, Singapore.
*Author to whom any correspondence should be addressed.

**E-mail:** wuzhe@nus.edu.sg

**Status**
Model predictive control (MPC) is a widely adopted control scheme in industrial applications for its ability to handle complex multivariate scenarios, address constraints, and optimize performance. Based on the current state information, MPC calculates a sequence of optimal control actions using predictions of the dynamical system, of which only the first control input is implemented. In the MPC framework, prediction accuracy plays an essential role in ensuring satisfactory performance. However, for chemical processes that typically exhibit complex dynamics and strong nonlinearity, it is difficult to develop a high-accuracy model using traditional first-principles methods, because complex physicochemical phenomena is often poorly understood. Unlike first-principles models, which are usually expressed in physical equations, machine learning methods aim to capture the dynamic behavior of nonlinear systems in a data-driven way with high accuracy and flexibility.

With the development of computational resources and open-source toolboxes, machine learning methods have received significant attention for addressing the modeling problem in chemical processes [1, 2, 3]. Specifically, recurrent neural networks (RNNs) and their variants are well suited for handling time-series data and have shown strong performance in modeling nonlinear systems. Moreover, RNN-based models have been effectively integrated into MPC frameworks as predictive models to accurately capture system dynamics and provide a robust foundation for MPC computations [4]. Machine learning methods have been integrated into general MPC, Lyapunov-based MPC, distributed MPC, and zone tracking MPC, with the aim of achieving different control objectives. In particular, the theoretical analysis and computational implementation of Lyapunov-based MPC using RNN models as predictive models have been investigated in [5, 6], demonstrating improved closed-loop performance compared to MPC using linear empirical models. ML-based MPC schemes have been successfully implemented in a wide range of chemical process applications, including but not limited to batch crystallization [7], reactors [8, 9], and steam methane reforming [10]. The key strengths of machine learning methods lie in their flexibility, efficiency, and accuracy [11]. However, the implementation of ML-based MPC in industrial chemical applications still faces several challenges, including data availability, controller integration, and in-depth analysis of black-box neural network models. These challenges will be discussed in the later section.

**Current and future challenges**
The implementation of ML-based MPCs faces both theoretical and practical challenges [12, 11]. On the theoretical side, a central concern is ensuring the accuracy and reliability of the learned model. The generalization error of ML models directly affects the fidelity of system dynamics representation, which in turn influences MPC stability, robustness, and constraint satisfaction. Establishing formal guarantees on closed-loop performance when using data-driven predictive models remains an active area of research. On the practical side, the computational burden of solving complex MPC optimization problems can become significant, especially for nonlinear models or when real-time decision-making is required. Extending ML-based MPC frameworks to large-scale, high-dimensional systems introduces additional challenges in scalability and solver efficiency. Data-related issues further compound these difficulties: chemical processes often provide only limited operational data, the available data may be noisy or incomplete. Addressing these challenges requires advances in both ML model development, such as incorporating physics-based priors or uncertainty quantification, and in MPC design, including efficient solvers, model reduction strategies, and robust control formulations capable of handling data and model imperfections. In this article, we will focus on the issue of data scarcity, which represents one of the most critical obstacles to the development and deployment of machine learning models for MPC applications.

**Advances in science and technology to meet challenges**
Various approaches have been proposed to address the data scarcity problem, among which transfer learning and physics-informed methods have shown strong potential for modeling chemical processes with limited data and integrating them into MPC schemes.

Physics-informed machine learning (PIML) aims to improve the accuracy and interpretability of the model by embedding domain knowledge of the process into the development of the ML model [13, 14]. Available physics models, even if partially known or uncertain, can be incorporated into machine learning models in various ways to enhance performance, such as customizing network architectures or modifying loss functions [15]. For instance, structural knowledge of complex chemical processes can be utilized to design neural network architectures, thereby improving modeling accuracy and reducing the need for high-quality data [?]. PIML has attracted significant attention for its ability to learn effectively with limited data and has been successfully applied to chemical processes. For example, [7] proposed a physics-informed recurrent neural network (PIRNN) for batch crystallization, and its integration into MPC demonstrated improved closed-loop performance compared with traditional methods.

Transfer learning is another promising approach to reduce data requirements by leveraging knowledge from existing models, and is particularly valuable for process scale-up. Unlike conventional ML methods that train models from scratch, transfer learning reuses knowledge from a source process with a similar configuration and sufficient data to develop a model for a nonlinear target process [16]. Specifically, to develop a transfer learning model for a chemical process (e.g., a chemical reactor), a source process is first selected, which can be the same reactor operating under different conditions, involving different reactions, or even a different type of reactor. Sufficient data collected from the source process enable training a high-accuracy source model, which then provides a strong initialization for the target model, leading to faster convergence, improved prediction accuracy, and reduced data requirements. This approach has been widely adopted across various domains for its ability to address both data scarcity and distribution mismatch. In chemical process applications, transfer learning-based frameworks have shown promising results in both open- and closed-loop settings, as demonstrated in [17, 18] for tackling complex modeling and control problems.

**Concluding remarks**
ML-based MPC offers powerful capabilities for modeling and controlling complex chemical processes, yet its adoption is hindered by challenges such as data scarcity, stability, and computational demands. Approaches like physics-informed machine learning and transfer learning can mitigate these issues by embedding domain knowledge and reusing existing models, thereby improving accuracy and reducing data requirements. Continued research integrating these methods with robust and scalable MPC designs will be key to enabling their practical industrial application.

**Acknowledgments**
Financial support from MOE AcRF Tier 1 FRC Grant (22-5367-A0001), Singapore, and NUS Start-Up Grant (A-0009486-03-00) is gratefully acknowledged.

**References**
[1] B. Bhadriraju, A. Narasingam, and J. S. Kwon. Machine learning-based adaptive model identification of systems: Application to a chemical process. *Chemical Engineering Research and Design*, 152:372–383, 2019.

[2] H. Zhang, V. Juraskova, and F. Duarte. Modelling chemical processes in explicit solvents with machine learning potentials. *Nature Communications*, 15:6114, 2024.

[3] N. Sharma and Y. Liu. A hybrid science-guided machine learning approach for modeling chemical processes: A review. *AIChE Journal*, 68:e17609, 2022.

[4] M. Ławryńczuk and K. Zarzycki. LSTM and GRU type recurrent neural networks in model predictive control: A review. *Neurocomputing*, 632:129712, 2025.

[5] Z. Wu, A. Tran, D. Rincon, and P. D. Christofides. Machine learning-based predictive control of nonlinear processes. Part I: Theory. *AIChE Journal*, 65:e16729, 2019.

[6] Z. Wu, A. Tran, D. Rincon, and P. D. Christofides. Machine-learning-based predictive control of nonlinear processes. Part II: Computational implementation. *AIChE Journal*, 65:e16734, 2019.

[7] G. Wu, W. T. G. Yion, K. L. N. Q. Dang, and Z. Wu. Physics-informed machine learning for MPC: Application to a batch crystallization process. *Chemical Engineering Research and Design*, 192:556–569, 2023.

[8] E. Terzi, F. Bonassi, M. Farina, and R. Scattolini. Learning model predictive control with long short-term memory networks. *International Journal of Robust and Nonlinear Control*, 31:8877–8896, 2021.

[9] Z. Wang, D. Yu, and Z. Wu. Real-time machine-learning-based optimization using input convex long short-term memory network. *Applied Energy*, 377:124472, 2025.

[10] Y. Wang, X. Cui, D. Peters, B. Çıtmacı, A. Alnajdi, C. G. Morales-Guio, and P. D. Christofides. Machine learning-based predictive control of an electrically-heated steam methane reforming process. *Digital Chemical Engineering*, 12:100173, 2024.

[11] M. R. Dobbelaere, P. P. Plehiers, R. Van de Vijver, C. V. Stevens, and K. M. Van Geem. Machine learning in chemical engineering: strengths, weaknesses, opportunities, and threats. *Engineering*, 7:1201–1211, 2021.

[12] Z. Wu, P. D. Christofides, W. Wu, Y. Wang, F. Abdullah, A. Alnajdi, and Y. Kadakia. A tutorial review of machine learning-based model predictive control methods. *Reviews in Chemical Engineering*, 41:359–400, 2025.

[13] G. E. Karniadakis, I. G. Kevrekidis, L. Lu, P. Perdikaris, S. Wang, and L. Yang. Physics-informed machine learning. *Nature Reviews Physics*, 3:422–440, 2021.

[14] C. Meng, S. Griesemer, D. Cao, S. Seo, and Y. Liu. When physics meets machine learning: A survey of physics-informed machine learning. *Machine Learning for Computational Science and Engineering*, 1:20, 2025.

[15] W. Bradley, J. Kim, Z. Kilwein, L. Blakely, M. Eydenberg, J. Jalvin, C. Laird, and F. Boukouvala. Perspectives on the integration between first-principles and data-driven modeling. *Computers & Chemical Engineering*, 166:107898, 2022.

[16] J. Jiang, Y. Shu, J. Wang, and M. Long. Transferability in deep learning: A survey. *arXiv preprint arXiv:2201.05867*, 2022.

[17] M. Xiao, C. Hu, and Z. Wu. Modeling and predictive control of nonlinear processes using transfer learning method. *AIChE Journal*, 69:e18076, 2023.

[18] M. S. Alhajeri, Y. M. Ren, F. Ou, F. Abdullah, and P. D. Christofides. Model predictive control of nonlinear processes using transfer learning-based recurrent neural networks. *Chemical Engineering Research and Design*, 205:1–12, 2024.

# Generative Design of Inorganic Materials

Kedar Hippalgaonkar[1,2,3]

[1]Institute for Functional Intelligent Materials, National University of Singapore, Singapore, Singapore.

[2]Institute of Materials Research and Engineering, A*STAR (Agency for Science), Singapore, Singapore.

[3]School of Materials Science and Engineering, Nanyang Technological University, Singapore, Singapore

## Introduction

Recent advances in AI have sparked excitement for an "AlphaFold moment" in materials discovery – a leap where algorithms can predict novel crystal structures as readily as AlphaFold predicts protein folds. In fact, we are witnessing early signs of this: DeepMind's *Graph Networks for Materials Exploration (GNoME)* model recently predicted 2.2 million presumably new inorganic crystals, including about 380,000 that are theoretically stable[1]. Such deep learning tools dramatically accelerate the exploration of materials by predicting which new compositions might form stable crystals[2]. However, discovering candidates on a computer is only half the battle – *synthesizability* remains a formidable challenge. As the DeepMind team noted, behind each "new, stable" crystal predicted can lie *months of painstaking experimentation* to synthesize it. This gap between computational prediction and laboratory realization underscores a core theme in generative design of materials: proposals must not only be stable on paper but also *feasible to create in practice*.

A key factor is that generative models must respect the fundamental physics and chemistry of crystals. One critical aspect is crystallographic symmetry. Real inorganic crystals almost always exhibit symmetrical arrangements of atoms; indeed, over 98% of known crystals possess symmetry beyond the trivial P1 (simple periodic lattice) group[2]. High symmetry often correlates with desirable properties (for example, non-centrosymmetric crystals can be piezoelectric[3]). Yet many early generative models frequently produced low-symmetry or even P1 structures that are rare in nature[3]. For instance, diffusion-based crystal generators like DiffCSP and CDVAE tend to generate a large fraction (30–40%) of outputs in the P1 space group[2]. Even advanced models that allow conditioning on symmetry, such as MatterGen (2023), recover the intended space group only ~20% of the time for highly symmetric targets[3]. Such physically implausible outputs (e.g. a high rate of low-symmetry crystals) indicate a mismatch with nature's "symmetry preference" and can hinder experimental realization. Clearly, *symmetry preservation* is essential – generative models need to "speak the native language" of crystal structures (space groups and Wyckoff positions)[4] so that their proposals are valid crystal prototypes rather than unphysical atomic arrangements.

Beyond symmetry and thermodynamic stability, a truly useful generative model must grapple with synthesizability – the practical reality of making a predicted crystal. This entails considering factors that go beyond a single DFT formation energy calculation. *Local phase stability* and *disorder* can decisively impact whether a material can be made. For example, a generative model

(WyCryst) recently suggested a new compound $Cu_3SnSe_4$ as a stable and unique crystal structure[2]. DFT computations placed this material on the convex hull (thermodynamically stable) with only ~30 meV/atom to the nearest competitor phase. Yet when researchers attempted to synthesize $Cu_3SnSe_4$ via rapid sintering techniques, the result was not the predicted phase but a related compound ($Cu_5SnSe_7$) dominating the product. In other words, even though $Cu_3SnSe_4$ is theoretically stable, the kinetics and local structural preferences during synthesis led to a different phase. This *experimental validation* case study highlights the importance of kinetic accessibility and phase competition: generative algorithms must account for the possibility that multiple polymorphs or off-stoichiometric phases exist nearby in energy, and that fast or non-equilibrium synthesis routes might be needed to obtain the desired phase. Similar challenges arise with disordered crystals (those with partial occupancies or mixed atomic sites). Many functional materials – from high-entropy alloys to solid solutions – exhibit compositional disorder or defects that are not captured by ideal, ordered crystal models. Generative design is now pushing into this realm by developing representations for disordered materials (e.g. encoding fractional site occupancies as special "null" atoms in the crystal representation)[5,6]. Incorporating disorder is crucial for proposing *synthesizable* materials because perfectly ordered crystals are often hard to achieve; slight off-stoichiometry or site mixing can dictate whether a phase forms or not.

In summary, the generative design of inorganic materials sits at the intersection of materials science and machine learning. The goal is to propose novel compounds with target properties ("materials by design"), but to succeed, these proposals must obey physical constraints and be guided by both thermodynamics and kinetics. The community has identified several key criteria to evaluate progress in this field[4,7]:

- **Validity** – Are generated structures physically plausible (e.g. no atomic overlaps and charge-balanced)? Most models enforce basic validity checks[7], so this is a minimal requirement.
- **Novelty & Uniqueness** – Are the generated crystals truly new and distinct? Models should not merely regurgitate known materials. We often measure the fraction of generated samples that are not present in the training data (unique) and not duplicates of each other[7]. Element-agnostic *template novelty* is used as a stricter measure, counting unique Wyckoff position arrangements irrespective of which elements occupy them[4,7].
- **Stability** – Are the proposed materials thermodynamically stable or at least metastable (within a threshold energy above the convex hull)? A common metric is the percentage of generated structures that have a formation energy within a certain cutoff (e.g. ≤0.08 eV/atom) of the hull, often verified by DFT relaxations[7]. The combined fraction of Stable, Unique, Novel (S.U.N.) structures reflects how many truly new and stable materials a model can discover[7].
- **Symmetry** – Do generated structures exhibit the symmetry distributions seen in real crystals? Important indicators include the fraction of outputs in non-trivial space groups (ideally minimizing the P1 fraction)[7] and how closely the frequency of each space group in the generated set matches those in nature[2]. The ultimate benchmark introduced recently is **S.S.U.N.**, the fraction of generated structures that satisfy *all* key criteria at once: **S**ymmetric (not P1), **S**table, **U**nique, and **N**ovel.

With these metrics, researchers can quantify the performance of generative models in a physically meaningful way. In the following sections, we review two state-of-the-art approaches that have pushed the field forward by explicitly incorporating symmetry into the generative process and addressing the above challenges: a diffusion-based model SymmCD (ICLR 2025) and a transformer-based model WyFormer (ICML 2025). We then discuss how these advances, along with integration of DFT and experimental feedback, are paving the way towards generative design of *synthesizable* inorganic materials.

## SymmCD: Symmetry-Preserving Diffusion Model

One recent approach to enforce crystallographic symmetry in generation is SymmCD (Symmetry-Preserving Crystal Diffusion), introduced by Levy *et al.* (2025)[3]. SymmCD is a diffusion probabilistic model designed to *build crystals by first building one fundamental piece*. It achieves this by decomposing a crystal into two parts during generation: (1) the asymmetric unit – a minimal set of atoms from which the entire crystal can be generated by symmetry operations, and (2) the symmetry operations themselves (the space group and the specific rotations/reflections that map the asymmetric unit onto the full structure). By learning a joint distribution over both components, SymmCD ensures that when it "unfolds" the asymmetric unit with the sampled symmetry operations, the result is a valid symmetric crystal (much like cutting a paper snowflake after folding the paper). Importantly, the symmetry operations are not hard-coded from a database; SymmCD uses a continuous representation (binary symmetry matrices) that generalizes across the 230 space groups, allowing it to share information between less-populated and more-populated symmetries in the training data.

**Competitive performance**: Despite enforcing stricter symmetry constraints, SymmCD manages to perform on par with prior generative models in terms of yielding stable materials, all while greatly reducing the incidence of nonsensical low-symmetry outputs. In their experiments on Materials Project data, the SymmCD model generated crystals that were diverse, valid, and had realistic symmetry distributions, addressing the major shortcomings of earlier methods. The authors report that SymmCD's rate of producing stable unique novel crystals is comparable to the best existing approaches, but with the crucial distinction that *virtually all* of SymmCD's outputs obey crystallographic symmetry by construction. In other words, SymmCD generates "symmetric, stable, and valid" crystals in one go. This is a significant improvement over baseline diffusion models like DiffCSP or the earlier CDVAE, which often needed a post hoc relaxation step that could alter symmetry (e.g. generated structures relaxing to a different, typically higher-symmetry phase or collapsing into known structures)[8]. By eliminating that inconsistency, SymmCD ensures the property one intends to design for (which might depend on symmetry, such as ferroelectricity requiring non-centrosymmetry) remains intact after generation[8]. Another advantage highlighted is efficiency: representing a crystal via its asymmetric unit significantly reduces the degrees of freedom, which makes the diffusion process faster. The SymmCD framework was noted to offer order-of-magnitude improvements in sampling speed and model size compared to prior 3D diffusion models, since it works in a reduced space without having to place every atom individually. This computational efficiency is not just academic – faster generation means one can sample larger candidate pools and integrate the model more easily with iterative workflows (like alternating generation and DFT validation).

In summary, SymmCD demonstrates that incorporating symmetry *explicitly* into generative modeling is a viable strategy that yields physically-plausible candidates. It lays the groundwork for future diffusion models to further incorporate other physical constraints (like partial occupancies or lattice dynamics) while maintaining competitive accuracy in generating stable compounds. The next model we discuss takes a different route to the same problem of symmetry preservation: instead of diffusion, it uses an autoregressive transformer with a symmetry-aware language of crystal building blocks.

## WyFormer: Wyckoff-Token Transformer Model

**WyFormer** (Wyckoff Transformer) is a generative model developed by Kazeev *et al.* (2025) that achieves symmetry preservation through a fundamentally discrete, group-theoretic approach. Rather than working in real-space atomic coordinates, WyFormer represents a crystal as a set of tokens capturing its space group and *Wyckoff positions* – the structured positions of atoms mandated by that space group's symmetry[2,4]. In essence, it treats crystal generation as analogous to composing a sentence, where the "grammar" is given by space group symmetry rules. The first token of a sequence specifies the space group, and subsequent tokens each represent an atomic site by combining an element with a site-symmetry label (e.g. a Wyckoff position like 4a, 8i, etc., which encodes multiplicity and symmetry of that site)[7]. This clever encoding means any sequence of tokens generated by the model corresponds to a symmetry-consistent crystal structure by construction – no post-relaxation can break the symmetry because the fractional coordinates, which are determined after token generation, are placed in symmetric orbits by design[8]. Since over 98% of stable crystals in Materials Project have a unique Wyckoff representation[7], implying that this tokenization does not lose generality: almost every real material can be described uniquely by a sequence of Wyckoff tokens. WyFormer uses a permutation-invariant *autoregressive Transformer* to generate these token sequences one by one, which allows it to model the complex conditional dependencies (for example, certain combinations of Wyckoff positions might be more likely for certain space groups or chemistries)[7]. The transformer is trained on a database of known crystal structures (e.g. the MP-20 dataset of ~45k stable compounds) with a training objective that encourages it to generate valid, novel structures.

**Performance and metrics**: By "speaking the crystal's language" of symmetry, WyFormer achieves outstanding results on the benchmarks discussed earlier. First, it nearly eliminates nonsymmetric outputs – the fraction of generated structures in space group P1 is about 1.44%, essentially at the level of the actual data (≈1.7% in Materials Project). This is a significant improvement over earlier generative models like DiffCSP, MatterGen or FlowMM, which, as noted, yielded P1 in 5–30% of cases. Controlling symmetry also has a ripple effect on stability: high-symmetry structures often correspond to deep minima in the energy landscape, whereas low-symmetry random structures are more likely to be high in energy or unstable. In head-to-head comparisons that included representative models (DiffCSP, an improved DiffCSP++, CrystalFormer, WyCryst VAE, etc.), WyFormer had one of the highest rates of generating stable, unique, novel materials. When 105 generated samples from each model were relaxed and evaluated with DFT, only ~22.4% of WyFormer's samples were confirmed stable (energy below the hull threshold) – a modest fraction, but notably all of these were symmetry-respecting, when combined with DiffCSP++ for final relaxation of atomic coordinates[7]. This combined approach outperformed all baseline models under the same DFT test, indicating a synergy between the fast symmetry-guided proposal of WyFormer and the coordinate-wise refinement of a diffusion model. In fact, the S.S.U.N. metric – the percentage of outputs that are simultaneously symmetric, stable, unique, and novel – reached about >20% for WyFormer. Meanwhile, older diffusion models often showed higher raw stable rates *only by rediscovering known materials*: e.g., DiffCSP had ~19.7% stable rate but also a very high training set overlap (low uniqueness). When it comes to genuinely novel and symmetric outcomes, WyFormer leads the pack.

Another relevant aspect of WyFormer is the diversity of new crystal prototypes it produces. Thanks to the element-agnostic template analysis, researchers found that WyFormer generated new symmetry templates (unique Wyckoff position arrangements not seen in training data)[4]. In one evaluation, it produced ~206 novel Wyckoff combinations out of 1000 samples – significantly more than most baselines, indicating it explores new structural motifs rather than just perturbing known ones. This is important because it means the model is expanding the design space in a meaningful way. The team validated some of these novel structures with extensive DFT calculations (10,000 DFT relaxations were performed across models for a thorough comparison). The result was that WyFormer's symmetric outputs were confirmed to be as thermodynamically stable as the best diffusion models' outputs, while also being more novel in structure[4]. In fact, its overall S.U.N. score matched that of MatterGen (a state-of-the-art diffusion model by Zeni et al., 2023[9]), and its S.S.U.N. was the highest among all methods evaluated. This shows that enforcing symmetry did not sacrifice the ability to find stable compounds – on the contrary, it helped in discovering *new stable compounds* that other models might miss by focusing on random asymmetric arrangements.

Finally, it's worth noting the practical advantage of WyFormer's approach: speed. Generating a crystal as a sequence of tokens is extremely fast with a trained transformer – on the order of *4000 structures per second on a single GPU*[4]. This is orders of magnitude faster than diffusion models, which require iterative refinement over hundreds of timesteps per sample. The high sampling speed makes WyFormer very attractive as a front-end generator that can propose thousands of candidates, which can then be quickly screened (via predictive models or rapid DFT) for stability and properties. In summary, WyFormer represents a major step forward in generative inorganic chemistry, combining group theory and deep learning to produce candidate materials that are *symmetry-correct, diverse, and as computationally accessible as generating text*. The model's success also spurred the release of open datasets and benchmarks – thousands of generated crystals from WyFormer and other models have been shared publicly to encourage reproducibility and further improvements.

Conclusion

Generative design of inorganic materials is emerging as a powerful alternative paradigm for materials discovery – one that complements traditional *Edisonian* trial-and-error with AI-driven creativity. The latest models have shown they can effectively navigate the vast search space of compounds and produce candidates that are not only novel but also meet essential criteria like stability and symmetry. However, as we push this field toward practical impact, several important themes have become clear:

- **Synthesizability is Key**: A hypothetical material is only valuable if it can be made. Generative models must go beyond predicting stable formulas on paper – they need to incorporate considerations of *phase stability across conditions, kinetic barriers, and disorder*. For instance, even with a favorable DFT stability prediction, making a phase experimentally could requires ultra-fast, non-equilibrium processing and still resulted in a different phase due to local energetics. Thus, models should be aware of competing phases and perhaps aim for candidates that are not just globally stable, but also *locally attainable* given real-world synthesis kinetics. Incorporating predictors of synthesis difficulty (e.g. phase diagrams, diffusion coefficients, or known metastable routes) into generative workflows is an open challenge.

- **Beyond Thermodynamics – Include Disorder and Metastability**: Many technologically relevant materials are metastable or entropically stabilized. Generative algorithms traditionally focus on thermodynamic ground states, but the *metastable landscape* is equally important. Future models might generate not just a single crystal structure, but a set of nearby configurations or disordered variants, to suggest pathways for stabilization (for instance, via dopants or strain). Encouragingly, early efforts like *Dis-GEN* are encoding partially occupied sites into generative models[2], allowing suggestions of solid solutions or defect-rich structures. Likewise, the notion of *phase space stability* – stability within a region of a phase diagram rather than at a single composition – could be used to prefer materials that are robust against minor composition shifts or disorder, which often occur in synthesis.

- **Symmetry and Physical Knowledge as Built-in Biases**: A recurring lesson is that encoding domain knowledge (like symmetry constraints) dramatically improves the quality of generated materials. Both SymmCD and WyFormer demonstrate how respecting crystal symmetry yields outputs that are more aligned with experimental reality, without diminishing the diversity of discoveries. We expect that other physical biases – e.g. electroneutrality, common structural motifs, Pauling's rules for ionic crystals, etc. – can be embedded in generative models to further improve realism. The ultimate AI-designed material should come not just with a formula and structure, but with "genes" of physical realism that make it a credible candidate for synthesis.

- **Rapid Iteration with Experiment**: Closing the loop with experimental validation is crucial[2]. Generative models can propose ideas at lightning speed, but those ideas must be tested. Advances in *high-throughput synthesis* and characterization are pivotal in this regard. Techniques like Rapid Joule Heating and Ultrafast Sintering can create materials on timescales of minutes,[10–12] offering a way to quickly attempt the synthesis of AI-generated compounds, including those that may only be accessible via non-equilibrium routes. When such experiments are performed, the feedback (success or failure, phase purity, observed microstructure) can loop back to refine the models. In our $Cu_3SnSe_4$ example, the outcome (formation of a competing phase) provides valuable training data about the role of kinetics and could guide the model to adjust its predictions or suggest a different composition to avoid that competitor. Establishing this iterative AI–experiment cycle will be vital to improve the *predictive power for synthesizability*.

- **Toward Functional Materials by Design**: Finally, while much of the generative work so far targets stability as the primary objective (a necessary starting point), the true promise of generative design lies in discovering materials with targeted *functional properties* – be it a superconductor with a high $T_c$, a battery cathode with fast diffusion, or a photovoltaic with optimal bandgap. Achieving this requires that models be guided by property evaluations, which in turn demands large datasets of materials properties. Currently, such data (e.g. piezoelectric moduli, ionic conductivity, catalytic activity) are far scarcer than crystal structure data. A call has been made to grow databases for functional properties and integrate them into generative model training[2]. Progress in this direction is underway (for example, conditioning generative models on bandgap or elastic moduli, as in some extensions of MatterGen), but much remains to be done. We anticipate that as property data and multi-objective optimization techniques improve, generative AI will move from simply proposing *stable* materials to proposing *useful* materials.

In conclusion, the generative design of inorganic materials is evolving from a curious exercise in AI to a sophisticated, physics-aware discipline. The community is learning that metrics like S.S.U.N. (symmetric,stable, unique, novel) are not just academic – they encapsulate the multi-faceted requirements any newly discovered material must meet[2]. Models like WyFormer and SymmCD have set new benchmarks by satisfying these requirements more completely than ever before. Yet, the *"best-laid schemes"* of computational and AI-driven algorithms can *"oft go awry"* (to borrow from poet Robert Burns) when confronted with the reality of the laboratory[2]. The solution is to further tighten the integration of theoretical generative models with experimental and physics-based feedback. By doing so – incorporating symmetry, disorder, kinetic factors, and property targets – we move closer to AI systems that can not only dream up new materials on a computer but also help bring them into existence in the real world. The ultimate success will be measured by discoveries of novel, synthesizable materials with extraordinary properties, discovered in silico and realized in situ, thus truly achieving the long-held dream of materials-by-design.

## Acknowledgements

K.H acknowledges the use of ChatGPT 5 from OpenAI to generate draft language. This research is supported by the National Research Foundation, Singapore under its AI Singapore Programme (AISG Award No: AISG3-RP-2022-028). K.H would also like to acknowledge funding from the MAT-GDT Program at A*STAR via the AME Programmatic Fund by the Agency for Science, Technology and Research under Grant No. M24N4b0034

## References:

(1) Merchant, A.; Cubuk, E. D. Millions of New Materials Discovered with Deep Learning. *Google DeepMind*. November 29, 2023. https://deepmind.google/discover/blog/millions-of-new-materials-discovered-with-deep-learning/.
(2) Hippalgaonkar, K. Generative Design of Inorganic Materials; AI4X Conference Presentation: Singapore, 2025.
(3) Levy, D.; Panigrahi, S. S.; Kaba, S.-O.; Zhu, Q.; Lee, K. L. K.; Galkin, M.; Miret, S.; Ravanbakhsh, S. SymmCD: Symmetry-Preserving Crystal Generation with Diffusion Models. *ArXiv Prepr. ArXiv250203638* **2025**.
(4) Hippalgaonkar, K. *Symmetry by Design: Wyckoff Transformer*. https://medium.com/@kedar.h/symmetry-by-design-wyckoff-transformer-6caa2399d801 (accessed 2025-03-11).
(5) Petersen, M. H.; Zhu, R.; Dai, H.; Aggarwal, S.; Wei, N.; Chen, A. P.; Bhowmik, A.; Lastra, J. M. G.; Hippalgaonkar, K. Dis-GEN: Disordered Crystal Structure Generation. *ArXiv Prepr. ArXiv250718275* **2025**.

(6) Kazeev, N.; Al-Maeeni, A. R.; Romanov, I.; Faleev, M.; Lukin, R.; Tormasov, A.; Castro Neto, A.; Novoselov, K. S.; Huang, P.; Ustyuzhanin, A. Sparse Representation for Machine Learning the Properties of Defects in 2D Materials. *Npj Comput. Mater.* **2023**, *9* (1), 113.

(7) Kazeev, N.; Nong, W.; Romanov, I.; Zhu, R.; Ustyuzhanin, A.; Yamazaki, S.; Hippalgaonkar, K. Wyckoff Transformer: Generation of Symmetric Crystals. *ArXiv Prepr. ArXiv250302407* **2025**.

(8) Zhu, R.; Nong, W.; Yamazaki, S.; Hippalgaonkar, K. WyCryst: Wyckoff Inorganic Crystal Generator Framework. *Matter* **2024**, *7* (10), 3469–3488.

(9) Zeni, C.; Pinsler, R.; Zügner, D.; Fowler, A.; Horton, M.; Fu, X.; Wang, Z.; Shysheya, A.; Crabbé, J.; Ueda, S.; others. A Generative Model for Inorganic Materials Design. *Nature* **2025**, *639* (8055), 624–632.

(10) Zhang, C.; Recatala-Gomez, J.; Aabdin, Z.; Jiang, Y.; Jiang, L.; Tan, S. Y.; Liu, H.; Qian, Y.; Lee, C. J. J.; Hachmioune, S.; others. Direct Joule-Heated Non-Equilibrium Synthesis Enables High Performing Thermoelectrics. *ArXiv Prepr. ArXiv250604447* **2025**.

(11) Wang, C.; Ping, W.; Bai, Q.; Cui, H.; Hensleigh, R.; Wang, R.; Brozena, A. H.; Xu, Z.; Dai, J.; Pei, Y.; others. A General Method to Synthesize and Sinter Bulk Ceramics in Seconds. *Science* **2020**, *368* (6490), 521–526.

(12) Choi, C. H. 'William'; Shin, J.; Eddy, L.; Granja, V.; Wyss, K. M.; Damasceno, B.; Guo, H.; Gao, G.; Zhao, Y.; Higgs III, C. F.; others. Flash-within-Flash Synthesis of Gram-Scale Solid-State Materials. *Nat. Chem.* **2024**, *16* (11), 1831–1837.

# AI as a research companion: Strengths, limits, and lessons from practice

Limsoon Wong

National University of Singapore, School of Computing, 13 Computing Drive, Singapore 117417. Email: dcswls@nus.edu.sg

**Abstract**

AI can store vast knowledge and manipulate formulas instantly, but it does not reason like a human scientist. Across mathematics, biology, and quantitative modeling, it performs mechanical reasoning well yet misapplies predictions and overlooks critical assumptions—making it a useful but fallible research companion, especially for interdisciplinary scientists who can leverage its breadth with careful oversight.

## 1. AI as a companion in scientific exploration

Artificial intelligence (AI) is becoming a familiar presence in scientific research. Beyond speeding up calculations or handling routine tasks, AI can act like a collaborator by recalling relevant knowledge, performing formulaic calculations, and providing summaries of large bodies of literature. In some cases, it can suggest plausible next steps or organize information in helpful ways.

However, AI is not an independent expert. It performs well at **mechanical formula manipulations and one-step reasoning** but struggles with conceptual leaps, integrating knowledge across fields, or drawing correct inferences when reasoning depends on combining information from multiple areas. In that sense, it resembles an "average scientist" who has read widely, can handle routine calculations, and recall facts accurately, but does not reason consistently or make truly original insights.

This perspective presents three stories drawn from my interactions with ChatGPT that illustrate these strengths and limitations. Each story shows a different facet of AI as a research companion: in mathematics, it can apply known facts to solve problems; in biology, it recalls information but misapplies it in predictive contexts; in reasoning with quantitative models, it manipulates formulas correctly but overlooks crucial real-world assumptions. By examining these examples, we can see where AI may support scientific work, where it can mislead, and how researchers might use it effectively in interdisciplinary and transdisciplinary settings.

## 2. Geometry and one-step reasoning

Consider this homework problem. A parallelogram PQRS has base PS equal to 5.5 cm, a height from PS to the opposite side QR equal to 3 cm, and a perpendicular from PQ to corner R equal to 5 cm. The task is to compute the perimeter of PQRS.

Two years ago, ChatGPT 3.0 gave the wrong answer. It noticed that the perpendicular from PQ to corner R could serve as the **height with respect to PQ**—that is, the height if PQ is viewed as the base—but it did not realize that the area calculated using PS and height 3 cm must equal the area calculated using PQ and height 5 cm. In other words, it failed to recognize that the **area of a parallelogram is preserved when the shape is rotated to view a different side as the base**. Without this insight, it could not solve for PQ and gave results that were entirely wrong.

ChatGPT 4.0 made the crucial connection by using this additional geometric fact. It understood that the area remains the same whether computed from base PS or PQ, set up the equation PS × 3 = PQ × 5, solved for PQ = 3.3 cm, and then computed the perimeter as 17.6 cm.

This shows that the AI's improvement came from acquiring an extra fact—the preservation of area under rotation—which enabled it to perform a **one-step reasoning** it could not do before. It did not demonstrate conceptual leaps or general integration across mathematics; its reasoning still relies on mechanical manipulations and applying known facts.

## 3. Biology and missed connections

Next, I tested ChatGPT on a biology problem that involves predicting **translation initiation sites (TIS)** in mRNA. The TIS is the location where the ribosome begins translating the mRNA into a protein, usually starting at an AUG codon, but which AUG is used depends on surrounding sequences and other signals.

I asked the same questions to ChatGPT 3.0 two years ago and ChatGPT 4.0 recently. Both recalled biological facts correctly but struggled to reason about predictions.

The first question concerned the nucleotide at **position minus three**, which means the nucleotide located three positions before the AUG start codon. Both versions correctly cited the Kozak consensus sequence (Kozak, 1984), which states that having a purine (A or G) at this position increases the likelihood that the ribosome will initiate translation at that AUG.

The second question asked whether the number of in-frame upstream AUG codons matters. Both recalled the ribosome scanning model, noting that ribosomes usually initiate at the **first AUG in a good context** (Kozak, 2002). However, both drew the wrong conclusion, claiming that more upstream AUGs make prediction harder. In reality, knowing the number of upstream AUGs makes prediction easier. Consider a simplified model where each AUG has probability p of being a TIS, the (n+1)th AUG can only succeed if all n upstream AUGs fail, making its probability at most $p \times (1 - p)^n$. Therefore, upstream AUG count is a strong predictive feature. Both versions got the biology right but misapplied it to prediction.

The third question concerned nearby in-frame downstream stop codons TAA, TAG, and TGA. Both versions of ChatGPT **correctly explained the reasoning**. They noted that if translation starts at that AUG, the ribosome would encounter the stop codon very soon after, producing a very short and likely nonfunctional peptide. Therefore, an AUG closely followed by a stop codon is unlikely to be a real TIS. This shows that the AI can recall biological mechanisms and explain them logically, even though it struggles on other predictive questions like upstream AUGs or codon bias.

The fourth question asked about downstream codons such as CTG, GAC, GAG, and GCC. Both identified these as normal coding codons but said they were irrelevant to TIS prediction. Both missed the cross-domain reasoning. In noncoding regions, codon frequencies are roughly uniform. In coding regions, evolution favors some codons over others because certain codons match more abundant tRNAs, are translated more efficiently, or reduce errors (Quax et al., 2015). This results in a **codon usage bias**, where some codons appear more frequently than others. As a result, downstream codon-frequency patterns can help distinguish real TIS in coding regions from false AUGs in noncoding regions.

Unlike in mathematics, AI showed no progress from version 3.0 to version 4.0. Both made the same mistakes, demonstrating that ChatGPT struggles to integrate knowledge across domains. It recalled facts correctly but failed to apply them in predictive contexts.

This story shows that AI has broad knowledge of biology and can reason correctly in some areas, such as the minus three nucleotide and stop codons, but it misses inferences when biology must be combined with predictive logic, such as upstream

AUGs and codon bias. There was no improvement across versions, as the same reasoning failures persisted.

## 4. Quantitative models and overlooked assumptions

Suppose a classifier has 90 percent sensitivity and 90 percent precision on a test set with 80 percent positives and 20 percent negatives. What happens in the real world, where 20 percent are positives and 80 percent are negatives?

ChatGPT 3.0 mixed up definitions and produced contradictory reasoning. ChatGPT 4.0 reasoned correctly that **sensitivity**, which measures the true positive rate, and **specificity**, which measures the true negative rate, remain the same because they are mathematically independent of prevalence. **Precision**, however, depends on prevalence; with more negatives, false positives rise relative to true positives, reducing precision (Altman & Bland, 1994). ChatGPT 4.0 calculated the new precision using algebra.

The subtle point that AI missed is that **applying these formulas correctly requires the assumption that the distribution of relevant covariates—such as age, sex, ancestry, or disease stage—remains similar between the test set and the real world** (Pepe, 2003). Sensitivity and specificity formulas are mathematically correct regardless of prevalence, but if these covariate distributions shift, the classifier's behavior may change in practice. ChatGPT 4.0 did not highlight this assumption or warn users to check it, focusing only on the formulas themselves.

This story shows that AI can handle quantitative reasoning and algebraic adjustments, but it **fails to alert users to critical real-world conditions that determine whether predictions will remain valid**.

## 5. Reflections on AI as a fallible but useful partner

These three stories illustrate both the promise and the limits of AI as a scientific partner. In mathematics, AI can apply known facts to solve problems using one-step reasoning. In biology, it misapplies knowledge and cannot reliably connect facts with predictive reasoning. In reasoning with quantitative models, it manipulates formulas correctly but overlooks key assumptions.

AI resembles an "average scientist." It is knowledgeable and performs well at **mechanical formula manipulations and one-step reasoning**, but it struggles to make conceptual leaps or integrate knowledge across domains. It sometimes produces useful ideas but also makes mistakes, misapplies knowledge, or draws incorrect conclusions. It can recall background knowledge and provide relevant information, but it requires human oversight. Its value lies in expanding our cognitive reach by surfacing ideas and hypotheses that humans can then test, refine, or correct.

I see a particular role for AI as a research partner for scientists working in interdisciplinary or transdisciplinary areas. Such scientists are often experts in one field but have only fragmented knowledge in others. They can leverage AI's broad knowledge base and "patience" to gather relevant background information, summarize concepts from other fields, and highlight points that might otherwise take a long time to assemble. A reflective scientist can then use this information to reason across disciplines, make conceptual links, and explore implications, while relying on their own expertise to evaluate and integrate the insights.

AI is not yet a reliable expert, but it is already a useful collaborator. With careful guidance, it may continue to mature as a research partner. For now, it is best seen as a companion whose strengths and weaknesses we must understand to make the most of its potential in scientific discovery.

**References**

D. G. Altman, J. M. Bland. Diagnostic tests 2: Predictive values. *BMJ*, 309(6947):102, 1994.

M. Kozak. Compilation and analysis of sequences upstream from the translational start site in eukaryotic mRNAs. *Nucleic Acids Research*, 12(2):857-872, 1984.

M. Kozak. Pushing the limits of the scanning mechanism for initiation of translation. *Gene*, 299(1):1-34, 2002.

M. S. Pepe. *The Statistical Evaluation of Medical Tests for Classification and Prediction*. Oxford University Press, 2003.

T. E. F. Quax, N. J. Claassens, D. Soell, J. van der Oost. Codon bias as a means to fine-tune gene expression. *Molecular Cell*, 59(2):149-161, 2015.

# Automated workflows for advanced materials properties

Lorenzo Bastonero[1] and Nicola Marzari[1,2,3,4,5,*]

[1] U Bremen Excellence Chair, Bremen Centre for Computational Materials Science, and MAPEX Center for Materials and Processes, University of Bremen, 28359 Bremen, Germany
[2] Theory and Simulation of Materials (THEOS), École Polytechnique Fédérale de Lausanne, 1015 Lausanne, Switzerland
[3] National Centre for Computational Design and Discovery of Novel Materials (MARVEL), École Polytechnique Fédérale de Lausanne, 1015 Lausanne, Switzerland
[4] PSI Center for Scientific Computing, Theory and Data (CSD), Paul Scherrer Institut, 5232 Villigen PSI, Switzerland
[5] Cavendish Laboratory (TCM), University of Cambridge, Cambridge CB3 0US, United Kingdom
* Author to whom any correspondence should be addressed.

**E-mail:** nicola.marzari@epfl.ch

## 1  Status

Computing advanced material properties from first-principles has been for a long time the domain of highly specialized computational scientists [1]. In addition to a thorough understanding of the underlying theory and its implementation, persistent numerical pitfalls and runtime failures in the codes themselves demanded detailed and code-specific expertise.

In the last two decades, the advent of high-throughput materials discovery pushed the development of complex computational infrastructures, such as AiiDA [2, 3], AFLOW [4, 5], Atomate [6, 7], ASE [8], to name a few, able to orchestrate and even persist complex computational workflows. These workflows can be run on remote high-performance resources, monitoring their state, retrieving and storing the results following FAIR (Findable, Accessible, Interoperable, and Reusable) principles [9], and taking autonomous actions such as recovering from known errors. Thanks to these advances, automated calculations can now been performed to obtain even complex properties - recent examples include infrared and Raman spectra [10], muon spin spectroscopies [11], and advanced electronic-structure methods [12, 13], where open-source graphical user interfaces allow to make these capabilities accessible to non-experts [14]. Not only is this greatly beneficial to collaborations between experimental and computational scientists, but it also simplifies routine calculations and provides fast access to advanced capabilities to the computational community for ever more complex tasks.

A paradigmatic example of the complex capabilities needed and the benefit of automated workflows can be seen in the training of machine-learning interatomic potentials (MLIPs): accurate training requires iterative procedures that combine exploration of materials potential energy surfaces, efficient and precise selection of relevant atomic configuration, extensive first-principles calculations, and refinement of the MLIP. Recently, complex "active learning" strategies have been implemented [15, 16], providing another important step forward to a fully automated approach to computational materials science.

One can now fully expect that automated workflows will drive ambitious scientific challenges, such the green technological transition on Earth, which demands the accelerated understanding, discovery, and design of complex materials properties, through a multidisciplinary and multiscale approach. A notable example is the optimization of carbon-free technologies for metal electrowinning on Mars, a case study which is representative of a scarcity-driven paradigm for the low-energy production of materials, and whose success would also transfer to major technological advancement on Earth [17]. This goal requires interoperable, multiscale workflows: accurate electronic ground states of transition-metal compounds, finite-temperature anharmonic free energies, corrosion phase-diagram, and transport properties. In Figure 1 we illustrate the different building blocks that will support this research in the near future. All these tasks can be performed by complex calculations managed by independent advanced workflows, automatically interconnected to one another.

## 2  Current and future challenges

Advanced workflow engines have transformed computational materials research, but many pipelines remain tightly coupled to specific atomistic codes, which hinges on the reusability and interoperability of existing workflows. Recently, a common workflow definition for computational materials science has been proposed for Python-based workflows [18, 19], leaving as next challenge not merely its technical standardization, but its community adoption at scale: shared semantics for tasks and datasets, stable public APIs for plugins that wrap diverse simulation codes, and disciplined metadata schemas that satisfy the FAIR principles. Broad adoption will depend on lowering the entry barrier for non-developers, providing robust reference implementations, and ensuring that abstractions remain expressive enough to cover domain-specific edge cases.

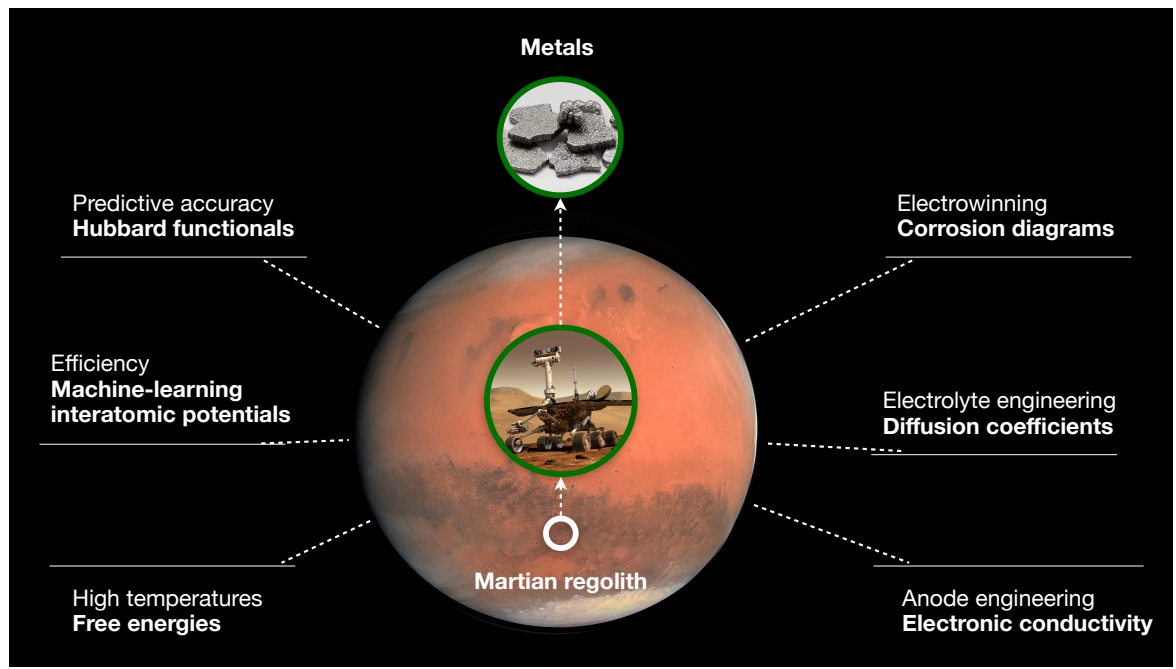

**Figure 1.** Artistic rendition of the automated building blocks required to optimize electrowinning technologies to extract metals from the Martian regolith. (Credits: Mars: ESA & MPS for OSIRIS Team. Rover: NASA/JPL/Cornell University. Iron chips: Alchemist-hp.)

As workflows become more complex and more heterogeneous, robust and reusable protocols are required to keep accuracy under control [20]. For instance, when training machine-learning potentials the first-principles datasets of energies, forces, and stresses, must be computed with consistent and tight computational parameters in order to avoid producing noisy potential energy surfaces. This not only applies to standard MLIPs, but also to future datasets which will provide more advanced materials properties, such as higher-order responses, including Born effective charges, dielectric and Raman tensors, as well as interatomic force constants matrices.

Verification across atomistic codes is an equally pressing challenge that has greatly advanced thanks to the development of common, universal workflows: having a single workflow manager driving multiple codes that implement the same physical capability — e.g., the equation of state with different DFT engines [21, 22] — enables the cross-validations that are essential for demonstrating numerical accuracy in computed properties. While these efforts have been accomplished for ground-state DFT [21, 22], the frontier now lies with more advanced properties, such as phonons, spectra, and transport.

Finally, validation and the methodological accuracy itself are the ultimate goal for predictive accuracy, and the automated calculation of the many properties required by contemporary materials science should not reduce the effort into fundamental theoretical developments. As much of the future research will be driven by artificial intelligence, it is important to remember that machine-learning models cannot be more accurate than the reference they are trained on.

## 3    Advances in science and technology to meet challenges

Large language models (LLMs) offer a forward-looking approach to respond to today's challenges and to advance the tailoring of automated workflows. In order to correctly train potentials, a key step forward would certainly be the widespread adoption of standardized computational input parameters after careful protocols have been devised. All the calculations performed by thousands of researchers should be joined to create community suites of datasets, spanning crystals, low-dimensional materials, and disordered phases, covering a broad range of properties, such as vibrational, electronic, and transport properties. Each suite should include parameterized protocols with target accuracies. Moreover, beyond the code generation, domain-specialized LLMs can translate scientists written-prompts into working workflows, and propose corrective actions with minimal human intervention. In the longer term, specific tuned LLMs that can interrogate databases, launch sandboxed calculations, and interpret results could become reliable co-pilots for streamlining advanced computational capabilities.

## 4  Concluding remarks

Automated workflows will be at the core of the AI revolution: they represent externalizable capababilities that can be driven reliably and robustly by universal APIs, driving both the streamline application and use of advanced computational capabilities and the widespread deployment of machine-learning approaches for materials design and discovery.

**Acknowledgments**

We acknowledge financial support by the Deutsche Forschungsgemeinschaft (DFG) under Germany's Excellence Strategy (EXC 2077, No. 390741603, University Allowance, University of Bremen) and Lucio Colombi Ciacchi, the host of the "U Bremen Excellence Chair Program". NM acknowledges financial support by the NCCR MARVEL, a National Centre of Competence in Research, funded by the Swiss National Science Foundation (grant number 205602).

**References**

[1] N. Marzari, A. Ferretti, and C. Wolverton. Electronic-structure methods for materials design. *Nature Materials*, 20(6):736–749, 2021. ISSN 1476-1122. doi: 10.1038/s41563-021-01013-3. URL https://doi.org/10.1038/s41563-021-01013-3.

[2] Sebastiaan P. Huber, Spyros Zoupanos, Martin Uhrin, Leopold Talirz, Leonid Kahle, Rico Häuselmann, Dominik Gresch, Tiziano Müller, Aliaksandr V. Yakutovich, Casper W. Andersen, Francisco F. Ramirez, Carl S. Adorf, Fernando Gargiulo, Snehal Kumbhar, Elsa Passaro, Conrad Johnston, Andrius Merkys, Andrea Cepellotti, Nicolas Mounet, Nicola Marzari, Boris Kozinsky, and Giovanni Pizzi. AiiDA 1.0, a scalable computational infrastructure for automated reproducible workflows and data provenance. *Scientific Data*, 7 (300):300, sep 2020. doi: 10.1038/s41597-020-00638-4.

[3] Martin Uhrin, Sebastiaan P. Huber, Jusong Yu, Nicola Marzari, and Giovanni Pizzi. Workflows in AiiDA: Engineering a high-throughput, event-based engine for robust and modular computational workflows. *Computational Materials Science*, 187:110086, feb 2021. doi: 10.1016/j.commatsci.2020.110086.

[4] Camilo E. Calderon, Jose J. Plata, Cormac Toher, Corey Oses, Ohad Levy, Marco Fornari, Amir Natan, Michael J. Mehl, Gus Hart, Marco Buongiorno Nardelli, and Stefano Curtarolo. The aflow standard for high-throughput materials science calculations. *Computational Materials Science*, 108:233–238, October 2015. ISSN 0927-0256. doi: 10.1016/j.commatsci.2015.07.019.

[5] Corey Oses, Marco Esters, David Hicks, Simon Divilov, Hagen Eckert, Rico Friedrich, Michael J. Mehl, Andriy Smolyanyuk, Xiomara Campilongo, Axel van de Walle, Jan Schroers, A. Gilad Kusne, Ichiro Takeuchi, Eva Zurek, Marco Buongiorno Nardelli, Marco Fornari, Yoav Lederer, Ohad Levy, Cormac Toher, and Stefano Curtarolo. aflow++: A C++ framework for autonomous materials design. *Computational Materials Science*, 217:111889, January 2023. ISSN 0927-0256. doi: 10.1016/j.commatsci.2022.111889. URL https://www.sciencedirect.com/science/article/pii/S0927025622006000.

[6] Kiran Mathew, Joseph H. Montoya, Alireza Faghaninia, Shyam Dwarakanath, Muratahan Aykol, Hanmei Tang, Iek-heng Chu, Tess Smidt, Brandon Bocklund, Matthew Horton, John Dagdelen, Brandon Wood, Zi-Kui Liu, Jeffrey Neaton, Shyue Ping Ong, Kristin Persson, and Anubhav Jain. Atomate: A high-level interface to generate, execute, and analyze computational materials science workflows. *Computational Materials Science*, 139:140–152, November 2017. ISSN 0927-0256. doi: 10.1016/j.commatsci.2017.07.030. URL https://www.sciencedirect.com/science/article/pii/S0927025617303919.

[7] Alex M. Ganose, Hrushikesh Sahasrabuddhe, Mark Asta, Kevin Beck, Tathagata Biswas, Alexander Bonkowski, Joana Bustamante, Xin Chen, Yuan Chiang, Daryl C. Chrzan, Jacob Clary, Orion A. Cohen, Christina Ertural, Max C. Gallant, Janine George, Sophie Gerits,

Rhys E. A. Goodall, Rishabh D. Guha, Geoffroy Hautier, Matthew Horton, T. J. Inizan, Aaron D. Kaplan, Ryan S. Kingsbury, Matthew C. Kuner, Bryant Li, Xavier Linn, Matthew J. McDermott, Rohith Srinivaas Mohanakrishnan, Aakash N. Naik, Jeffrey B. Neaton, Shehan M. Parmar, Kristin A. Persson, Guido Petretto, Thomas A. R. Purcell, Francesco Ricci, Benjamin Rich, Janosh Riebesell, Gian-Marco Rignanese, Andrew S. Rosen, Matthias Scheffler, Jonathan Schmidt, Jimmy-Xuan Shen, Andrei Sobolev, Ravishankar Sundararaman, Cooper Tezak, Victor Trinquet, Joel B. Varley, Derek Vigil-Fowler, Duo Wang, David Waroquiers, Mingjian Wen, Han Yang, Hui Zheng, Jiongzhi Zheng, Zhuoying Zhu, and Anubhav Jain. Atomate2: modular workflows for materials science. *Digital Discovery*, 4(7): 1944–1973, 2025. ISSN 2635-098X. doi: 10.1039/d5dd00019j.

[8] Ask Hjorth Larsen, Jens Jørgen Mortensen, Jakob Blomqvist, Ivano E Castelli, Rune Christensen, Marcin Dułak, Jesper Friis, Michael N Groves, Bjørk Hammer, Cory Hargus, Eric D Hermes, Paul C Jennings, Peter Bjerre Jensen, James Kermode, John R Kitchin, Esben Leonhard Kolsbjerg, Joseph Kubal, Kristen Kaasbjerg, Steen Lysgaard, Jón Bergmann Maronsson, Tristan Maxson, Thomas Olsen, Lars Pastewka, Andrew Peterson, Carsten Rostgaard, Jakob Schiøtz, Ole Schütt, Mikkel Strange, Kristian S Thygesen, Tejs Vegge, Lasse Vilhelmsen, Michael Walter, Zhenhua Zeng, and Karsten W Jacobsen. The atomic simulation environment – a Python library for working with atoms. *Journal of Physics: Condensed Matter*, 29(27):273002, June 2017. ISSN 1361-648X. doi: 10.1088/1361-648x/aa680e.

[9] Mark D. Wilkinson et al. The FAIR guiding principles for scientific data management and stewardship. *Scientific Data*, 3(160018):160018, mar 2016. doi: 10.1038/sdata.2016.18.

[10] Lorenzo Bastonero and Nicola Marzari. Automated all-functionals infrared and Raman spectra. *npj Computational Materials*, 10(1):1–12, March 2024. ISSN 2057-3960. doi: 10.1038/s41524-024-01236-3. URL https://www.nature.com/articles/s41524-024-01236-3.

[11] Ifeanyi J. Onuorah, Miki Bonacci, Muhammad M. Isah, Marcello Mazzani, Roberto De Renzi, Giovanni Pizzi, and Pietro Bonfà. Automated computational workflows for muon spin spectroscopy. *Digital Discovery*, 4(2):523–538, 2025. ISSN 2635-098X. doi: 10.1039/d4dd00314d.

[12] Junfeng Qiao, Giovanni Pizzi, and Nicola Marzari. Automated mixing of maximally localized Wannier functions into target manifolds. *npj Computational Materials*, 9(1):1–9, October 2023. ISSN 2057-3960. doi: 10.1038/s41524-023-01147-9. URL https://www.nature.com/articles/s41524-023-01147-9.

[13] Lorenzo Bastonero, Cristiano Malica, Eric Macke, Marnik Bercx, Sebastiaan Huber, Iurii Timrov, and Nicola Marzari. First-principles hubbard parameters with automated and reproducible workflows. *npj Computational Materials*, 11(1), June 2025. ISSN 2057-3960. doi: 10.1038/s41524-025-01685-4.

[14] Xing Wang, Edan Bainglass, Miki Bonacci, Andres Ortega-Guerrero, Lorenzo Bastonero, Marnik Bercx, Pietro Bonfà, Roberto De Renzi, Dou Du, Peter N. O. Gillespie, Michael A. Hernández-Bertrán, Daniel Hollas, Sebastiaan P. Huber, Elisa Molinari, Ifeanyi J. Onuorah, Nataliya Paulish, Deborah Prezzi, Junfeng Qiao, Timo Reents, Christopher J. Sewell, Iurii Timrov, Aliaksandr V. Yakutovich, Jusong Yu, Nicola Marzari, Carlo A. Pignedoli, and Giovanni Pizzi. Making atomistic materials calculations accessible with the aiidalab quantum espresso app, 2025.

[15] Yuanbin Liu, Joe D. Morrow, Christina Ertural, Natascia L. Fragapane, John L. A. Gardner, Aakash A. Naik, Yuxing Zhou, Janine George, and Volker L. Deringer. An automated framework for exploring and learning potential-energy surfaces. *Nature Communications*, 16(1), August 2025. ISSN 2041-1723. doi: 10.1038/s41467-025-62510-6.

[16] Davide Bidoggia, Nataliia Manko, Maria Peressi, and Antimo Marrazzo. Automated training of neural-network interatomic potentials, 2025.

[17] The Martian Mindset. 2025. URL https://www.uni-bremen.de/en/the-martian-mindset.

[18] Jan Janssen, Janine George, Julian Geiger, Marnik Bercx, Xing Wang, Christina Ertural, Jörg Schaarschmidt, Alex M. Ganose, Giovanni Pizzi, Tilmann Hickel, and Jörg Neugebauer. A python workflow definition for computational materials design. *Digital Discovery*, 2025. ISSN 2635-098X. doi: 10.1039/d5dd00231a.

[19] Frédéric Suter, Tainã Coleman, İlkay Altintaş, Rosa M. Badia, Bartosz Balis, Kyle Chard, Iacopo Colonnelli, Ewa Deelman, Paolo Di Tommaso, Thomas Fahringer, Carole Goble, Shantenu Jha, Daniel S. Katz, Johannes Köster, Ulf Leser, Kshitij Mehta, Hilary Oliver, J.-Luc Peterson, Giovanni Pizzi, Loïc Pottier, Raül Sirvent, Eric Suchyta, Douglas Thain, Sean R. Wilkinson, Justin M. Wozniak, and Rafael Ferreira da Silva. A terminology for scientific workflow systems. *Future Generation Computer Systems*, 174:107974, January 2026. ISSN 0167-739X. doi: 10.1016/j.future.2025.107974.

[20] Gabriel de Miranda Nascimento, Flaviano José dos Santos, Marnik Bercx, Davide Grassano, Giovanni Pizzi, and Nicola Marzari. Accurate and efficient protocols for high-throughput first-principles materials simulations, 2025.

[21] Sebastiaan P. Huber, Emanuele Bosoni, Marnik Bercx, Jens Bröder, Augustin Degomme, Vladimir Dikan, Kristjan Eimre, Espen Flage-Larsen, Alberto Garcia, Luigi Genovese, Dominik Gresch, Conrad Johnston, Guido Petretto, Samuel Poncé, Gian-Marco Rignanese, Christopher J. Sewell, Berend Smit, Vasily Tseplyaev, Martin Uhrin, Daniel Wortmann, Aliaksandr V. Yakutovich, Austin Zadoks, Pezhman Zarabadi-Poor, Bonan Zhu, Nicola Marzari, and Giovanni Pizzi. Common workflows for computing material properties using different quantum engines. *npj Computational Materials*, 7(136):136, aug 2021. doi: 10.1038/s41524-021-00594-6.

[22] Emanuele Bosoni, Louis Beal, Marnik Bercx, Peter Blaha, Stefan Blügel, Jens Bröder, Martin Callsen, Stefaan Cottenier, Augustin Degomme, Vladimir Dikan, Kristjan Eimre, Espen Flage-Larsen, Marco Fornari, Alberto Garcia, Luigi Genovese, Matteo Giantomassi, Sebastiaan P. Huber, Henning Janssen, Georg Kastlunger, Matthias Krack, Georg Kresse, Thomas D. Kühne, Kurt Lejaeghere, Georg K. H. Madsen, Martijn Marsman, Nicola Marzari, Gregor Michalicek, Hossein Mirhosseini, Tiziano M. A. Müller, Guido Petretto, Chris J. Pickard, Samuel Poncé, Gian-Marco Rignanese, Oleg Rubel, Thomas Ruh, Michael Sluydts, Danny E. P. Vanpoucke, Sudarshan Vijay, Michael Wolloch, Daniel Wortmann, Aliaksandr V. Yakutovich, Jusong Yu, Austin Zadoks, Bonan Zhu, and Giovanni Pizzi. How to verify the precision of density-functional-theory implementations via reproducible and universal workflows. *Nature Reviews Physics*, 6(1):45–58, November 2023. ISSN 2522-5820. doi: 10.1038/s42254-023-00655-3.

# Artificial Intelligence driven computational workflows as accelerators of innovative advanced materials design and engineering

Dorye Luis Esteras Córdoba[1], Andrei Tomut[1], Alba Quiñones Andrade[1], José-Hugo Garcia[1] and Stephan Roche [1,2]

[1] Catalan Institute of Nanoscience and Nanotechnology (ICN2), CSIC and BIST, Campus UAB, Bellaterra, 08193 Barcelona, Spain
[2] ICREA, Institució Catalana de Recerca i Estudis Avançats, 08070 Barcelona, Spain

E-mail: stephan.roche@apeironai.eu

**Status**

The discovery of novel materials has always been a fundamental driver of technological innovation, shaping human progress from the earliest tools to today's strategic sectors such as energy conversion, storage, and electronics. Computational methods have historically played a central role in this predictive revolution. Early electronic-structure approaches, such as Hartree–Fock, laid the groundwork, though they were computationally prohibitive for realistic systems. The advent of Density Functional Theory (DFT) brought a practical balance between accuracy and efficiency, unlocking the study of increasingly complex solids, surfaces, and interfaces. Over subsequent decades, methodological innovations, including linear-scaling techniques and multiscale modelling, further expanded the range and complexity of systems accessible to simulation, elevating computation to a critical complement of experimental research.

The combination of DFT and deep learning has started to reshape computational materials science [1-3]. Due to the computationally costly nature of DFT calculations, many researchers now turn to machine learning as a means to overcome the demands in terms of length and time scales of the phenomena in DFT calculations [4]. Recent frameworks, such as DeepH, demonstrate how deep neural networks can accurately and efficiently represent DFT Hamiltonians, thereby reducing computational cost. This balance between efficiency and precision enables the study of defects, disorder, interfaces, and heterostructures that were previously too demanding [3]. However, one major obstacle remains: training neural networks typically demands large amounts of high-quality data [3,5].

Real-world materials exhibit phenomena across multiple spatial and temporal scales, from quantum electron correlations and lattice dynamics to mesoscale defects and microstructural evolution. Capturing this complexity requires more than isolated methods: brute-force scaling of established approaches often leads to prohibitive computational costs, while the growing demand for high-throughput screening and vast chemical space exploration demands efficient, scalable strategies that transcend traditional single-method paradigms.

A central response is the construction of integrated computational workflows for materials discovery, design, and engineering. These workflows connect quantum-level calculations with atomistic simulations, and further with mesoscale and device-level models, creating a systematic multiscale bridge. Such integration is indispensable to capture materials' complexity in a reproducible, automated manner. Yet, building these workflows introduces significant challenges: orchestrating heterogeneous codes, overcoming theory-specific limitations, and ensuring accessibility to scientists across disciplines.

In this landscape, Artificial Intelligence is redefining what workflows can achieve. AI-assisted systems can intelligently orchestrate diverse methods, accelerate property prediction, and guide the exploration of vast structural and chemical spaces.

**Current and future challenges**

Despite the development of sophisticated frameworks for materials simulation, a universally adopted solution remains elusive. The primary challenges stem from the inherent complexity of simulation tasks and a persistent lack of standardization across different codes and methodologies. The current landscape is dominated by powerful but philosophically distinct ecosystems. For instance, the AiiDA framework prioritizes rigorous, automated data provenance to ensure reproducibility, but this comes at the cost of a steep learning curve.[6, 7] In contrast, the FireWorks engine, integral to the Materials Project, is often coupled with Automate and offers greater flexibility for high-throughput dynamic workflows, though it place a greater responsibility on the user to enforce strict provenance conventions [8, 9, 10] This fragmentation creates significant obstacles to interoperability and reusability, often compelling researchers to develop custom parsers or ad-hoc scripts to bridge the gap between tools.

Moreover, to unlock their full predictive power, these workflows must also deliver observables that can be quantitatively compared with experiments**.** This is where the integration of the LSQUANT code (www.lsquant.org) becomes transformative. LSQUANT enables the calculation of experimentally relevant properties (including optical, electronic, transport, dielectric, and magnetic observables) directly on large-scale models. Its linear-scaling quantum transport and electronic-structure methods make it possible to study disordered systems, complex heterostructures, and amorphous phases with unprecedented fidelity [11]. This capability is critical not only for validating theoretical predictions but also for exploring materials that defy crystalline order. For example, amorphous boron nitride (a-BN) has recently demonstrated ultralow dielectric constants suitable for next-generation interconnect technologies [12].

Finally, although artificial intelligence appears as a key game changer, it requires extremely large volumes of data which are quite expensive for most of accurate scientific simulations such as LSQUANT or DFT. Placing the need for efficient training and validation methods that use optimal available datasets as additional central challenges. The issue of data efficiency in material workflows is not confined to materials science. Many other computational areas face similar obstacles. Generating a large volume of data requires time, computing resources, and effort. Therefore, active learning is a promising approach to reduce the training datasets and support accurate or even improved predictions [13].

**Advances in science and technology to meet challenge**

Recent advances in Artificial Intelligence, particularly large language models (LLMs), offer a viable path toward their solution [14, 15]. An LLM-based framework could function as a dynamic, adaptive engine, translating high-level natural language commands into executable, multi-step simulations. For instance, an AI agent could take a prompt to find a material's band structure and then run a molecular dynamics simulation, and in response, automatically query a database, generate input files for a DFT code, parse the results, and use the output to initialize the subsequent simulation with a different software package.[16] This workflow-enhanced AI approach is critical because it combines the adaptability of modern AI with the transparency, control, and reproducibility of traditional scientific workflows. It avoids the "black box" problem of pure AI generation by preserving the logic and data provenance of each computational step, ensuring that the results are not only accelerated but also scientifically rigorous and verifiable.[17]

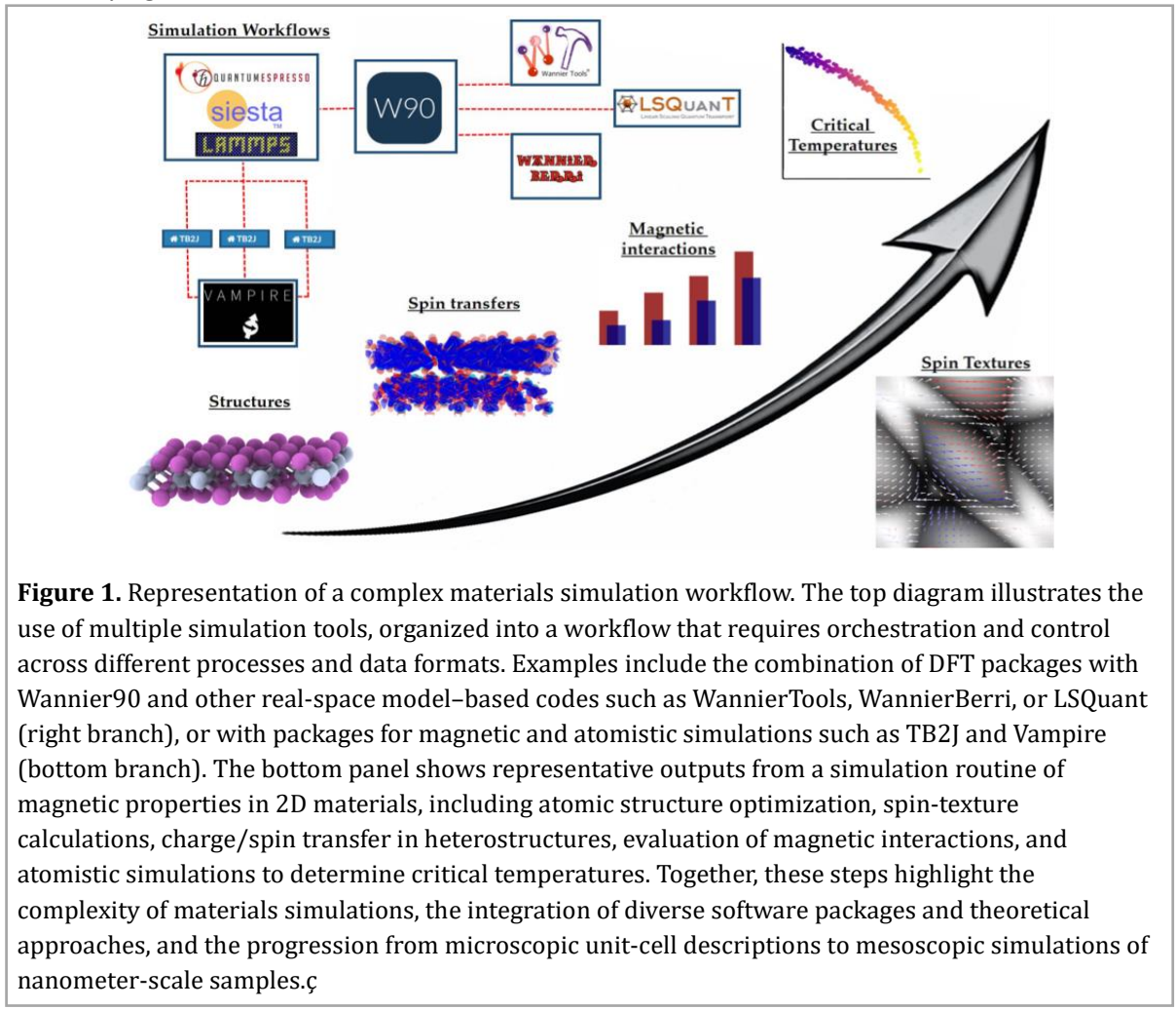

**Figure 1.** Representation of a complex materials simulation workflow. The top diagram illustrates the use of multiple simulation tools, organized into a workflow that requires orchestration and control across different processes and data formats. Examples include the combination of DFT packages with Wannier90 and other real-space model–based codes such as WannierTools, WannierBerri, or LSQuant (right branch), or with packages for magnetic and atomistic simulations such as TB2J and Vampire (bottom branch). The bottom panel shows representative outputs from a simulation routine of magnetic properties in 2D materials, including atomic structure optimization, spin-texture calculations, charge/spin transfer in heterostructures, evaluation of magnetic interactions, and atomistic simulations to determine critical temperatures. Together, these steps highlight the complexity of materials simulations, the integration of diverse software packages and theoretical approaches, and the progression from microscopic unit-cell descriptions to mesoscopic simulations of nanometer-scale samples.ç

LLMs as well as foundation material models can be optimally trained using active learning, a technique that has emerged as a valuable strategy to reduce both the data requirements and the computational load in AI-assisted workflows. Examples include active learning strategies based on Bayesian optimization, which have achieved improvements in efficiency by selecting the most informative candidates and minimizing the number of DFT calculations required [18].

Bayesian active learning also provides a reliable methodology for building effective Hamiltonian parameters. This enables the study of super-large-scale atomic structures without the complications and approximations of conventional parameterization [19].

By selecting the most informative data points for training, active learning can optimize the use of expensive computational resources while maintaining predictive accuracy. In practice, this workflow has been applied to the design and optimization of organic light-emitting diodes, reducing the number of physics-based simulations needed to estimate material properties [20].

**Concluding remarks**

Artificial intelligence is changing very fast the way we do computational materials science, opening the door to predictive and automated workflows that can connect different scales. Density functional theory and linear-scaling quantum transport have been essential steps, and with machine learning and active learning we are already reaching systems that before were too complex or too big to study. Still, the challenges are evident. The diversity of codes and the lack of standardisation create barriers, and the cost of generating high-quality datasets continues to be a limitation.

In this context, LSQUANT is especially relevant because it links theory with experimentally measurable properties, making possible the study of disordered and amorphous materials with high fidelity. Some strategies are already on the table, but generation and benchmarking of reliable datasets remain key. At the same time, algorithms and training methods will need to keep evolving to make the best use of them.

In the future, combining AI with scalable quantum methods and active learning will not only accelerate discovery, but also ensure results are reproducible and directly connected to experiments, which is what finally matters for advancing materials design and engineering.

**Acknowledgements**

All authors acknowledge support from PID2019- 106684GB-I00 also funded by MCIN/AEI/10.13039/501100011033/FEDER, UE, as well as PID2022-138283NB-I00 funded by MCIU/AEI/10.13039/501100011033 and SGR and DLE funded by Generalitat de Catalunya. ICN2 is funded by the CERCA Programme from Generalitat de Catalunya, and is currently supported by the Severo Ochoa Centres of Excellence programme, Grant CEX2021-001214-S, both funded by MCIN/AEI/10.13039.501100011033. This work has received funding from the European Union's Horizon Europe research and innovation programme – European Research Council Executive Agency under grant agreement no. 101078370 – AI4SPIN. Disclaimer: Funded by the European Union. Views and opinions expressed are however those of the author(s) only and do not necessarily reflect those of the European Union or the European Research Council Executive Agency. Neither the European Union nor the granting authority can be held responsible for them.

**References**

1. Malica C, Novoselov K S, Barnard A S, Kalinin S V, Spurgeon S R, Reuter K, Alducin M, Deringer V L, Csányi G and Marzari N 2025 Artificial intelligence for advanced functional materials: exploring current and future directions *J. Phys. Mater.* **8** 021001
2. Chávez-Angel E, Eriksen M B, Castro-Alvarez A, Garcia J H, Botifoll M, Avalos-Ovando O, Arbiol J and Mugarza A 2025 Applied artificial intelligence in materials science and material design *Adv. Intell. Syst.* **7** 2400986
3. Li H, Wang Z, Zou N, Ye M, Xu R, Gong X, Duan W and Xu Y 2022 Deep-learning density functional theory Hamiltonian for efficient *ab initio* electronic-structure calculation *Nat. Comput. Sci.* **2** 367–377
4. del Rio B G, Phan B and Ramprasad R 2023 A deep learning framework to emulate density functional theory *npj Comput. Mater.* **9** 158
5. Ren P, Xiao Y, Chang X, Huang P Y, Li Z, Gupta B B, Chen X and Wang X 2021 A survey of deep active learning *ACM Comput. Surv.* **54** 9 180
6. Pizzi G, Cepellotti A, Sabatini R, Marzari N and Kozinsky B 2016 Comput. Mater. Sci. 111 218–230
7. Huber S P, Zoupanos S, Uhrin M, Talirz L, Kahle L, Häuselmann R, Gresch D, Müller T, Yakutovich A V, Andersen C W, Ramirez F F, Adorf C S, Gargiulo F, Kumbhar S, Passaro E, Johnston C, Merkys A, Cepellotti A, Mounet N, Marzari N, Kozinsky B and Pizzi G 2020 AiiDA 1.0, a scalable computational infrastructure for automated reproducible workflows and data provenance Sci. Data 7 300
8. Jain A, Ong S P, Chen W, Medasani B, Qu X, Kocher M, Brafman M, Petretto G, Rignanese G M, Hautier G, Gunter D and Persson K A 2015 FireWorks: a dynamic workflow system designed for high-throughput applications *Concurrency Computat.: Pract. Exper.* **27** 5037–5059

9. Mathew K, Montoya J H, Faghaninia A, Dwarakanath S, Aykol M, Tang H, Chu I, Smidt T, Bocklund B, Horton M, Dagdelen J, Wood B, Liu Z, Neaton J, Ong S P, Persson K and Jain A 2017 Atomate: a high-level interface to generate, execute, and analyze computational materials science workflows *Comput. Mater. Sci.* 139 140–152

10. Ong S P, Richards W D, Jain A, Hautier G, Kocher M, Cholia S, Gunter D, Chevrier V L, Persson K A and Ceder G 2013 Python Materials Genomics (pymatgen): a robust, open-source python library for materials analysis *Comput. Mater. Sci.* **68** 314–319

11. Fan Z, Garcia J H, Cummings A W, Barrios-Vargas J E, Panhans M, Harju A, Ortmann F and Roche S 2021 Linear scaling quantum transport methodologies *Phys. Rep.* **903** 1–69

12. Hong S, Lee C S, Lee M H, Lee Y, Ma K Y, Kim G, Yoon S I, Ihm K, Kim K J, Shin T J, Kim S W, Jeon E, Jeon H, Kim J Y, Lee H I, Lee Z, Antidormi A, Roche S, Chhowalla M, Shin H J and Shin H S 2020 Ultralow-dielectric-constant amorphous boron nitride *Nature* **582** 511–514

13. Raghavan K, Papadimitriou G, Jin H, Mandal A, Kiran M, Balaprakash P and Deelman E 2025 Advancing anomaly detection in computational workflows with active learning *Future Gener. Comput. Syst.* **166** 107608

14. Yildiz O and Peterka T 2025 arXiv:2412.10606

15. Fan S, Cong X, Fu Y, Zhang Z, Zhang S, Liu Y, Wu Y, Lin Y, Liu Z and Sun M 2024 arXiv:2411.05451

16. Campbell Q, Cox S, Medina J, Watterson B and White A D 2025 arXiv:2502.09565

17. Specification of www.apeironai.eu
18. Doan H A, Agarwal G, Qian H, Counihan M J, Rodríguez-López J, Moore J S and Assary R S 2020 Quantum chemistry-informed active learning to accelerate the design and discovery of sustainable energy storage materials *Chem. Mater.* **32** 5330–5342
19. Ma X, Chen H, He R, Yu Z, Prokhorenko S, Wen Z, Zhong Z, Íñiguez-González J, Bellaiche L, Wu D and Yang Y 2025 Active learning of effective Hamiltonian for super-large-scale atomic structures *npj Comput. Mater.* **11** 70
20. Abroshan H, Chandrasekaran A, Winget P, An Y, Kwak S, Brown C T, Morisato T and Halls M D 2022 Active learning for the design of novel OLED materials *SID Symp. Dig. Tech. Pap.* Book 2, Session 66: OLED Materials I, Paper 66-3 (Society for Information Display)